\pdfoutput=1
\documentclass[12pt]{report}
\usepackage[a4paper, total={6in, 8in}]{geometry}
\usepackage{comment}
\usepackage{cancel}
\usepackage{epigraph}
\AtBeginDocument{}

\usepackage{amsfonts}
\usepackage{lmodern}
\usepackage{dsfont}
\usepackage{amssymb}
\usepackage{graphicx}
\usepackage{multicol}
\usepackage{framed}
\usepackage{amsmath}
\usepackage{bm}
\usepackage{xfrac}
\usepackage{siunitx}
\usepackage{caption}
\usepackage[version=4]{mhchem} 
\usepackage{shtabularlines}
\usepackage{float}
\usepackage{subfigure}
\usepackage{lipsum}
\usepackage{mathtools}
\usepackage{cuted}
\usepackage{url}
\usepackage{hyperref}
\usepackage[utf8]{inputenc}
\usepackage[english]{babel}
\usepackage[dvipsnames,table]{xcolor}
\usepackage{array}
\newcolumntype{P}[1]{>{\centering\arraybackslash}p{#1}}
\usepackage{multirow}
\usepackage{csquotes}

\usepackage[
  backend=bibtex,
  style=numeric-comp, 
  doi=false, 
  url=false, 
  eprint=false 
]{biblatex}
\hypersetup{
    colorlinks=true,
    citecolor=Cyan,
    linkcolor=Maroon,
}

\usepackage{fancyhdr}
\DeclareUnicodeCharacter{2217}{*}
\begin{document}


\begin{titlepage}
	\centering
        {\huge\bfseries The Casimir Effect in Non-Abelian Gauge Theories on the Lattice\par}
	\vspace{2cm}
	
	{\Large \textbf{Blessed Arthur Ngwenya} \par}
	\vspace{1cm}
	{\large supervised by\par
	\textbf{Associate Professor W.A.\ Horowitz}\\
            \large (University of Cape Town)\par
            \large and\par
	\textbf{Professor A.K.\ Rothkopf}\\
            \large (University of Stavanger)\par }
	
	\vspace{1cm}
    \includegraphics[width=0.35\textwidth]{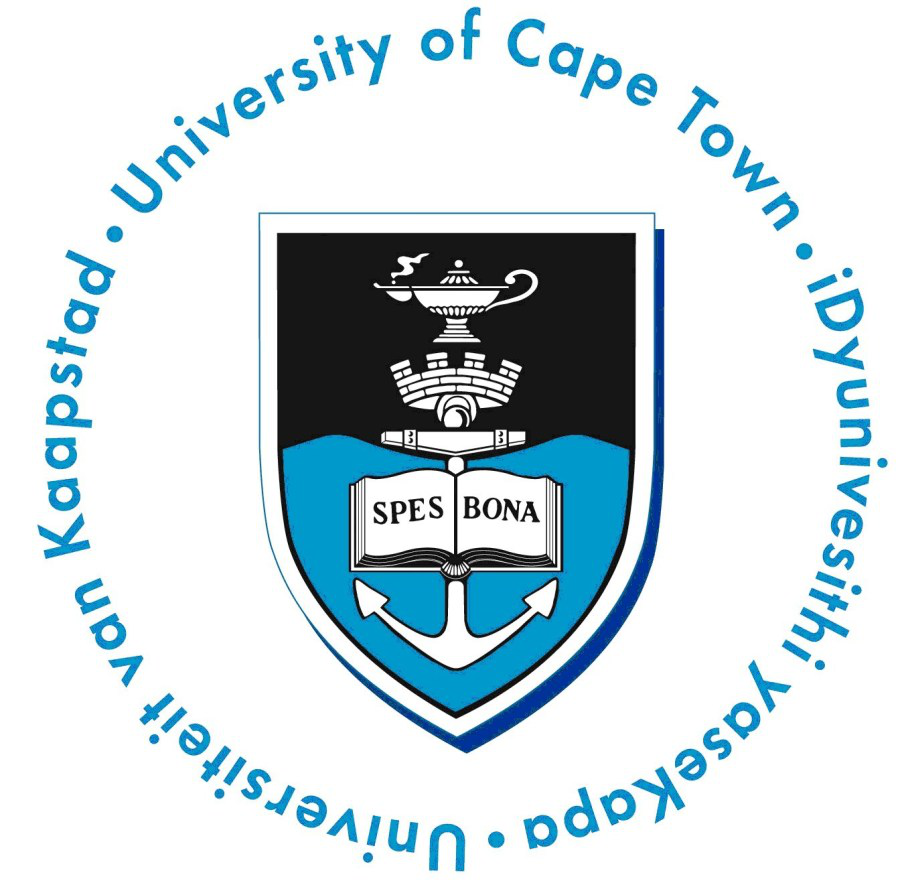}\par\vspace{0.2cm}
	{\scshape\large University of Cape Town \par}
	
    \vspace{1cm}
	{\large Thesis submitted in fulfilment of the requirements for the degree of\par
	Doctor of Philosophy in Physics\par
	\vspace{0.7cm}
	June, 2024\par}
\end{titlepage}




\begin{abstract}
We present non-perturbative results of the Casimir potential in non-abelian gauge theories in (2+1)D and (3+1)D for SU(2) and SU(3) in the confined and deconfined phases. For the first time, geometries beyond parallel plates in (3+1)D SU($N_c)$ are explored, and we show that the Casimir effect for the symmetrical and asymmetrical tube and box is attractive. The result for the tube is contrary to the weakly coupled, massless non-interacting scalar field theory result where a repulsive Casimir force, described by a negative slope of the potential, is measured in this geometry. We propose various techniques that can be used to account for the energy contributions from creating the boundaries in the geometry of a tube and box where the size of the faces forming the walls of the geometries changes with separation distance. We show that increasing the temperature from a confined to a deconfined phase does not alter the measured potential and we motivate for this observation by showing that the masses of the particles in the Casimir interactions are lower than the lowest glueball mass, $M_{0^{++}}$ in the pure gauge ground state, thus suggesting that the region inside the geometries is a boundary-induced deconfined phase. Through the measured expectation value of the Polyakov loop, we propose that the Casimir effect for the asymmetrical tube in the large separation distance limit $R\to \infty$ should be equivalent to that of the parallel plate geometry at smallest separation distance, while the asymmetrical box in the same limit should be equivalent to the symmetrical tube at the smallest separation distance. 

\end{abstract}          
\clearpage
\begin{center}
    \thispagestyle{empty}
    \vspace*{\fill}
    Dedicated to all the dreamers who haven't made it.
    \vspace*{\fill}
\end{center}

    \thispagestyle{empty}
    \vspace*{\fill}
    \epigraph{\centering During my lifetime I have dedicated myself to this struggle of the African people. I have fought against white domination, and I have fought against black domination. I have cherished the ideal of a democratic and free society in which all persons live together in harmony and with equal opportunities. It is an ideal which I hope to live for and to achieve. But if needs be, it is an ideal for which I am prepared to die.}{Nelson Mandela, \textit{Statement from the Dock at the Opening of the Defence Case in the Rivonia Trial, Pretoria Supreme Court, 20 April 1964}}
    \vspace*{\fill}        
\clearpage
\pagestyle{empty}
\begin{center}
    \vspace*{\fill}
    \textbf{Acknowledgements}\\
    \vspace{1cm}
    Words cannot express my gratitude to my supervisors, Associate Professor W.A.\ Horowitz (University of Cape Town) and Professor A.K.\ Rothkopf (University of Stavanger) for their invaluable insights and guidance throughout this work. Alexander played a crucial role in my PhD recovery process after some time spent in corporate and ensuring that we successfully see this project into completion. I have worked under the close supervision and mentorship of Will since I was a second year student at the University of Cape Town in 2016 and he has played a pivotal role in shaping my interest and understanding of high energy nuclear physics research. It is through his support, and dedication to the transformation of the South African Physics academic landscape that I have been able to pursue this path.\\
    \vspace{0.5cm}
    
    \noindent
    I am immensely grateful for the support that I received from my family and friends throughout this work. I am especially grateful to my mother, whose own hard-work and sacrifices have led me to this moment. Mother, I am truly honored to have the opportunity to occupy spaces enabling me to represent your hopes and dreams, and those of our historically disenfranchised communities. To my sister, Victoria Mbalenhle Ngwenya, whose energy and optimism I continue to draw strength from, I hope that not just this qualification, but the endeavor it took to break through the socioeconomic barriers as a first generation graduate to get to this point inspires you in your own journey as you pave your own path in the world. A special thank you to Simvuyele Rose Mdekazi, for all her love, support and encouragement.\\
    \vspace{0.5cm}
    
    \noindent
    I would also like to thank the members of our research group at the University of Stavanger, Rasmus Normann Larsen and Gaurang Parkar with whom I spent a lot of time in conversation with regarding various aspects of this project. The time spent engaging on various aspects of physics research, particularly in Lattice QCD and their own research topics has been instrumental to my own growth in the field.\\
    \vspace{0.5cm}

    \noindent
    My appreciation also goes to the faculty and staff at the University of Cape Town (mostly in the Department of Physics and the Department of Mathematics and Applied Mathematics), whose assistance on academic, administrative and personal matters over the years has been invaluable. I would also like to thank the staff at iThemba Laboratories for Accelerator Based Science (LABS), especially Candice Saaiman who has been handling all my academic travel arrangements. I am also thankful to Rene Kotze at the National Institute for Theoretical and Computational Sciences (NITheCS) who has been magical on the administrative front.\\
    \vspace{0.5cm}

    \noindent
    Special thanks to the staff at Brookhaven National Laboratory (BNL), especially Professor Raju Venugopalan and the staff at the European Organization for Nuclear Research (CERN), especially Professor Urs Wiedemann who have hosted me at the lab on numerous occasions.\\
    \vspace{0.5cm}

    \noindent
    I am also thankful to the PhD (and Postdoc) cohort that I overlapped with at the University of Stavanger with whom a lot of enlightening conversations were had, within and outside our own fields of expertise. To name but a few, Gerhard Ungersbäck, Jimmy Huy Tran, Minh Chi To, Parisa Roshaninejad, Daniil Krichevskiy, Magdalena Britt Eriksson, Oleg Komoltsev, Vegard Undheim, Abhijit Bhat Kademane, Divyarani Chandrababu Geetha, Mar Saiz Aparicio, Anna Maria Raukh and Jonas Elias El Gammal. I would be remiss in not mentioning my small community of friends in the city of Stavanger who have made life abroad a worthwhile experience. Their presence and insights have been instrumental in navigating the complexities of living abroad.\\
    \vspace{0.5cm}

    \noindent
    Having spent some time working in corporate while simultaneously pursuing my doctoral research, I must also express my deepest appreciation to my former managers, Kamil Wlazly and Felix Pezold at IXM S.A.\ for their support.\\
    \vspace{0.5cm}
    
    \noindent
    Notably, I would like to acknowledge SA-CERN, The National Research Foundation (NRF) of South Africa and the Erasmus+ programme for their generous financial contributions towards this work. In addition, I thank the South African Centre for High Performance Computing (CHPC) and the University of Stavanger's computing cluster for the use of their computational infrastructure.\\
    \vspace{0.5cm}

    \noindent
    I also extend gratitude to members of the doctoral committee who have provided me with insightful feedback following their review of the thesis.\\
    \vspace{0.5cm}

    \noindent
    While I have received a lot of support throughout this work, it has not been in isolation. This thesis reflects not only my efforts, but also the collective support and influence of everyone who has touched my life. Therefore, I would like to express my deepest gratitude to everyone who has walked this journey with me at various stages of my life leading up to this moment. The contributions you have made to my personal growth have certainly played a huge role in shaping the person I have become.\\
    \vspace{0.5cm}

    \noindent
    Thank you all. You'll always be famous!
    \vspace*{\fill}
\end{center}
\clearpage
\pagestyle{fancy}


\pagenumbering{roman} 
\tableofcontents            
\listoffigures              
\listoftables
\clearpage

\pagenumbering{arabic} 
\chapter{Introduction}
\label{Chapter 1}

\noindent
In this thesis, we seek to address the question of how the QCD vacuum energy is modified by the presence of physical boundaries of varying geometries. We study this from a lattice perspective through the computation of the energy density of a non-abelian field inside various geometries. While the relevant degrees of freedom are glueballs, i.e.\ colour singlet composites of gluons in pure gauge theory, we explore whether the system remains confined in the presence of boundary conditions. In the introduction, we briefly discuss quantum fluctuations as a manifestation of Heisenberg's uncertainty principle, and the resulting divergent zero-point energy. Then we go through the various developments of the Casimir effect in abelian gauge theory as a first finite quantification of the differences in the zero-point energy and the resulting physical phenomena. Lastly, we discuss the Casimir effect in non-abelian gauge theories using recently formulated first-principle non-perturbative approaches.\\

\section{Zero-Point Quantum Fluctuations}
\label{Chapter 1.1}
In quantum systems, \textit{quantum fluctuations} in the field manifest themselves in the form of \textit{virtual particles}, which rapidly\footnote{The constraint in the the lifetime of virtual particles is imposed by the energy-time uncertainty principle in order to ensure energy conservation, $E^2=(pc)^2 + (mc^2)^2$ within observable timescales.} appear and disappear in empty-space. Such quantum fluctuations are intrinsic to the formulation of quantum mechanics in the second quantisation which facilitates the presence of many-body interactions. Heisenberg's uncertainty principle\footnote{Which is a direct consequence of the Dirac quantisation condition between canonically conjugate variables, e.g.\ $[\hat{x},\hat{p}]=i\hbar$.} provides a description of quantum fluctuations,
\begin{equation}
    \sigma_x \sigma_p \geq \frac{\hbar}{2}, \quad \sigma_E \sigma_t \geq \frac{\hbar}{2},
\end{equation}
and describes the inherent limit to the precision on the, e.g., simultaneous measurement of a particle's position and momentum. Thus the momentary appearance of virtual particles does not lead to a net violation of energy conservation. \\

\noindent
This net energy conservation can be understood from Einstein's mass-energy equivalence, $E=mc^2$, describing the rest energy as well as the creation and annihilation of particles through the temporary conversion of energy into mass. Accordingly, the momentary appearance of these virtual particles is within the constraints of the energy-time uncertainty principle. Their representation in Feynman diagrams is internal lines signalling intermediate particle interaction states that do not violate the required energy-momentum conservation at the vertices (initial and final states) of the particle interactions.\\

\noindent
According to the uncertainty principle, a quantum particle placed at a local minimum of a potential does not remain motionless there, else the uncertainty principle is violated. This is contrary to the classical case, where a marble placed at the bottom of a well remains there indefinitely, governed by Newton's laws of motion. In quantum field theory, particles are excitations of the underlying fields, thus quantum fields in a vacuum at every point in space are always fluctuating and do not maintain a specific constant value. \\

\noindent
These fluctuating fields produce an infinite amount of energy in free space, the \textit{zero-point energy}. This can be understood by considering the quantum harmonic oscillator with quantised energy,
\begin{equation}
    E_n = \left( n+ \frac{1}{2} \right) \hbar \omega,
\end{equation}
where the ground state energy, $E_0$ does not vanish. In quantum field theory, assigning a ground state to each normal mode of the real scalar field (i.e.\ the vacuum energy is an infinite sum of uncoupled harmonic oscillators, where each oscillator corresponds to each frequency) results in ultraviolet divergences\footnote{The divergent energy is a source of the old cosmological constant problem \cite{Burgess:2013ara}, where the large theoretically predicted value of the vacuum energy density in QFT is incomparable with the observed small value of the cosmological constant.},
\begin{equation}
    E_0 = \frac{\hbar}{2} \sum\limits_{j} \omega_j,
\end{equation}
where $j$ is the quantum number of the normal modes and $\omega_j$ represents all possible frequencies.\\

\noindent
In quantum electrodynamics (QED), this non-vanishing ground state energy of an electromagnetic field with a sea of virtual photons can also be understood from the perspective of the field commutation relations. In general, the commutator of the field operators \cite{Hannabuss:1999ex}, 
\begin{equation}
    [\hat{E}_i(x_1), \hat{B}_j(x_2)] = -i\hbar\mu_0\epsilon_{ijk} \delta_k(x_1-x_2) \neq 0,
\end{equation}
does not vanish and the presence of quantum fluctuations of the vacuum electromagnetic field is consistent with the uncertainty principle. \\

\noindent
In principle, this infinite amount of energy in free space arising in QFT is not a physically measurable quantity. While the presence of an infinite amount of energy in free space may sound counter-intuitive, a multitude of experimental results have been consistent with QFT predictions on the presence of vacuum energy. This energy is instead measured by imposing specific boundary conditions on the electromagnetic fields in free space, which modify the vacuum properties and allows for measurement of \textit{zero-point energy differences}. These energy differences are described by the \textit{Casimir effect}, which we discuss in the following section. \\


\section{The Casimir Effect}
\label{Chapter 1.2}
\subsection{Abelian Gauge Theories}
The Casimir effect \cite{Casimir:1948dh, Casimir:1947kzi} provides experimental evidence for the existence of zero-point fluctuations which can be theoretically quantified. It was first derived in 1948 by Hendrik Casimir for a parallel plate configuration in quantum electrodynamics. It states that if two identical neutral ideally conducting plates are placed at some distance $d$ apart in a vacuum (i.e.\ the ground state of QED ), the plates would experience an attractive force towards each other\footnote{This was the initial definition based on the outcome of Casimir's parallel plate configuration. A more general definition states that the spectrum of virtual particles and the vacuum zero-point energy is modified by the presence of boundaries (as a generalisation to other geometries).}. The magnitude of the force of attraction per unit area of the plates (pressure) is,
\begin{equation}
    F(d) = -\hbar c\frac{\pi^2}{240} \frac{1}{d^4},
    \label{eqn:casimir_original}
\end{equation}
which is small, but non-zero\footnote{The Casimir effect is a universal quantum phenomenon, and in this abelian theory, describes the fluctuations of a virtual photon gas. In the classical limit (classical electrodynamics) where $\hbar \to 0$, this effect vanishes.}. For example for plates placed $\sim 1$ $\mu$m apart, the resulting Casimir force is $\sim 10^{-3}$ Nm$^{-2}$.\\

\noindent
Casimir explained that the presence of the neutral conducting plates in the vacuum imposes boundary conditions on the stationery modes (standing waves) of the electromagnetic field in-vacuum. In free space, electromagnetic modes span an infinite range of wavelengths\footnote{Their sum is divergent, but the difference of the sums at varying plate separation distance is well-defined and subtracts the infinite vacuum energy.}; however, between the plates, the long wavelengths ($\lambda >d$) modes get \textit{frozen out} (i.e., cannot be accommodated). The freezing out occurs because the `optical cavity' formed between the plates limits the number of modes (based on the wavelength) that can exist within the cavity. The available modes between the plates is thus less than the number of modes in free space (outside the plates).\\

\noindent
While the long wavelengths are frozen-out between the plates, the very short wavelengths (up to the order of ultraviolet wavelengths) are not restricted by the presence of the plates and can exist between the two regions. As a result, the zero-point energy of the modes with short wavelengths are independent of the separation distance between the plates, hence the requirement for the additive renormalisation, often in the form of a subtraction procedure. Therefore, the contributing modes between the plates have wavelengths comparable or smaller to the separation distance, $d$.\\

\noindent
The limited number of modes between the plates correspond to a lower vacuum energy density in that region, leading to a pressure difference and subsequently the attractive force. When the plates are pulled apart (i.e.\ the separation distance between them is increased), the normal modes of `longer' wavelengths are no longer frozen out. The energy density in the region between the plates increases and the magnitude of the attractive force is reduced.\\

\noindent
Apart from the parallel plate geometry, a more commonly used Casimir geometry is that of a large plate and a conducting spherical shell of radius $r >> d$, where $d$ is the separation distance. This geometry has experimental motivations introduced by the perfect alignment difficulties when placing two large plates at distances on the order of microns apart. The resulting Casimir-Polder force obtained from employing the proximity force approximation\footnote{In the proximity force approximation \cite{Blocki:1977zz}, a general geometry such as those involving a sphere is locally approximated plane-plane geometry where the plane-wave basis can be applied.} is \cite{Bordag:2001qi},
\begin{equation}
    F(d) = -\hbar c\frac{\pi^3r}{360} \frac{1}{d^3}.
\end{equation}

\noindent
This approximation is used because the theoretical formulation of geometries with curved surfaces is nontrivial in comparison to its parallel plate counterpart and continues to be of interest to theorists. See, for example Ref.\ \cite{Bimonte:2017ahs} for a discussion on the improvement of the proximity force approximation and the reduction to the associated errors. We refer the reader to Ref.\ \cite{hartmann2018casimir,Bimonte:2017ahs,Chan:2001zzb} and references therein for a detailed discussion on this geometry and the corresponding experimental studies. Most importantly, we note the difficulties in the formulation of certain geometries to study the Casimir effect, both analytically and experimentally.\\

\noindent
In its initial formulation, the Casimir effect was only understood for perfectly conducting metals plates at zero temperature. Around the mid-1950's, it was generalised to materials with arbitrary dielectrics and rough surfaces at finite temperature by Evgeny Lifshitz. The short-distance interaction between materials (atomic bodies) occurs through Van der Waal's/London forces which depend on atomic polarizability \cite{genet2004electromagnetic}. This generalisation opened room for Casimir studies under physical conditions. The resulting force between dielectric macroscopic bodies has the form \cite{Lifshitz:1956zz},
\begin{equation}
    F(d) = -\hbar c\frac{\pi^2}{720} \frac{1}{d^3} \frac{(\epsilon_0 - 1)^2}{(\epsilon_0 + 1)^2} \varphi (\epsilon_0),
\end{equation}
where $\epsilon_0$ is the dielectric constant and $\varphi (\epsilon_0)$ is a tabulated function. The ideal conductor behaviour in the zero temperature limit obtained by Casimir in Eq.\ (\ref{eqn:casimir_original}) is recovered in the limit $\epsilon_0 \to \infty$.\\

\noindent
At finite temperatures, the fields will exhibit thermal fluctuations around the field expectation values in thermal equilibrium. The thermal fluctuations result in radiation pressure which adds thermal corrections to the Casimir force \cite{Mitter:1999hu, Decca:2003td,Ghisoiu:2010fga}. The thermal wavelength is given by, $\lambda_T = \hbar c/(2\pi k_BT)$ and $\lambda_T \simeq 1.2$ $\mu$m at room temperature. Hence the relevant scale where thermal fluctuations become important is at separation distances, $d \gtrsim 1$ $\mu$m  where the wavelength of such fluctuations can fit inside the cavity of the configuration. \\

\noindent
The Lifshitz formulation also predicted that there is a geometrical element to the Casimir effect, where some configurations result in a repulsive force. An example of such a configuration with a repulsive potential is the double spherical cavity where an inner conducting spherical shell of radius $r$ is surrounded by an outer spherical shell with radius, $R\to \infty$. The zero-point energy for the vacuum of the inner shell was derived\footnote{The resulting expression from the initial derivation is very complex and employs some special functions.} in the late 1960's and later improved through numerical approximation \cite{Boyer:1968uf,Balian:1977qr}, 
\begin{equation}
    \langle E \rangle = C\frac{\hbar c}{2r}, \quad F=-\frac{\partial E}{\partial r} >0
\end{equation}
where $C=0.092353$ \cite{Milton:1999ge}. Note the positive sign of the Casimir force, essentially indicating that the spherical conducting shell tends to expand. This provides motivation for our exploration of different geometries.\\

\noindent
Subsequent studies in the abelian gauge also show additional interesting results, such as the implausible Scharnhorst effect \cite{Scharnhorst:1990sr,deClark:2016mvw}, where radiative corrections computed in perturbation theory for the parallel plate geometry lead to photons between the plates (moving perpendicular to the plates) travelling faster than in-vacuum photons with speed, $c$. Such a theoretical prediction could suggest that imposing the Casimir boundaries on the QED vacuum amplifies some frequencies in the medium cavity between the plates. This effect remains unobserved experimentally, hence the result could be plagued by theoretical errors which introduce the need to explore non-perturbative methods to compute these radiative corrections in field theories.\\ 

\noindent
Recent studies of the abelian theory have been performed in lattice field theory in an attempt to improve the inaccuracies in the measured Casimir effect due to e.g., thermal corrections in perturbative analytic methods for general geometries which are usually formulated using simplifications such as fixed boundary conditions. See, for example Ref.\ \cite{Bordag:2001qi} where Green's functions are used to describe the boundary effects on vacuum fluctuations in quantum field theory. Initial lattice studies focused on how to describe the boundary conditions for the geometries of interest in the lattice formalism since the vacuum fluctuations spectrum is dependent on them. One such boundary condition is the Chern-Simons action (boundary condition)\footnote{The Casimir effect between parallel plates using the Chern-Simons boundary condition in QED has also been computed analytically, see Ref.\ \cite{Markov:2006js}, and is consistent with the lattice results. In addition, the Chern-Simons boundaries have also been applied to compact QED \cite{Pavlovsky:2009kg}. } described in Ref.\ \cite{Pavlovsky:2009kg} for a parallel plates in (3+1)D \textit{non-compact}\footnote{In non-compact QED, the gauge fields, $A_{\mu}$ are treated as continuous variables, i.e., they are real-valued and unbounded. The lattice action is consistent with the continuum action given by the first term in Eq.\ (\ref{eqn:chern_simons}).} QED. \\

\noindent
The Chern-Simons action for electromagnetic fields in a (3+1)-dimensional spacetime region $V$ is given by the Maxwell action with an additional Chern-Simons surface action term \cite{Bordag:1999ux},
\begin{eqnarray}
    S = -\frac{1}{4}\int_V d^4x F_{\mu\nu}F^{\mu\nu} - \frac{a}{2}\int_{S} d^3x \epsilon^{ijk}A_i\partial_jA_k,
    \label{eqn:chern_simons}
\end{eqnarray}
where $S$ is the boundary and $a$ is a real parameter. The volume term is used to generate the wave equation for $A_{\mu}$ in a suitable gauge and the surface term is used to generate the boundary conditions. We refer the reader to Ref.\ \cite{Bordag:1999ux, Markov:2006js, Pavlovsky:2009kg} for extensive discussions on these boundary conditions.\\
    
\noindent
These lattice simulations have been extended to \textit{compact}\footnote{In compact QED, the gauge fields, $A_{\mu}$ are periodic and bounded on the compact manifold, U(1). The lattice action is given by a sum over the plaquette variables, $S[\theta]=\beta\sum_P(1-\cos \theta_P)$ \cite{Chernodub:2016owp}.} QED which exhibits non-perturbative properties that are consistent with non-abelian gauge theories. We refer the reader to Ref.\ \cite{Chernodub:2016owp, Chernodub:2017mhi} for the case of parallel conducting wires in (2+1)D and Ref.\ \cite{Chernodub:2022izt} for parallel conducting plates in (3+1)D compact U(1) gauge theory using chromoelectric boundary conditions\footnote{On the lattice, this corresponds to increasing the lattice coupling at the boundaries by a Lagrange multiplier $\lambda$, $\beta_P(\lambda)=\beta+\lambda$, $\beta = 1/g^2a^{(4-D)}$ such that the tangential field component vanishes, i.e.\ $\theta_P \to \mathds{1}$. }. In these studies, the resulting action describes photon dynamics which characterise the perturbative effects and the formation of abelian monopole dynamics describing the non-perturbative effects. The monopoles lead to mass-gap generation which screens the Casimir potential at large separation distances similarly to strongly interacting theories. At short separation distances, the monopole density between the boundaries is diminished into a dilute gas of monopole-antimonopole pairs and the region between the wires/plates goes through an induced deconfining phase transition.\\

\noindent
We have so far discussed the historical theoretical progressions in abelian gauge theory, however, experimental studies of the Casimir effect in various geometries have been performed and continually improved in recent years with the technological advances allowing for precision measurements of micro-nano scale forces\footnote{It is worth mentioning that given the small (i.e.\ micro-nano range) magnitude of the Casimir force, Casimir measurements are of interest to the gravitational interactions community for the possible detection of any deviations from the expected gravitational interactions at such small distances and possibly posing arguments for a fifth fundamental force \cite{Onofrio:2006mq, Antoniadis:2011zza}.}. Such studies point out the physical significance of zero-point energies and pave way for its applications in technological devices. See the reviews in Ref.\ \cite{Klimchitskaya:2009cw,Bordag:2001qi,Bordag:2009zz,Milton:2004ya} for some discussions on the experimental front and theoretical developments alike.\\

\subsection{Non-Abelian Gauge Theories}
\noindent
Extensive studies of the Casimir effect have been performed with abelian gauge fields because the weakly coupled theory is well-understood and the boundary conditions associated with electromagnetic fields can be experimentally controlled with great precision. In addition, radiative corrections to the Casimir energy due to interactions of virtual photons with virtual fermions can be calculated using perturbative techniques. The Casimir effect has also been studied in non-abelian gauge theories, which are relevant in the description of the strong and weak interactions.\\ 

\noindent
Yang-Mills theories are more complicated than their abelian counterparts, which is partly attributed to their non-perturbative confining nature, they are non-linear and self interacting. The geometrical boundaries imposed on the confining gauge fields are not necessarily equivalent to lattice studies in a \textit{finite closed volume}\footnote{We refer the reader to Ref.\ \cite{Horowitz:2021dmr} for a study of finite system size corrections to the equation of state.}. In the latter, as the lattice volume approaches zero, the infrared cut-off may become higher than the ultraviolet scale and the theory becomes weakly coupled due to asymptotic freedom\footnote{In the finite volume geometry where the lattice volume is decreased, Yang-Mills theories become deconfining at zero temperature.}, while this can not necessarily be generalised to the Casimir set-up where at least one spatial direction is unbounded.\\

\noindent
In the preceding subsection, we discussed the computation of the abelian Casimir effect in lattice simulations. Our approach in the non-abelian theory will also be implemented on the lattice, where numerous other QCD observables have been computed non-perturbatively. The lattice implementation involves the regularisation of QCD in Euclidean space and discretising the fields on a hypercubic lattice. The fermionic fields are placed on the sites and gluonic fields are placed on the links between the sites and their interactions are simulated in thermal equilibrium. We dedicate the second chapter of this thesis to lattice gauge theory formalism.\\

\noindent
We highlight the formulation and findings of Ref.\ \cite{Chernodub:2018pmt}, taken as a starting point for our studies. In that work, the Casimir effect in non-abelian gauge theory is studied on the lattice for perfectly conducting static parallel wires in (2+1)D SU(2) at zero temperature ($N_s=N_{\tau}=32$) using chromoelectric boundaries. The resulting attractive Casimir potential is described by the empirically chosen fitting function,
\begin{equation}
    V_{\text{Cas}}^{\text{fit}} (R) = -(N_c^2-1)\frac{\zeta(3)}{16\pi} \frac{1}{R^{(\nu+2)}\sigma^{(\nu+1)}} e^{-M_{\text{Cas}}R},
    \label{eqn:vcas_fit}
\end{equation}
with fit parameters, $\nu$ and $M_{\text{Cas}}$, where $\sigma$ is the string tension and $N_c$ is the number of colours. The parameter $\nu$ describes the anomalous scaling dimension of the potential at short separation distances compared to the expected tree-level behaviour,
\begin{equation}
    V_{\text{Cas}}^{\text{tree}} (R) = -(N_c^2-1)\frac{\zeta(3)}{16\pi} \frac{1}{R^2},
    \label{eqn:vcas_tree}
\end{equation}
describing the Casimir energy of $(N_c^2-1)$ non-interacting copies of a monopole-free $U(1)$ gauge theory analogous to a free scalar field with Dirichlet boundary conditions \cite{Ambjorn:1981xw}. Similarly to the finite volume geometry and the formulation in compact QED, this Casimir geometry is bound to induce a smooth deconfining phase transition between the wires. We explore the presence of this smooth deconfining phase transition for other geometries.\\

\noindent
The vacuum zero temperature Yang-Mills theory exhibits non-perturbative dynamic mass-gap generation\footnote{The mass gap describes the spontaneously broken chiral symmetry and lack of a first principle description of how the bound states (hadron) masses are generated from lighter/massless quarks and gluons through strong interactions.} in both (2+1)D and (3+1)D, similarly to compact QED. This non-perturbative mass-gap generation results in an effective screening of the Casimir potential at large separation distances quantified by the parameter $M_{\text{Cas}}$ (the Casimir mass) in the exponential of Eq.\ (\ref{eqn:vcas_fit}). The mass-dependent exponential decay of the Casimir potential is consistent with the Casimir energy decay for a massive scalar particle obtained in a gauge-invariant Hamiltonian formulation of (2+1)D non-abelian gauge theories \cite{Karabali:2018ael}. \\ 

\noindent
Interestingly, the resulting Casimir mass corresponding to the zero-energy of the (2+1)D gauge theory with perfect conductor parallel wires \cite{Chernodub:2018pmt},
\begin{equation}
    M_{\text{Cas}}^{\infty} = 1.38(3)\sqrt{\sigma} < M_{0^{++}} = 4.718(43)\sqrt{\sigma}
\end{equation}
is less than the lowest $0^{++}$ glueball mass in (2+1)D SU(2) gauge theory \cite{Teper:1998te}. In principle, the lowest glueball mass, $M_{0^{++}}$ should be the lowest mass in the system, hence this result suggests that the dominant degrees of freedom in the non-abelian theory Casimir interaction are a lighter gluonic state, possibly forming a `light-gluon' plasma due to the boundary-induced deconfining phase transition in the region between the wires. \\

\noindent
This low Casimir mass has also been obtained in an alternative study employing analytical methods to show that the propagator of the gauge-invariant gluon field in SU(2) corresponds to the propagator of a massive scalar field in (2+1)D, with a mass corresponding to the Casimir mass \cite{Karabali:2018ael}. The same study also shows that the (2+1)D Casimir mass has physical implications for the (3+1)D theory. It is shown that the two-dimensional theory Casimir mass is equal to the high temperature magnetic mass due to screening effects in the three-dimensional theory. Hence studies of the Casimir effect in (2+1)D gauge theories can be used as a probe of the magnetic screening mass of a pure gauge QCD plasma. These masses are not at the centre of this work, but come at no additional cost and will be stated.\\ 

\noindent
In the previous subsection, we discussed the geometrical aspects of the Lifshitz formulation of the Casimir force in abelian gauge theory, where the resulting potential is attractive or repulsive depending on the geometry under consideration. This result has also been presented for a weakly coupled, massless non-interacting scalar field theory with Dirichlet boundary conditions in Ref.\ \cite{Mogliacci:2018oea}, where finite volume effects on the equation of state are studied. This study suggests that the resulting zero-temperature Casimir pressure is attractive for a parallel plate and box configurations, and repulsive for a tube configuration in the abelian theory. Understanding whether such a Casimir force geometric dependence exists in non-abelian gauge theories is of interest in this work.\\

\noindent
The Casimir effect remains of interest in various fields of physics, and we refer the reader to Ref.\ \cite{Milton:2004ya, Milton:2008st} and references therein for comprehensive discussions. In motivation of our study of the Casimir effect in non-abelian gauge theories, we highlight its importance in quantum field theory. In quantum chromodynamics, the Casimir effect has been pivotal in building our understanding of the phenomenologically relevant MIT bag model \cite{Chodos:1974pn,DeGrand:1975cf,Chodos:1974je} of hadrons which describes hadron spectroscopy, confinement and other hadron properties.\\ 

\noindent
In Casimir terms, the bag model is similar to the QED spherical shell model discussed in the previous subsection. The main difference in the QCD bag model is that we are only interested in the contributions of gluon field modes in the interior of the shell/bag due to confinement. The model assumes that quarks are confined in a sphere of radius $R$ but can move freely within the sphere and obey the free Dirac equation. Gluons are also confined in the sphere, they are described by field configurations which can fluctuate but the gluons are not free to move similarly to the quarks. The hadron bag energy is approximated by the sum of the Casimir energy, the bag volume energy\footnote{This energy term guarantees quark confinement in the bag.}, single quark kinetic energies and the gluon exchange energy \cite{DeGrand:1975cf}. The hadron mass spectrum is then extracted from the bag energy by numerically solving the function relating the bag energy to the hadron mass \cite{Bernotas:2012gm}.\\

\noindent
The exact functional form of the Casimir energy term in the bag model has been subject to scrutiny \cite{Milton:1982iw, Leseduarte:1996ah}. The reason is that the phenomenological value of the Casimir energy parameter differs both in sign and magnitude from its theoretical value \cite{Milton:1982iw}. As a result, the Casimir energy term is removed in some modified versions of the bag model \cite{Bernotas:2004bn}. It was shown in Ref.\ \cite{Bernotas:2012gm} that including improved values of the zero-point (Casimir) energy in the \textit{centre-of-mass-motion corrections}\footnote{These corrections are associated with the modelling of the bag as an independent particle shell, leading to energy contributions from the centre of mass motion confined inside the bag.} \cite{Halprin:1982pb} of the bag model results in improved magnetic moments and light hadron masses relative to experimental data. Hence our understanding of the Casimir effect in non-abelian gauge theories remains relevant.\\

\noindent
In this thesis, we study the Casimir effect in non-abelian gauge theories in (2+1)D and (3+1)D for the gauge groups SU(2) and SU(3) using lattice techniques. Inspired by the first-principle numerical simulations of Ref.\ \cite{Chernodub:2018pmt}, we apply chromoelectric boundary conditions on the lattice to formulate new as of yet unexplored types of geometries. We provide extended results of the Casimir potential in (2+1)D gauge theories for parallel conducting wires, then provide new results for the Casimir potential in (3+1)D for a parallel conducting plate configuration, a symmetrical and asymmetrical tube and box respectively.\\

\noindent
Moreover, we explore the Casimir energy dependence on the temperature given that Yang-Mills theories contain a phase transition where the theory deconfines into a plasma of gluons with respect to the temperature scale (and at high densities). As such, we compute the Polyakov loop (deconfinement order parameter) as we move from the confined to the deconfined phase. In order to understand the gluodynamics of the restricted gluon modes inside the configuration volume, we compare the Polyakov loop of the fields on the interior and exterior (vacuum) of each configuration.\\

\noindent
The outline of the thesis is as follows: In chapter (\ref{chapter:lattice_formalism}), we describe the formalism of lattice gauge theory, motivating for lattice QCD methods and setting up the pure gauge Wilson action. We describe an unexpected step-size reduction effect on the convergence of plaquette averages when using the Metropolis update algorithm. In chapter (\ref{chapter:strings_and_booundaries}), we discuss periodic boundary conditions used in our lattice volume, and chromoelectric boundary conditions used to study the Casimir effect. We also present Padé fitting functions to the QCD string tension, which we use to introduce a physical scale to our lattice measurements. Such fits allow for the interpolation of the string tension with respect to the inverse coupling, as well as extrapolation to the continuum limit. In chapter (\ref{chapter:The Casimir Effect: Fields and Symmetries}), we discuss an approximation to the electromagnetic field strength tensor using single plaquettes on the lattice. We then derive the symmetry relations of the electromagnetic field strength tensor components for the wires, plates and tube geometries. Lastly, we present numerical results for the field components in the various geometries showing the field suppression at the boundaries and the resulting field symmetries. The main results of the thesis are provided in chapter (\ref{chapter:The Casimir Potential}) and include the Casimir energy for the geometries of parallel wires in (2+1)D, parallel plates, symmetrical and asymmetrical tube and box in (3+1)D. These are followed by the Polyakov loop results and the test for Casimir temperature dependence in chapter (\ref{chapter:The Polyakov Loop and Deconfinement}). Our results are then summarised in chapter (\ref{chapter:Conclusions}), and appendices of the generator matrices and rotation transformations are provided.\\

\chapter{Lattice Gauge Theory}
\label{chapter:lattice_formalism}


\noindent
In this chapter, we briefly introduce the formulation of non-abelian pure gauge theory on the lattice using the Wilson action. We start by looking at the gluonic action and how it relates to the lattice plaquette variables and the electromagnetic field strength tensor. Then we discuss the implementation of the Metropolis algorithm in updating the gauge/link variables. We show that there is a step-size dependence of the Metropolis updates on the convergence of lattice measurements in non-abelian theory which is not well documented in literature and is not observed in classical field theory. This step-size dependence varies based on the method used to generate group elements and is independent of the distribution used. We also touch briefly on the Hamiltonian Monte Carlo (HMC) algorithm and the effect of topology in Monte Carlo algorithms. Lastly, we discuss the analysis of lattice measurements including thermalisation, auto-correlations and Jackknife errors. \\

\section{QCD on the Lattice}
\label{qcd_on_the_lattice}

\noindent
Quantum chromodynamics (QCD) in the weakly coupled regime (high energies and large momentum transfers between quarks) has been accurately studied using perturbative techniques made possible by the asymptotic freedom feature of the theory \cite{Grozin:2008yd}. We refer the reader to Ref.\ \cite{Deur:2016tte,Sumino:2014ipa} for some reviews on perturbative QCD and discussions alike. On the other hand, these perturbative methods also show that the strong coupling constant, $\alpha_s = g^2/4\pi >1$ below $\Lambda_{QCD} \sim 200$ MeV, rendering such methods unusable at low energies due to the lack of convergence of the underlying series expansion, even for scales around $1$ GeV.\\

\noindent
The inability to apply perturbative methods in studying the strongly coupled regime (low energies) of QCD introduces the need for non-perturbative approaches, and thus provides a strong motivation for the lattice method \cite{Ratti_2018,davoudi2014formal}. The lattice method is a regularisation of QCD in Euclidean space, thus fermionic and gluonic fields are discretised on a hypercubic lattice with spacing, $a$. The quark fields are placed on sites and gauge fields are placed on the links between these sites, then their interactions are simulated in thermal equilibrium \cite{Ratti_2018}.\\

\noindent
These non-perturbative calculations are performed on the lattice by numerically evaluating the QCD path integral in Euclidean space,
\begin{eqnarray}
    \langle \mathcal{O} \rangle = \frac{1}{Z} \int \mathcal{D} A \mathcal{D} q \mathcal{D} \Bar{q} \mathcal{O}[A,q,\Bar{q}] e^{-S[A,q,\Bar{q}]},
    \label{eqn:qcd_path_integral}
\end{eqnarray}
where $\mathcal{O}$ is an observable given by a product of gauge ($A$), quark ($q$) and antiquark ($\Bar{q})$ fields and $S$ is the Euclidean QCD action. The expectation value is computed over all possible field configurations using Monte Carlo sampling. As a result, the only limitations to practical lattice calculations are computational resources and algorithm efficiency, which introduce statistical and systematic uncertainties \cite{Tanabashi:2018oca}.\\

\noindent
The statistical uncertainties are a result of the use of Monte Carlo importance sampling\footnote{Importance sampling involves using a weighted average of random draws from the proposal distribution $q(x)$ (which is easier to sample) in order to approximate the estimator of the mean of the target distribution $p(x)$ \cite{tokdar}.} in the evaluation of the path integral in Eq.\ (\ref{eqn:qcd_path_integral}). There are various sources for the systematic uncertainties such as the discretisation of the QCD action since $a\neq 0$. The Symanzik effective action has been used to reduce these discretisation errors \cite{Symanzik:1983dc}. Some errors arise due to finite volume (of the lattice box) effects where the values of physical quantities, e.g.\ quark masses differ from the infinite volume values. See Ref.\ \cite{ParticleDataGroup:2012pjm, Alexandrou:2019lfo} for further discussions on systematic errors such as the continuum limit extrapolation, chiral extrapolation, operator matching etc. \\

\noindent
In the continuum theory, the strong coupling constant, $\alpha_s$, is dependent on the \textit{QCD scale parameter}, $\Lambda_{QCD}$ \cite{Deur:2016tte}, which is an emergent dimensionful scale arising from quantum fluctuations. In the lattice regularisation scheme, this scale is related to the inverse physical lattice spacing $a$, i.e.\ $\Lambda_{UV} = 1/a$, which acts as a scale setting parameter. Thus choosing the lattice coupling constant, $\beta\equiv 2N_c/g^2$, corresponds to fixing the lattice spacing, which in turn gives us the ultraviolet cut-off of the theory. As a result, it is important to pick the lattice spacing such that it is much smaller than any physical length scale of the system being investigated. This recovers the Euclidean space rotational symmetries broken by the lattice discretisation.\\

\noindent
We obtain physical results of the continuum theory in lattice calculations through extrapolation to the continuum limit (i.e., $a \to 0$). Lattice calculations in this limit of the physical lattice spacing are not computationally feasible as they require increased number of lattice grid points. A good estimation of discretisation errors becomes important in keeping this extrapolation under control, and the extrapolation methods generally perform better when data from various lattice spacings is used \cite{Luscher:1998pe}.\\

\begin{figure}[!htb]
\begin{center}
\includegraphics[scale=0.5]{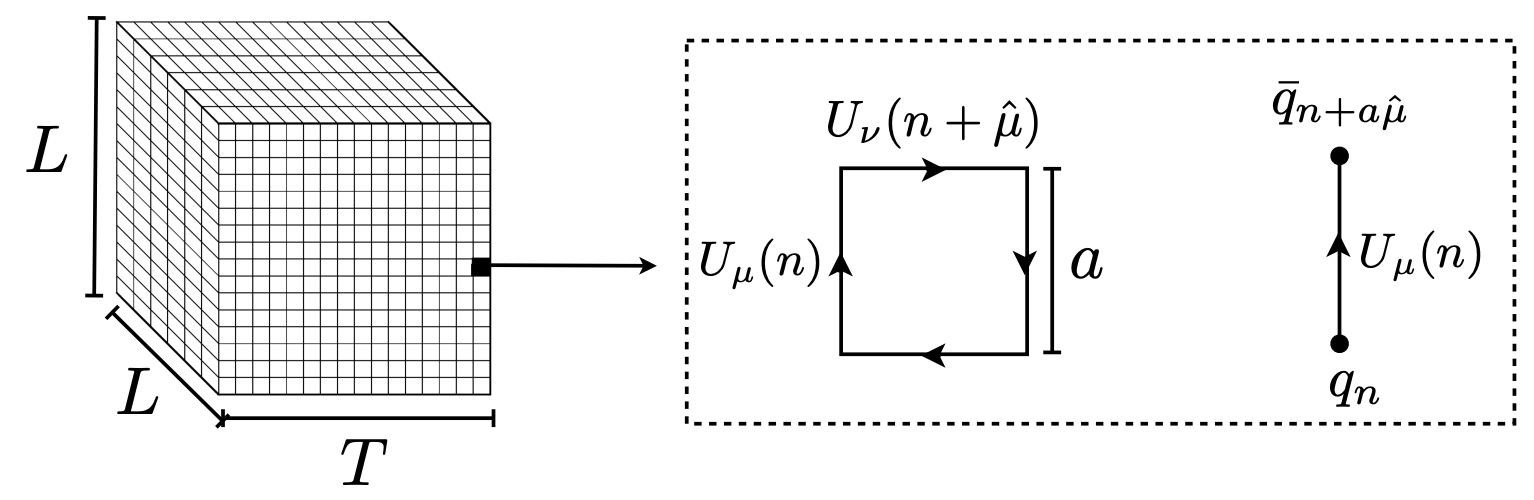}
\caption{\label{Fig1} A ($2+1$) dimensional cubic lattice \cite{davoudi2014formal}.}
\end{center}
\end{figure}

\noindent
A ($2+1$)-dimensional cubic lattice is shown in Fig.\ (\ref{Fig1}), showing the quarks, $q_n$ on the sites and gauge fields, $U_{\mu}(n)$ on the links. In the figure, $L$ and $T$ are the spatial and temporal lengths, respectively. The middle frame shows a single plaquette, while the right frame represents a Wilson line, and we will discuss these in detail shortly. The physical spatial and temporal sizes are $L=aN_s$ and $T=aN_t$, respectively, where $a$ is the physical lattice spacing and $N_s$ and $N_t$ are the number of lattice points in the spatial and Euclidean temporal directions.\\

\noindent
In general, the QCD action appearing in Eq.\ (\ref{eqn:qcd_path_integral}) can be split into three parts; the gluonic part, describing the propagation and self-interaction of gluonic fields; the fermionic part, which describes the quark fields and the interaction part, describing the coupling of quarks and gluons. This study is performed in pure gauge theory, and thus we restrict
ourselves to the pure gluonic part of the action.\\

\noindent
The gluonic action depends on the spin-$1$ bosonic gauge fields (the gluons), $A_{\mu}(n)$, where $n$ is the position argument (e.g., lattice site). These are traceless hermitian matrices that live in the Lie algebra of SU($3$) and constitute a vector field with Lorentz index $\mu$. We require gauge invariance of the action, which means the gauge fields themselves need to be gauge covariant. Using the definition of the covariant derivative, the gauge fields transform as follows,
\begin{eqnarray}
    A_{\mu}(n) \to A_{\mu}(n)^{'} &=& \Omega(n)A_{\mu}(n)\Omega(n)^{\dag} + i(\partial_{\mu}\Omega(n))\Omega(n)^{\dag},
    \label{eqn:gauge_transform}
\end{eqnarray}
where $\Omega(n)$ are SU($N_c$) group elements at a lattice site, i.e., unitary $N_c\times N_c$ with unit determinant. \\

\noindent
Such local gauge transformations only result in unphysical local phase changes in the field strength tensor, which in turn cancel in the contribution to the action and do not change the physical nature of the theory. When viewed from a geometrical perspective, one thinks of gauge fields as connections on the principal bundle enabling infinitesimal parallel transport of the fields relating different fibres \cite{Vakar:2021qsl}. Then the gauge transformations describe the corresponding change in the choice of coordinates on the fibres.\\

\noindent
Since the gauge fields live in the Lie algebra, su($3$) in the continuum formulation, they can be expressed by the group generators, $T_j$ (i.e., Gell-Mann matrices),
\begin{eqnarray}
    \label{eqn:gauge_field}
    A_{\mu}(n) = \sum\limits_{j=1}^{8} A^{(j)}_{\mu}(n) T_j,
\end{eqnarray}
while in the lattice representation, they are instead represented by SU$(3)$ group elements\footnote{This representation ensures the preservation of gauge invariance and simplifies lattice computations.} given by,
\begin{eqnarray}
    \label{eqn:link_variable}
    U_{\mu}(n) &=& \text{exp}\left[ia A_{\mu}(n) \right] \sim P\text{ exp}\left(i \int_{C_{xy}}A\cdot ds \right),\\
                &\sim& 1 + iaA_{\mu}(n) + \mathcal{O}(a^2),
\end{eqnarray}
and are the elementary link variables between lattice sites, where $a$ is the lattice spacing. When discretising the continuum action, gauge invariance is explicitly chosen by calculating in link variables, which ensures that gauge invariance is preserved on the finite grid. In moving from the continuum theory to the lattice, the integral appearing in the gauge transporter in Eq.\ (\ref{eqn:link_variable}) is approximated by the product of the path length, $a$ (lattice spacing), and the field value, $A_{\mu}(n)$, at the starting point.\\

\noindent
The link variables play the role of the gauge transporter from one point, $n$, to another. The colour trace of a product of these gauge fields forming a closed loop (\textit{Wilson loop}) describes a gauge invariant physical observable. We require physical observables to be gauge invariant. Thus the expectation value of the observables is unaffected by the gauge fixing choice. However, individual link variables have a vanishing expectation value, which leads to all other gauge invariant observables vanishing \cite{Gattringer:2010zz}. On the lattice, gauge fixing only simplifies some calculations, as opposed to its essential and integral role in perturbative calculations \cite{Giusti:2001xf}.\\

\noindent
In lattice terms, the product of four link variables forming a square is called the \textit{plaquette},
\begin{eqnarray}
     U_{\mu\nu}(n) &=&  U_{\mu}(n)U_{\nu}(n+\hat{\mu})U_{-\hat{\mu}}(n +\hat{\mu}+\hat{\nu})U_{-\hat{\nu}}(n+\hat{\nu}),\\
                    &=&  U_{\mu}(n)U_{\nu}(n+\hat{\mu})U_{\mu}(n+\hat{\nu})^{\dag}U_{\nu}(n)^{\dag},
                    \label{eqn:plaquette}
\end{eqnarray}
and is shown on the middle frame of Fig.\ (\ref{Fig1}). The plaquette variable is path dependent, traversing a single $1\times 1$ square loop in a specific plane. The sum of all possible plaquettes on the lattice describes the Wilson gauge action, with each plaquette counted with a single orientation. The Wilson action is given by,
\begin{eqnarray}
    \label{eqn:wilsonaction}
    S_G \left[ U_{\mu}(n) \right] \equiv \frac{\beta}{N} \sum\limits_{n\in\Lambda} \sum\limits_{\mu<\nu} \text{Re Tr} \left[ 1 -  U_{\mu\nu}(n) \right],
\end{eqnarray}
where the sum is performed over all lattice points containing a plaquette and the Lorentz indices. The pre-factor, $\beta$ is the inverse coupling,
\begin{eqnarray}
    \beta \equiv \frac{2N_c}{g_s^2},
\end{eqnarray}
and provides the physical scale on the lattice.\\

\section{Metropolis Algorithm}
\label{sec:Metropolis Algorithm}
The purpose of an update algorithm is to use a proposal probability distribution to generate a set of field configurations which approximate a target distribution. The way an update algorithm moves from one field configuration, to another (also called a Monte Carlo update step) is through a stochastic process (i.e., Markov process) following the specified proposal distribution. The Markov process\footnote{The Markov process is a stochastic process where the state at $\tau_{n+1}$ depends only on the state at $\tau_{n}$ and not the entire sequence of states that preceded it.} starts at some initial configuration, which is either an identity element (cold start) or a set of random numbers drawn from a specified distribution (hot start). A good choice for the starting configuration reduces the computation time.\\ 

\noindent
The Markov process then samples the distribution, visiting all configurations, but those with a high probability or large Boltzmann factor exp$(-S)$ more frequently and ultimately approaching a unique stationary distribution. A stationery distribution always exists because we assume a finite state (configuration) space, else it would not exist when sampling an infinite space, see for example Ref.\ \cite{shoesmith1986huygens}. The sequence of states or field configurations formed by the Markov process constitutes the Markov chain.\\

\noindent
The basis of any Markov chain process is that it has to satisfy the \textit{detailed balance} condition requiring a symmetry in the transition between configurations,
\begin{eqnarray}
\label{eqn:detailed_balance}
    T(U \to U')P(U') = T(U' \to U)P(U),
\end{eqnarray}
where $T$ is the transition probability between subsequent configurations (i.e., at $n-1$ and at $n$ along the chain), and $P(U)$ is the probability that the system is in the configuration $U$. One can show that the stationery probability distribution is a fixed point of the Markov process, and once reached, the system stays there. This introduces the need for thermalisation steps prior to using any measurements from the Markov process to compute observables. To test whether the system has thermalised (equilibrium distribution is reached), one can compute the expectation value of the observable of interest as a function of the Monte Carlo time.\\

\noindent
The Metropolis algorithm \cite{reiher1966hammersley} is a local update algorithm allowing us to change only a single link variable $U_{\mu}(n) \to U_{\mu}(n)'$ at each update. Our goal is the equilibrium probability,
\begin{eqnarray}
    P(U) \propto \text{exp}(-S[U_{\mu}(n)]),
\end{eqnarray}
where $S[U]$ is the Wilson gauge action given in Eq.\ (\ref{eqn:wilsonaction}). The single link variable that we want to change is connected to four plaquettes in $(2+1)$D and six plaquettes in $(3+1)$D. We show this link and associated plaquettes for the former case in Fig.\ (\ref{fig:link_updates}), and these are the only plaquettes that are affected when the local configuration change is made. The local contribution to the action of the above-mentioned plaquettes is given by,
\begin{eqnarray}
    \label{eqn:wilsonactionlocal}
    S \left[ U_{\mu}(n)' \right]_{loc} = \frac{\beta}{N}\text{Re Tr} \left[ p\cdot 1 -  U_{\mu}(n)' A \right],
\end{eqnarray}
where $p$ is the number of locally affected plaquettes, $A$ is a sum of the other link variables forming the plaquette (black link variables in Fig.\ (\ref{fig:link_updates}), called \textit{staples}) that are not being changed during the update, but form part of the plaquette that are affected by the update. The corresponding sum of all the staples is given by,
\begin{multline}
    A = \sum\limits_{\hat{\mu} \neq \hat{\nu}} \left[ U_{\nu}(n+\hat{\mu})U_{\mu}(n+\hat{\nu})^{\dag}U_{\nu}(n)^{\dag} \right.\\ \left. + U_{\nu}(n+\hat{\mu} - \hat{\nu})^{\dag}U_{\mu}(n-\hat{\nu})^{\dag}U_{\nu}(n-\hat{\nu}) \right], \hspace{1.5cm}
\end{multline}
where the sum considers plaquettes in the forward (first term) and backward (second term) directions. Here $\hat{\mu}$ refers to the direction of the link variable being updated, while $\hat{\nu}$ refers to the other directions in Euclidean space.\\

\noindent
The candidate link variable $U_{\mu}(n)'$ that is proposed in the update step is a group element. The main technique that is applied to obtain this candidate configuration is to multiply the current configuration $U_{\mu}(n)$ by a random element of SU($N_c$) close to unity,
\begin{eqnarray}
    \label{gaugeupdate}
    U_{\mu}(n)' &=& \Omega U_{\mu}(n),
\end{eqnarray}
where $\Omega$ is the random element of SU($N_c$) close to unity. The random element, $\Omega$ needs to be selected with equal probability as $\Omega^{-1}$ in order to achieve a symmetric selection probability to ensure detailed balance, which is a criterion to achieve convergence to the target distribution. The problem then simplifies to computing the matrices $\Omega$ which are random elements of SU($N_c$). There are various methods in which one can generate group elements of SU($N_c$) close to unity, and we will briefly discuss three of these methods in the following subsection.\\

\noindent
Once the candidate link variable has been computed according to Eq.\ (\ref{gaugeupdate}), we then calculate the local change in action (i.e., for the affected plaquettes),
\begin{eqnarray}
    \Delta S_{loc} &=& S \left[ U_{\mu}(n)' \right]_{loc} - S \left[ U_{\mu}(n) \right]_{loc},\\
    \label{eqn:deltaS}
    &=& -\frac{\beta}{N}\text{Re Tr} \left[ ( U_{\mu}(n)' - U_{\mu}(n)') A \right].
\end{eqnarray}
The last step of the algorithm is the accept-reject step, where we draw a random number $k \in [0,1)$ from a uniform distribution and accept the proposed link variable $U_{\mu}(n)'$ with probability,
\begin{equation}
    P(U \to U') = 
\begin{cases}
    1,          &  \Delta S_{loc} \leq 0\\
    \text{exp}(-\Delta S_{loc}),    & \Delta S_{loc} > 0,
    \label{eqn:accept_reject}
\end{cases}
\end{equation}
i.e., $k \leq \text{exp}(-\Delta S_{loc})$, else reject.\\

\subsection{Computing a Random SU($N_c$) Element }
\label{subsec:groupelement}
\noindent
The Metropolis algorithm is used to update the gauge configuration, $U_{\mu}(n) \to U_{\mu}(n)'$ in each sweep\footnote{A sweep is a complete update cycle where each lattice site $n$ is visited and potentially updated at least once based on the outcome of the accept-reject step in Eq.\ (\ref{eqn:accept_reject}).}. We have have shown that in order to obtain the candidate link variable $ U_{\mu}(n)'$, one needs to generate a random group element $\Omega$. We now discuss three methods that can be used to generate this random group element. These methods used to generate a random element of SU($N_c$) which we discuss here form basis for the discussion in section (\ref{subsec:stepsize}) where we show the effect of the step-size used in the update of the random group element $\Omega$.\\

\subsubsection{First Approach:}
The first approach that we follow to generate the candidate update matrices is a direct sampling of the gauge algebra which seeks to reduce the computational cost of this step. It is analogous to the approach discussed in Ref.\ \cite{Weigert:1997mf}, where one computes an SU($N_c$) group element by taking the complex exponentiation of a linear combination of the algebra elements in their fundamental representation, i.e.,
\begin{eqnarray}
    \label{eqn:exponentiatedgen}
    \Omega = \text{exp}\left(i \sum\limits_{j=1}^{m} w^j T_j \right) \equiv \text{exp}(iQ),
\end{eqnarray}
where $m$ is the number of generators, $m=2$ for SU(2) and $m=8$ for SU(3) and $T_j$ are the generators. The $w^j$ are random numbers generated from either a uniform distribution with $w^j \in (-\varepsilon,\varepsilon)$ or normal distribution, where $\varepsilon$ corresponds to the standard deviation. This parameter determines the extent in which you traverse the generator space, and will be discussed further in the second approach.\\ 

\noindent

\noindent
We implement this approach in the following steps:
\begin{itemize}
    \item Instead of taking the complex exponentiation of a linear combination of generators, one only generates a single random number, $w$ and computes the group element using Eq.\ (\ref{eqn:exponentiatedgen}) without the sum. Thus only taking the complex exponentiation of a single generator.
    \item Given that we apply many update steps, we can select a different generator each time. Thus one needs to draw a second random number $d \in (1,m)$ from a uniform distribution, which allows you to sample the different generator directions on each sweep.
\end{itemize}
While Eq.\ (\ref{eqn:exponentiatedgen}) suggests stepping into random directions with random “length”, $w^j$ in generator space, here we step along one generator direction each time with a different random “length”. Choosing different directions at subsequent update steps, in total leads to the same result, i.e., randomly traversing the whole generator space.\\

\noindent
A similar, but alternative method of generating a group element is proposed in Ref.\ \cite{Curtright:2015iba}, allowing one to express the group element as a second order polynomial in a hermitian generating matrix. The polynomial coefficients consist of elementary trigonometric functions that depend on the group parameter and the determinant of the hermitian generating matrix.\\

\subsubsection{Second Approach:}
The second approach is discussed in detail in Ref.\ \cite{Gattringer:2010zz}, and we will summarise it for SU($2$) and SU($3$). In SU($2$), based on the Pauli representation of SU(2) matrices, one starts by generating four uniformly distributed random numbers $x_i \in (-\sfrac{1}{2}, \sfrac{1}{2})$. An SU(2) group element can be parameterised by these four real numbers which satisfy the normalisation condition in Eq.\ (\ref{eqn:ran_normalisation}) such that they lie on the surface of a four-dimensional hyper-sphere (3-sphere). A uniform distribution is used to ensure that the resulting group elements are evenly spread over the group manifold. The symmetric interval eliminates bias in the generated matrices and ensures detailed balance.\\

\noindent
One also needs to pick a second number, $\varepsilon$. Where $|\varepsilon| <1$ is the parameter tuned to determine the average step-size of the random walks, thus informs the acceptance rate and the spread of the generated update matrices around the unit element. The average step-size, $\Delta_{MC}$ is given by, 
\begin{eqnarray}
    \label{eqn:metrostepsize}
    \Delta_{MC} &=&  \frac{\int_{-a}^{a} \rho (x) |x| dx}{\int_{-a}^{a} \rho (x) dx},
\end{eqnarray}
where $\rho(x)$ is the probability density function, commonly a normal distribution with standard deviation $\sigma = \varepsilon$ or uniform distribution with $a=\varepsilon$. If $\rho(x) = 1$, i.e., uniform distribution, then $\Delta_{MC} = |\varepsilon|^2$. \\

\noindent
The update matrix is then given by,
\begin{eqnarray}
    \label{eqn:updateapproach1}
    \Omega = x_0\cdot\mathds{1} + i \bm{x}\cdot\bm{\sigma}, \quad \text{det}[\Omega] = \sum\limits_{j=0}^{3}x_j^2 = 1,  \label{eqn:ran_normalisation}\\
    x_0 = \text{sign}(r_0)\sqrt{1 - \varepsilon^2}, \quad \bm{x}= \varepsilon\bm{r/|r|},
    \label{eqn:updateapproach_epsilon}
\end{eqnarray}
where $\bm{\sigma}$ corresponds to the summed generators.\\

\noindent
The SU(3) update matrices are computed by generating three SU(2) matrices using Eq. (\ref{eqn:updateapproach1}), and embedding these in $3\times 3$ matrices according to,
\begin{eqnarray}
    R = \begin{pmatrix}
    r_{00} & r_{01} & 0\\
    r_{10} & r_{11} & 0\\
    0 & 0 & 1
    \end{pmatrix}, \quad S = \begin{pmatrix}
                            s_{00} & 0 & s_{01}\\
                            0 & 1 & 0\\
                            s_{10} & 0 & s_{11}
                            \end{pmatrix}, \quad T = \begin{pmatrix}
                                                        1 & 0 & 0\\
                                                        0 & t_{00} & t_{01}\\
                                                        0 & t_{10} & t_{11}
                                                        \end{pmatrix},
\end{eqnarray}
then taking the product of these three matrices,
\begin{eqnarray}
    \Omega=RST.
\end{eqnarray}

\noindent
A slight modification of this approach that we also consider is to replace $\varepsilon$ in Eq.\ (\ref{eqn:updateapproach_epsilon}) by $\varphi \in (-\varepsilon, \varepsilon)$. That is, instead of picking a fixed value $\varepsilon < 1$ to guide our step-size, we generate and use a second random number $\varphi \in (-\varepsilon, \varepsilon)$ from a uniform or normal distribution. We will discuss the step-size effect of this modification in the following subsection.\\

\subsubsection{Third Approach:}
The last approach that we look at is a slight modification of the second approach. One generates four random values, $r_j$ following a normal distribution with mean zero and standard deviation, $\varepsilon \leq 1$. The $r_j$ are normalised to unity, then the new group element is obtained as follows,
\begin{eqnarray}
    \Omega = r_0\cdot\mathds{1} + i\sum_{j=1}^m r_j\sigma_j,
\end{eqnarray}
where $\sigma_j$ are the generators of the Lie Algebra, and $m$ is the number of generators. Generally, one picks the update matrices for the candidate link, $U$ such that they are close to the previous link. However, one can also explore the effect of picking a completely random element, i.e., we take $U_{\mu}(n)' = \Omega.$\\

\subsection{Metropolis Algorithm Step-size Dependence for SU($N_c$) }
\label{subsec:stepsize}
\noindent
One challenge with the Metropolis algorithm is picking a step-size (i.e., the parameter $\varepsilon$ discussed in the previous subsection) which informs the efficiency of the algorithm. This is a system-specific parameter, however, an acceptance rate of $\sim 30-50\%$ is deemed optimal, thus one tries to tune the step-size such that it matches it. An interesting comment from \cite{Gattringer:2010zz} is that ``One sees that this topic moves from science to art" as there are no clear guiding strategies. The simplest form of the `art' would be to start by testing different step-sizes in comparison to their corresponding acceptance rates before taking any measurements. \\ 

\noindent
Another option is to write a program that changes the step-size at each update step in order to maintain the desired acceptance rate \cite{graves2011automatic}. This approach is relevant in instances where the step-size strongly depends on the domain of the proposal probability distribution, $\rho(x)$. That is, a very large step-size (low acceptance rate) where $\rho(x)$ is highly variable or a very small step-size (high acceptance rate) where $\rho(x)$ is relatively flat. Improving the step-size selection is subject to ongoing research, see Ref. \cite{marnissi2020majorize, graves2011automatic, swendsen2011maximum}.\\

\begin{figure}[!htb]
    \centering
    \subfigure[Plaquette Averages]{{\includegraphics[scale=0.5]{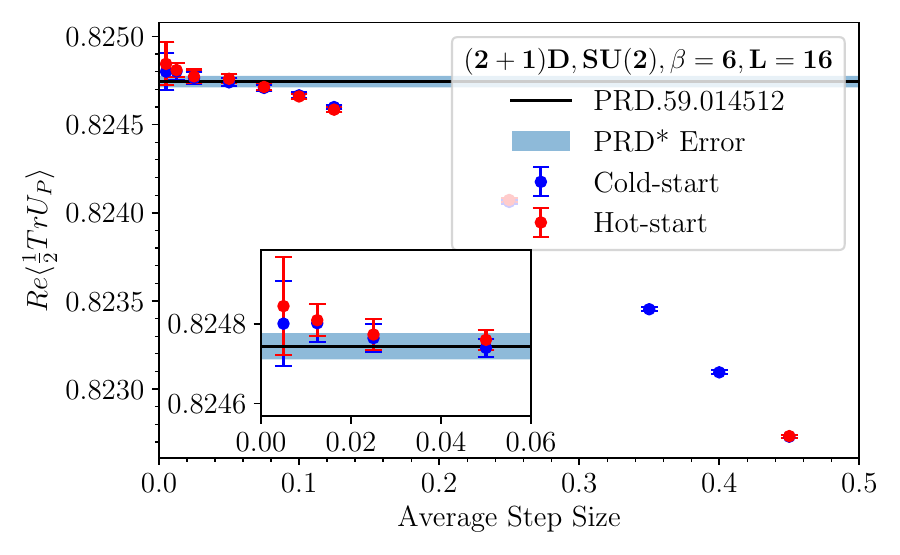} }}%
    \hspace{-0.4cm}
    \subfigure[Plaquette Average Ratio]{{\includegraphics[scale=0.5]{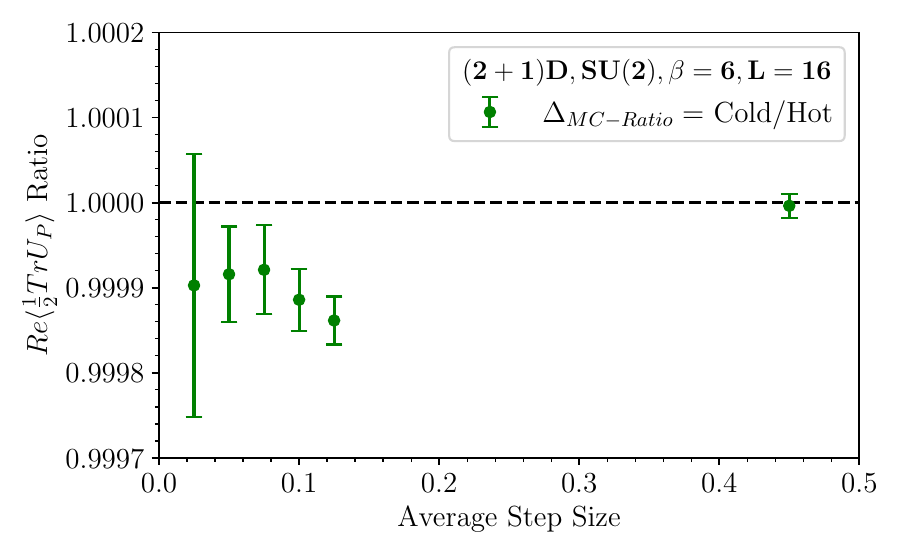} }}%
    \caption{Step-size dependence of plaquette averages for a hot and cold start using the first approach with a uniform distribution in $(2+1)$D SU($2$) for $\beta = 6$ and $L=16$.}%
    \label{fig:stepsize_starts}%
\end{figure}

\noindent
The consequence of a smaller step-size is that the configuration space is explored much slower, resulting in large \textit{thermalisation times} depending on the starting configuration. In addition, a small step-size leads to a high acceptance rate which results in long \textit{auto-correlation} times, drastically increasing the statistical errors in the measurement. In order to overcome these large errors, one would need to increase the Monte-Carlo sample size resulting in an inefficient algorithm. On the other hand, if the step-size is too large, then the configuration space is explored more broadly, leading to a very small acceptance rate. \\

\noindent
Irrespective of the step-size that one picks for the Metropolis algorithm, it is expected that for large enough Monte-Carlo time, the configurations converge (within uncertainty) of the  stationary distribution. In this section, we show that although this is the case in scalar field theory, the statement does not in general hold for the gauge group SU($N_c$), possibly due to an asymmetry in the proposals leading to a modification of the acceptance probability. Given that the plaquette average is the cheapest quantity to compute on the lattice, we use Eq.\ (\ref{eqn:metrostepsize}) to show how the plaquette average in different pure gauge theory systems varies with step-size. We show a comparison to the plaquette averages measured using a combination of the heatbath algorithm and over-relaxation in Ref.\ \cite{Teper:1998te} for $(2+1)$D and Ref.\ \cite{Athenodorou:2021qvs} for $(3+1)$D.\\

\begin{figure}[!htb]
\begin{center}
\includegraphics[scale=0.7]{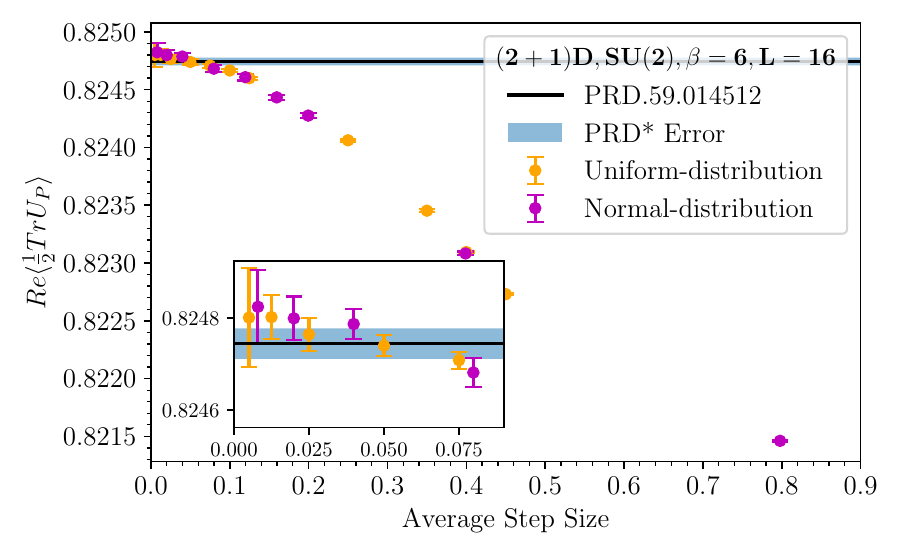}
\caption{Step-size dependence of plaquette averages for different distributions using the first approach with a cold start in $(2+1)$D SU($2$) for $\beta = 6$ and $L=16$.}
\label{fig:stepsize_distributions}
\end{center}
\end{figure}

\noindent
Note that we have computed the plaquette average using uncorrelated measurements, with statistical errors obtained from the Jackknife technique, thus should be reflective of the true error in the measurements. In Fig.\ (\ref{fig:stepsize_starts}), we show that for gauge group SU($2)$, the measured plaquette averages only converge to the literature result in Ref.\ \cite{Teper:1998te} in the limit as one approaches smaller step-sizes. In the literature, e.g., Ref.\ \cite{Gattringer:2010zz}, it is suggested that a value of $\varepsilon < 1$ should suffice. However, the fact that there is a $\sim 0.25\%$ difference between measured plaquette averages at $\varepsilon = 0.9$ (bigger step-size) and $\varepsilon = 0.1$ (smaller step-size) comes as a surprise, and perhaps suggests that this choice of $\varepsilon$ should not be generalised.\\

\noindent
We also test whether the starting configuration (hot or cold) converges to the same plaquette average for different step-sizes. We obtain a ratio of one, showing that the starting configuration does not affect the measured plaquettes (as expected) because of the convergence of the Markov chain to an equilibrium distribution over long Monte Carlo times as discussed earlier in the chapter. This check is to ensure that our update algorithm does not suffer from topological freezing \cite{Boyd:1997nt, Schaefer:2010hu}, which is a common issue for Monte Carlo algorithms \cite{Albandea:2021lvl}. We discuss this problem briefly later in this chapter.\\

\begin{figure}[!htb]
\begin{center}
\includegraphics[scale=0.7]{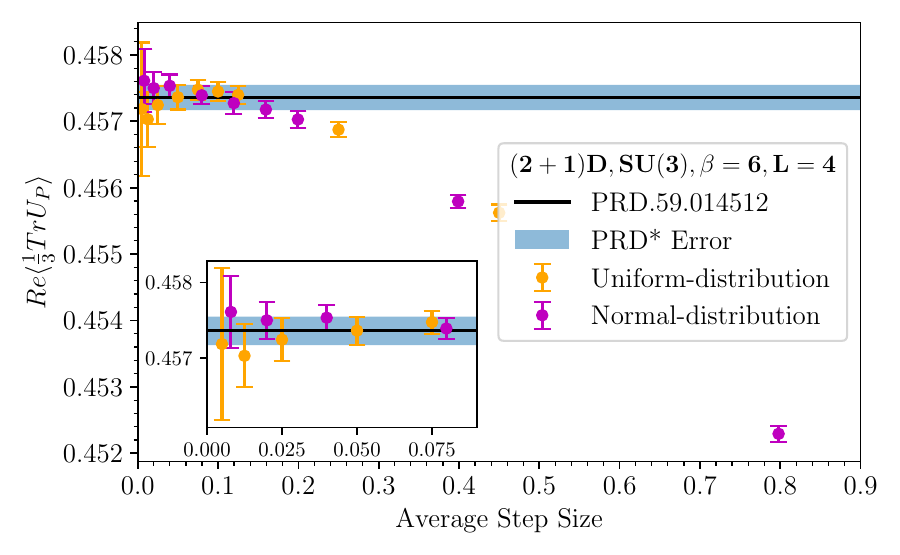}
\caption{Step-size dependence of plaquette averages for different distributions using the first approach with a cold start in $(2+1)$D SU($3$) for $\beta = 6$ and $L=4$.}
\label{fig:stepsize_distributions_su3}
\end{center}
\end{figure}

\noindent
In the meantime, let us proceed by discussing the comparison of the step-size effect on a uniform and normal proposal distributions as shown in Fig.\ (\ref{fig:stepsize_distributions}). In this plot, we show that the behavior of the Markov chain is less sensitive to the shape of the proposal distribution due to the symmetry imposed by the detailed balance condition i.e., Eq.\ (\ref{eqn:detailed_balance}). This is evident from the shape of the plaquette average dependence on the step-size (for both a uniform and normal distribution), which follows approximately the same curve and plateaus around the same $\Delta_{MC}$.\\

\noindent
Recall that the Lie group, SU($2$) can be expressed as a subgroup of SU($3$) by either taking the first three generators of SU($3$) and accounting for the extra factors of ($+1$) when taking the trace or by changing the representation. We show in Fig.\ (\ref{fig:stepsize_distributions_su3}) that the step-size dependence observed in SU($2$) is consistent in SU($3$), thus not just a SU($2$) group artefact. Given that this effect is not observed in scalar field theories, this would suggest that it is related to the sampling of non-abelian groups.\\

\begin{figure}[!htb]
    \centering
    \subfigure[Plaquette Averages]{{\includegraphics[scale=0.5]{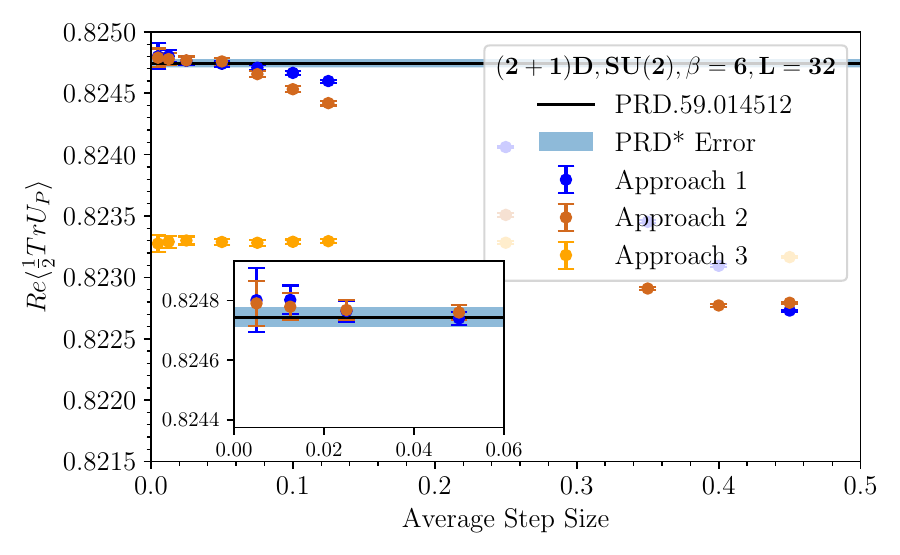} }}%
    \hspace{-0.4cm}
    \subfigure[Plaquette Average Ratio]{{\includegraphics[scale=0.5]{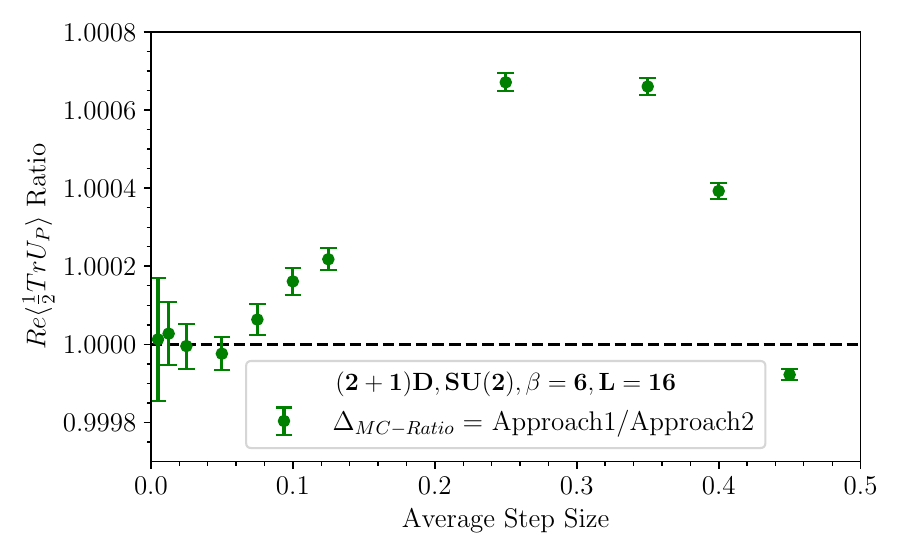} }}%
    \caption{Step-size dependence of plaquette averages for the first, second and third approach using a uniform distribution with a cold start in $(2+1)$D SU($2$) for $\beta = 6$ and $L=16$.}%
    \label{fig:approach1_2}%
\end{figure}

\noindent
In principle, the various ways in which one computes the update matrices should be equivalent (within errors). In Sec.\ (\ref{subsec:groupelement}), we discussed three such approaches and we now compare the different approaches in  Fig.\ (\ref{fig:approach1_2}). The first two methods converge to the same plaquette average with decreasing step-size, however they show a clear systematic difference as can be seen from the ratio plot on the right frame. On the other hand, the third method shows no explicit step-size dependence as seen from the orange data-points on the left frame. While this is not at the centre of this work, we highlight it as a possible subject of investigation to improve algorithm efficiency.\\

\begin{figure}[!htb]
    \centering
    \subfigure[Plaquette Averages]{{\includegraphics[scale=.5]{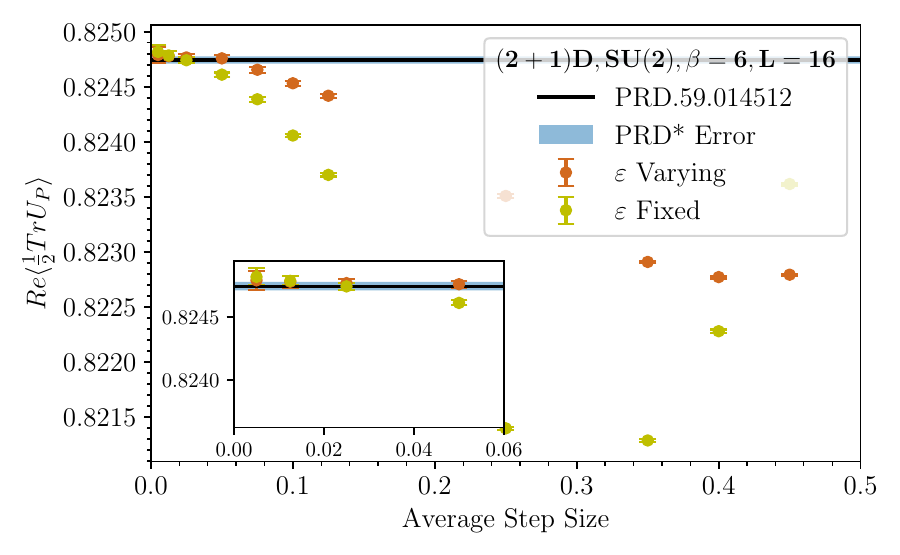} }}%
    \hspace{-0.4cm}
    \subfigure[Plaquette Average Ratio]{{\includegraphics[scale=.5]{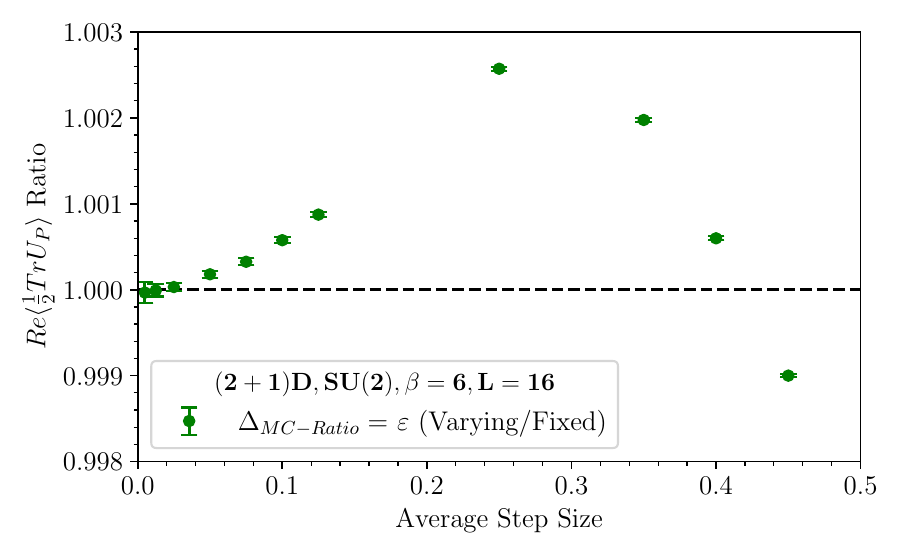} }}%
    \caption{Step-size dependence of plaquette averages for the second approach with fixed and varying $\varepsilon$ using a uniform distribution with a cold start in $(2+1)$D SU($2$) for $\beta = 6$ and $L=16$.}%
    \label{fig:approach2_epsilon}
\end{figure}

\begin{figure}[!htb]
    \centering
    \subfigure[Plaquette Averages]{{\includegraphics[scale=0.5]{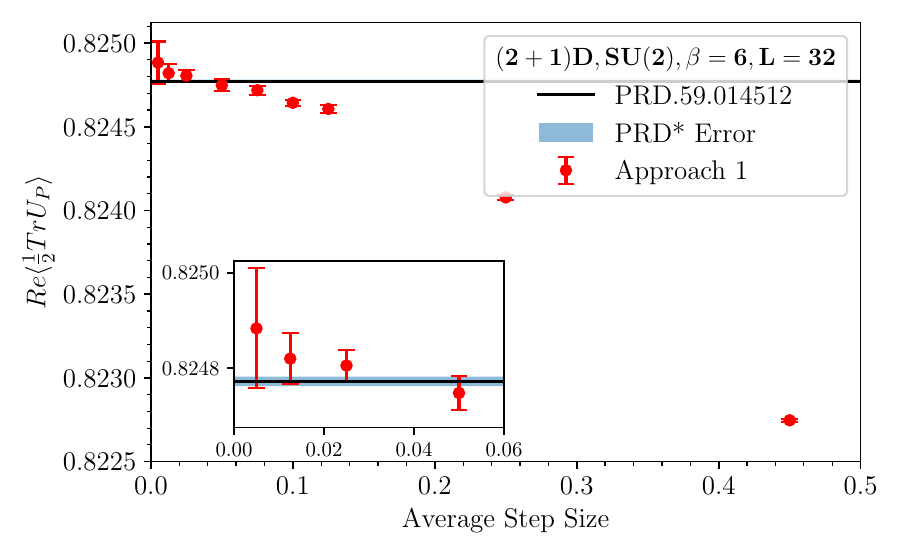} }}%
    \hspace{-0.4cm}
    \subfigure[Plaquette Average Ratio]{{\includegraphics[scale=0.5]{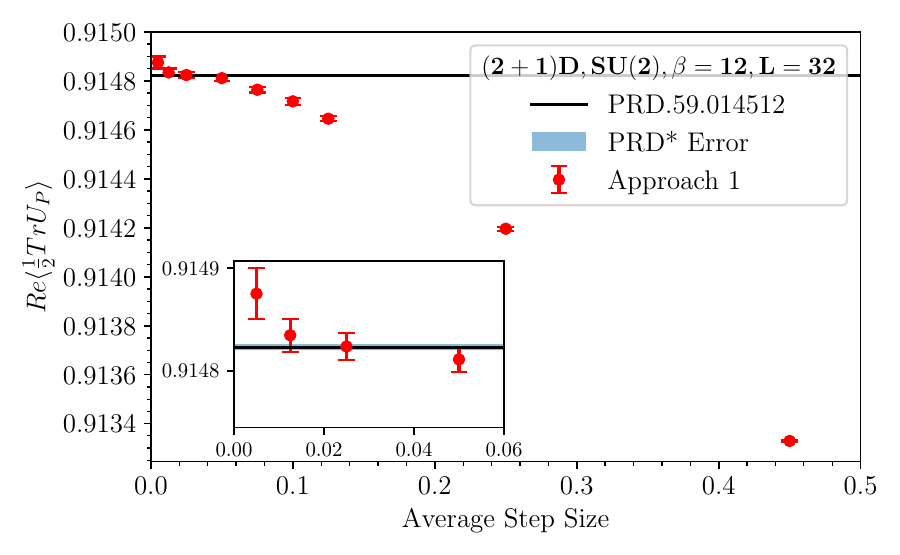} }}%
    \caption{Step-size dependence of plaquette averages in $(2+1)$D SU($2$) for $\beta = 6$ and $\beta = 12$ using the first approach with a uniform distribution with a cold start at lattice size $L=32$.}%
    \label{fig:approach1_beta}
\end{figure}

\noindent
As mentioned earlier, the second approach can be slightly modified to fix the variable, $\varepsilon$ to a constant or draw it from a uniform distribution. We show the resulting step-size dependence in Fig.\ (\ref{fig:approach2_epsilon}), and both approaches converge to the same value (at different rates) with reducing step-size, but varying $\varepsilon$ is generally closer to the heatbath plus over-relaxation result. For completeness, Fig.\ (\ref{fig:approach1_beta}) shows the step-size dependence at varying $\beta$ and larger lattice size. It is clear that the convergence with step-size is not a coupling nor lattice size artefact, but rather intrinsic to the group sampling formalism.\\

\noindent
The goal of this thesis is to study the Casimir effect, where the lattice observables of interest are not the plaquette averages, but the plaquette action. We now look at the effect of the step-size in the measured plaquette action, by comparing the averages at step-sizes of $\Delta_{MC} = 0.25$ and $\Delta_{MC} = 0.025$. The former corresponds to plaquette averages outside the range of the result in Ref.\ \cite{Teper:1998te}, while the latter corresponds to plaquette averages within (or at the least closest to) the same result. See the example given in Fig.\ (\ref{fig:stepsize_starts}). Note that this discussion involves concepts such as the string tension, which are expanded on in the following chapter and are not relevant for the understanding of the current discussion, thus can be taken at face value.\\

\begin{figure}[!htb]
    \centering
    \subfigure[Plaquette Averages]{{\includegraphics[scale=0.5]{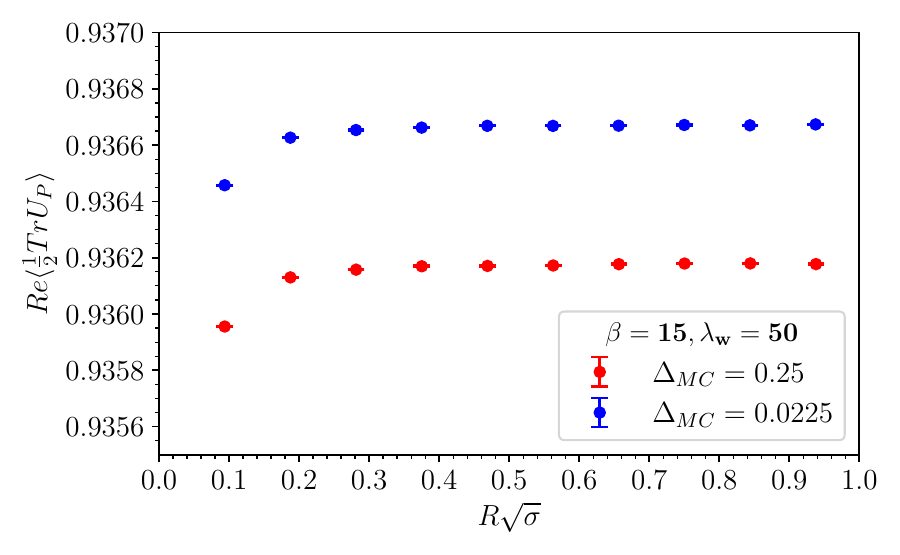} }}%
    \hspace{-0.4cm}
    \subfigure[Plaquette Average Ratio]{{\includegraphics[scale=0.5]{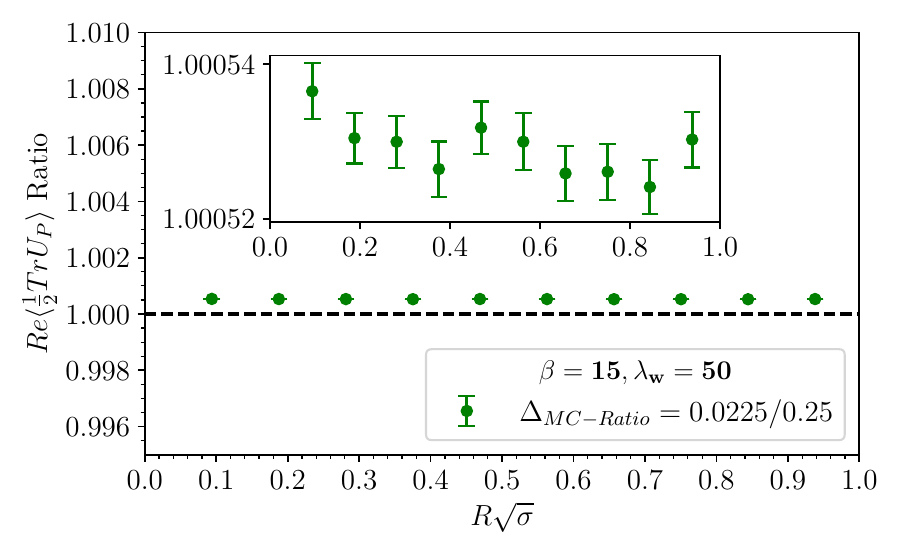} }}%
    \caption{Plaquette averages for the $yt$ plaquettes as a function of the string tension at two different values of $\Delta_{MC}$ in $(2+1)$D SU($2$) for $\beta = 15$, $L=32$ and $\lambda_w=50$.}%
    \label{fig:plaquettemc}%
\end{figure}

\begin{figure}[!htb]
    \centering
    \subfigure[Action Averages]{{\includegraphics[scale=0.5]{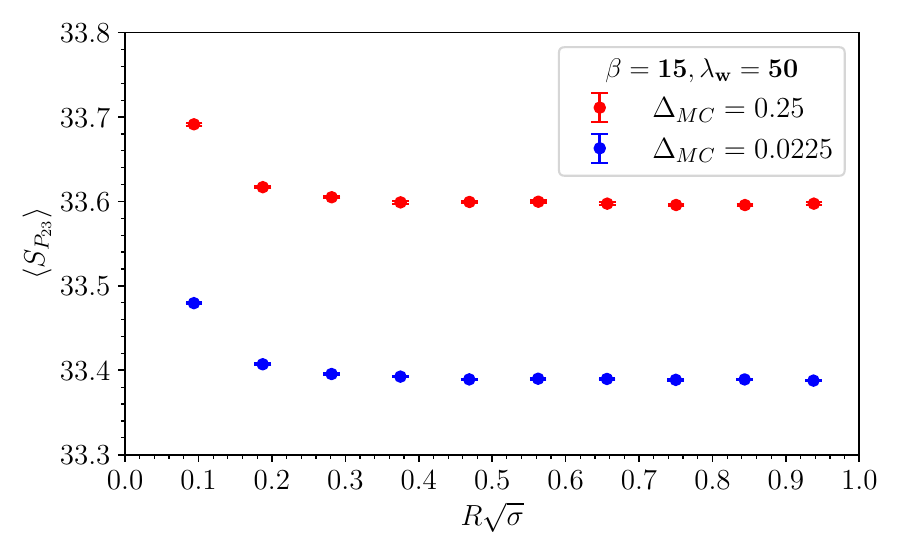} }}%
    \hspace{-0.4cm}
    \subfigure[Action Averages Ratio]{{\includegraphics[scale=0.5]{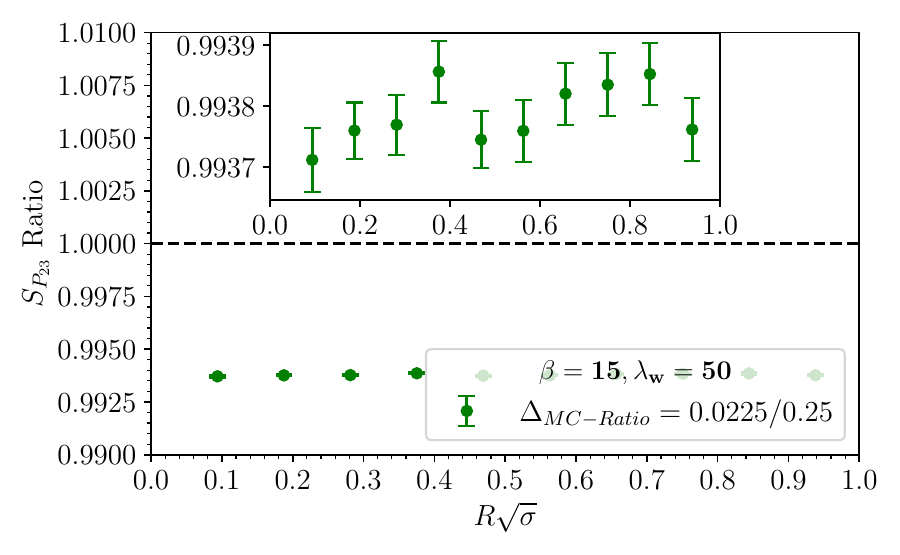} }}%
    \caption{Average action for the $yt$ plaquettes as a function of the string tension at two different values of $\Delta_{MC}$ in $(2+1)$D SU($2$) for $\beta = 15$, $L=32$ and $\lambda_w=50$.}%
    \label{fig:plaquettemcratio}%
\end{figure}

\noindent
In Fig.\ (\ref{fig:plaquettemc}), we show the step-size effect on the $yt$ plaquette average as a function of the string tension (which we use to introduce a physical scaling to our lattice measurements), which shows a difference of $\sim 0.05\%$ for the step-sizes compared. This, in turn results in a difference of $\sim 0.3\% > 0.05\%$ in the measured average action for the $yt$ plaquettes as shown in Fig.\ (\ref{fig:plaquettemcratio}). The emphasis here, is that when one picks the step-size for the Metropolis update, checking the convergence of plaquette averages alone is not enough. Instead, one should pick a step-size based on the convergence of their preferred observable of interest on the lattice (in this case, the average action for the $yt$ plaquette). \\

\section{Hamiltonian Monte Carlo Algorithm}

\noindent
The Metropolis algorithm is an exact update method where one only updates a single link variable at each sweep (i.e.\ local updates) and the proposal depends on the current state of the system. It suffices for our purposes of pure gauge studies, but can become very inefficient in studying the full QCD action with fermionic fields. This inefficiency stems from the challenge of local update algorithms for systems with long range correlators where changes in the system need not be localised. There exists an ensemble of more efficient algorithms, some of which perform global updates at each sweep and others where the gauge field changes are determined by the action and not picked randomly. \\

\noindent
The Hamiltonian Monte Carlo (also referred to as the Hybrid Monte Carlo in some texts) approach is one such algorithm where the proposal is non-local, but rather explores the system's state-space more efficiently. The Markov time numerical evolution of the gauge fields in this algorithm is informed by molecular dynamics and classical equations of motion,
\begin{eqnarray}
    H[Q,P] &=& \frac{1}{2}P^2 + S[Q],\\
    \Dot{P} &=& -\frac{\partial H }{\partial Q} = -\frac{\partial S}{\partial Q},\\
    \Dot{Q} &=& -\frac{\partial H }{\partial P} = P,
\end{eqnarray}
where $H$ is the Hamiltonian, $Q$ are elements of the Lie algebra defining the link variables in Eq.\ (\ref{eqn:exponentiatedgen}) and,
\begin{eqnarray}
    P_{\mu}(n) = \sum\limits_{j=1}^{m} P^j_{\mu}(n) T_j,
\end{eqnarray}
are also elements of the algebra defined by $m$ (number of generators) real `momentum' variables, $P^j_{\mu}(n)$.\\

\noindent
Since the Hamiltonian is a constant of motion, the path followed by the gauge field configurations is a hypersurface of constant energy in phase space and the resulting update needs to be ergodic. Such a numerical evolution (leapfrog evolution) of the classical equations naturally introduce a discrete step-size, $\varepsilon$ which needs to be carefully chosen, and comes with numerical errors $\mathcal{O}(\varepsilon^2)$. Note that this formulation necessitates the computation of derivatives of the action, and consequently derivatives with respect to the elements of the algebra.\\

\begin{figure}[!htb]
\begin{center}
\includegraphics[scale=0.7]{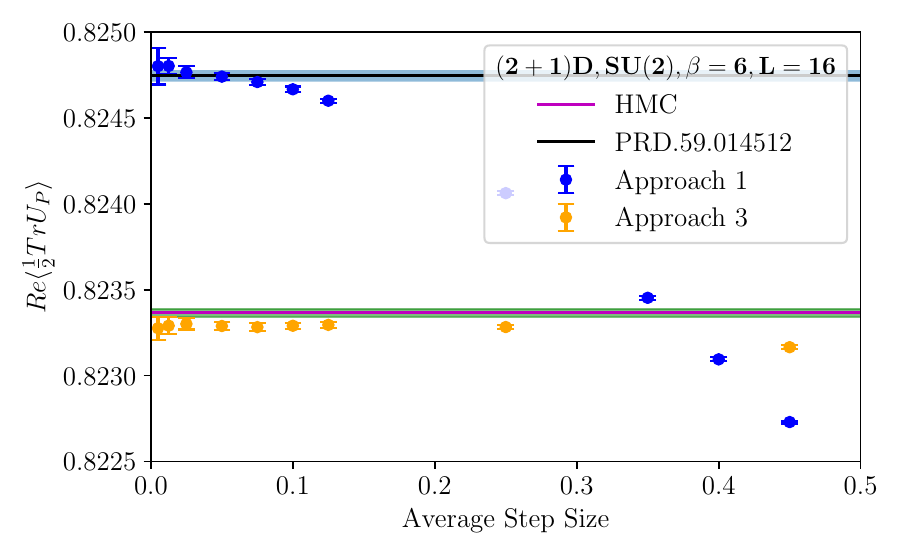}
\caption{Step-size dependence of plaquette averages for different algorithms in $(2+1)$D SU($3$) for $\beta = 6$ and $L=16$.}
\label{fig:stepsize_hmc}
\end{center}
\end{figure}

\noindent
The Wilson plaquette is composed of four link variables, such derivatives can be calculated from the product rule or employing some clever techniques, see Ref.\ \cite{Gattringer:2010zz}. There exists other hybrid algorithms where the classical evolution is perturbed with a noise term that introduces quantum fluctuations \cite{Duane:1986iw}, similarly to stochastic differential equations. An extensive discussion on the formulation of the Hamiltonian Monte Carlo algorithm can be found in Ref.\ \cite{Gattringer:2010zz}, and we briefly mention it here as an alternative approach that we have explored to make sense of the observed step-size dependence in numerical results obtained using Metropolis update steps.\\

\noindent
We show a comparison of plaquette averages measured using the Metropolis algorithm (data-points), the Hamiltonian Monte Carlo algorithm (purple line), a well as a combination of the heatbath algorithm and over-relaxation (black line) in Fig.\ (\ref{fig:stepsize_hmc}). All three algorithms provide equivalent numerical results up-to second digit.  In the case of the Metropolis algorithm, we compare the first approach (blue data-points) of generating candidate update elements, which we showed to converge to the result of Ref.\ \cite{Teper:1998te} (heatbath + over-relaxation) with decreasing step-size. This approach overlaps with the HMC result and is therefore equivalent at some step-size, $0.3<\Delta_{MC}<0.4$.\\

\noindent
We also show the third approach (orange data-points), which lies very close to the HMC result and surprisingly shows no explicit step-size dependence. It is unclear at this point what the exact source of this step-size dependence is, but it clearly does not result in a deviation from other available algorithms based on our observations in Fig.\ (\ref{fig:stepsize_hmc}). It may be important for studies performing precision measurements to take into account these step-size effects. However, such effects are negligible for our purposes and we generate our candidate update matrices using the first approach for the remainder of this thesis. \\

\section{Topology on the Lattice}
In section (\ref{subsec:stepsize}), we discussed the observed step-size effect on the measured plaquette averages when the field configurations are generated using the Metropolis update algorithm. While it is inconclusive that such effects are related to topological aspects of traversing the phase space, we briefly discuss lattice topology and highlight some common issues with Monte Carlo algorithms.\\

\subsection{Topological Sectors}
\noindent
Topology plays a significant role in our understanding of lattice QCD, mostly through the explanation of the properties of mathematical objects such as groups that are used in the formulation of gauge theories. In the integral formulation of gauge theories in Euclidean space, the field configuration space may be decomposed into disjoint subsets. The set of all continuously deformed configurations live in the same subset called \textit{topological sectors} \cite{rajaraman1987introduction,coleman1988aspects}. \\

\noindent
An example of the occurrence of topological sectors is a single quantum scalar particle moving on the real line $\mathbb{R}^1$, e.g., along the $x$-axis. When formulated in Euclidean space with periodic boundary conditions, the endpoints of the path (boundaries) as $x\to \pm \infty$ become indistinguishable and can be identified as a single point at $\infty$. The resulting path followed by the particle is a compactification of the real projective line $\mathbb{R}^1 \cup \{\infty\}$ into an $S^1$-sphere (circle).\\

\noindent
The elements, $U$ of the circle group (i.e., $S^1$) can be parametrised by the angle measure, $\theta$, also describing the exponential map of the group,
\begin{equation}
    U = e^{i\theta} = \cos \theta + i\sin \theta,
\end{equation}
where $\theta$ has period $2\pi$ as required for the trigonometric functions. This means that in our transformation of the real line, $\mathbb{R}^1$ into a circle, the endpoints have to fulfil the condition, $\text{lim}_{x \to \infty} \theta (x) = 2\pi n$ for some integer $n$. One can rewrite the resulting group elements including the period,
\begin{equation}
    U^{(n)}(\theta) = e^{in\theta} = \cos n\theta + i\sin n\theta.
\end{equation}
Topological sectors arise because any unique mappings $U^{(n)}$ and $U^{(m)} \in U(1)$ with $n\neq m$ of the elements from the spatial real line to the circle group cannot be continuously deformed irrespective of their angle, since this will break the periodicity condition at the endpoints requiring discrete deformations only. Thus each topological sector is labelled by the integer $n$, also called the \textit{winding number} and describes the number of angles in the real domain space that maps to the same element $U \in$ U(1) in the target complex space.\\

\noindent
An observable $\mathcal{O}$ of our quantum scalar particle system moving along the $S^1$-sphere has an expectation value given by the path integral,
\begin{equation}
    \langle \mathcal{O} \rangle = \frac{1}{Z} \int \mathcal{D} \psi \text{ exp}(-S[\psi]) \mathcal{O}[\psi],
\end{equation}
summing over all closed paths $\psi(t) \in S^1$, where $Z$ is the partition function and $t\in[0,2\pi]$. The set of all possible paths is divided into disjoint subsets (topological sectors) and the winding number is given by the \textit{topological charge},
\begin{equation}
    Q = \frac{1}{2\pi} \int_0^{2\pi R} dt \Dot{\psi},
\end{equation}
which is a conserved quantity.\\

\noindent
While we have given an overview of the simple example of how topological sectors may arise in one dimension, a generalisation can be made following a similar argument on the four-dimensional torus and the gauge group SU(3) on the $S^3$-sphere. We refer the reader to Ref.\ \cite{Mazur:2021zgi, Jahn:2019nmd, Bietenholz:2012sh} for a detailed discussion of topology in the context of lattice QCD. The idea is to find classical fields which minimise the Euclidean Yang-Mills vacuum action, which in turn requires the gluon field strength tensor to vanish at spacetime infinity. A vanishing field strength tensor implies that the gauge fields themselves need to vanish, i.e., $A_{\mu}(n)=0$. Using the gauge transform definition in Eq.\ (\ref{eqn:gauge_transform}), we find that the gauge fields need to approach a pure gauge at spacetime infinity,
\begin{equation}
    \lim_{r \to \infty} A_{\mu}(n) = A_{\mu}(n)^{pg} = i(\partial_{\mu}\Omega(n))\Omega(n)^{\dag},
    \label{eqn:pure_gauge}
\end{equation}
A proof that using these pure gauge fields results in a vanishing gluon field strength tensor is given in Ref.\ \cite{Mazur:2021zgi}. The resulting topological charge is given by \cite{Mages:2015scv, vanBaal:1982ag},
\begin{eqnarray}
\label{eqn:su3_topological_charge}
    Q &=& \int_\mathcal{M} d^4x q(x),\\
    q(x) &=& \frac{1}{32\pi^2} \epsilon_{\mu \nu \rho \sigma} \text{Tr}[F_{\mu \nu}F_{\rho \sigma}],
\end{eqnarray}
where $q(x)$ is the topological charge density. In the lattice regularisation scheme, the gluonic topological charge can be measured from the components of the field strength tensor as follows,
\begin{eqnarray}
    Q_{top} &=& a^4 \sum \limits_{x} q(x),\\
    q(x) &=& \frac{1}{8\pi^2} \left[ F_{01}(x)F_{23}(x) + F_{02}(x)F_{31}(x) + F_{03}(x)F_{12}(x)\right],
\end{eqnarray}
where we have defined the field strength components in terms of the plaquette variables in Eq.\ (\ref{eqn:field_components}).\\ 

\noindent
The lattice gluonic topological charge, $Q \notin \mathbb{Z}$ due to the action discretisation errors and is dominated by short-range ultraviolet fluctuations of the gauge fields which appear in Monte Carlo simulations, thus very high gluon momenta \cite{Gruber:2013efo}. In order to isolate the relevant degrees of freedom of the long-range structure of the topological charge, the ultraviolet noise is eliminated from the gauge fields before measuring the topological charge through smoothing methods \cite{Gruber:2013efo, deForcrand:2006my, Teper:1997am, APE:1987ehd, Bernard:1999kc}. We define the \textit{topological susceptibility},
\begin{eqnarray}
    \chi_{top} = \frac{1}{V} \left( \langle Q^2 \rangle - \langle Q \rangle^2  \right),
\end{eqnarray}
as the variance of the topological charge, normalised by the lattice Euclidean volume, and it describes the fluctuations in the topological charge or equivalently, the width of the topological charge distribution. See Ref.\ \cite{Jahn:2019nmd, Moore:2017ond, HotQCD:2018pds, GrillidiCortona:2015jxo, Gorghetto:2018ocs, Taniguchi:2016tjc} for further discussions on the topological susceptibility in different temperature regimes and number of flavours.\\

\subsection{Topological Freezing in Monte Carlo Algorithms}
\noindent
In general, any group elements with different winding numbers are in separate topological sectors and it is impossible to continuously deform two transformations $U_{\mu}(n) \in$ SU(3) with different topological charges. However, in the case of gauge fields, tunneling effects occur between different topological sectors and are mediated by \textit{instantons} \cite{Vandoren:2008xg, Jahn:2019nmd, Gross:1980br, Forkel:2000sq}. Such tunnelling phenomena break the pure gauge condition given in Eq.\ (\ref{eqn:pure_gauge}) and result in a non-zero action \cite{Mazur:2021zgi} in the transition region known as the \textit{sphaleron barrier}.\\

\noindent
The sphaleron barrier is an energy barrier with a height of the order $\mathcal{O}(\Lambda_{QCD}/\alpha_s)$ \cite{Petreczky:2016vrs}, and is studied on the lattice through the analysis of the distribution of topological charge configurations and identifying regions where the topological charge changes. Instanton-like gauge field configurations are identified similarly. In the continuum theory, instantons are classical solutions to the Euclidean equations of motion with finite action \cite{Jahn:2019nmd}. Since different topological sectors are disparate in the continuum, they are only connected by configurations with infinite action, thus cannot be continuously deformed.\\

\noindent
Given the infinite action requirement for tunnelling in the continuum, one would expect that generating gauge field configurations on the lattice using Monte Carlo algorithms should result in such configurations getting trapped in a single topological sector. However, while this is probable, it is not entirely the case since the continuum infinite action is translated to a large (but finite) action on the lattice, making it possible for instantons to tunnel through different topological sectors.\\

\begin{figure}[!htb]
\begin{center}
\includegraphics[scale=0.3]{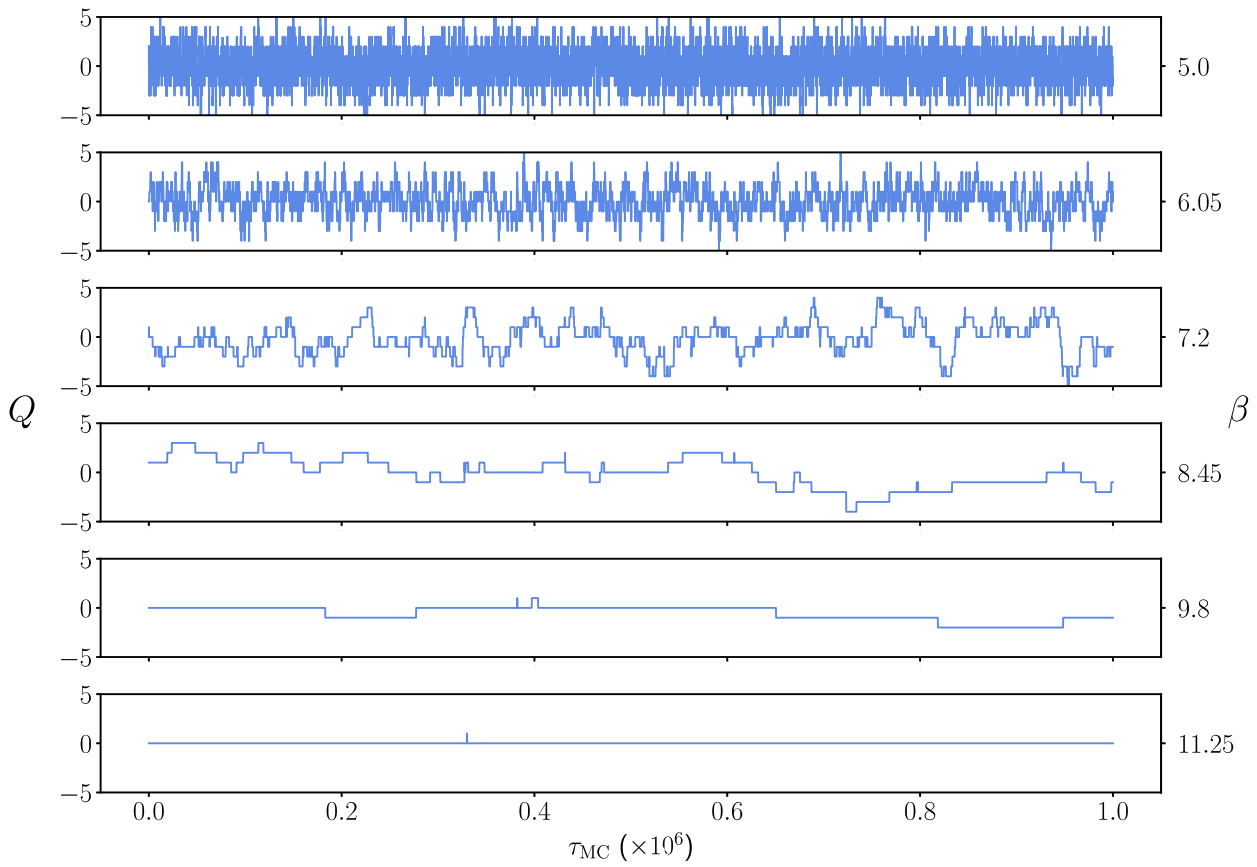}
\caption{Topological charge history along a Markov chain of HMC configurations at different couplings in $U(1)$ \cite{Albandea:2021lvl}.}
\label{fig:topological_charge}
\end{center}
\end{figure}

\noindent
While we have discussed that the infinite action barrier in the continuum does not result in poor exploration of topological sectors on the lattice, there are other lattice artefacts that do. During a Monte Carlo simulation, the generated gauge field configurations can get stuck in a single topological sector or evolve very slowly across configurations with different topological charges \cite{Alles:1996vn, DelDebbio:2004xh}. This phenomenon is known as \textit{topological freezing} and can occur as a result of finite-size effects or smaller lattice spacings \cite{Luscher:2010we,Luscher:2010iy}, or due to algorithm inefficiencies. In Fig.\ (\ref{fig:topological_charge}), we show an example of topological freezing at small lattice spacings in a $U(1)$ gauge theory with configurations obtained from the HMC algorithm in Ref.\ \cite{Albandea:2021lvl}.\\

\noindent
One consequence of topological freezing is that it results in very long autocorrelation times \cite{Luscher:2010we, DelDebbio:2004xh, Schaefer:2010hu} of the measured observable, resulting in statistical difficulties and leading to enhanced errors. See Ref.\ \cite{Florio:2020itn, Luscher:2011kk,Albandea:2021lvl,Bonanno:2020hht} for discussions on approaches used to attempt the reduction of autocorrelation times. We discussed the step-size effect on the Metropolis algorithm update step in section (\ref{subsec:stepsize}), and we show the effect of step-size reduction on the autocorrelation times in Fig.\ (\ref{fig:autocorrelations}) for (3+1)D SU(3). The observed long autocorrelation times with decreasing step-size can be associated with a slower sampling of the configuration space, a behaviour consistent with topological freezing which is due to a difficulty in overcoming the action barriers.\\

\noindent
It is worth mentioning that while we have shown behaviour reminiscent of topological freezing for (2+1)D SU($N_c$) non-abelian gauge theories in section (\ref{subsec:stepsize}), we do not anticipate that step-size effects are due to topological freezing. Testing whether such effects persist (as anticipated) in (3+1)D is left for future explorations, but could be of interest because instantons are absent or rather, instanton effects are suppressed in (2+1)D. Thus the concept of instantons in this case does not have the same significance as in the (3+1)D counterpart. This can be attributed to different confinement dynamics, topological obstructions and the absence of some dynamical features in (2+1)D theories. We refer the reader to Ref.\ \cite{Dunne:1998qy} for detailed discussions. In full QCD with fermionic degrees of freedom, instantons play a crucial role in the fermion mass generation, and this is not a concern in our pure gauge studies.\\

\section{Measuring Observables}
\subsection{Thermalisation}
\begin{figure}[!htb]
\begin{center}
\includegraphics[scale=0.35]{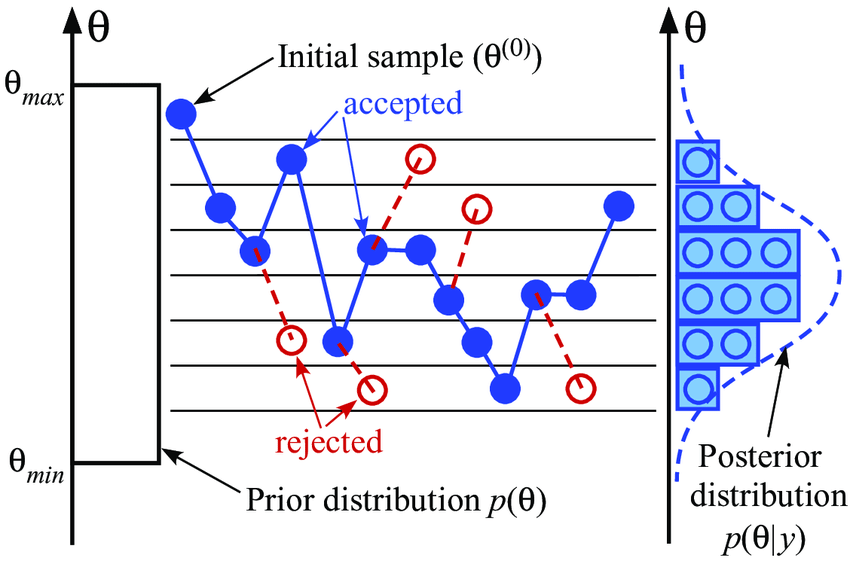}
\caption{Illustration of a Markov chain Monte Carlo process \cite{app10010272}.}
\label{fig:mcmc}
\end{center}
\end{figure}

\noindent
Monte Carlo methods are based on repeated random sampling of a specified prior probability distribution to obtain some variables, $r_i$ used as input parameters. This so called Markov process (repeated sampling) forms a Markov chain, see Section (\ref{sec:Metropolis Algorithm}) for a detailed description. In a Markov chain, the generated configuration at position $n$, depends only on the preceding configuration at position ($n-1$) along the Markov chain, thus the configurations are correlated. An example of a Markov process is shown in Fig.\ (\ref{fig:mcmc}) \cite{tomicmcmc}, where the chain moves from left to right sampling a uniform distribution.\\

\noindent
As discussed in Section (\ref{sec:Metropolis Algorithm}), from the initial point ($n=0$ or $\theta^{(0)}$) of the Markov process, the system undergoes a thermalisation process before the stationery distribution is reached. All measurements taken during the thermalisation process are not useful in the computation of the observable of interest and need to be discarded, else they can skew the expectation value. This is an essential step before studying auto-correlation times, because we are only concerned with the auto-correlations of measurements that will be used in computing the expectation value of our observable of choice.\\

\begin{figure}[!htb]
    \centering
    \subfigure[Vacuum Normalised Action]{{\includegraphics[scale=0.5]{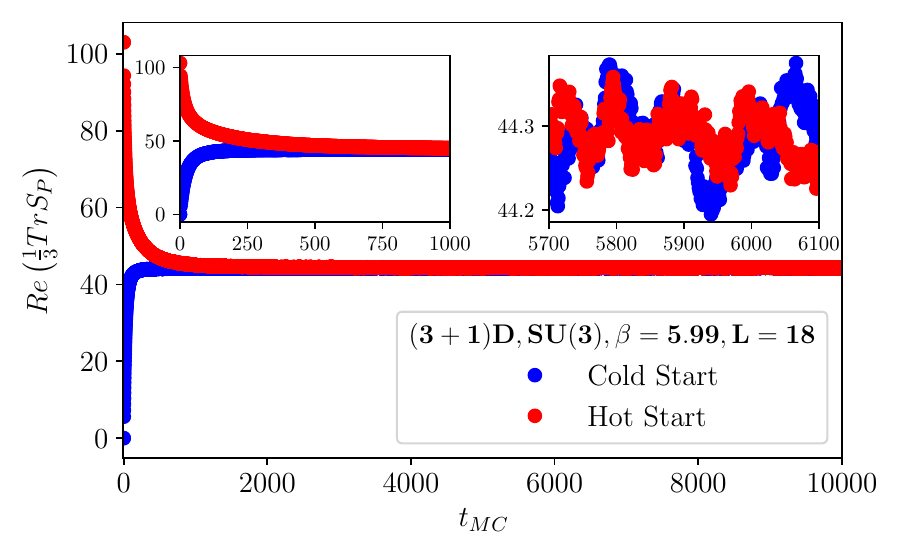} \label{subfig:thermalisationa} }}%
    \hspace{-0.4cm}
    \subfigure[Vacuum Normalised Action Averages]{{\includegraphics[scale=0.5]{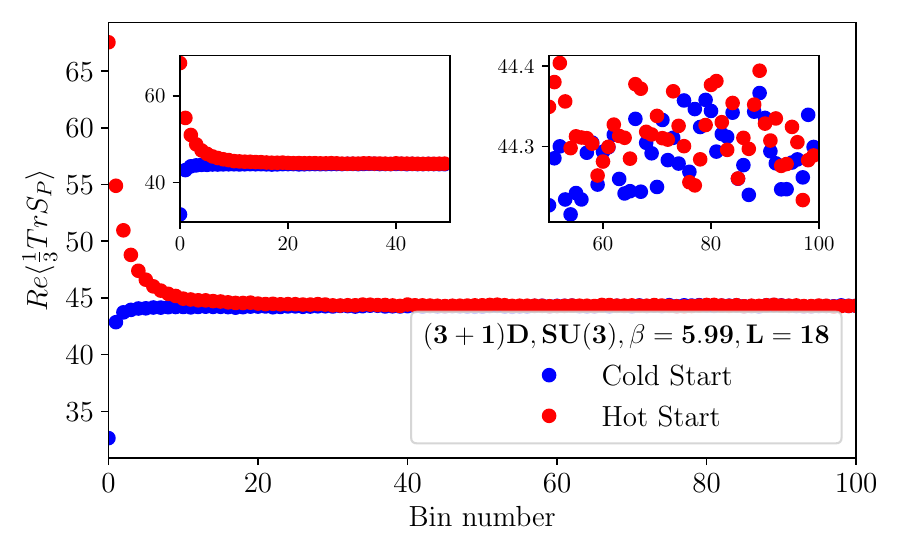} \label{subfig:thermalisationb} }}%
    \caption{Thermalisation of the action for a cold and hot start in $(3+1)$D SU($3$) for $\beta = 5.99$, $L=18$.}%
    \label{fig:thermalisation}%
\end{figure}

\noindent
In Fig.\ (\ref{fig:thermalisation}), we show the thermalisation steps of the spatial plaquette in $(3+1)$D SU($3$) for $\beta = 5.99$, $L=18$ (note that the temporal plaquette thermalises similarly because we are using an isotropic lattice). We sample a normal distribution following the Metropolis algorithm with $\Delta_{MC} \sim 0.8$, starting from a cold and hot start. Figure\ (\ref{subfig:thermalisationa}) shows a comparison of these thermalisation steps in Monte Carlo time. We observe that the cold start thermalises faster than the hot start, as can be seen from the left inset-figure. The right inset-figure shows that over long Monte Carlo times, both starts converge to the same result.\\

\noindent
In general, we test for thermalisation as shown in Fig.\ (\ref{subfig:thermalisationb}) where the normalised action measurements along Monte Carlo time are binned with a reasonable bin-size (similar to the Jackknife bin-size). In this case, we have used a bin-size of one hundred measurements. Then one computes the expectation value of the observable on each bin of these normalised action measurements and checks where it plateaus. The data in bins at low Monte Carlo times with averages that deviate (because the system has not yet reached thermal equilibrium) from the average of thermalised measurements at large Monte Carlo times is then discarded in this thermalisation step.\\

\subsection{Auto-correlations}
\noindent
Now that we have addressed thermalisation, in order to quantify how strongly correlated any two measurements on the Markov chain are, we compute the \textit{auto-correlation time}. The autocorrelation time is a measure of how many Monte Carlo steps it takes to obtain statistically independent measurements. An autocorrelation time of $\sim 1$ indicates that the measurements are highly correlated. If correlated measurements are not discarded, standard error estimating methods which assume that the measurements are statistically independent will underestimate the true variance of the measured observable. The goal is to get the auto-correlation time to $\sim 0$, by skipping some elements of the chain (the \textit{auto-correlation length}) between measurements.\\ 

\noindent
Recall that the \textit{unbiased estimator} for the mean and the corresponding variance for the observable $O$ from $N$ measurements are given by,
\begin{eqnarray}
    \langle O \rangle = \frac{1}{N} \sum \limits_{i=1}^N O_i, \quad \text{and} \quad \sigma^2_{O} = \frac{1}{N-1} \sum \limits_{i=1}^N [ O_i - \langle O \rangle ]^2,
    \label{eqn:naive_mean}
\end{eqnarray}
where the above error estimate truly holds provided the thermalised configurations used in computing the error are statistically uncorrelated. However, it also provides a decent error estimate if the configurations are only `slightly' correlated. Now, let us consider a situation where we are dealing with correlated measurements, as is the case in our MCMC implementation.\\

\noindent
Let us start by defining the auto-correlation function $\Gamma_t$ and the normalised auto-correlation function $\rho_t$ \cite{Joseph_2020},
\begin{eqnarray}
    \Gamma_t = \frac{1}{(N-t-1)} \sum \limits_{i=1}^{N-t} [ O_i - \langle O_i \rangle ] [ O_{i+t} - \langle O_{i+t} \rangle ] , \quad \text{and} \quad \rho_t =\frac{\Gamma_t}{\Gamma_0},
\end{eqnarray}
where $t$ is the so called lag-time (i.e., number of configurations between measurements along the Markov chain) allowing us to compare the running average to the first and last ($N-t$) measurements and $\Gamma_0 = \sigma^2_{O}$. Note that in the absence of auto-correlations, the auto-correlation function vanishes as expected.\\ 

\noindent
The normalised auto-correlation function exhibits exponential asymptotic behavior as the lag-time becomes large and can be approximated by fitting a linear combination of exponentials \cite{Gattringer:2010zz},
\begin{eqnarray}
    \rho_t =\frac{\Gamma_t}{\Gamma_0} \sim \sum \limits_{j=1}^{n} b_j e^{-t/\tau_j},
\end{eqnarray}
where $n$ is the number of exponential terms, $b_j$ are coefficients determined from the fit and normalised to unity, and $\tau_j$ is also a fit parameter, where the asymptotically leading term, $\tau_n$ is called the \textit{exponential auto-correlation time}. One then uses the value of $\tau_n$ to discard correlated measurements.\\

\noindent
An alternative to this approach (which avoids dealing with the intricacies of fitting functions), is to instead compute the \textit{integrated auto-correlation time}, $\tau_{int}$ defined as,
\begin{eqnarray}
    \tau_{int} = \frac{1}{2} + \sum \limits_{t=1}^{M} \rho_t, \quad \text{with} \quad M \geq 4\tau_{int} + 1.
    \label{eqn:intcorr}%
\end{eqnarray}
See Ref.\ \cite{Westbroek:2017tym} for a detailed derivation of this result. The sum in Eq.\ (\ref{eqn:intcorr}) should be taken along the entire length of the Markov chain, however, this becomes computationally costly and $\rho_t$ becomes unreliable (e.g., negative values and some noise). Thus finding the right length to truncate can be nontrivial, and various techniques have been proposed. We have adopted the approach in Ref.\ \cite{Joseph_2020}, where one truncates at some chain length, $M$ satisfying the given condition.\\

\begin{figure}[!htb]
\begin{center}
\includegraphics[scale=.7]{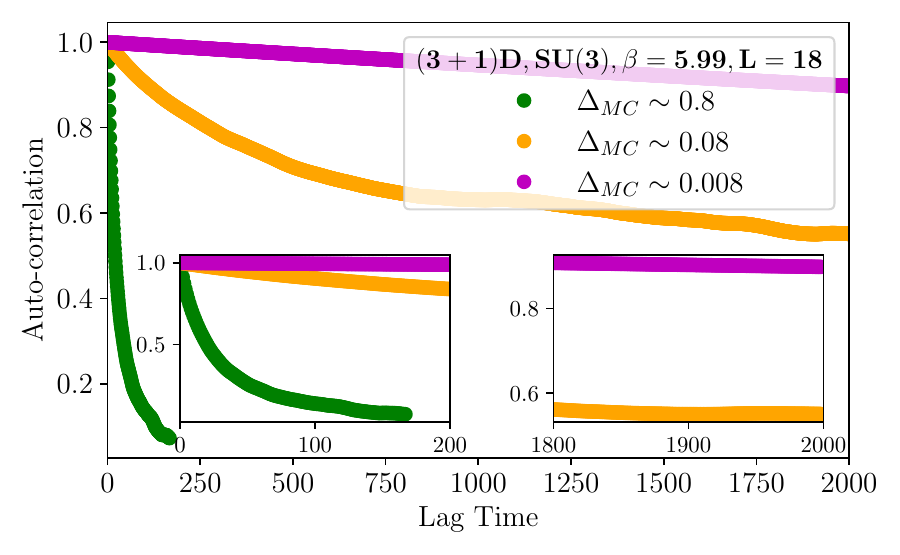}
\caption{Auto-correlations on the action measurements in $(3+1)$D SU($3$) for $\beta = 5.99$, $L=18$ using different step-sizes sampling a normal distribution.}
\label{fig:autocorrelations}
\end{center}
\end{figure}

\noindent
The corresponding error for the integrated auto-correlation time is given by \cite{Joseph_2020},
\begin{eqnarray}
    \delta \tau_{int} = \tau_{int} \sqrt{\frac{4M +2}{N}},
\end{eqnarray}
where $M$ is our auto-correlation length. The statistical error in the observable, $O$ accounting for the auto-correlations in the data is then given by,
\begin{eqnarray}
    \delta O =  \frac{\sigma_O}{\sqrt{N}} \sqrt{2\tau_{int}}.
\end{eqnarray}
This is one way of accounting for the auto-correlations in the error analysis in order to ensure that the statistical errors in the measurements are not underestimated. In this thesis, we explicitly compute the autocorrelation time of our measurements and use this analysis to discard the correlated measurements, $M$. Then we apply the Jackknife error technique discussed in the next subsection.\\

\noindent
Lastly, we revisit the discussion in Sec.\ (\ref{subsec:stepsize}), on the step-size ($\Delta_{MC}$) dependence of the Metropolis algorithm and the effect that it has on the auto-correlations in the measurements. The reduction in the step-size results in a higher acceptance rate as we sample around the same region in the phase space and the proposed state is very close to the current state. This, in turn leads to very strong auto-correlations in the measurements. We show, in Fig.\ (\ref{fig:autocorrelations}), how rapidly the auto-correlation time of measured action grows with a reduction in step-size. In our subsequent analysis, we have chosen the autocorrelation length according to Eq.\ (\ref{eqn:intcorr}).\\

\subsection{Jackknife Errors}
\label{subsec:jackknife}

\noindent
In the last subsection, we discussed one approach of tackling error analysis associated with lattice observables obtained through correlated Markov chain configurations. This involved the computation of auto-correlation times, a potentially costly process for lattice observables. There are various alternative methods used to obtain an estimate of the correlation in Monte Carlo measurements. These include the bootstrap approach, some data blocking methods and resampling methods such as the Jackknife procedure, which we discuss next.\\ 

\noindent
Starting with our Markov chain (sample of configurations) of the lattice observable of interest, $O_i$ (in our case, this is the Wilson action), the corresponding unbiased estimator of the mean and variance are given by Eq.\ (\ref{eqn:naive_mean}). This description of the variance (i.e., estimator) is true for uncorrelated and slightly correlated measurements. Thus it does not result in a complete underestimation of the errors, which would occur if one simply took the variance of the distribution.\\

\noindent
The Jackknife method involves a resampling procedure where one builds an ensemble of estimators from the original sample of configurations, omitting different configurations on the sample each time. One picks a bin width, $B$ and divides the original sample of configurations of size, $N$ into $N_B = N/B$ blocks. The measurements in each block need to be uncorrelated, thus one condition for the choice of the bin width is that, $B > M$, where $M$ is the auto-correlation length (see the previous subsection). Each block consists of the number of measurements that are omitted from the original sample each time the jackknife estimator is computed. The estimator of the mean, $\Tilde{O}$ is given by,
\begin{eqnarray}
    \Tilde{O}_k = \frac{1}{N} \sum \limits_{i=1}^{N-B} \left( O_i - \sum \limits_{k=1}^{B} O_{(k-1)B+i} \right),
\end{eqnarray}
where $k\in [1, N_B]$, and each estimator consists of ($N-B$) measurements. The corresponding variance of the estimators is,
\begin{eqnarray}
    \sigma^2_{\Tilde{O}} &=& \frac{N_B-1}{N_B} \sum \limits_{k=1}^{N_B} \left( \Tilde{O}_k - \langle O \rangle \right)^2.
\end{eqnarray}
More detailed discussions on the Jackknife technique can be found in Ref.\ \cite{shao2012jackknife,wolff2004monte,Westbroek:2017tym}.\\

\noindent
In summary, we use the Wilson action with the Metropolis algorithm to sample the field configurations. We take a minimum of $5\times 10^3$ thermalisation steps and discard a minimum of $50$ configurations between measurements. The autocorrelation time is computed for each dataset and the autocorrelation lengths are chosen carefully according to Eq.\ (\ref{eqn:intcorr}). We discard $n>M$ measurements such that we only use statistically independent measurements. Where possible, for example in the case of $(2+1)$D, more than $10^4$ configurations have been used. However, we do not perform any continuum extrapolations and a relatively small number of configurations is sufficient for our calculations and we expect good accuracy due to their statistical independence.\\ 


\chapter{The String Tension and Boundary Conditions}
\label{chapter:strings_and_booundaries}

\noindent
In this chapter, we introduce the string tension, which sets the overall physical scale in our lattice measurements. We perform Padé fits of string tensions as a function of the bare inverse coupling, $\beta$ and mean-field-tadpole improved inverse coupling, $\beta_I$ in (2+1)D and (3+1)D SU(2) and SU(3), respectively. We then discuss periodic boundary conditions and the associated four-dimensional torus geometry used in our lattice volume. Lastly, we look at the electric-type boundary conditions that we use to impose physical configurations of interest in our Casimir effect studies in non-abelian pure gauge theory. \\

\section{String Tension Fits}
\label{section:string_tension}
\noindent
We start our discussion by looking at the `renormalisation' of lattice gauge theory in order to obtain physical quantities. In pure gauge theory, this renormalisation is usually done through the introduction of a known dimensionful quantity i.e.\ by studying the force between two static quarks \cite{Necco:2001xg, Sommer:1993ce}, which then sets the overall scale in lattice measurements. The static QCD potential in (3+1)D can be approximated by \cite{Gattringer:2010zz},
\begin{eqnarray}
    V(r) = A - \frac{B}{r} + \sigma r,
    \label{eqn:hq_potential}
\end{eqnarray}
where the constant $A$ is an overall normalisation of the energy, the middle term is the Coulomb part which describes short-range interactions and the last term is the linearly rising term describing long-range interactions. The corresponding force between two static quarks is,
\begin{equation}
    F(r) = -\frac{dV}{dr} = -\frac{B}{r^2} - \sigma.
    \label{eqn:hq_force}
\end{equation}

\noindent
The motivation for the presence of Coulomb-type interactions in the potential comes from taking the small coupling limit, i.e., $g\to 0$, of the continuum gluonic action. Looking at the field dependence of gluonic action, see Eq.\ (\ref{eqn:field_strength_def}), where the commutator term is rescaled by a factor of $1/g$, in this limit, the field strength tensor reduces to the abelian form. The Coulomb-type interactions in the gluonic action are then deduced from our understanding of QED-type interactions \cite{Bohm:2001yx, Roepstorff:1994ga, Griffiths:2008zz, Zee:2003mt}.\\

\noindent
The linear term is a direct manifestation of quark confinement in colour-neutral hadrons and points out the linear rise in energy with increasing separation distance between the quark-antiquark pair. A lattice formulation derivation of this linearly rising potential is given in Ref.\ \cite{Gattringer:2010zz} by taking the strong coupling expansion of the Wilson loop. It is known from predictions of the string model of hadron confinement \cite{Nambu:1974zg} that colour flux tubes (also called strings) form between colour charges because the self-interaction of gluons results in the fields being squeezed into narrow tubes.\\

\noindent
Strings/flux-tubes thus act as mediators of the strong interactions between the quarks. When the separation distance between quarks is increased, the energy stored in the flux tubes (strings of gluonic fields) increases linearly and results in a constant tension along the string. Therefore, the \textit{string tension} describes the energy per unit length of a flux tube formed between a quark-antiquark pair and is defined in the large separation distance limit as
\begin{eqnarray}
    -\sigma = \lim_{r\to \infty} F(r).
\end{eqnarray}

\noindent
Confinement is addressed in the non-perturbative regime of QCD, thus lattice calculations (which are inherently non-perturbative) have been used to perform studies allowing for the extraction of the string tension at different energy scales \cite{Sommer:1993ce}. On the other hand, these lattice studies are supported by indirect evidence from hadron sprectra measured experimentally in collider studies since the string tension affects the spectrum and properties of hadrons. An example of this is found in hadron spectroscopy studies, where the mass of hadrons shows a linear dependence on the total spin \cite{Perkins:1982xb}. \\

\noindent
We refer the reader to Ref.\ \cite{Sommer:2014mea} and references therein for various string tension measurements available in literature. We focus on the results of \cite{Teper:1998te} in the (2+1)D theory and \cite{Athenodorou:2021qvs} in its (3+1)D counterpart. The primary reason for these choices is that we seek to extend their investigations on functional forms that can be used to interpolate and possibly extrapolate (to the continuum) these string tension measurements as functions of the inverse coupling. Such functions are of interest because lattice measurements of the string tension are only available at discrete values of $\beta$ over some finite range.\\

\begin{figure}[!htb]
    \centering
    \subfigure[Bare Coupling]{{\includegraphics[scale=0.48]{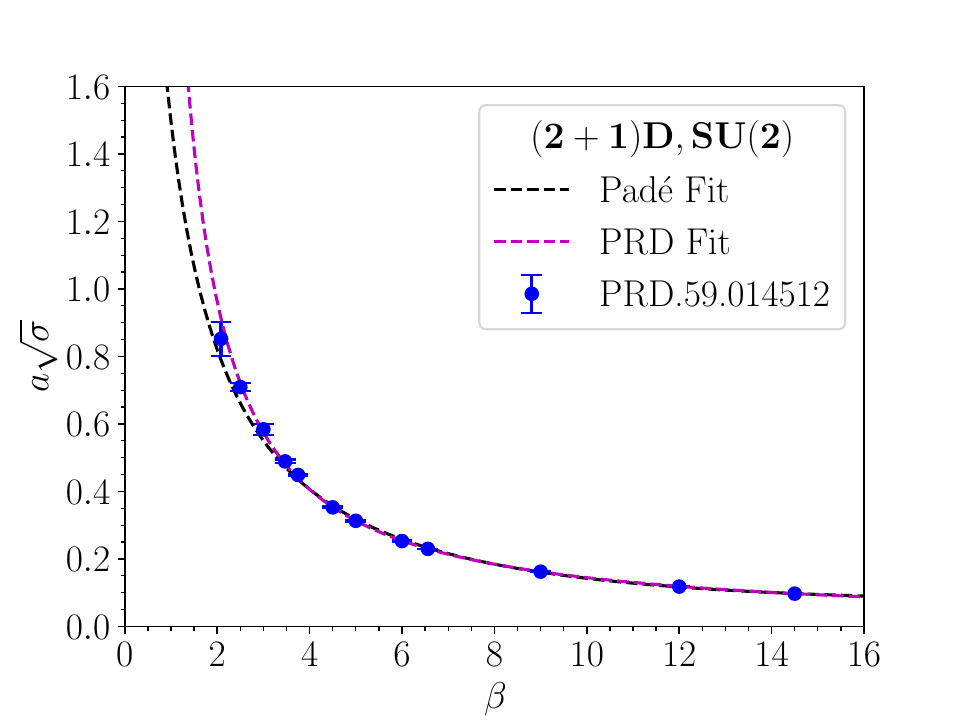} \label{fig:pade_2dsu2beta_original} }}%
    \hspace{-0.8cm}
    \subfigure[Tadpole Improved Inverse Coupling]{{\includegraphics[scale=0.48]{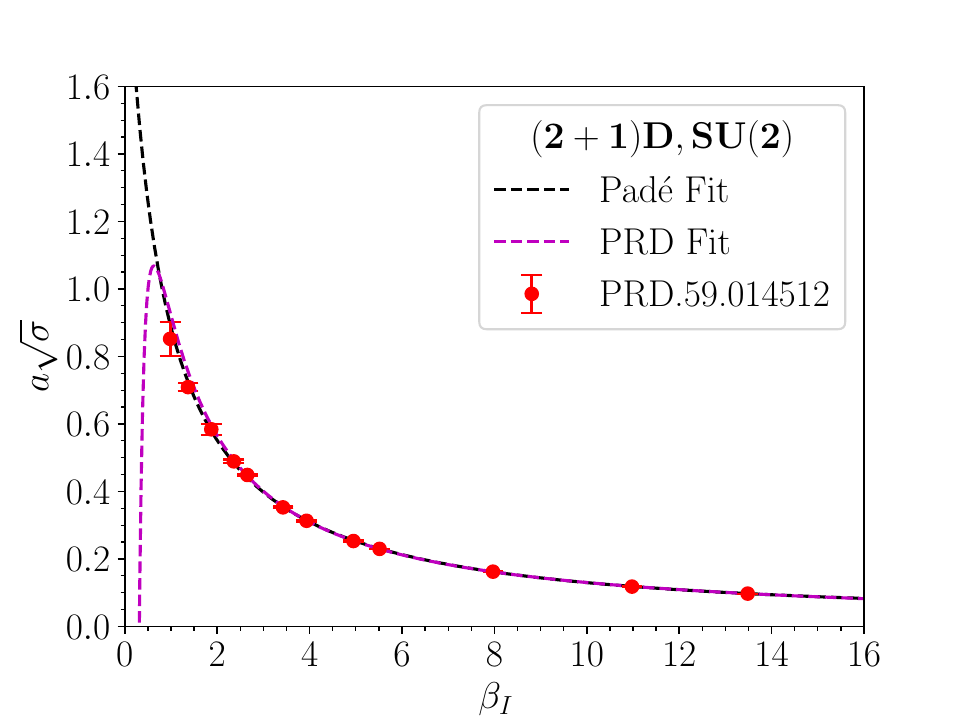} \label{fig:pade_2dsu2betaI_original} }}%
    \caption{String tension Padé fits in $(2+1)$D SU(2) using the bare and mean-field-tadpole improved inverse coupling.}%
    \label{fig:pade_2dsu2}
\end{figure}

\begin{figure}[!htb]
    \centering
    \subfigure[Bare Coupling]{{\includegraphics[scale=0.48]{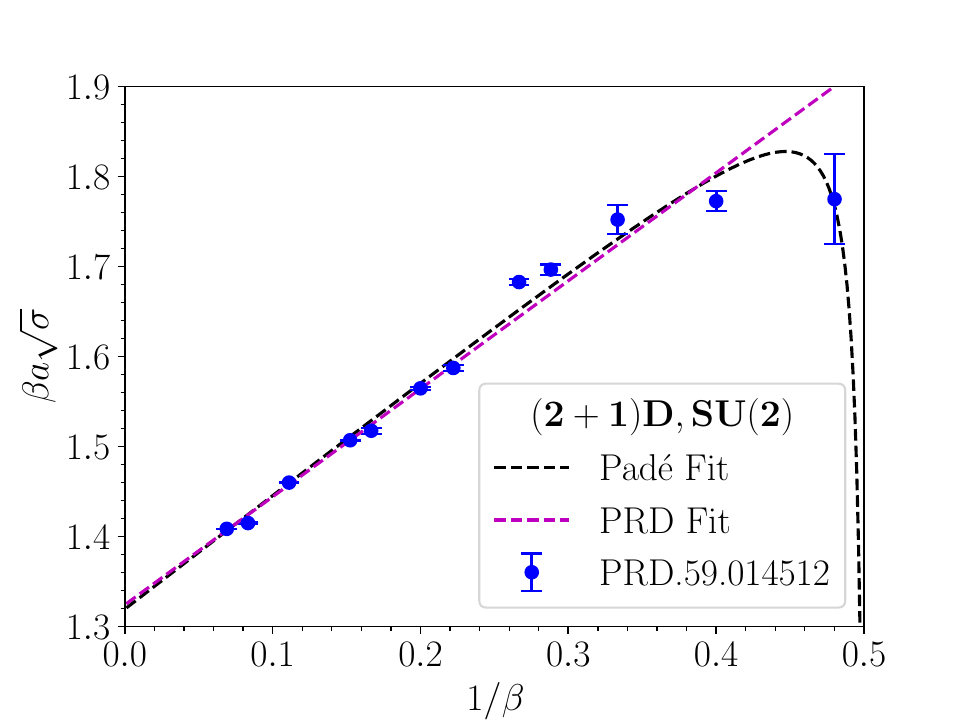} \label{fig:pade_2dsu2beta_prd} }}%
    \hspace{-0.8cm}
    \subfigure[Tadpole Improved Inverse Coupling]{{\includegraphics[scale=0.48]{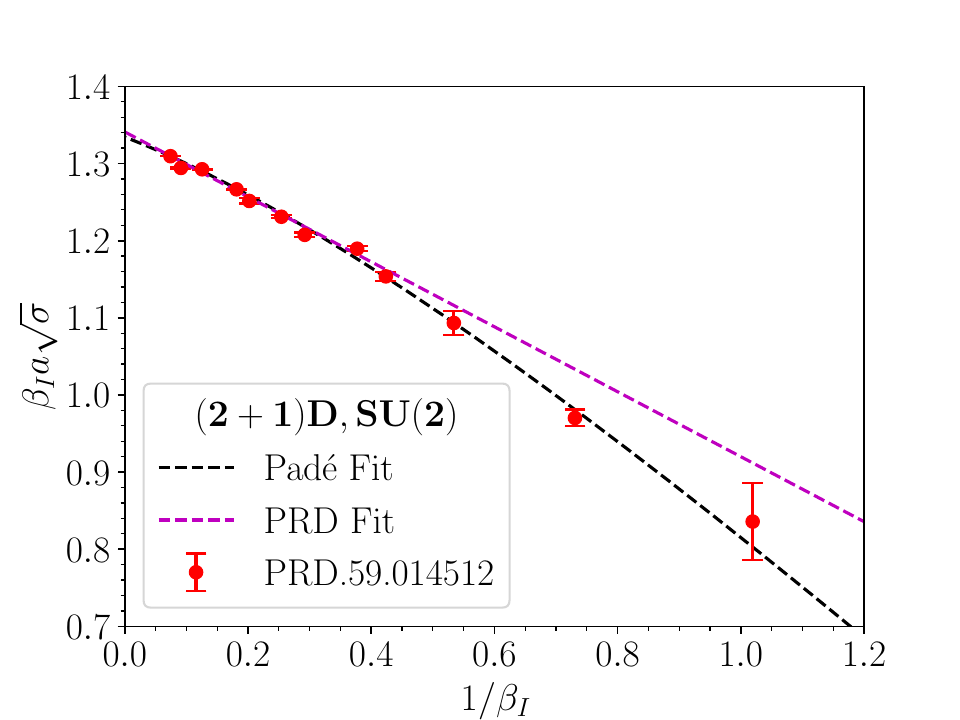} \label{fig:pade_2dsu2betaI_prd} }}%
    \caption{Linearised string tension Padé fits in $(2+1)$D SU(2) using the bare and mean-field-tadpole improved inverse coupling.}%
\end{figure}

\begin{figure}[!htb]
    \centering
    \subfigure[Bare Coupling]{{\includegraphics[scale=0.48]{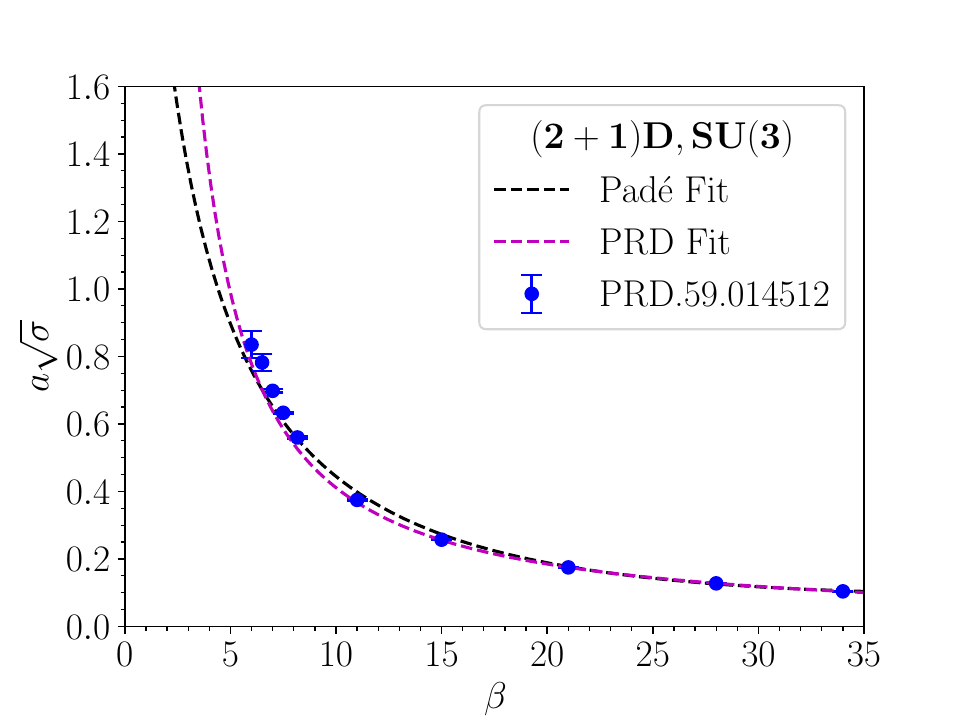} \label{fig:pade_2dsu3beta_original} }}%
    \hspace{-0.8cm}
    \subfigure[Tadpole Improved Inverse Coupling]{{\includegraphics[scale=0.48]{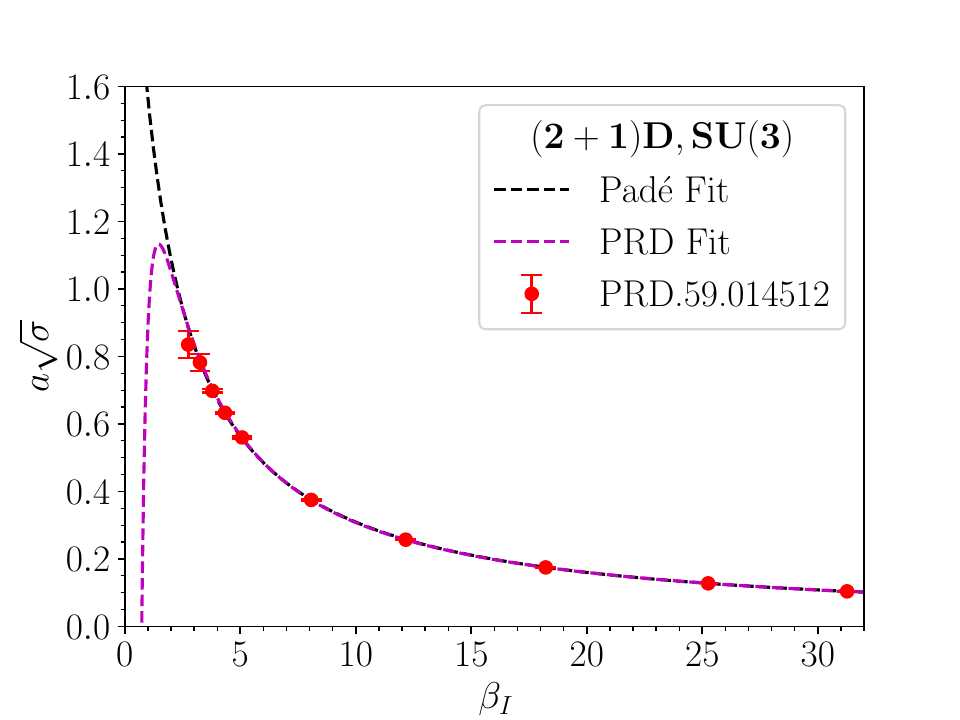} \label{fig:pade_2dsu3betaI_original} }}%
    \caption{String tension Padé fits in $(2+1)$D SU(3) using the bare and mean-field-tadpole improved inverse coupling.}%
\end{figure}

\begin{figure}[!htb]
    \centering
    \subfigure[Bare Coupling]{{\includegraphics[scale=0.47]{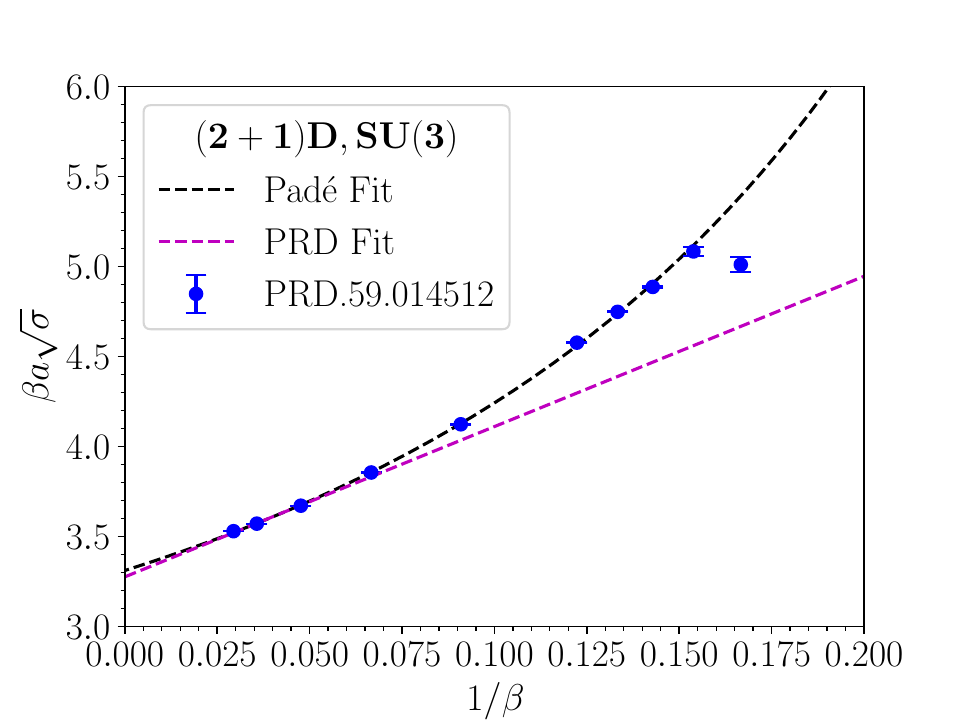} \label{fig:pade_2dsu3beta_prd} }}%
    \hspace{-0.7cm}
    \subfigure[Tadpole Improved Inverse Coupling]{{\includegraphics[scale=0.47]{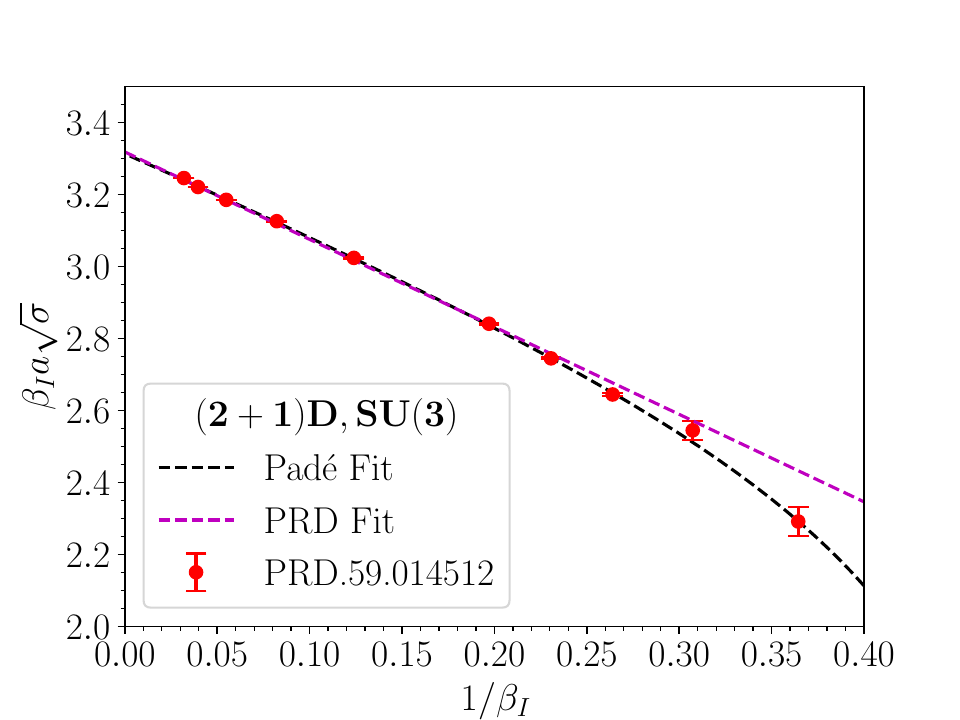} \label{fig:pade_2dsu3betaI_prd} }}%
    \caption{Linearised string tension Padé fits in $(2+1)$D SU(3) using the bare and mean-field-tadpole improved inverse coupling.}%
    \label{fig:pade_2dsu3}
\end{figure}

\noindent
In the (2+1)D theory, the string tension is extracted in Ref.\ \cite{Teper:1998te} from the calculated mass of the lightest state of a static quark-antiquark pair with a periodic flux tube of length $aL$ winding around a spatial torus according to \cite{Teper:1998te},
\begin{equation}
    am_P(L) = a^2\sigma L - \frac{\pi}{6L} + ...,
\end{equation}
where $m_P$ is the flux tube mass, $a$ the lattice spacing and $L$ the lattice spatial extent. The second term is the universal string correction \cite{Badalian:2002rc, Shifman:2007rc, Greensite:2003bk}  arising from quantum fluctuations in the gluon field when the separation distance between the quarks is comparable to the size of the flux tube. We show these string tension values as data-points in Fig.\ (\ref{fig:pade_2dsu2} - \ref{fig:pade_2dsu3}), where the dashed lines correspond to fitting functions that we will now describe.  \\

\noindent
Given that the lattice data for the string tension is only available at discrete points along the $\beta$-axis, a fitting function is required to obtain the string tension for a wider range of $\beta$ values as well as to extrapolate to the continuum. The corresponding fitting function that is used to perform the continuum extrapolation in Ref.\ \cite{Teper:1998te} is of the form
\begin{equation}
    \beta a\sqrt{\sigma} = c_0 + \frac{c_1}{\beta}, \quad c_0 = 2N_c\frac{\sqrt{\sigma}}{g^2},
\end{equation}
which is an expected functional form in the limit $\beta \to \infty$ based on the expected mass scale ($g^2$) dependence of $\sqrt{\sigma}$ \cite{Teper:1998te}. The constants $c_0$ and $c_1$ are the fit parameters and the $\mathcal{O}(1/\beta)$ term is a correction term. The resulting fits up to $\mathcal{O}(1/\beta)$ are \cite{Teper:1998te}, 
\begin{eqnarray}
    \beta a \sqrt{\sigma} &=& 1.324(12) + \frac{1.20(11)}{\beta}, \quad \text{SU}(2), \\
     \beta a \sqrt{\sigma} &=& 3.275(24) + \frac{8.35(61)}{\beta}, \quad \text{SU}(3),
\end{eqnarray}
and perform well for $\beta \geq 4.5$ in SU(2) and $\beta \geq 15$ in SU(3). The aforementioned paper also provides optimal parameters for $\mathcal{O}(1/\beta^2)$ fits which are applicable for a wider $\beta$ range and shows that these fits suffer from intrinsic systematic errors rendering them unreliable for the continuum extrapolation. Such errors arise due to the lattice bare coupling.\\

\noindent
The lattice bare coupling, $\beta$ is known to perform poorly in defining the running coupling especially at small $\beta$ \cite{Lepage:1996jw, Lepage:1992xa} due to lattice artifacts from the discretisation, finite volume effects, non-perturbative effects as well as uncertainties arising from renormalisation schemes that can lead to different running couplings. As a result, it requires careful treating in order to obtain a well defined running coupling. Such treatment has come in the form of improved lattice actions \cite{Wilson:1983xri}, finite size scaling \cite{Engels:1989fz} and the use of non-perturbative renormalisation schemes \cite{DeGrand:1996ri}.\\

\noindent
One such improvement of the bare coupling is the mean field tadpole improved coupling \cite{Parisi:1980pe, Lepage:1996jw},
\begin{equation}
    \beta_I = \beta \left\langle \frac{1}{N_c} \text{Tr} [U_P] \right\rangle,
\end{equation}
drawing its physical motivations from perturbation theory, in that the tadpole improvement scheme improves the convergence of perturbative expansions to larger distances up to $\sim 0.5$ fm \cite{Lepage:1992xa, Alford:1995hw}.\\

\noindent
It is shown in Ref.\ \cite{Teper:1998te} that using the improved coupling reduces the systematic errors in the fits and gives the following fit parameters,
\begin{eqnarray}
    \beta_I a \sqrt{\sigma} &=& 1.341(7) - \frac{0.421(50)}{\beta_I}, \quad \text{SU}(2), \\
     \beta_I a \sqrt{\sigma} &=& 3.318(12) - \frac{2.43(22)}{\beta_I}, \quad \text{SU}(3),
\end{eqnarray}
with good fits for $\beta \geq 3.0$ in SU(2) and reasonable fits up to $\beta \geq 6.0$ in SU(3). These fits employing the improved coupling perform better even without the inclusion of higher order terms, which show strongly suppressed contributions at $\mathcal{O}(1/\beta^2)$. Please see Ref.\ \cite{Teper:1998te} for a detailed discussion on the $\beta_I$ improvement. \\

\noindent
As can be seen from Fig.\ (\ref{fig:pade_2dsu2} - \ref{fig:pade_2dsu3}), we provide additional fits utilising a Padé approximant of order $\mathcal{O}(2,1)$ in comparison to the fits already discussed. The corresponding fitting function for the string tension is,
\begin{equation}
    \label{eqn:pade_fit}
    a\sqrt{\sigma} (\beta) = \frac{a_0 + a_1\beta + a_2\beta^2}{1 + b_1\beta},
\end{equation}
where we have used the argument, $\beta$, to refer to either the bare inverse coupling, $\beta$, or the mean-field-tadpole improved coupling, $\beta_I$, and $(a_0, a_1, a_2, b_1)$ are the fit parameters. While we do not provide a physical motivation for this functional form, this choice is rather intuitive and its application will become apparent in (3+1)D where the bare coupling decreases logarithmically with the lattice scale $a$ and not linearly as we have seen in (2+1)D. \\

\noindent
More general motivations for the use of Padé approximants include their versatility to represent different functional forms, complemented by a wide range of applications in physics problems. See, for example, Ref.\ \cite{Leung:2000kg,Samuel:1995jc,Samuel:1994bw} for applications in QCD. They can perform better than truncated Taylor series which usually converge locally around a point, and they also exhibit faster conversion rates. In addition, the natural pole structure means that these fits can be easily extended to functions with singularities. \\

\noindent
One difficulty with the use of rational functions for fitting is that unphysical poles may arise. Such singularities can be mitigated by performing a Borel-Padé analysis \cite{Spada:2018iyp,baker1975essentials}. Another issue is the number of fit parameters which grows with the order, see for example, that we are moving from two fit parameters in Ref.\ \cite{Teper:1998te} to four parameters in our Padé implementation in Eq.\ (\ref{eqn:pade_fit}). An increase in the number of fitting parameters can result in increased fitting errors, resulting in extrapolating difficulties. A common trick is to pick the order of the denominator such that it is at least one less than the order of the numerator. \\

\noindent
We provide the (2+1)D string tension Padé fit parameters in Table (\ref{tab:pade_2dsun_beta}) for the bare coupling and Table (\ref{tab:pade_2dsun_betaI}) for the tadpole improved coupling. In both cases, our results are consistent with Ref.\ \cite{Teper:1998te} in the continuum extrapolation. The added advantage of these Padé fits is that they produce better interpolation results at low coupling for both $\beta$ and $\beta_I$. Due to the number of fit parameters and associated errors, we do not attempt to extrapolate to or make any quantitative statements on the strong coupling regime.\\

\begin{table}[!htb]
    \centering
        {\rowcolors{2}{green!80!yellow!50}{green!70!yellow!40}
        \begin{tabular}{ |P{0.8cm}|P{1.8cm}|P{1.8cm}|P{1.8cm}|P{1.8cm}|P{1.8cm}|P{1.8cm}|  }
        \hline
        \hline
        $\mathbf{N_c}$ & \bf{Fit type} & $\mathbf{a_0}$ & $\mathbf{a_1}$ & $\mathbf{a_2}$ & $\mathbf{b_1}$ & \bf{Figure} \\
        \hline
        2 & $\beta$ & 6.0 & -0.298196 & -0.011415 & 2.820948 & \ref{fig:pade_2dsu2beta_original} \\
        2 & $1/\beta$ & 1.319173 & -1.354793 & -2.545725 & -1.986305 & \ref{fig:pade_2dsu2beta_prd}\\
        \hline
        3 & $\beta$ & 4.0 & -0.126544 & 0.002101 & 0.562759 & \ref{fig:pade_2dsu3beta_original} \\
        3 & $1/\beta$ & 3.310344 & -1.671054 & 5.0 & -2.471226 & \ref{fig:pade_2dsu3beta_prd}\\
        \hline
        \end{tabular}}
    \caption{String tension Padé fitting parameters in (2+1)D SU($N_c$) using the bare coupling.}
    \label{tab:pade_2dsun_beta}
\end{table}

\begin{table}[!htb]
    \centering
        {\rowcolors{2}{green!80!yellow!50}{green!70!yellow!40}
        \begin{tabular}{ |P{0.8cm}|P{1.8cm}|P{1.8cm}|P{1.8cm}|P{1.8cm}|P{1.8cm}|P{1.8cm}|  }
        \hline
        \hline
        $\mathbf{N_c}$ & \bf{Fit type} & $\mathbf{a_0}$ & $\mathbf{a_1}$ & $\mathbf{a_2}$ & $\mathbf{b_1}$ & \bf{Figure} \\
        \hline
        2 & $\beta_I$ & 2.158519 & -0.034008 & 0.001277 & 1.401243 & \ref{fig:pade_2dsu2betaI_original} \\
        2 & $1/\beta_I$ & 1.334442 & 0.740851 & -0.617788 & 0.787799 & \ref{fig:pade_2dsu2betaI_prd}\\
        \hline
        3 & $\beta_I$ & 2.763801 & -0.014082 & 0.000245 & 0.757265 & \ref{fig:pade_2dsu3betaI_original} \\
        3 & $1/\beta_I$ & 3.313017 & -8.409216 & 3.737772 & -1.852322 & \ref{fig:pade_2dsu3betaI_prd}\\
        \hline
        \end{tabular}}
    \caption{String tension Padé fitting parameters in (2+1)D SU($N_c$) using the mean-field-tadpole improved coupling.}
    \label{tab:pade_2dsun_betaI}
\end{table}

\noindent
We now move our attention to the confining string tension in the (3+1)D theory, with a focus on the results of Ref.\ \cite{Athenodorou:2021qvs} where the string tension is estimated from the Nambu-Goto formula \cite{Nambu:1974zg, Arvis:1983fp,Athenodorou:2010cs,Aharony:2013ipa},
\begin{equation}
    E_k(l) = \sigma_k l \left( 1 - \frac{2\pi}{3\sigma_k l^2} \right)^{1/2},
\end{equation}
which describes the ground state energy of a flux tube winding around a spatial torus. The flux tube length is given by $l$ and $k$ is the units of the fundamental flux, where $k=1$ corresponds to the flux tube with the fundamental flux. The string tension is, by definition, the energy per unit length of a long flux tube formed between the quarks; thus the asymptotic string tension is extracted in the large $l$ limit,
\begin{equation}
    \sigma_k = \lim_{l \to \infty} \frac{E_k(l)}{l}.
    \label{eqn:nambu_goto}
\end{equation}

\begin{figure}[!htb]
    \centering
    \subfigure[Bare Coupling]{{\includegraphics[scale=0.48]{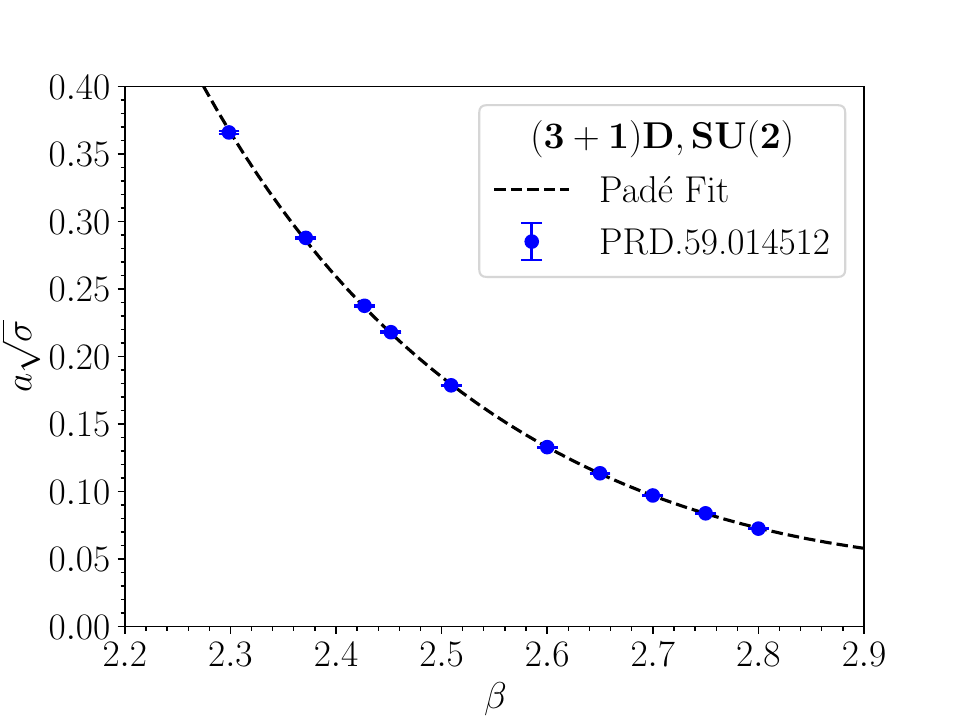} \label{fig:pade_3dsu2beta_original} }}%
    \hspace{-0.8cm}
    \subfigure[Tadpole Improved Inverse Coupling]{{\includegraphics[scale=0.48]{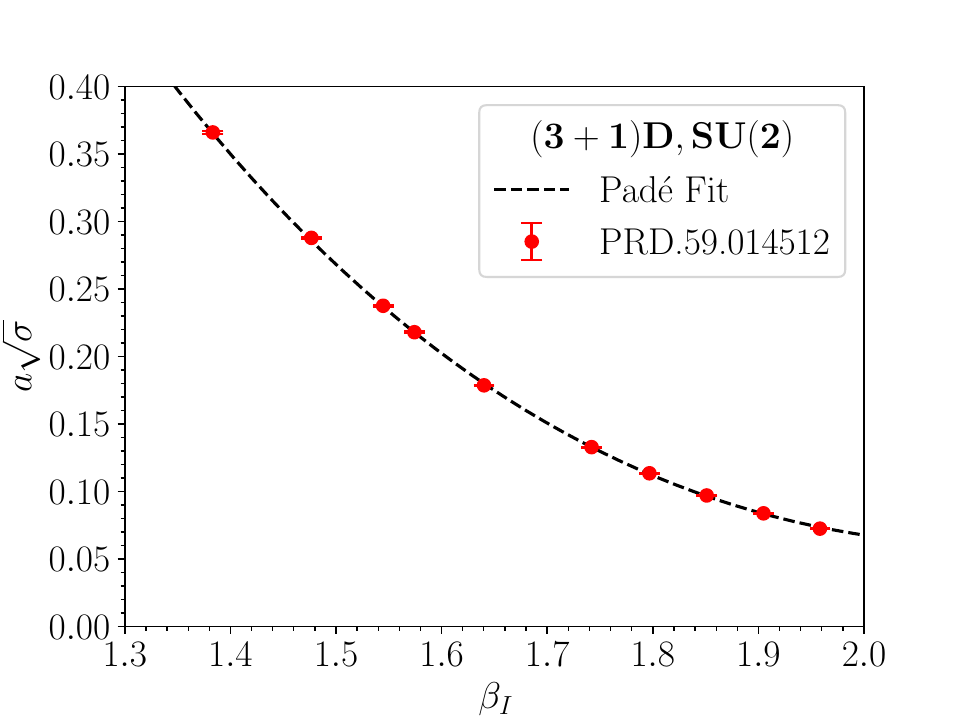} \label{fig:pade_3dsu2betaI_original} }}%
    \caption{String tension Padé fits in $(3+1)$D SU(2) using the bare and mean-field-tadpole improved inverse coupling.}%
    \label{fig:pade_3dsu2}%
\end{figure}

\begin{figure}[!htb]
    \centering
    \subfigure[Bare Coupling]{{\includegraphics[scale=0.48]{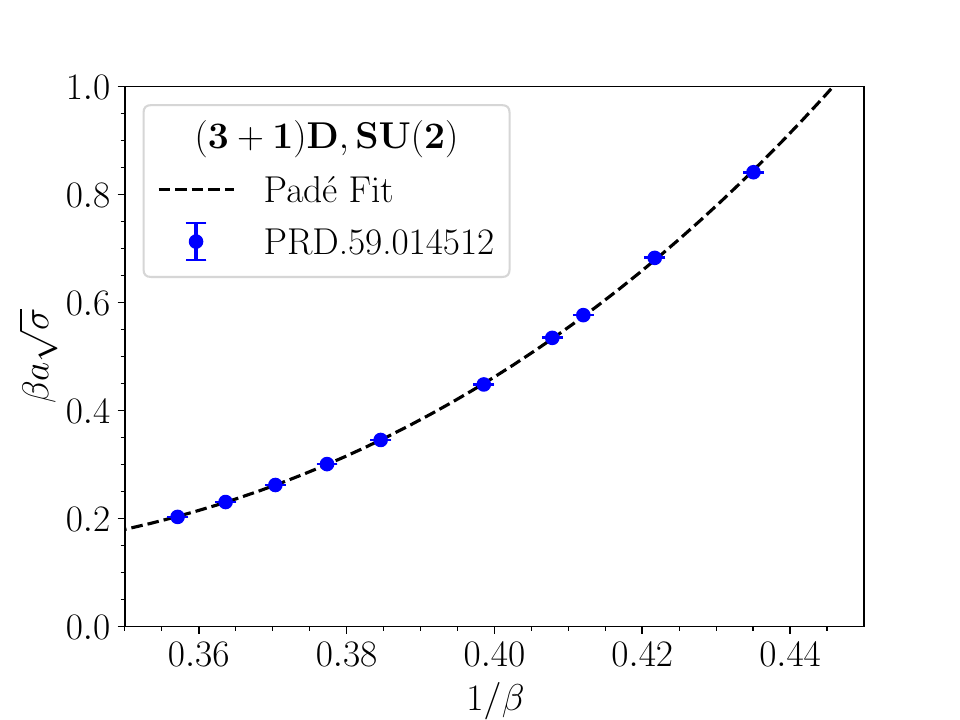} \label{fig:pade_3dsu2beta_prd} }}%
    \hspace{-0.8cm}
    \subfigure[Tadpole Improved Inverse Coupling]{{\includegraphics[scale=0.48]{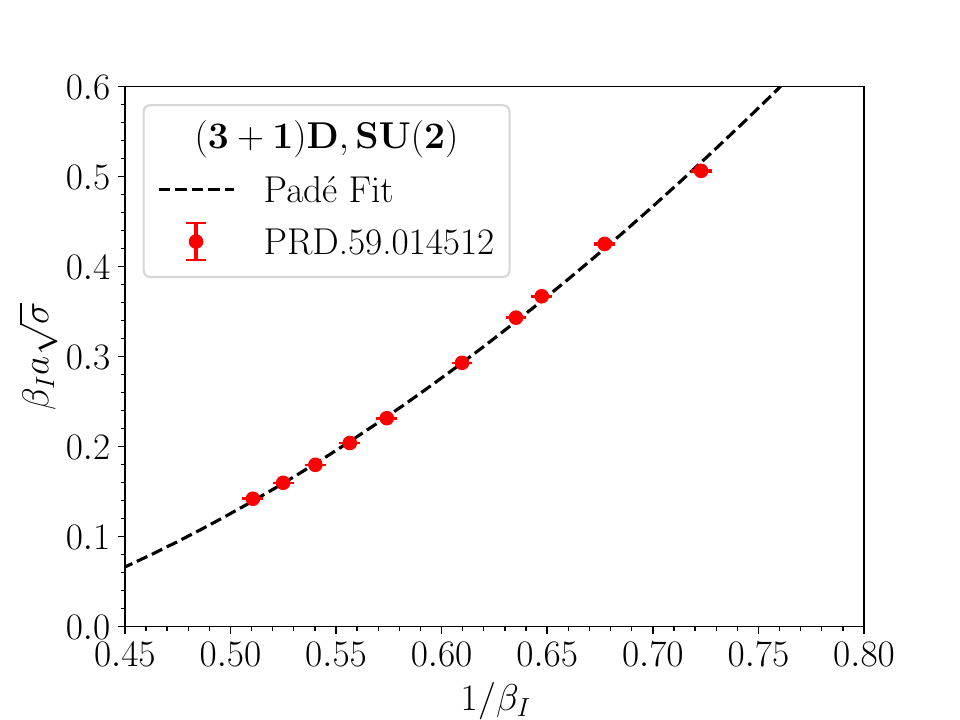} \label{fig:pade_3dsu2betaI_prd}  }}%
    \caption{Linearised string tension Padé fits in $(3+1)$D SU(2) using the bare and mean-field-tadpole improved inverse coupling.}%
\end{figure}

\begin{figure}[!htb]
    \centering
    \subfigure[Bare Coupling]{{\includegraphics[scale=0.48]{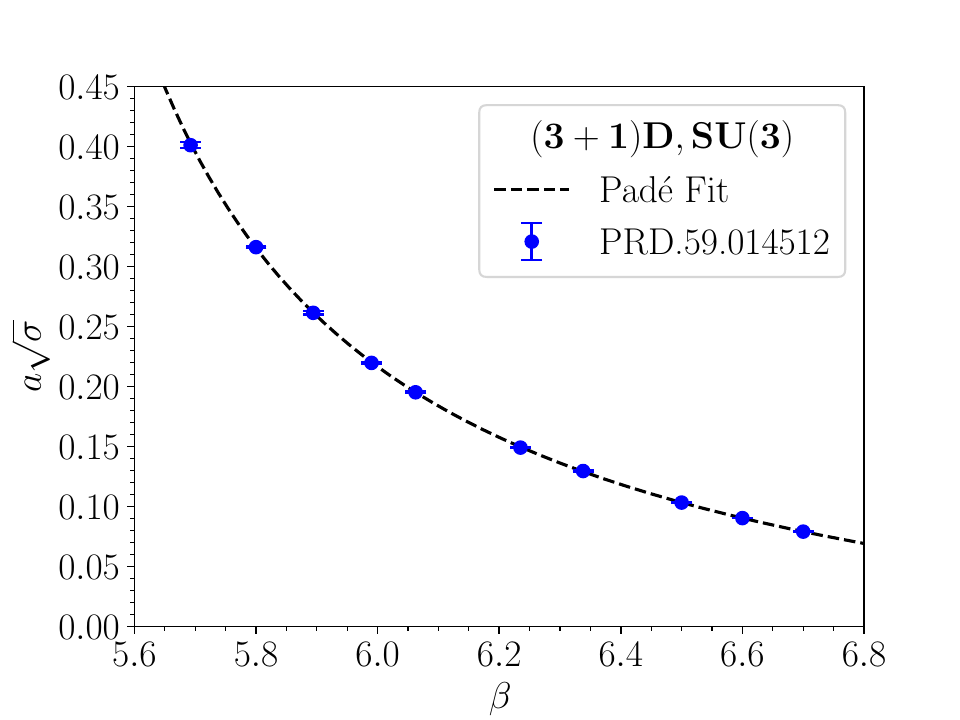} \label{fig:pade_3dsu3beta_original} }}%
    \hspace{-0.8cm}
    \subfigure[Tadpole Improved Inverse Coupling]{{\includegraphics[scale=0.48]{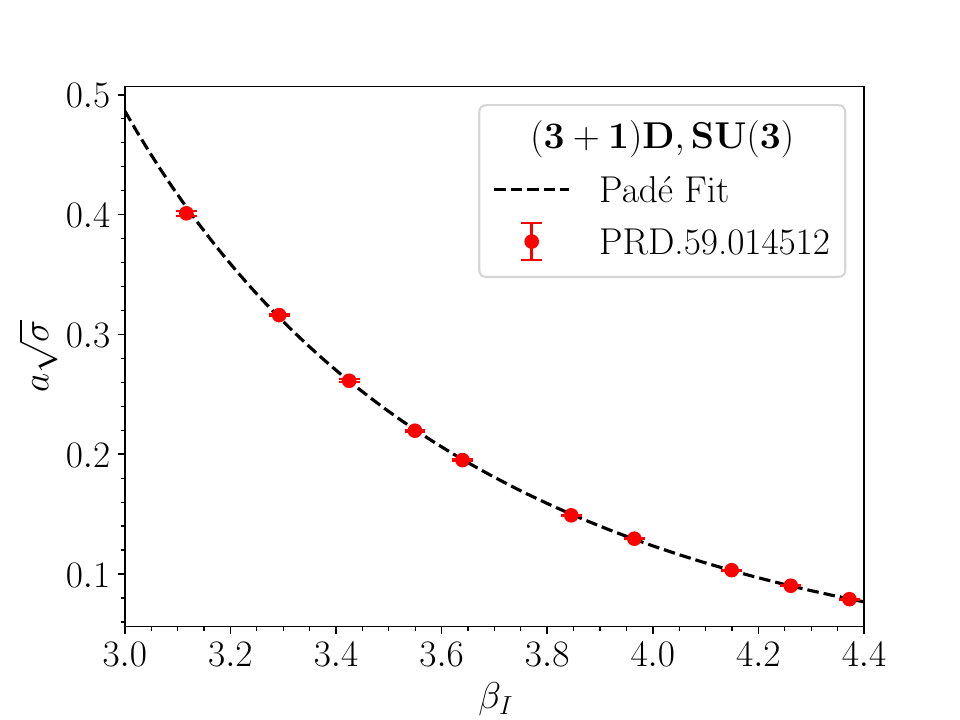} \label{fig:pade_3dsu3betaI_original} }}%
    \caption{String tension Padé fits in $(3+1)$D SU(3) using the bare and mean-field-tadpole improved inverse coupling.}%
\end{figure}

\begin{figure}[!htb]
    \centering
    \subfigure[Bare Coupling]{{\includegraphics[scale=0.47]{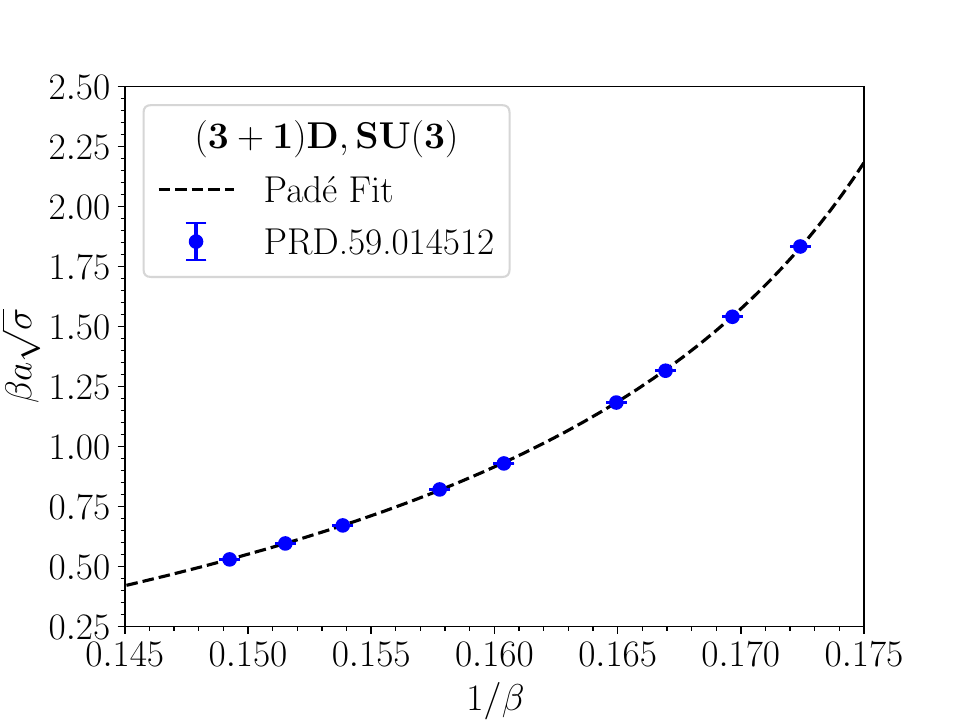} \label{fig:pade_3dsu3beta_prd} }}%
    \hspace{-0.7cm}
    \subfigure[Tadpole Improved Inverse Coupling]{{\includegraphics[scale=0.47]{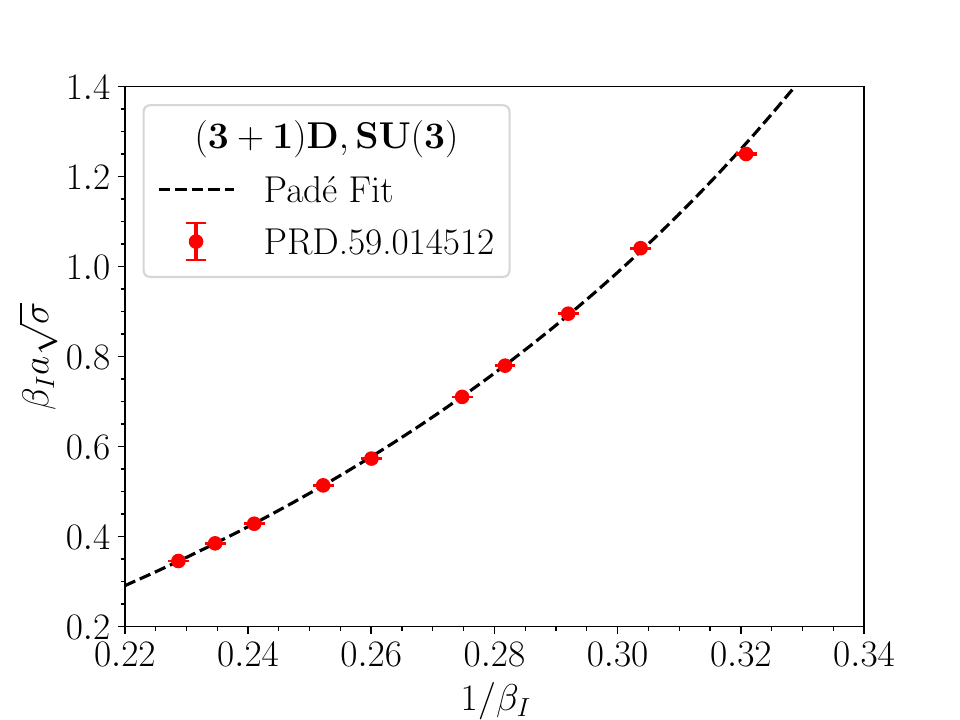} \label{fig:pade_3dsu3betaI_prd} }}%
    \caption{Linearised string tension Padé fits in $(3+1)$D SU(3) using the bare and mean-field-tadpole improved inverse coupling.}%
    \label{fig:pade_3dsu3}%
\end{figure}

\noindent
The relevant scale for the flux tube energy given in Eq.\ (\ref{eqn:nambu_goto}) is the flux tube length, $l$, and the expression holds in the infinite transverse volume limit. However, in the lattice formulation, the calculation is performed on a finite $l\times l$ transverse volume torus, which introduces finite volume errors. It is shown in Ref.\ \cite{Athenodorou:2021qvs} that such errors in the fundamental flux energy are insignificant for $l\sqrt{\sigma} \gtrsim 3$ . We employ the string tension data from lattice sizes  $l\sqrt{\sigma} \sim 4$ for gauge groups SU(2) and SU(3) which we show as data-points in Fig.\ (\ref{fig:pade_3dsu2} - \ref{fig:pade_3dsu3}).\\

\noindent
In the interest of the reader, we provide some references \cite{Teper:1998te, Athenodorou:2021qvs, Lucini:2004eq,Lucini:2001ej,Allton:2008ty} that contain discussions on string tension calculations for SU($N_c$) gauge theories in the large-$N_c$ limit. The SU($N_c \to \infty$) theories are of interest to the community because they offer theoretical simplifications which provide insight into QCD dynamics \cite{Lucini:2001ej}. In addition, there are lattice calculations that  have shown that SU(3) and SU($N_c \to \infty$) are \textit{similar theories} for both pure gauge \cite{Lucini:2001ej} and with fermionic degrees of freedom present (quenched theories) \cite{Bali:2013kia}. One may note that perhaps the simpler SU($N_c \to \infty$) theory can teach us some things about the more nontrivial, yet more physically relevant SU(3). \\

\noindent
It is of interest to have expressions for $a\sqrt{\sigma}$ for a continuous range of $\beta$ values for reasons already discussed in the (2+1)D case. Such interpolating functions are provided in Ref.\ \cite{Athenodorou:2021qvs}, drawing physical motivation from weak-coupling perturbation theory and only applicable for the Wilson plaquette action. We have attempted to use these interpolating functions with the provided fit parameters, but could not find optimal results. We have also performed Padé fits for these string tensions in (3+1)D and these are shown in Fig.\ (\ref{fig:pade_3dsu2} - \ref{fig:pade_3dsu3}). The corresponding fit parameters are given in Table (\ref{tab:pade_3dsun_beta}) for the bare coupling and Table (\ref{tab:pade_3dsun_betaI}) for the tadpole improved coupling.\\

\begin{table}[!htb]
    \centering
        {\rowcolors{2}{green!80!yellow!50}{green!70!yellow!40}
        \begin{tabular}{ |P{0.8cm}|P{1.8cm}|P{1.8cm}|P{1.8cm}|P{1.8cm}|P{1.8cm}|P{1.8cm}|  }
        \hline
        \hline
        $\mathbf{N_c}$ & \bf{Fit type} & $\mathbf{a_0}$ & $\mathbf{a_1}$ & $\mathbf{a_2}$ & $\mathbf{b_1}$ & \bf{Figure} \\
        \hline
        2 & $\beta$ & -1.903865 & 1.231340 & -0.203532 & -0.611210 & \ref{fig:pade_3dsu2beta_original} \\
        2 & $1/\beta$ & 2.724228 & -16.59037 & 26.06294 & -1.102794 & \ref{fig:pade_3dsu2beta_prd}\\
        \hline
        3 & $\beta$ & -0.090927 & 0.005699 & 0.000675 & -0.191639 & \ref{fig:pade_3dsu3beta_original} \\
        3 & $1/\beta$ & -0.966538 & 11.13525 & -25.99295 & -5.226978 & \ref{fig:pade_3dsu3beta_prd}\\
        \hline
        \end{tabular}}
    \caption{String tension Padé fitting parameters in (3+1)D SU($N_c$) using the bare coupling.}
    \label{tab:pade_3dsun_beta}
\end{table}

\begin{table}[!htb]
    \centering
        {\rowcolors{2}{green!80!yellow!50}{green!70!yellow!40}
        \begin{tabular}{ |P{0.8cm}|P{1.8cm}|P{1.8cm}|P{1.8cm}|P{1.8cm}|P{1.8cm}|P{1.8cm}|  }
        \hline
        \hline
        $\mathbf{N_c}$ & \bf{Fit type} & $\mathbf{a_0}$ & $\mathbf{a_1}$ & $\mathbf{a_2}$ & $\mathbf{b_1}$ & \bf{Figure} \\
        \hline
        2 & $\beta_I$ & 5.448942 & -5.0 & 1.183582 & 0.854934 & \ref{fig:pade_3dsu2betaI_original} \\
        2 & $1/\beta_I$ & 0.114700 & -1.435460 & 3.0 & 0.346613 & \ref{fig:pade_3dsu2betaI_prd}\\
        \hline
        3 & $\beta_I$ & -1.088703 & 0.380935 & -0.035350 & -0.514273 & \ref{fig:pade_3dsu3betaI_original} \\
        3 & $1/\beta_I$ & -0.335459 & 1.633556 & 3.0 & -1.898821 & \ref{fig:pade_3dsu3betaI_prd}\\
        \hline
        \end{tabular}}
    \caption{String tension Padé fitting parameters in (3+1)D SU($N_c$) using the mean-field-tadpole improved coupling.}
    \label{tab:pade_3dsun_betaI}
\end{table}

\noindent
We show that these Padé fits perform very well in the interpolation of the string tension and the logarithmic fall-off of the string tension in (3+1)D motivates this fitting functional form. Despite the strong physical motivation for the use of the improved coupling, $\beta_I$, for such interpolating functions, good fits are still obtained using the bare coupling. Such interpolations suffice for our coupling range of interest, and we do not attempt to extrapolate this string tension data either to the continuum limit nor the strong coupling regime.\\

\noindent
In the succeeding chapters, where we do not use lattice units, we will express all physical quantities in units of the string tension, which we use as our energy scale. The main reason for this choice is that the string tension is a fundamental and universal scaling variable in pure gauge theory that reflects the QCD dynamics of confinement \cite{Sommer:2014mea}. While it is common to express quantities in physical units of energy such as GeV for comparison to phenomenology, one has to be careful of the subtleties between pure gauge theories (gluonic degrees of freedom) and QCD (including dynamical fermions). We shall revisit this discussion in chapter (\ref{chapter:The Polyakov Loop and Deconfinement}).\\

\section{Boundary Conditions}
\label{section:boundary_conditions}
\noindent
We perform our Casimir studies on a `zero temperature' lattice ($N_s=N_{\tau}$) with periodic boundary conditions in all spatial and temporal directions. This implies that the spacetime manifold, $\mathcal{M}$ of our cubic lattice is a four-dimensional torus. The link variables defined in Eq.\ (\ref{eqn:link_variable}) are the dynamical degrees of freedom of the gauge fields and imposing periodic boundary conditions requires that
\begin{eqnarray}
    U_{\mu}(n) = U_{\mu}(n + L),
\end{eqnarray}
where $L$ is the lattice size.\\

\noindent
On the torus in $(3+1)$D SU(3), the field space is split into disconnected topological sectors each with quantised integer values of the topological charge \cite{Mages:2015scv, vanBaal:1982ag}, given in Eq.\ (\ref{eqn:su3_topological_charge}). Periodic boundary conditions are advantageous due to translational invariance and they minimise boundary effects as well as finite volume errors in lattice calculations. On the downside, the torus manifold introduces computational/algorithm efficiency challenges mostly associated with strong autocorrelations when computing topological properties of the fields \cite{Mages:2015scv, Schaefer:2010hu, LSD:2014yyp}. \\

\noindent
In general, lattice results should be independent of the choice of boundary conditions in the infinite volume limit. However, in order to obtain well-defined observables, the action must be a single-valued function and the corresponding fields must be periodic up to the symmetries of the action \cite{Mehen:2005fw}. See Ref.\ \cite{Christ:2019sah, Li:2018sfo, Mages:2015scv, Luscher:2012av, Bruno:2014lra, Mehen:2005fw} for discussions employing other commonly used boundary conditions on the lattice such as; open, anti-periodic, C-type, twisted and Dirichlet boundary conditions. Each of these have their own added advantages based on the system that is under consideration.\\

\noindent
Performing studies of the Casimir effect requires us to impose boundaries on the lattice. While this is a non-trivial exercise for most physical cases based on the geometry and material of the physical object of interest, it is understood that for ideal situations, one may impose boundary conditions that are consistent with a material made of either a perfect magnetic or electric conductor \cite{lindell2005perfect,Wolski:2011fy}. The former implies that the normal magnetic field and tangential electric field components vanish at the conductor's boundary, and we discuss the latter in more detail. \\

\noindent
In the case of electric-type boundary conditions in abelian gauge theory, the material is made up of a perfect conductor (i.e., infinite electrical conductivity, implying zero electrical resistance). According to Ohm's law, the current density of a material is directly proportional to the electric field, i.e., $\Vec{J} = \sigma \Vec{E}$, where $\sigma$ is the conductivity. Given the requirement that $\sigma \to \infty$ inside a perfect conductor, this implies that the electric field inside a perfect conductor is zero, else the current density would diverge. According to Faraday's law, 
\begin{eqnarray}
    \nabla \times \Vec{E} = - \frac{\partial \vec{B}}{\partial t}= 0;
\end{eqnarray}
thus the tangential component of the electric field has to be continuous across a surface. Since we have $\Vec{E} =0$ inside a perfect conductor, then the electric field at the surface of the conductor can only be normal to it, i.e. $E_{||} = 0$.\\ 

\noindent
Perfect conductors are also equipotential surfaces since there is zero scalar potential difference between any two points, and the electric vector potential is time-independent, $\Vec{E}= -\nabla V - \frac{\partial \vec{A}}{\partial t} = 0.$ The time-independent nature of the vector potential implies that the path integral of the vector potential around any closed loop along the surface of a perfect conductor does not change with time. This integral is given by $\oint \vec{A} \cdot d\Vec{l} = \int \vec{B} \cdot dA$ (where $dA$ is the area element) and implies that the magnetic field inside a perfect conductor should remain constant irrespective of any external changes. \\

\noindent
Gauss' law for magnetism, $\nabla \cdot \Vec{B} = 0$, suggests that the normal component of $\Vec{B}$ is continuous across a surface since the surface integral, $\oint \vec{B} \cdot dS = 0$. If we make the assumption of starting with a magnetic field of zero inside a conductor, a property of superconductors, then the magnetic field at the surface of a perfect conductor can only be tangential to it, $B_{\perp} = 0$. See Ref.\ \cite{mcdonaldelectromagnetic} for a literature review on the historical conceptual development of perfect conductors. We summarise these implied perfect electric conductor boundary conditions at the surface below:
\begin{equation}
    E_{||} = 0, \quad B_{\perp} = 0.
    \label{eqn:fields_boundary}
\end{equation}

\noindent
These electric-type boundary conditions have been used to study the Casimir effect in a simple $(2+1)$D geometry of parallel chromoelectric wires in a U(1) Maxwellian gauge theory \cite{Chernodub:2016owp} and in SU(2) non-abelian gauge theory \cite{Chernodub:2018pmt}. On the lattice, these boundary conditions are enforced by the following condition \cite{Chernodub:2018pmt};
\begin{equation}
    \beta_P = 
\begin{dcases}
    \beta, & P\notin S\\
    \lambda_w \beta,              & P\in S,
\end{dcases}
\label{eqn:boundary_condition}
\end{equation}
requiring the coupling at the surface of the perfect conductor material to be different from the vacuum inverse coupling, where the subscript $P$ refers to plaquette and $S$ refers to the worldsurfaces or worldvolume of the conductor material. The inverse coupling is defined as
\begin{equation}
    \beta \equiv 
\begin{dcases}
    \frac{2N_c}{ag^2}, & \text{(2+1)D}\\[5pt]
    \frac{2N_c}{g^2}, & \text{(3+1)D},
\end{dcases}
\label{eqn:inverse_coupling}
\end{equation}
where there's an extra factor of $1/a$ in the (2+1)D case because the coupling, $g^2$, in (2+1)D gauge theories is a dimensionful parameter with dimensions of mass, which sets the mass scale for the theory \cite{Teper:1998te}; the coupling in (3+1)D gauge theories is dimensionless.\\

\noindent
The boundary condition in Eq.\ (\ref{eqn:boundary_condition}) implies that a different coupling is applied on the plaquettes that lie on the worldsurfaces of the wires $(2+1)$D or worldvolume of the plates in $(3+1)$D, respectively. In order for the tangential component of the chromoelectric field on each conductor to vanish (perfect conductor condition), one takes the limit, $\lambda_w \to \infty$. On the lattice, this implies that the expectation value of the plaquettes lying along the worldsurfaces of the conductor material approaches the identity, i.e.\ $U_{P} \to \mathds{1}$. This coupling condition means that when computing the average trace of the plaquettes of a given configuration, the plaquettes on the worldsurfaces of the physical geometry of interest are weighted more. \\

\chapter{The Casimir Effect: Fields and Symmetries}
\label{chapter:The Casimir Effect: Fields and Symmetries}

\noindent
In this chapter, we introduce the main topic of this thesis, which is understanding the Casimir effect for different geometries such as; parallel wires, parallel plates, a symmetrical and asymmetrical tube and box. We look at the lattice definition of the electromagnetic field strength tensor and how it is approximated by the Wilson plaquettes in Euclidean space up to $\mathcal{O}(a^4)$. This leads to an expression for the Euclidean energy density. Lastly, we numerically compute the field strength tensor components and explore the associated rotational symmetries for the various geometries of interest to extract the equivalent field relations under rotation transformations. We highlight the effect of imposing the chromoelectric boundary condition on the measured field components at the boundaries.\\

\section{The Field Strength Tensor}
\noindent
The Casimir potential in non-abelian theory is numerically computed on the lattice by evaluating the energy of vacuum fluctuations in the gluonic fields described by the local expectation value of the Euclidean energy density, the $T^{00}$ component of the energy-momentum tensor $T^{\mu\nu}$. In chapter (\ref{qcd_on_the_lattice}), we introduced the Wilson gauge action, which is described by the product of four link variables forming a plaquette. We now look at how the Wilson plaquette variable appearing in the gluonic action approximates the electromagnetic field strength tensor. We do this by expanding Eq.\ (\ref{eqn:plaquette}) as a product of exponentials of gauge fields as follows,\\
\begin{eqnarray}
     U_{\mu\nu}(n) &=&  U_{\mu}(n)U_{\nu}(n+\hat{\mu})U_{\mu}(n+\hat{\nu})^{\dag}U_{\nu}(n)^{\dag}\\
                    &=& \text{exp}\left[iaA_{\mu}(n)\right] \times \text{exp}\left[iaA_{\nu}(n+\hat{\mu})\right]\nonumber \\ && \times \text{exp}\left[iaA_{\mu}(n+\hat{\nu})^{\dag}\right] \times \text{exp}\left[iaA_{\nu}(n)^{\dag}\right]\\
                    &=& \text{exp}\left(  iaA_{\mu}(n) + iaA_{\nu}(n+\hat{\mu}) -\frac{a^2}{2}\left[A_{\mu}(n), A_{\nu}(n+\hat{\mu})\right]\nonumber \right.\\ && -iaA_{\mu}(n+\hat{\nu}) -iaA_{\nu}(n) -\frac{a^2}{2}\left[A_{\mu}(n+\hat{\nu}), A_{\nu}(n)\right]\nonumber \\ && +\frac{a^2}{2}\left[A_{\nu}(n+\hat{\mu}), A_{\mu}(n+\hat{\nu})\right]  +\frac{a^2}{2}\left[A_{\mu}(n), A_{\nu}(n)\right]\nonumber \\ && \left. +\frac{a^2}{2}\left[A_{\mu}(n), A_{\mu}(n+\hat{\nu})\right] +\frac{a^2}{2}\left[A_{\nu}(n+\hat{\mu}), A_{\nu}(n)\right] \right),
                    \label{eqn:plaquette_expanded}
\end{eqnarray}
where the Baker-Campbell-Hausdorff formula \cite{hausdorff1906symbolische},
\begin{eqnarray}
\label{eqn:campbell}
    \text{exp}(A) \text{exp}(B) = \text{exp}\left(A + B + \frac{1}{2}[A,B]+... \right),
\end{eqnarray}
has been used to expand the product of exponentiated non-commuting gauge fields matrices into a single exponential of a linear combination of the gauge fields.\\

\noindent
The gauge fields with shifted arguments can be Taylor expanded,
\begin{eqnarray}
    A_{\mu}(n+\hat{\nu}) = A_{\mu}(n) + a\partial_{\nu}A_{\mu}(n) + \mathcal{O}(a^2),
\end{eqnarray}
where the forward finite differences method is employed for the derivative terms. Keeping up to $\mathcal{O}$(a) terms, and substituting this expression to Eq.\ (\ref{eqn:plaquette_expanded}), we find the following,
\begin{eqnarray}
    \label{eqn:plaquette_derivatives}
     U_{\mu\nu}(n)  &=& \text{exp}\left( ia^2( \partial_{\mu}A_{\nu}(n) - \partial_{\nu}A_{\mu}(n) + i[A_{\mu}(n), A_{\nu}(n)]) + \mathcal{O}(a^3) \right)\\
                    &=& \text{exp}\left(ia^2F_{\mu\nu}(n) + \mathcal{O}(a^3) \right)\\
                    \label{eqn:plaquette_taylor}
                    &=& 1 + ia^2F_{\mu\nu}(n) - \frac{a^4}{2}\left(F_{\mu\nu}(n)\right)^2 + \mathcal{O}(a^6),
\end{eqnarray}
where we have used the continuum definition of the field strength tensor generalised from the definition in electrodynamics,
\begin{eqnarray}
    F_{\mu\nu}(x) = -i\left[D_{\mu}(x), D_{\nu}(x)\right] = \partial_{\mu}A_{\nu}(x) - \partial_{\nu}A_{\mu}(n) + i\left[A_{\mu}(n), A_{\nu}(n)\right].
    \label{eqn:field_strength_def}
\end{eqnarray}
In comparison to the definition in electrodynamics, the Yang-Mills field strength tensor contains a non-vanishing commutator term of the the gauge fields, which gives rise to gluon self-interactions in QCD, which are in turn responsible for colour confinement.\\

\noindent
It is clear from Eq.\ (\ref{eqn:plaquette_taylor}) that integrating around the Wilson loop does not result in a unit expectation value, which implies that the connection, described by the gauge fields has curvature \cite{Makeenko:2009dw}. The field strength tensor provides a local measure of this curvature. Omitting the imaginary terms, we can then express the real components of the square of the electromagnetic field strength tensor (in lattice units) in terms of the plaquette variables,
\begin{eqnarray}
\label{eqn:field_components}
    a^4[F_{\mu\nu}(n)]^2 = 2\left(1 - U_{\mu\nu}(n) \right).
\end{eqnarray}

\noindent
Using the signature, $g_{\mu\nu} = (+,-,-,-) \leftrightarrow (t,x,y,z)$ in Minkowski spacetime, the contra-variant (3+1)D field-strength tensor is given by
\begin{equation}
\label{eqn:Minkowski_field_strength}
F_{\mu\nu}=
\begin{bmatrix}
0 & E_x & E_y & E_z\\
-E_x & 0 & -B_z & B_y\\
-E_y & B_z & 0 & -B_x\\
-E_z & -B_y & B_x & 0
\end{bmatrix} ,
\end{equation}
and the corresponding energy-momentum tensor associated with the Yang-Mills Lagrangian,
\begin{eqnarray}
    \mathcal{L}_{YM} = -\frac{1}{4} F^a_{\mu\nu} F^{\mu\nu,a} ,
\end{eqnarray}
is given by \cite{Yang:1954ek}
\begin{eqnarray}
    T^{\mu\nu} = -F^{\mu\alpha,a} F^{\mu,a}_{\alpha} + \frac{1}{4} \eta^{\mu\nu} F^{a}_{\alpha\beta} F^{\alpha\beta,a},
\end{eqnarray}
where $F_{0i} = E_i$, $F_{ij} = \epsilon_{ijk}B_k$ and $\frac{1}{4}F_{ij}F_{ij} = \frac{1}{2}B_iB_i$. The resulting energy density in Minkowski spacetime is
\begin{eqnarray}
    T^{00} = \frac{1}{2}(\Vec{E^2} + \Vec{B^2}).
\end{eqnarray}

\noindent
We work on a Euclidean lattice using the coordinate system notation, $(x,y,z,t)$ to simplify our lattice calculations. We thus need to transform the field-strength tensor in Eq.\ (\ref{eqn:Minkowski_field_strength}) into this new coordinate system. We perform this coordinate transformation by applying rotational matrices with an angle, $\theta =\pi/2$, starting with the $xy$ plane ($zt$ fixed), then the $yz$ plane ($xt$ fixed) and lastly the $zt$ plane ($xy$ fixed). The resulting field-strength tensor in the notation, $(x,y,z,t)$ is
\begin{equation}
F_{\mu\nu}^{'} = R_{\theta} F_{\mu\nu} R_{\theta}^{-1} =
\begin{bmatrix}
0 & -B_z & B_y & E_x\\
B_z & 0 & -B_x & E_y\\
-B_y & B_x & 0 & E_z\\
-E_x & -E_y & -E_z & 0
\end{bmatrix} .
\end{equation}
After applying a Wick rotation to Euclidean space, the electric field components acquire a negative sign, resulting in the following Euclidean field strength tensor:
\begin{equation}
F_{\mu\nu}^E = 
\begin{bmatrix}
0 & -B_z & B_y & -E_x\\
B_z & 0 & -B_x & -E_y\\
-B_y & B_x & 0 & -E_z\\
E_x & E_y & E_z & 0
\end{bmatrix} .
\label{eqn:euclidean_fieldstrength}
\end{equation}

\noindent
In our Euclidean lattice formulation, comparing the tensor, $F_{\mu\nu}^E$ to Eq.\ (\ref{eqn:field_components}), one deduces the following correspondence between the field strength and plaquette components:
\begin{itemize}
    \centering
    \item $U_{01} = U_{xy} \leftrightarrow B_{z} = F^E_{01}$
    \item $U_{02} = U_{xz} \leftrightarrow B_{y} = F^E_{02}$
    \item $U_{13} = U_{yz} \leftrightarrow B_{x} = F^E_{13}$
    \item $U_{03} = U_{xt} \leftrightarrow E_{x} = F^E_{03}$
    \item $U_{13} = U_{yt} \leftrightarrow E_{y} = F^E_{13}$
    \item $U_{23} = U_{zt} \leftrightarrow E_{z} = F^E_{23}$ , 
\end{itemize}
where the spatial and temporal components of the plaquettes correspond to magnetic and electric fields, respectively.\\

\noindent
The full expression for the energy density in Euclidean space whose local expectation value describes the energy of vacuum fluctuations of the gluon field becomes
\begin{eqnarray}
    T^{00}_E &=& \frac{1}{2}\left(\Vec{B^2} - \Vec{E^2}\right)\\
    \label{eqn:euclidean_energy_density}
            &=& \frac{1}{2}\left(B_x^2 + B_y^2 + B_z^2 - E_x^2 - E_y^2 - E_z^2\right)
\end{eqnarray}
and is numerically computed by taking the field expectation values of the individual components of all spatial and temporal plaquettes on the Euclidean lattice,
\begin{eqnarray}
    T^{00}_E &=& \frac{1}{2}\left[ \langle B_x^2 \rangle + \langle B_y^2 \rangle + \langle B_z^2 \rangle - \langle E_x^2 \rangle - \langle E_y^2 \rangle - \langle E_z^2 \rangle \right].
\end{eqnarray}

\noindent
In the following sections of this chapter, we discuss the field configurations for various geometries of interest for our Casimir effect studies to map out the effect of imposing chromoelectric boundaries. This numerical result of the field components is normalised by their in-vacuum expectation values in the absence of boundaries (i.e., ultraviolet subtraction). We use this normalisation because we are interested in seeing the effect of the chromoelectric boundary condition on the field components. Whereas, in the following chapter where we discuss the Casimir effect (and need to exclude the boundary effects), we normalise by subtracting the energy at $R_0\to\infty$ to account for the energy contributions of the boundaries. Based on our choice of periodic boundary conditions, we will take $R_0=L/2$, which is the furthest possible separation distance before the Casimir geometry mirrors itself\footnote{This description holds in the case of parallel wires/plates and can be easily visualised on a torus. For the tube and box, one can think of this as the distance in which the number of degrees of freedom inside/outside is the same.}.\\ 

\section{Geometry and Symmetries: Wires in (2+1)D}
\label{section:Geometry and Symmetries: Wires in $(2+1)$D}

\begin{figure}[!htb]
\begin{center}
\includegraphics[scale=0.8]{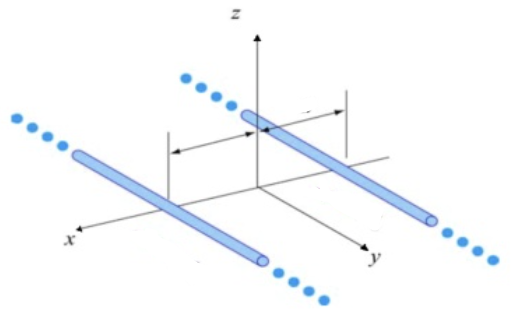}
\caption{Two parallel wires at $x_0$ and $x_1$ separated by a distance $R$.}
\label{fig:parallel_wires}
\end{center}
\end{figure}

\noindent
We start with the case of two chromoelectric wires placed in parallel along the $\hat{y}$ direction in (2+1)D, a distance $R$ apart. This case is discussed extensively in Ref.\ \cite{Chernodub:2018pmt} for SU(2) using the Wilson plaquette action with Hybrid Monte Carlo updates. We reproduce this study for the same action, using Metropolis updates and we extend the study to SU(3). The geometry of this set-up is shown in Fig.\ (\ref{fig:parallel_wires}), where $z$ corresponds to the Euclidean time direction. \\

\begin{figure}[!htb]
\begin{center}
\includegraphics[scale=.65]{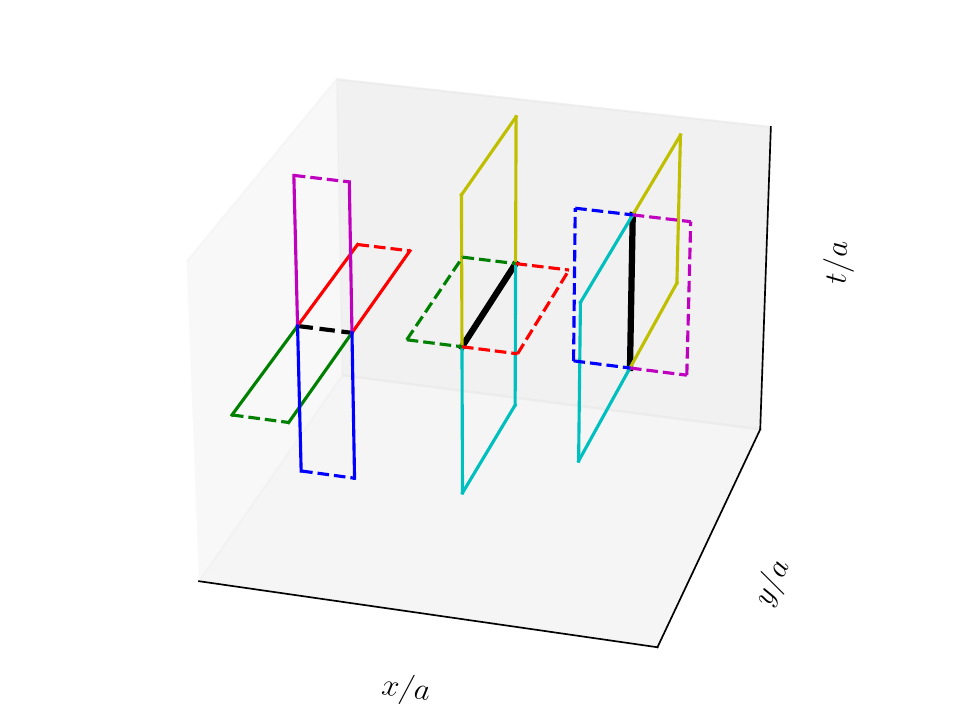}
\caption{Update of link variables in the directions $\mu=0$ (left), $\mu=1$ (middle) and $\mu=2$ (right) in a $(2+1)$D cubic lattice with chromoelectric wires.}
\label{fig:link_updates}
\end{center}
\end{figure}

\noindent
In Euclidean space, the parallel wires form worldsurfaces along the $yt$-plane and have the geometrical appearance of two parallel sheets. This implies that only the plaquettes lying on the $yt$ plane, i.e., $U_{12}$, where $\mu=1$ and $\nu=2$ in the notation $(x,y,t)$ have vanishing tangential fields based on the boundary condition in Eq.\ (\ref{eqn:boundary_condition}). The resulting plaquettes in the geometry of the wires are depicted in Fig.\ (\ref{fig:link_updates}) showing forward and backwards plaquettes during the update of link variables along the directions $\mu = 0,1,2$ (from left to right shown as thick black lines). The same colours correspond to the same plaquette, e.g.\ red is the forward $xy$ plaquette at $\mu=0$ and forward $yx$ plaquette at $\mu=1$. \\

\noindent
In Fig.\ (\ref{fig:link_updates}), solid lines represent the link variables sitting on the world-surface of the wires and dotted lines represent link variables outside the wires' world-surfaces. Only the plaquettes formed \textit{entirely} by solid links have vanishing tangential fields. However, recall that the  change in the local action depends not only on the individual plaquettes, but also on the neighbouring staples, see Eq.\ (\ref{eqn:deltaS}). This means that the staples sitting close to the world-surfaces of the wires also experience the effect of the presence of the chromoelectric wires. We will revisit this discussion shortly when looking at the field components at and around the world-surfaces of the wires. \\

\noindent
In the present case, the Euclidean field-strength tensor in the notation $(x,y,t)$ is obtained by applying two $\theta =\pi/2$ rotations to the Minkowski expression in Eq.\ (\ref{eqn:Minkowski_field_strength}) excluding the $F_{i3}$ and $F_{3i}$ components. The first rotation is the $xy$ plane along the $t$-axis, followed by a rotation of the $yt$ plane along the $x$-axis. The resulting Euclidean field strength is
\begin{equation}
F_{\mu\nu}^{E_{2D}} =
\begin{bmatrix}
0 & -B_z & E_x\\
B_z & 0 & E_y\\
-E_x & -E_y & 0
\end{bmatrix} .
\end{equation}

\noindent
Looking at the remaining components in this field strength tensor, and comparing it to the four-dimensional expression in Eq.\ (\ref{eqn:euclidean_energy_density}), the corresponding expression for the (2+1)D energy density is
\begin{eqnarray}
    \label{eqn:2deuclidean_energy_density}
    T^{00}_{E_{2D}} &=& \frac{1}{2}(B_z^2 - E_x^2 - E_y^2).
\end{eqnarray}

\begin{figure}[!htb]
\begin{center}
\includegraphics[scale=.65]{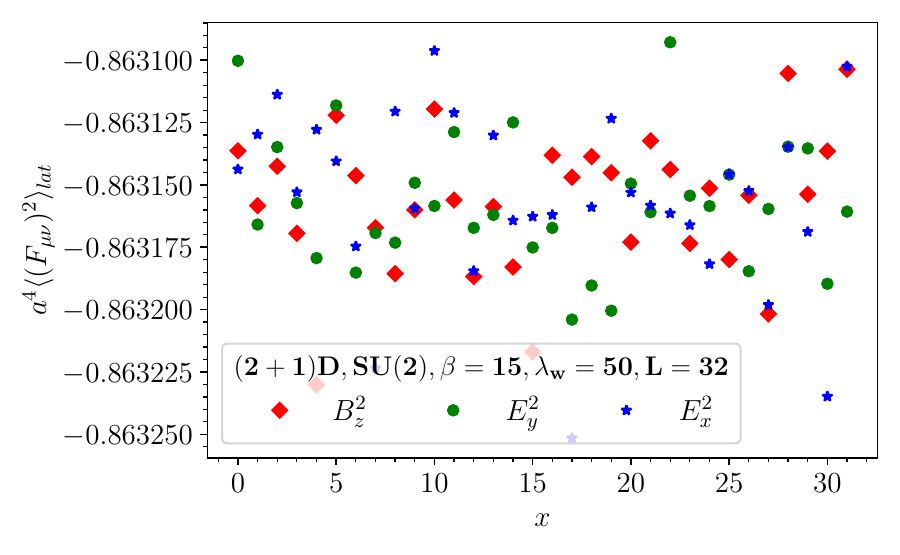}
\caption{Vacuum squared field strength tensor components expectation values in $(2+1)$D SU(2) along the $x$-axis for $\beta=15$.}
\label{fig:vacuum_fields_2DSU2}
\end{center}
\end{figure}

\noindent
It is sufficient to use Eq.\ (\ref{eqn:2deuclidean_energy_density}), taking expectation values of all the field components to compute the energy density on the lattice. However, the computation can be simplified by exploring the symmetries of the system. Based on the geometry (two parallel wires along the $\hat{y}$ direction), it is clear that the system is invariant under any $\pi/2$ rotations about the $x$-axis. Applying such a rotation,
\begin{equation}
F_{\mu\nu}^{F_{2D}} = R_{yz,\theta} F_{\mu\nu}^E R_{yz,\theta}^{-1} =
\begin{bmatrix}
0 & E_x & B_z\\
-E_x & 0 & E_y\\
-B_z & -E_y & 0
\end{bmatrix} ,
\end{equation}

\begin{eqnarray}
    \text{F}_{2D}: \quad \langle B_z^2 \rangle = \langle E_x^2 \rangle,
    \label{eqn:fields_F2D}
\end{eqnarray}
results in an equivalence of the tangential gauge fields fluctuations of the components $B_z$ and $E_x$. Note that such a symmetry is only valid due to the chosen geometry. This rotational symmetry would also hold if applied about the $y$-axis with wires placed parallel to the $x$-axis, and would result in an equivalence between $B_z$ and $E_y$.\\

\noindent
We now look at the lattice results of these field strength tensor components at the position of the wires and the surrounding region. These results follow directly from Eq.\ (\ref{eqn:field_components}) in lattice units. In the absence of the wires, the vacuum field components are shown in Fig.\ (\ref{fig:vacuum_fields_2DSU2}) fluctuating around approximately equal expectation values, as expected for a zero temperature lattice. In principle, one can not differentiate between the contributions of each component in this case.\\

\begin{figure}[!htb]
\begin{center}
\includegraphics[scale=.65]{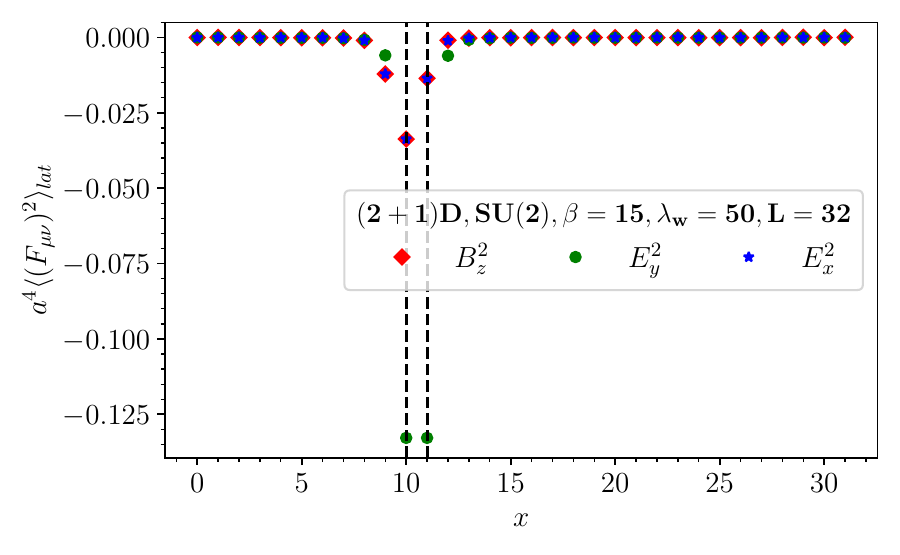}
\caption{Expectation values of the squared field strength tensor components in $(2+1)$D SU(2) along the $x$-axis orthogonal to the wires placed a distance $R_{lat}=1$ apart.}
\label{fig:fields_2DSU2_R1}
\end{center}
\end{figure}

\noindent
The wires are introduced in Fig.\ (\ref{fig:fields_2DSU2_R1}) at a shortest possible distance apart, $R_{lat}=1$. We normalise the fluctuations of the field components by subtracting the vacuum field expectation values. It is clear that at the position of each wire at $x_0=10$ and $x_1=11$, the electric field components parallel to the wires, $E_y$ are significantly suppressed relative to other components. This is in accordance with the chromoelectric boundary condition in use, given in Eq.\ (\ref{eqn:fields_boundary}) and the geometry. \\

\begin{figure}[!htb]
    \centering
    \subfigure[Distance $R_{lat}=2$]{{\includegraphics[scale=0.5]{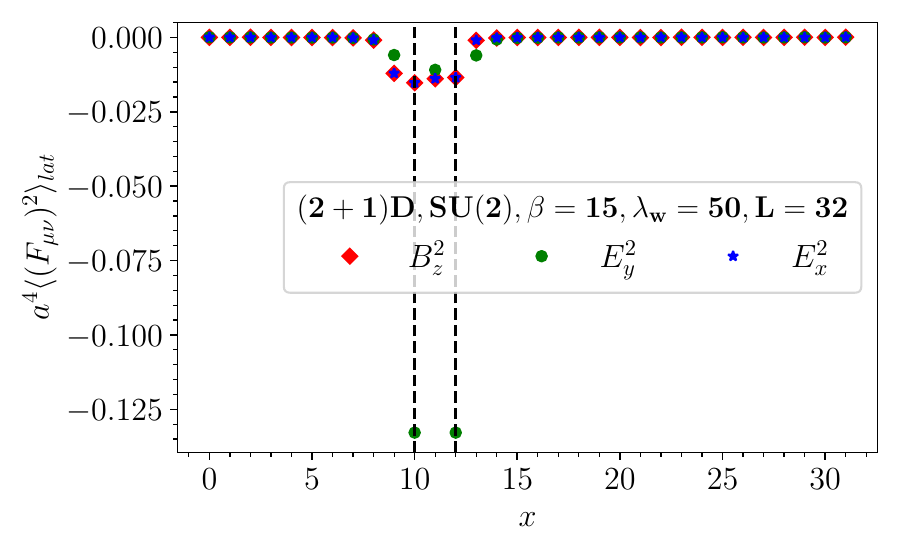} }}%
    \hspace{-0.45cm}
    \subfigure[Distance $R_{lat}=10$]{{\includegraphics[scale=0.5]{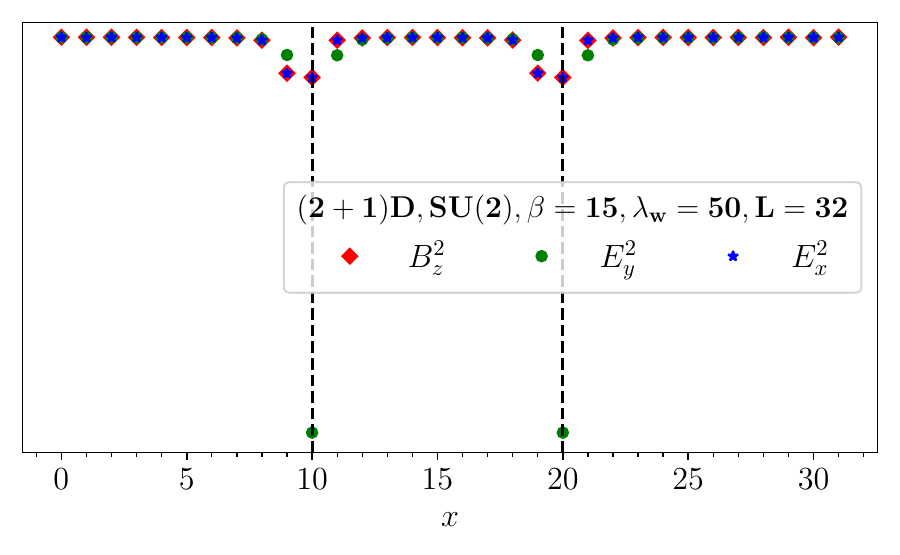} }}%
    \caption{Expectation values of the squared field strength tensor components in $(2+1)$D SU(2) along the $x$-axis orthogonal to the wires placed a distance $R$ apart.}%
    \label{fig:fields_2DSU2_R2_R10}%
\end{figure}

\noindent
While the suppression of the fluctuations in the $E_y$ fields is expected directly from the boundary conditions, we also observe that the remaining field components, i.e., $E_x$ and $B_z$ also react to the presence of the wires and are slightly suppressed. This follows from the formulation of the local action, where updating a single link variable affects the four plaquettes attached to it, see Fig.\ (\ref{fig:link_updates}) and Eq.\ (\ref{eqn:wilsonactionlocal}). It is for the same reason that we also observe a suppression of the field fluctuations close to (but not exactly) at the position of the wires. Our results are qualitatively consistent with the study of parallel wires in Ref.\ \cite{Chernodub:2016owp} on U(1) gauge theory at different couplings and permittivity.\\

\noindent
In Fig.\ (\ref{fig:fields_2DSU2_R2_R10}), we look at how the qualitative shape of the field fluctuations profiles changes as one moves the wires further apart\footnote{The position of the wires is represented by vertical dashed lines.}. The first observation is that the amount in which the $E_y$ fields are suppressed is independent of the position of the wires, as expected. Secondly, the $E_x$ and $B_z$ fields are suppressed slightly more when the wires are much closer together than when they are further apart because of the contributions from the backward plaquettes.\\

\begin{figure}[!htb]
\begin{center}
\includegraphics[scale=.65]{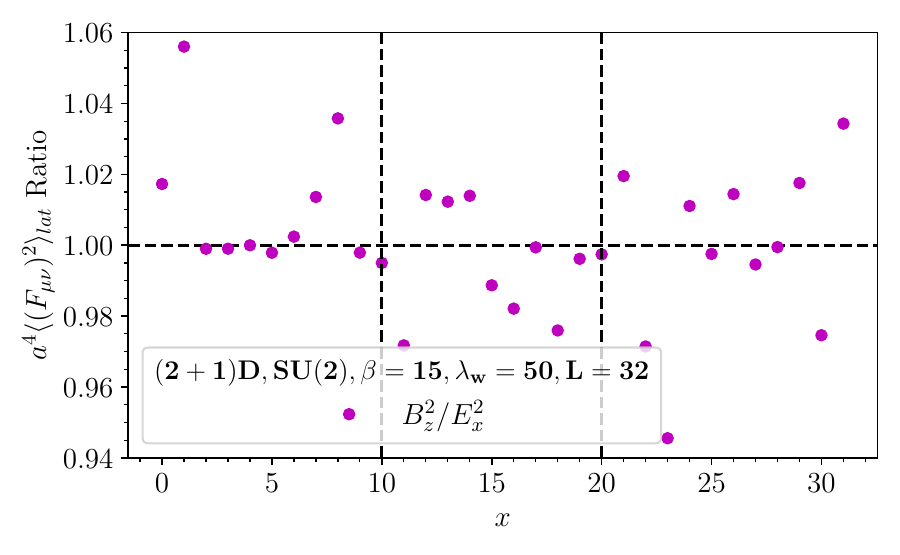}
\caption{Expectation values of the squared field strength tensor components in $(2+1)$D SU(2) along the $x$-axis orthogonal to the wires placed a distance $R_{lat}=10$ apart.}
\label{fig:fields_2DSU2_R10_ratio}%
\end{center}
\end{figure}

\noindent
However, this backward plaquette contribution will have no effect on the pressure of the system as the two components, $E_x$ and $B_z$ cancel. Also note that there is a clear symmetry in the orientation of the $E_y$ field fluctuations around the wires because these field components are artificially suppressed through the boundary condition. This is accompanied by an asymmetry in the $E_x$ and $B_z$ fields around the wires. This asymmetry is a consequence of the naive forward finite differences approach in taking the derivatives,
\begin{eqnarray}
    \partial_{\mu}A_{\nu}(x) = \frac{1}{a_{\mu}}\left(A_{\nu}(x + a_{\mu}\hat{\mu}) - A_{\nu}(x) \right),
\end{eqnarray}
in the Wilson field-strength formulation given in Eq.\ (\ref{eqn:plaquette_derivatives}). The asymmetry due to the forward differences approach is subject to ongoing research within the lattice community; see for example Ref.\ \cite{Rothkopf:2021jye} for an expansive discussion.\\

\noindent
In the meantime, we close the discussion on $(2+1)$D SU(2) field fluctuations by looking at a quantitative comparison of the $E_x$ and $B_z$ fluctuations. We showed in Eq.\ (\ref{eqn:fields_F2D}) that the expectation value of the square of the field fluctuations in these two components are equal due to the symmetry in the geometry. Although this is already qualitatively clear from the previous figures, given that we do not show the uncertainties in the measurements for neatness, we solidify this symmetry statement through the ratio of $E_x$ and $B_z$. We show this ratio in Fig.\ (\ref{fig:fields_2DSU2_R10_ratio}) for $R_{lat}=10$ and we see that it fluctuates closely around one, limited only by statistics.\\

\begin{figure}[!htb]
\begin{center}
\includegraphics[scale=.65]{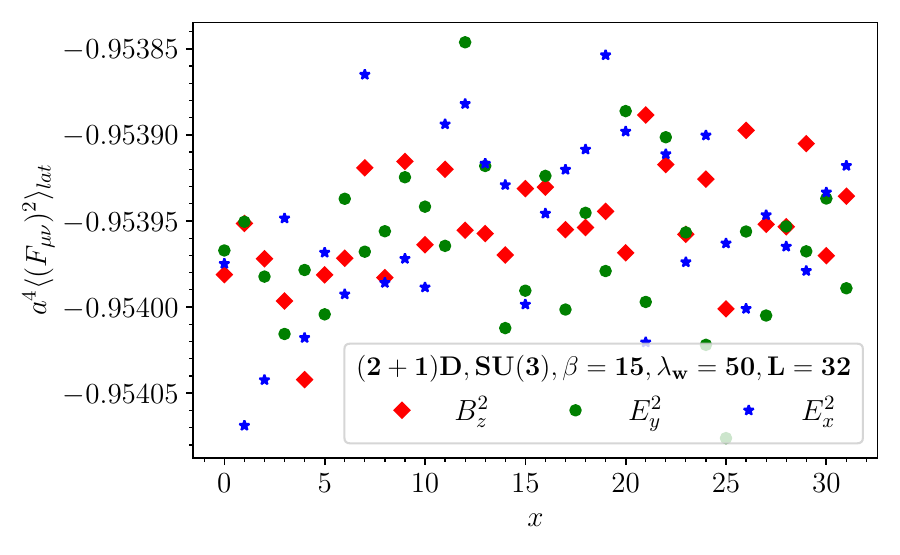}
\caption{Vacuum squared field strength tensor components expectation values in $(2+1)$D SU(3) along the $x$-axis for $\beta=15$.}
\label{fig:vacuum_fields_2DSU3}
\end{center}
\end{figure}
\begin{figure}[!htb]
\begin{center}
\includegraphics[scale=.65]{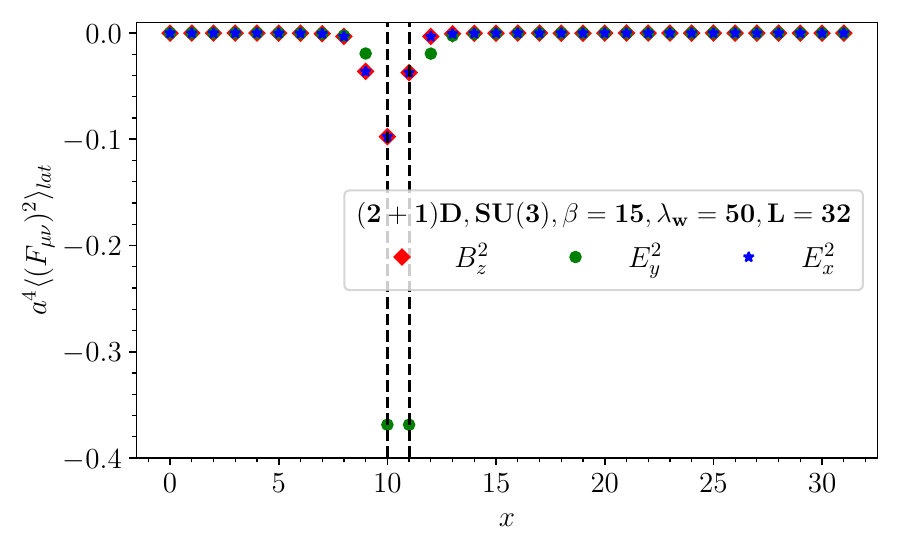}
\caption{Expectation values of the squared field strength tensor components in $(2+1)$D SU(3) along the $x$-axis orthogonal to the wires placed a distance $R_{lat}=1$ apart.}
\label{fig:fields_2DSU3_R1}
\end{center}
\end{figure}
\begin{figure}[!htb]
\begin{center}
\includegraphics[scale=.65]{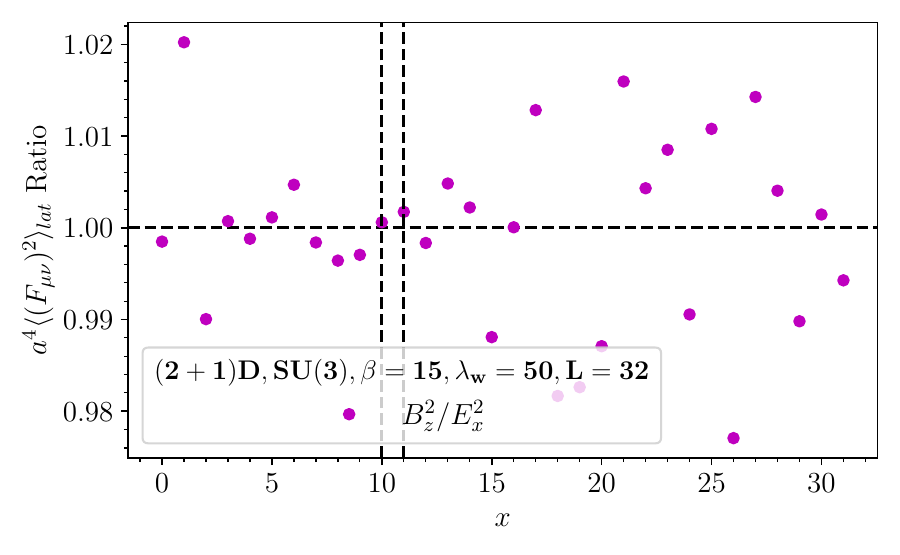}
\caption{Expectation values of the squared field strength tensor components in $(2+1)$D SU(3) along the $x$-axis orthogonal to the wires placed a distance $R_{lat}=1$ apart.}
\label{fig:fields_2DSU3_R1_ratio}
\end{center}
\end{figure}
\begin{figure}[!htb]
    \centering
    \subfigure[Distance $R_{lat}=2$]{{\includegraphics[scale=0.5]{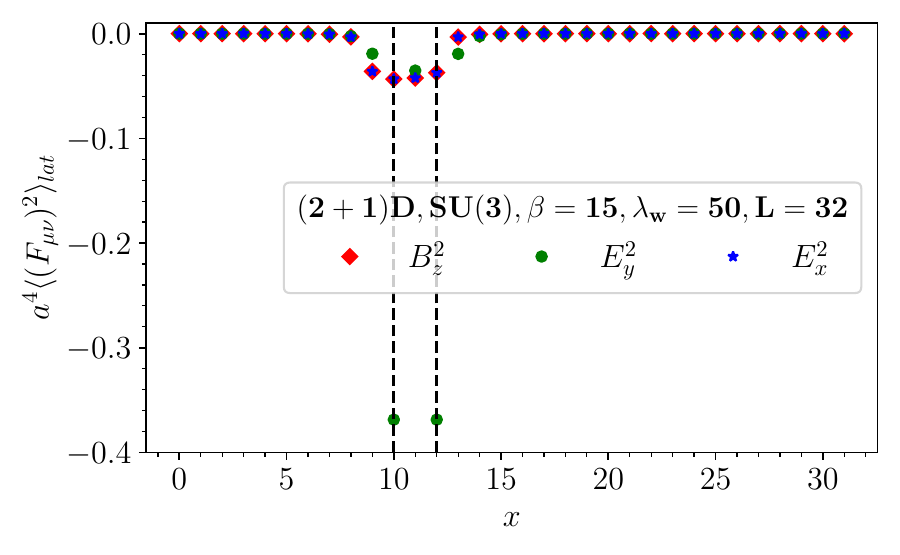} }}%
    \hspace{-0.45cm}
    \subfigure[Distance $R_{lat}=10$]{{\includegraphics[scale=0.5]{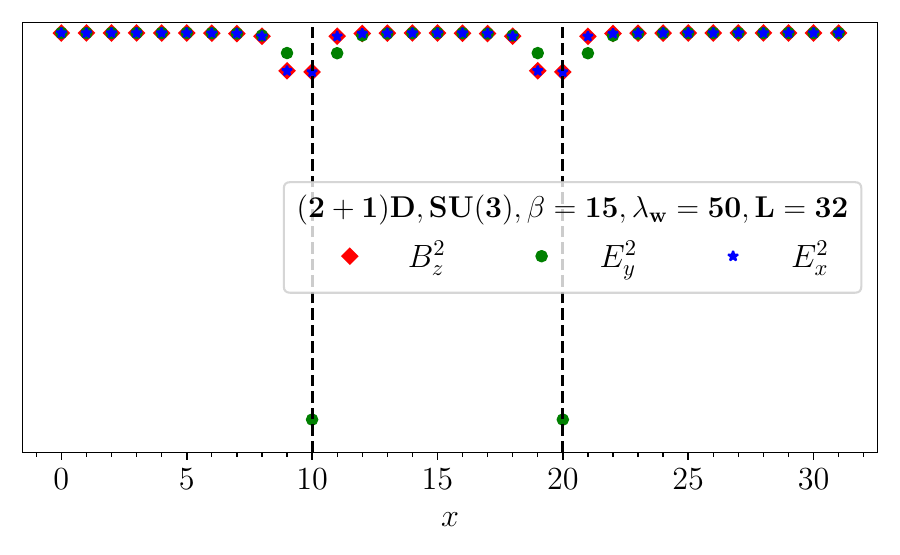} }}%
    \caption{Expectation values of the squared field strength tensor components in $(2+1)$D SU(2) along the $x$-axis orthogonal to the wires placed a distance $R$ apart.}%
    \label{fig:fields_2DSU3_R2_R10}%
\end{figure}

\noindent
In concluding this subsection, the following plots in Fig.\ (\ref{fig:vacuum_fields_2DSU3} - \ref{fig:fields_2DSU3_R2_R10}) show the field configurations for the parallel wires set-up in $(2+1)$D SU(3). The qualitative features of the fields at the position of the wires and around the wires are universal and are consistent with our SU(2) discussions. The only difference is that the increased number of the degrees of freedom results in an enhanced suppression of all the field components at the worldsheets of the wires and the surrounding region. This increased suppression, in turn, should result in an increased pressure experienced by the worldsheets of the wires and we will quantify this increase in pressure in the following chapter.\\

\section{Geometry and Symmetries: Plates}
\label{sec:Geometry and Symmetries: Plates}

\begin{figure}[!htb]
\begin{center}
\includegraphics[scale=.65]{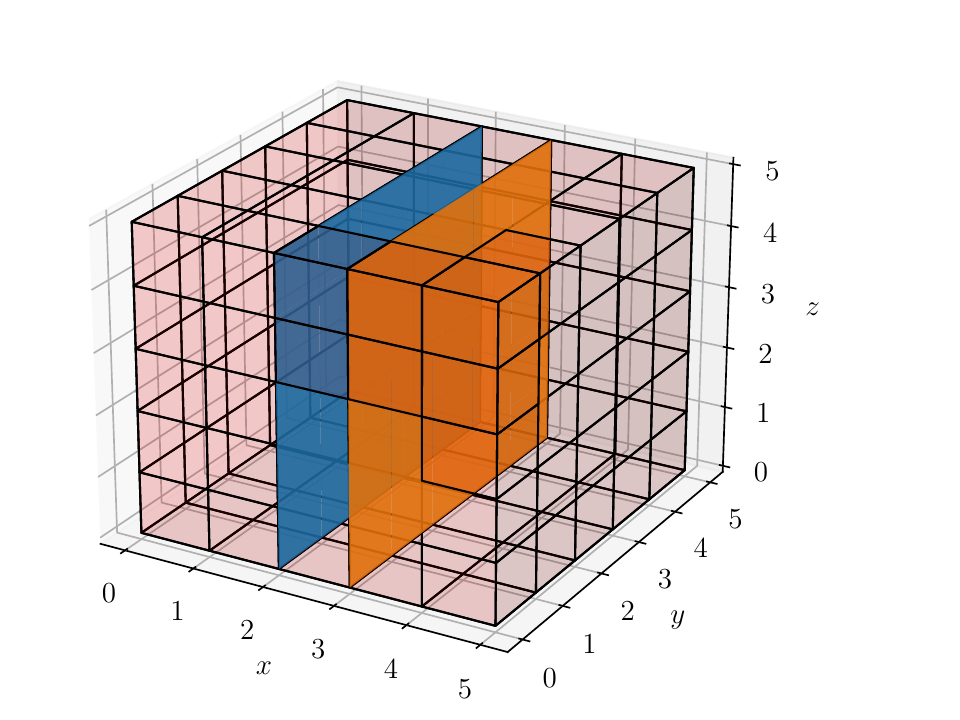}
\caption{Geometry of chromoelectric boundary conditions on parallel plates in a $(3+1)$D cubic lattice.}
\label{fig:geometry_plates}
\end{center}
\end{figure}

\noindent
In the case of (3+1)D non-abelian gauge theory, we start with the parallel chromoelectric plates configuration shown in Fig.\ (\ref{fig:geometry_plates}). The plates are placed at $\hat{x}=x_0$ and $\hat{x}=x_1$, a distance $R$ apart. They extend throughout the lattice spatial extents (i.e., they have `infinite' spatial extent) in the $\hat{y}$ and $\hat{z}$ directions with lattice area $L^2$. As opposed to the parallel wires in (2+1)D where the gauge field boundary conditions are formulated on the worldsurfaces of the wires, we now formulate such boundary conditions on the worldvolume of the plates.\\

\noindent
In order to explore the symmetries of the field strength tensor, we first note that all applicable rotations need to keep the $\hat{x}$-axis (perpendicular to the plates) fixed in order to preserve symmetries based on the geometry. This leaves us the freedom of only three (of six) possible rotations in Euclidean space: $yz$ plane with the $xt$ plane fixed, $yt$ plane with the $xz$ plane fixed and $zt$ plane with the $xy$ plane fixed. The three resulting field strength tensors showing the rotated $\bf E$ and $\bf B$ field components are given in Eq.\ (\ref{eqn:fields_F1} - \ref{eqn:fields_F3}).\\

\begin{equation}
F_{\mu\nu}^{F1} = R_{yz,\theta} F_{\mu\nu}^E R_{yz,\theta}^{-1} =
\begin{bmatrix}
0 & B_y & B_z & -E_x\\
-B_y & 0 & -B_x & -E_z\\
-B_z & B_x & 0 & E_y\\
E_x & E_z & -E_y & 0
\end{bmatrix} 
\label{eqn:fields_F1}
\end{equation}

\begin{equation}
F_{\mu\nu}^{F2} = R_{yt,\theta} F_{\mu\nu}^E R_{yt,\theta}^{-1} =
\begin{bmatrix}
0 & -E_x & B_y & B_z\\
E_x & 0 & E_z & -E_y\\
-B_y & -E_z & 0 & -B_x\\
-B_z & E_y & B_x & 0
\end{bmatrix} 
\label{eqn:fields_F2}
\end{equation}

\begin{equation}
F_{\mu\nu}^{F3} = R_{yz,\theta} F_{\mu\nu}^E R_{yz,\theta}^{-1} =
\begin{bmatrix}
0 & -B_z & -E_x & -B_y\\
B_z & 0 & -E_y & B_x\\
E_x & E_y & 0 & -E_z\\
B_y & -B_x & E_z & 0
\end{bmatrix} .
\label{eqn:fields_F3}
\end{equation}

\noindent
Performing a direct comparison of each of the resulting field strength components to the (3+1)D Euclidean field strength tensor in Eq.\ (\ref{eqn:euclidean_fieldstrength}), it is clear that the rotations result in the following equivalence relations between the electric and magnetic field components\footnote{We will use the labels F1 for the symmetry resulting from rotating the Euclidean field strength tensor about the $xt$ plane, F2 for the symmetry from rotating about the $xz$ plane and F3 for the symmetry from rotating about the $xy$ plane.}:
\begin{eqnarray}
    \label{eqn:F1}
    \text{F}1: \quad \langle B_z^2 \rangle = \langle B_y^2 \rangle \quad \text{and} \quad \langle E_z^2 \rangle = \langle E_y^2 \rangle\\
    \label{eqn:F2}
    \text{F}2: \quad \langle B_z^2 \rangle = \langle E_x^2 \rangle \quad \text{and} \quad \langle B_x^2 \rangle = \langle E_z^2 \rangle\\
    \text{F}3: \quad \langle B_y^2 \rangle = \langle E_x^2 \rangle \quad \text{and} \quad \langle B_x^2 \rangle = \langle E_y^2 \rangle
    \label{eqn:F3}
\end{eqnarray}

\begin{figure}[!htb]
\begin{center}
\includegraphics[scale=.65]{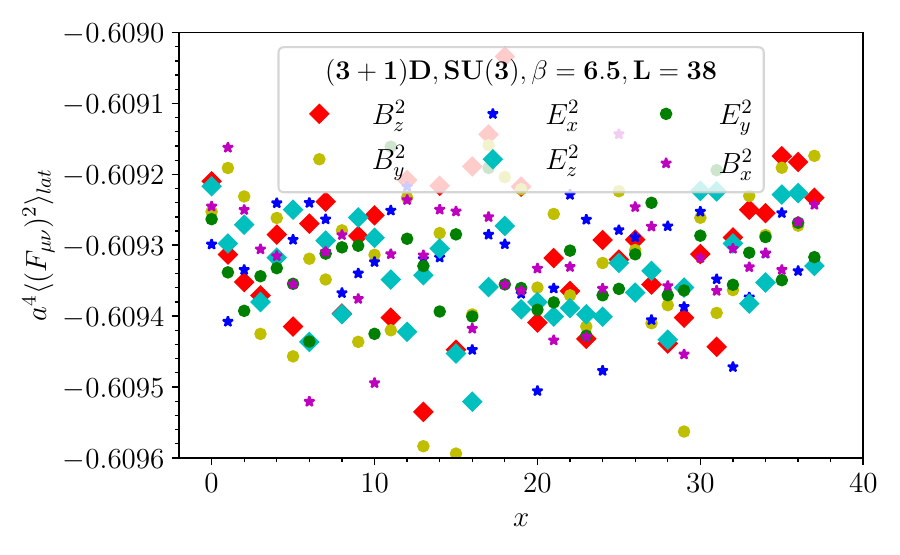}
\caption{Vacuum squared field strength tensor components expectation values in $(3+1)$D SU(3) along the $x$-axis for $\beta=6.5$.}
\label{fig:vacuum_fields_3DSU3}
\end{center}
\end{figure}

\noindent
In Fig.\ (\ref{fig:vacuum_fields_3DSU3}), we show the vacuum field expectation values in (3+1)D SU(3) for inverse coupling $\beta =6.5$. These fields correspond to a `zero temperature' theory and are computed on an isotropic lattice of size $N=38^4$. As expected for an isotropic lattice, the vacuum field expectation values are equal\footnote{and within statistical uncertainties, which are not shown for neatness.} (up to three digits in this case) and fluctuate around a common value. The magnitude of the errors is consistent with that of the plaquettes which the fields depend on, and these errors are given in chapter (\ref{chapter:lattice_formalism}). \\

\begin{figure}[!htb]
    \centering
    \subfigure[Distance $R_{lat}=1$]{{\includegraphics[scale=0.5]{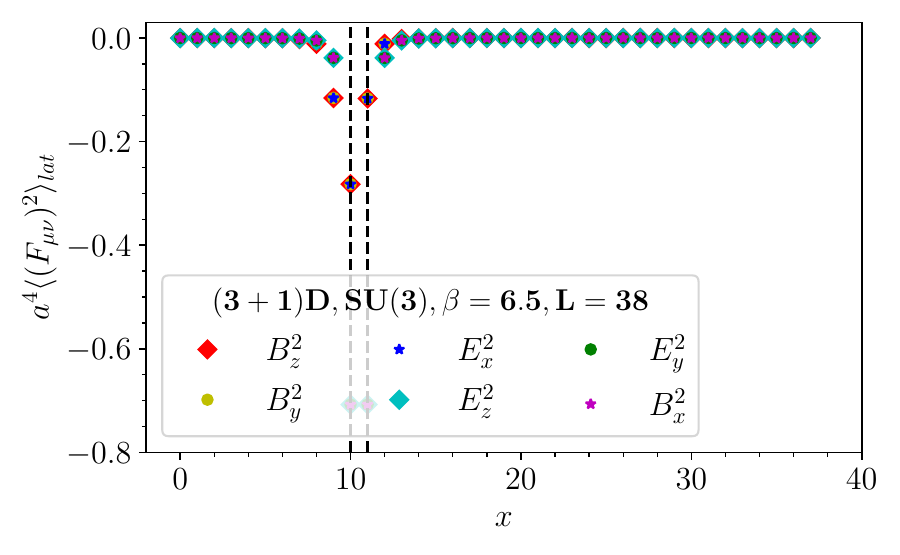} }}%
    \hspace{-0.45cm}
    \subfigure[Distance $R_{lat}=2$]{{\includegraphics[scale=0.5]{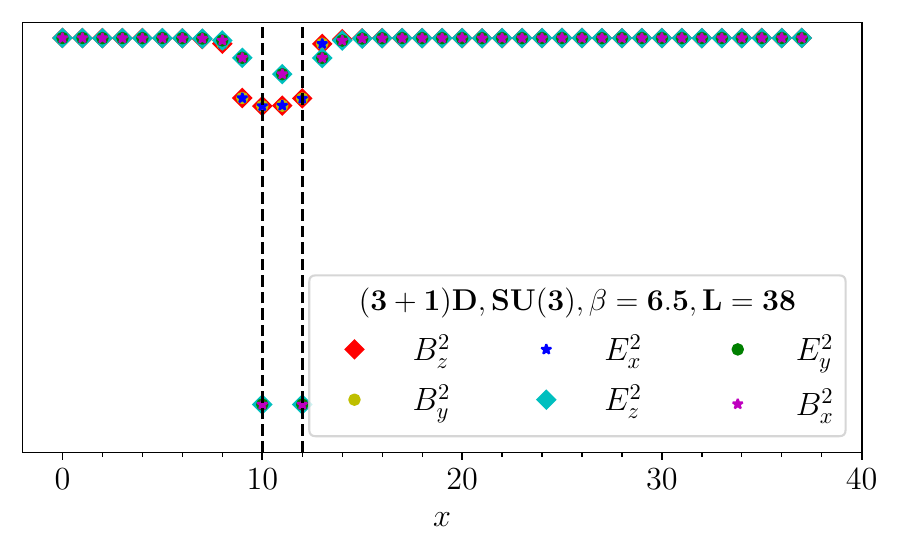} }}%
    \caption{Expectation values of the squared field strength tensor components in $(3+1)$D SU(3) along the $x$-axis orthogonal to the plates placed a distance $R$ apart.}%
    \label{fig:fields_3DSU3Plates_R1_R2}%
\end{figure}

\begin{figure}[!htb]
\begin{center}
\includegraphics[scale=.65]{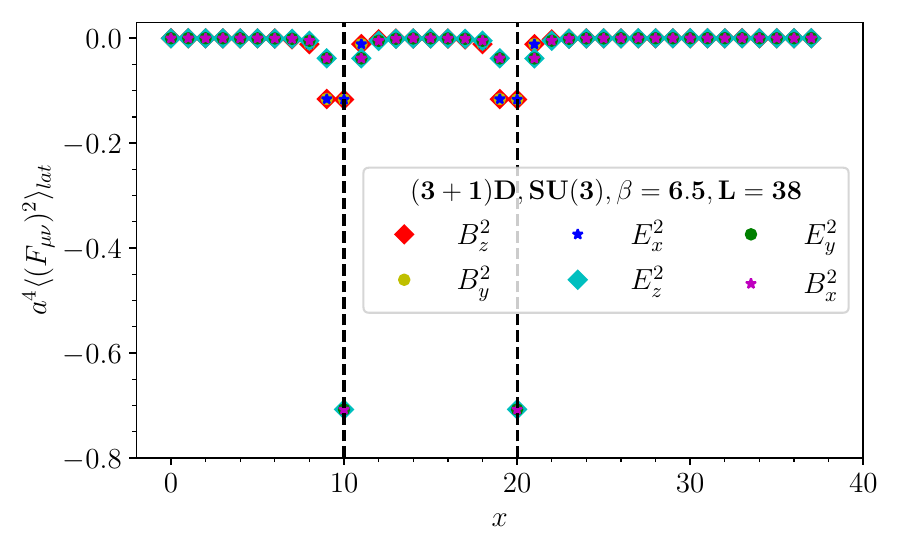}
\caption{Expectation values of the squared field strength tensor components in $(3+1)$D SU(3) along the $x$-axis orthogonal to the plates placed a distance $R_{lat}=10$ apart.}
\label{fig:fields_3DSU3Plates_R10}
\end{center}
\end{figure}

\noindent
The field expectation values in the presence of parallel chromoelectric plates\footnote{The position of the plates is represented by vertical dashed lines.} are given in Fig.\ (\ref{fig:fields_3DSU3Plates_R1_R2} - \ref{fig:fields_3DSU3Plates_R10}). The field profile is consistent with our observations in (2+1)D SU($N_c$) gauge theories in the presence of chromoelectric wires discussed in the previous subsection. In order to compare this pure gauge case to its (2+1)D SU(3) counterpart, we compare these results to the qualitative features of the field configurations in Fig.\ (\ref{fig:fields_2DSU3_R1} - \ref{fig:fields_2DSU3_R2_R10}). The enhanced magnitude in the suppression of the fields at and around the positions of the plates can be attributed to an increase in the boundary size and available degrees of freedom in the (3+1)D theory associated with an additional spatial dimension\footnote{In (2+1)D, each link variable is attached to four plaquettes, whereas in (3+1)D, each link variable is attached to six plaquettes.}. \\

\noindent
We now turn our attention to the calculated field symmetries, F1-F3 in Eq.\ (\ref{eqn:F1} - \ref{eqn:F3}) that result from the rotational symmetries of our parallel plate geometry. The numerically measured field expectation values show that $\langle B_z^2 \rangle = \langle B_y^2 \rangle = \langle E_x^2 \rangle$ and $\langle E_z^2 \rangle = \langle E_y^2 \rangle = \langle B_x^2 \rangle$, both consistent with our symmetry calculations. Lastly, the profile of the fields as we move the plates further apart is a good representation of the change in the number of modes (thus the pressure experienced by the plates) with respect to the separation distance.\\

\section{Geometry and Symmetries: Tube}
\label{sec:Geometry and Symmetries: Tube}

\begin{figure}[!htb]
    \centering
    \subfigure[Symmetric Tube]{{\includegraphics[scale=0.48]{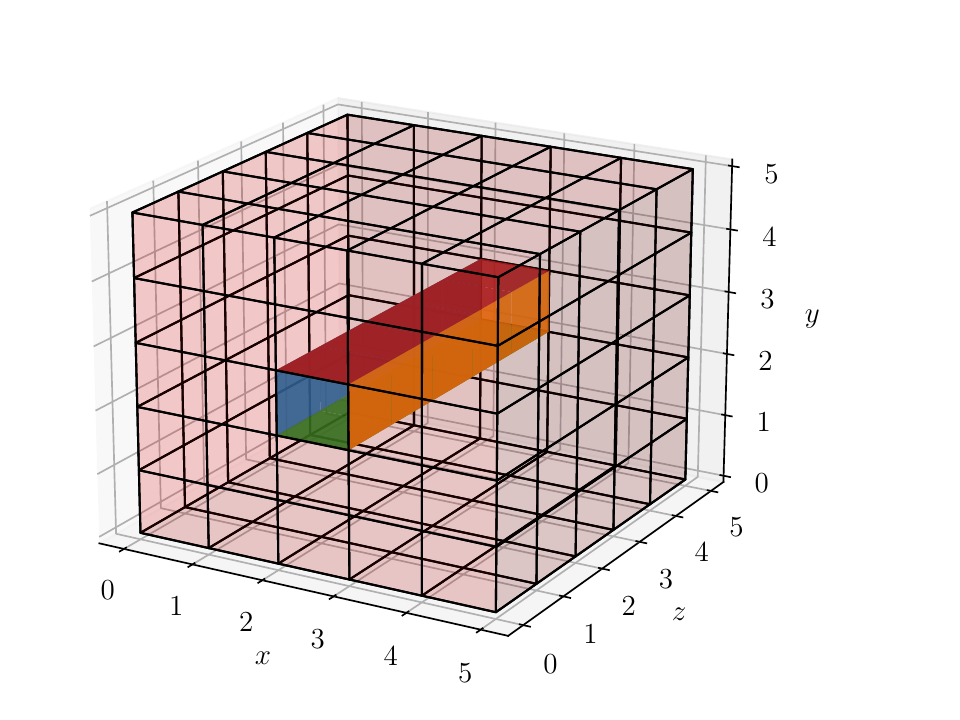} \label{fig:geometry_symmetric_tube}}}%
    \hspace{-1cm}
    \subfigure[Asymmetric Tube]{{\includegraphics[scale=0.48]{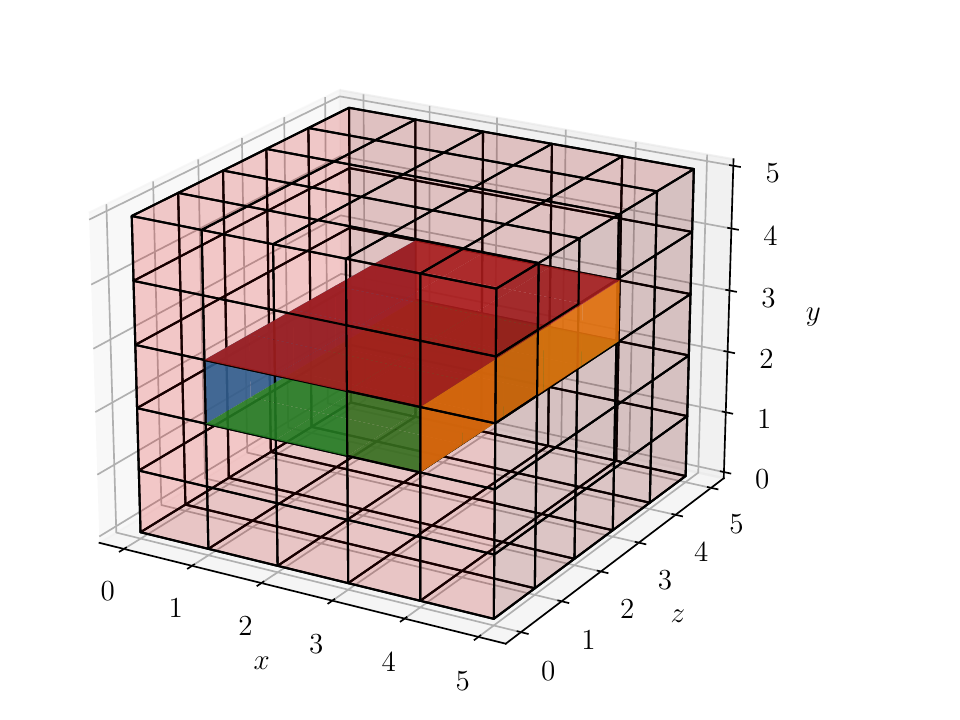} \label{fig:geometry_asymmetric_tube}}}%
    \caption{Geometry of chromoelectric boundary conditions on a tube in a $(3+1)$D cubic lattice.}%
    \label{fig:geometry_tube}%
\end{figure}

\noindent
We now consider the geometry of a tube that is elongated in the $\hat{z}$ direction. That is, it has finite width in the $\hat{x}$ and $\hat{y}$ directions, but covers the entire lattice extent, $L$ in $\hat{z}$. In general, we consider the Casimir effect of two cases, the first is a symmetrical tube shown in Fig.\ (\ref{fig:geometry_symmetric_tube}), where the hollow part is a square (i.e., the width and height sides are separated by the same distance $R_x=R_y=R$). The second is an asymmetrical tube shown in Fig.\ (\ref{fig:geometry_asymmetric_tube}), where the hollow part is rectangular with the $\hat{y}$ sides separated by a \textit{fixed} distance $R_y = 1$ and the $\hat{x}$ sides separated by a distance $R_x = R$. Because the field configurations are boundary dependent and their qualitative features do not vary much between the two geometries, we only discuss the fields for a symmetrical tube.\\

\noindent
Such geometries are obtained by delicately enforcing the gauge field boundary conditions on the plaquettes that form the required shape, with a careful treatment of the forward and backward plaquettes on the edges\footnote{and corners in the case of a box which we discuss in the next subsection.}. The Casimir-type geometries that can be explored on the lattice are limited by the hypercubic formulation due to discretisation errors, critical slowing down, finite volume effects, and other issues that arise in trying to model more complicated geometries e.g.\ a 2-sphere.\\ 

\begin{figure}[!htb]
    \centering
    \subfigure{{\includegraphics[scale=0.5]{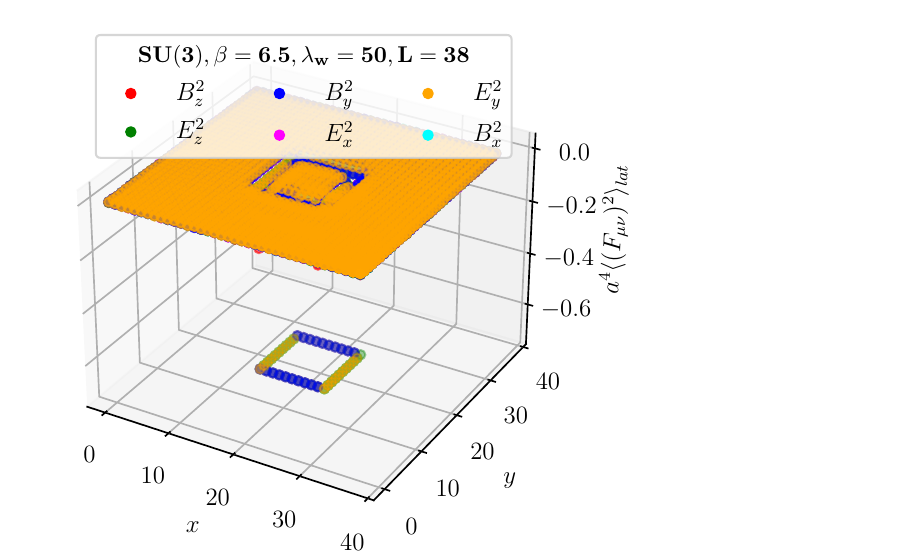} }}%
    \hspace{-2cm}
    \subfigure{{\includegraphics[scale=0.5]{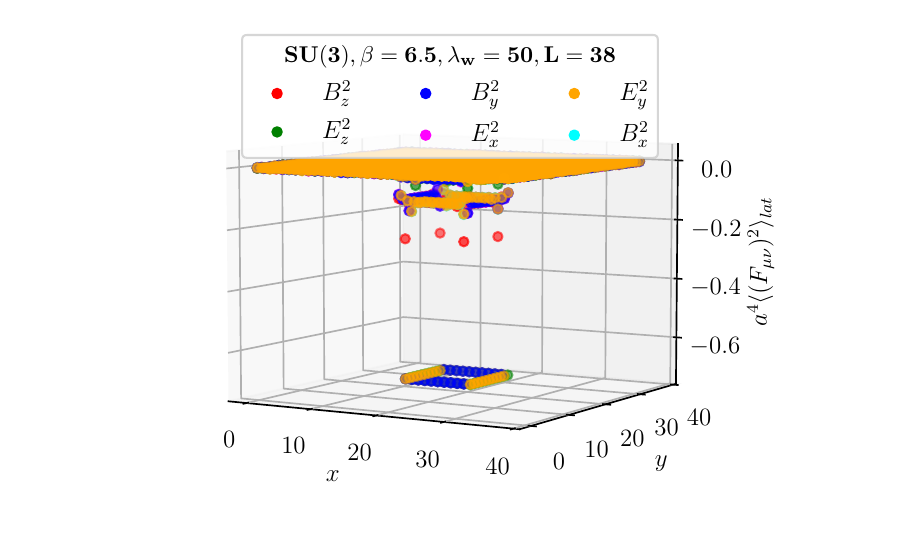} }}%
    \caption{Three dimensional expectation values of the squared field strength tensor components in $(3+1)$D SU(3) for a symmetrical tube with side lengths $R_{lat}=10$ and faces at $x=y=14$ and $x=y=24$.}%
    \label{fig:fields3D_tube_all}%
\end{figure}

\noindent
In terms of the rotational symmetries for the tube (both symmetric and asymmetric), notice that we now have fixed `chomoelectric plates' along both the $\hat{x}$ and $\hat{y}$ axis. Thus rotations that leave the system invariant should keep both these axes fixed. This leaves us the freedom to perform only one (of six) possible rotations, which is the $zt$ plane with the $xy$ plane fixed. The resulting field strength tensor corresponds to the F3-symmetry given in Eq.\ (\ref{eqn:F3}).\\

\noindent
We show a three-dimensional visualisation of the field configurations for a symmetrical tube with side lengths $R_x = R_y =10$ in Fig.\ (\ref{fig:fields3D_tube_all}). The elongated $\hat{z}$ direction has been integrated out since the fields should be the same along this direction. The left frame shows that all field components fluctuate around zero in vacuum, far from where the tube is positioned. Taking a closer look, one sees that as we get closer to the tube\footnote{from both the outside and the inside.}, the surrounding plaquettes start experiencing the gauge field contributions from the boundaries of the tube and the fields are slightly suppressed. This is seen on the right frame where the magnitude of suppression gets stronger as you approach the face of the tube.\\ 

\begin{figure}[!htb]
    \centering
    \subfigure{{\includegraphics[scale=0.5]{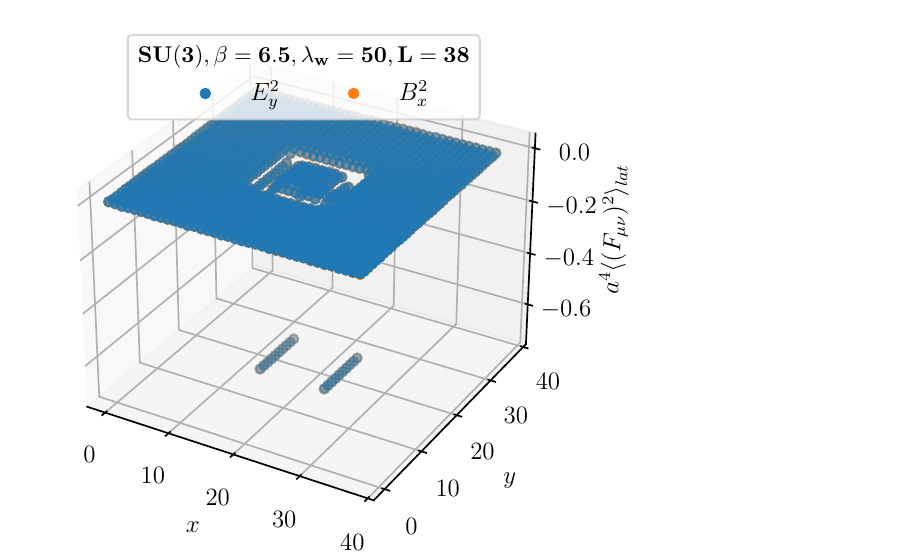} }}%
    \hspace{-2cm}
    \subfigure{{\includegraphics[scale=0.5]{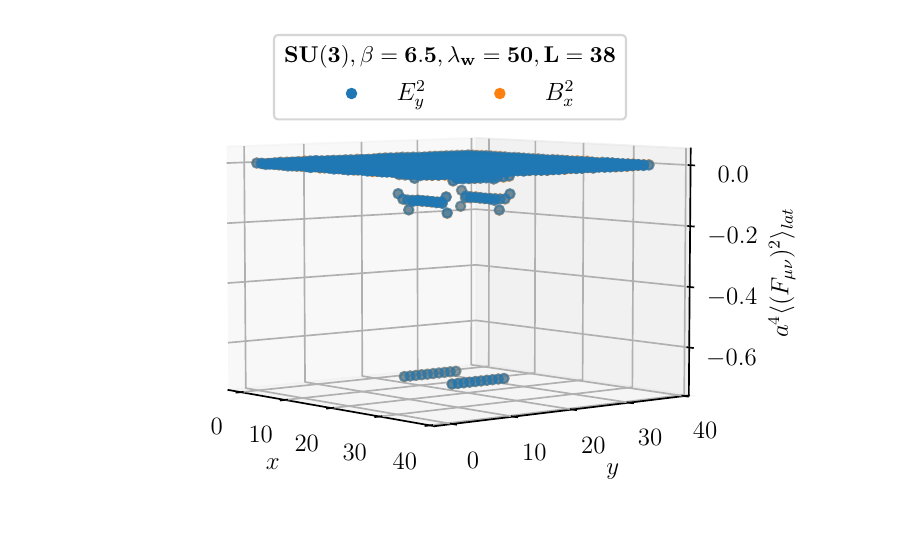} }}%
    \caption{Three dimensional expectation values of the $E_y^2$ and $B_x^2$ field components in $(3+1)$D SU(3) for a symmetrical tube with side lengths $R_{lat}=10$ and faces at $x=y=14$ and $x=y=24$.}%
    \label{fig:fields3D_tube_BxEy}
\end{figure}

\begin{figure}[!htb]
    \centering
    \subfigure{{\includegraphics[scale=0.5]{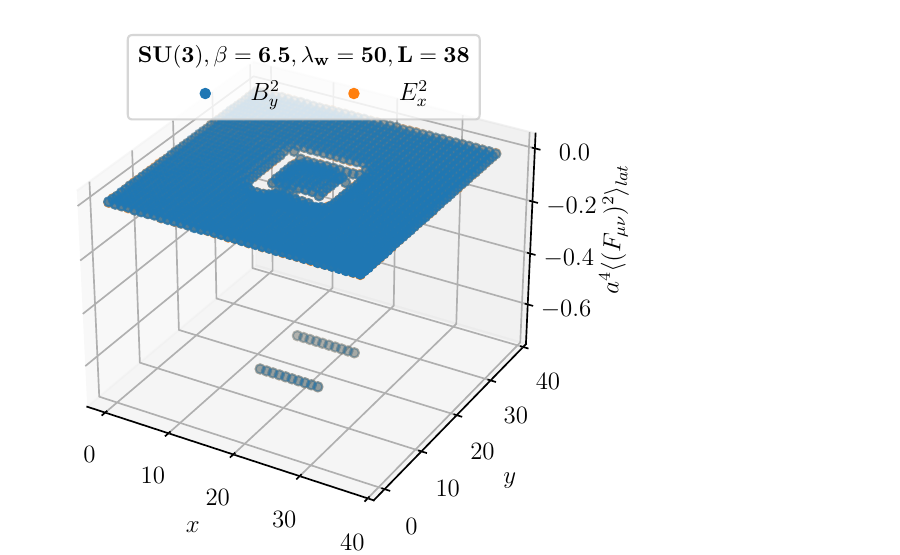} }}%
    \hspace{-2cm}
    \subfigure{{\includegraphics[scale=0.5]{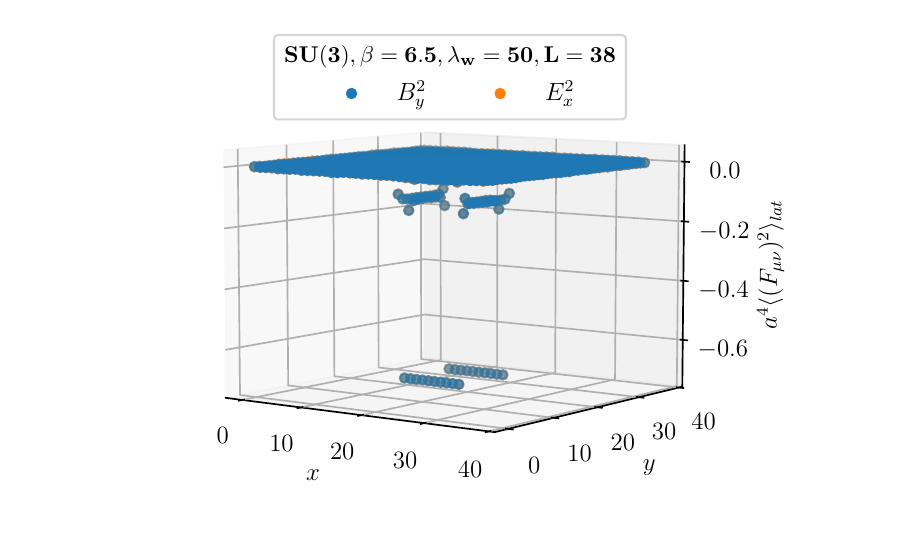} }}%
    \caption{Three dimensional expectation values of the $B_y^2$ and $E_x^2$ field components in $(3+1)$D SU(3) for a symmetrical tube with side lengths $R_{lat}=10$ and faces at $x=y=14$ and $x=y=24$.}%
    \label{fig:fields3D_tube_ByEx}
\end{figure}

\begin{figure}[!htb]
    \centering
    \subfigure{{\includegraphics[scale=0.5]{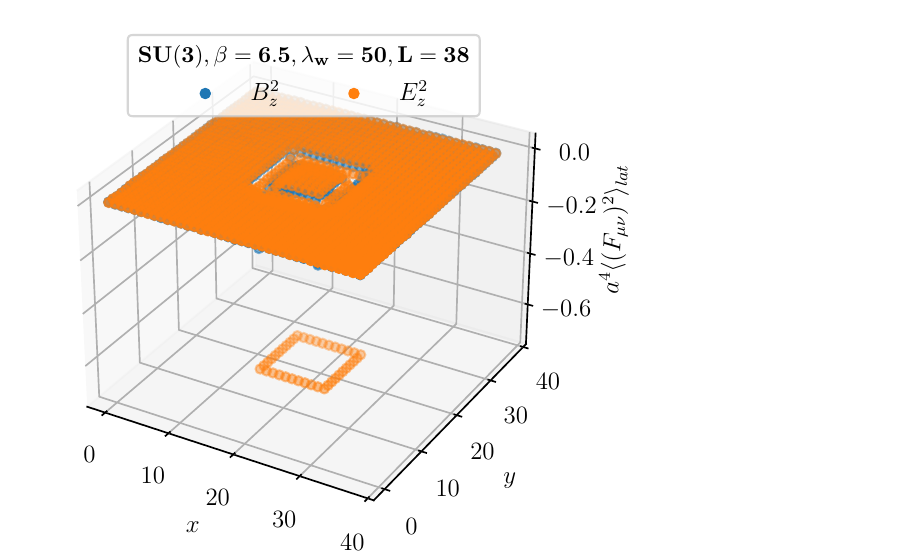} }}%
    \hspace{-2cm}
    \subfigure{{\includegraphics[scale=0.5]{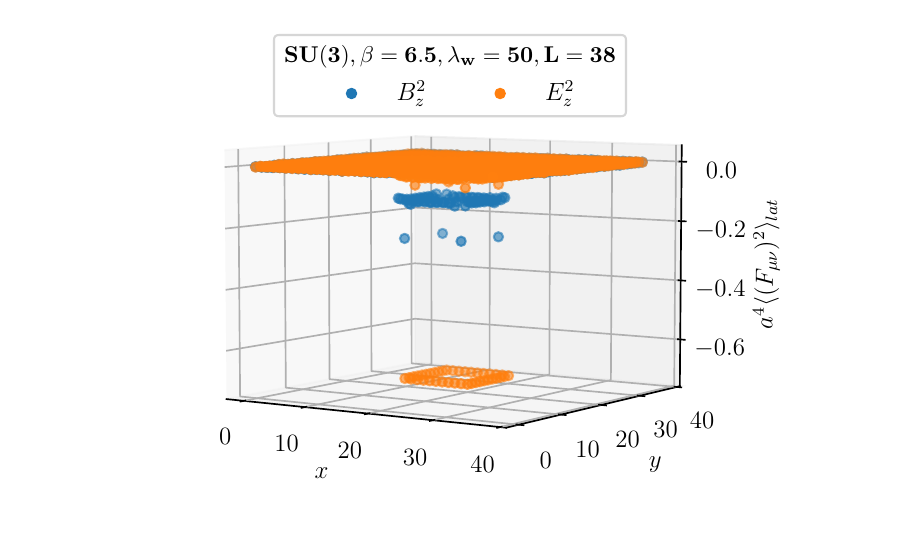} }}%
    \caption{Three dimensional expectation values of the $B_z^2$ and $E_z^2$ field components in $(3+1)$D SU(3) for a symmetrical tube with side lengths $R_{lat}=10$ and faces at $x=y=14$ and $x=y=24$.}%
    \label{fig:fields3D_tube_BzEz}
\end{figure}

\noindent
On the face of the tube, the fields experience maximal suppression, which manifests as a cavity in field configuration space. While it is difficult to clearly discern all six field components as they appear on the legend, it is evident that these field components are equal as per the F3-symmetry. The only exceptions to this are the $\hat{z}$ components of the electric and magnetic fields, whose behaviour is made clear on the right frame. To provide better visual clarity, in Fig.\ (\ref{fig:fields3D_tube_BxEy} - \ref{fig:fields3D_tube_BzEz}), we show the equivalent field components.\\

\noindent
It is clear from Fig.\ (\ref{fig:fields3D_tube_BxEy}) that $ \langle B_x^2 \rangle = \langle E_y^2 \rangle$ and Fig.\ (\ref{fig:fields3D_tube_ByEx}) that $\langle B_y^2 \rangle = \langle E_x^2 \rangle$. Note the different orientations of the suppressed fields (i.e., different field components suppressed on different sides of the tube) showing that the effect of the artificially modified gauge fields on the surface of the tube depends on the position along the tube based on the affected plaquettes. Lastly, we see in Fig.\ (\ref{fig:fields3D_tube_BzEz}) that $\langle B_z^2 \rangle \neq \langle E_z^2 \rangle$, which are the remaining components contributing to the energy density of the system.\\

\begin{figure}[!htb]
    \centering
    \subfigure[$l=2$ Above Top Face of Tube]{{\includegraphics[scale=0.5]{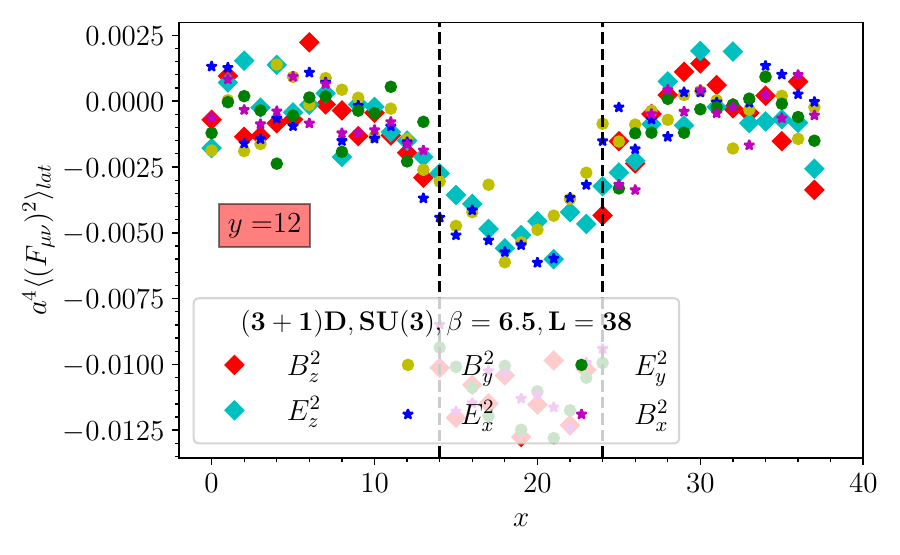} }}%
    \hspace{-0.45cm}
    \subfigure[$l=1$ Above Top Face of Tube]{{\includegraphics[scale=0.5]{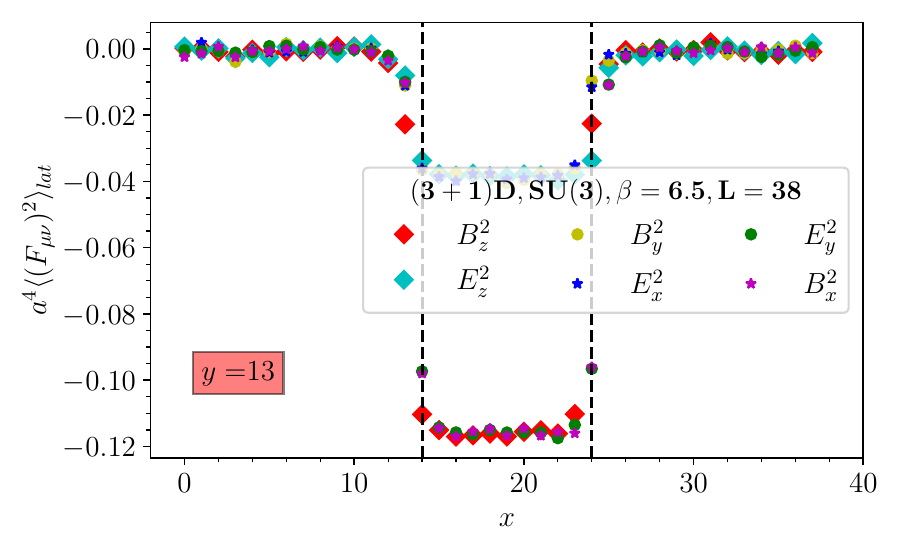} }}%
    \hspace{-0.45cm}
    \subfigure[$l=0$ On Top Face of Tube]{{\includegraphics[scale=0.5]{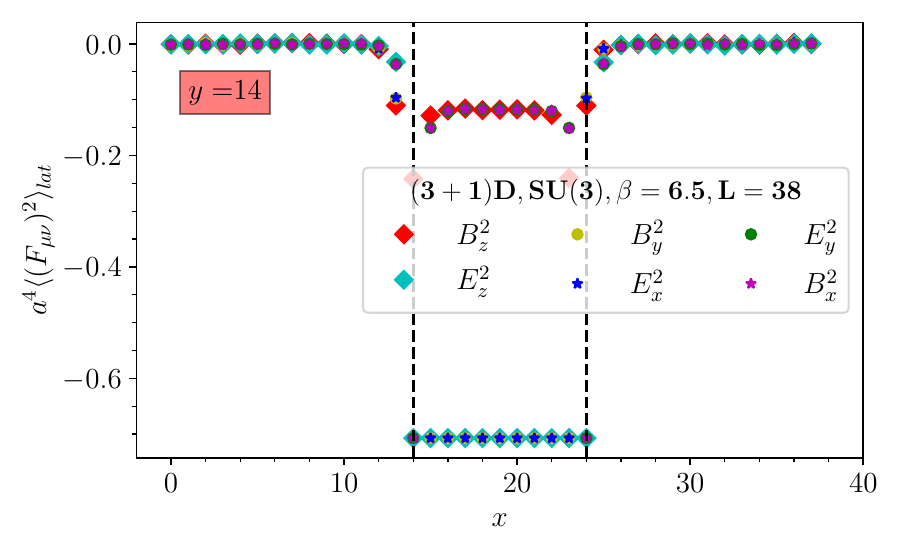} }}%
    \hspace{-0.45cm}
    \subfigure[$l=1$ Below Top Face of Tube]{{\includegraphics[scale=0.5]{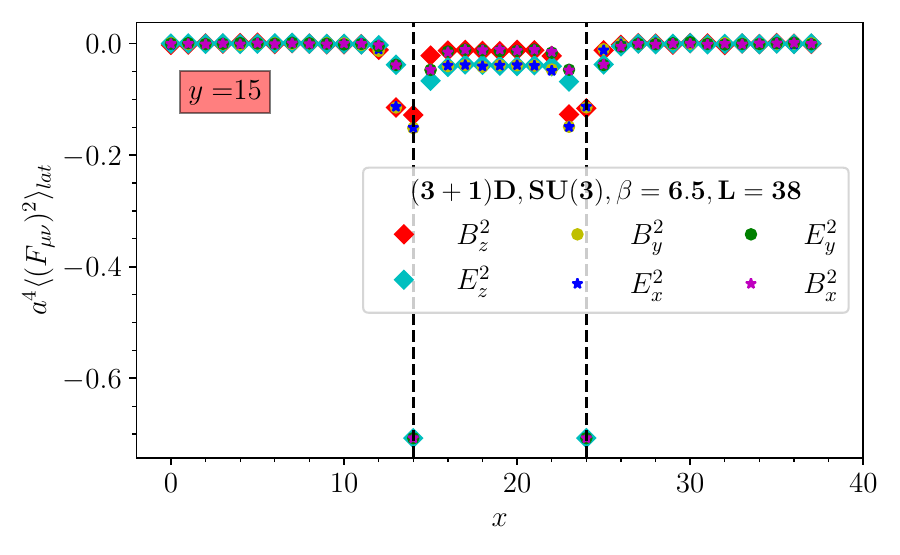} }}%
    \caption{Cross sections of expectation values of the squared field strength tensor components in $(3+1)$D SU(3) for a symmetrical tube with side lengths $R_{lat}=10$ around the top face at $y=14$ for all $x$.}%
    \label{fig:fields3D_tube_csout}
\end{figure}

\noindent
We have seen that the $\hat{x}$ and $\hat{y}$ electric and magnetic field components are suppressed equally, however, this is not true for the $\hat{z}$ components where the suppression of the electric fields is greater than that of the magnetic fields. We also observe a distinguishable contribution to the $B_z^2$ components coming from each of the tube's vertices and these contributions can be attributed to the $xy$ plaquettes on the hollow ends of the tube. While these plaquettes are not artificially modified in the construction of the tube geometry, they are attached to link variables that are affected.\\

\begin{figure}[!htb]
    \centering
    \subfigure[$l=2$ Above Bottom Face of Tube]{{\includegraphics[scale=0.5]{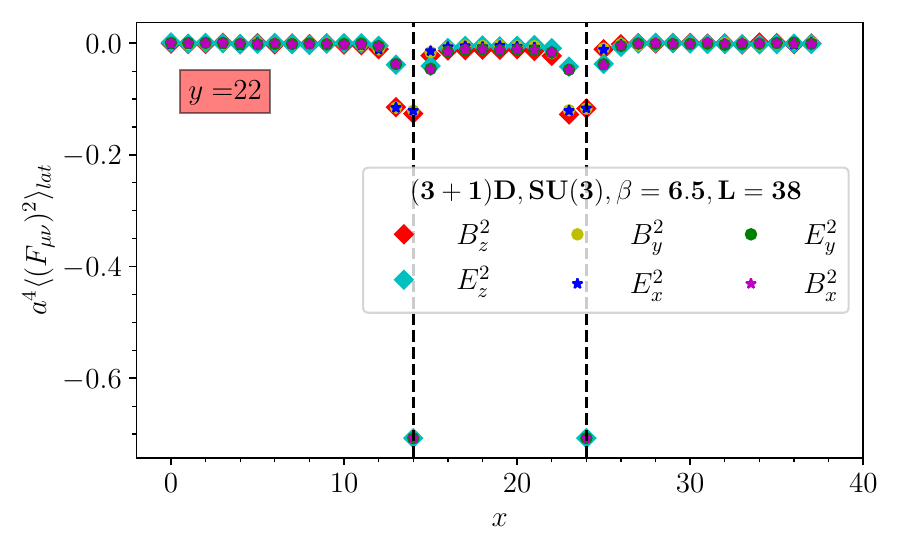} }}%
    \hspace{-0.45cm}
    \subfigure[$l=1$ Above Bottom Face of Tube]{{\includegraphics[scale=0.5]{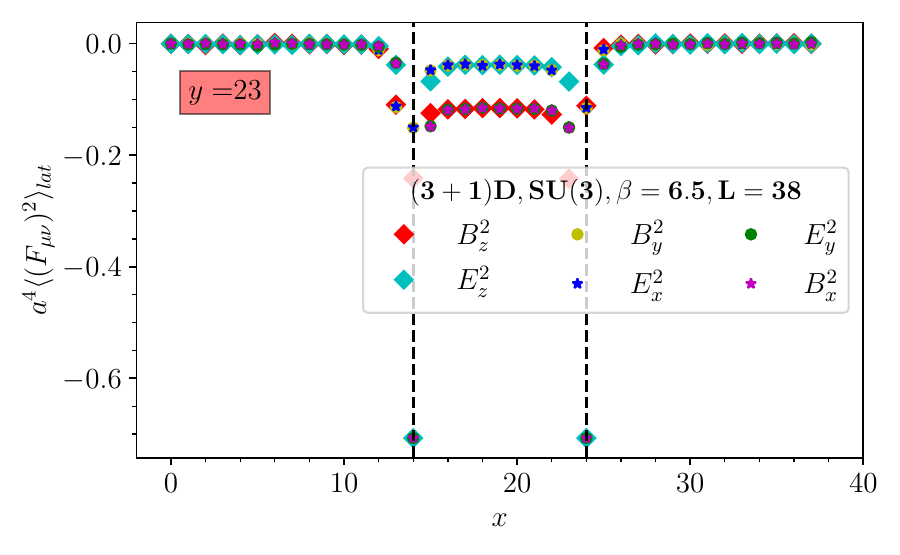} }}%
    \hspace{-0.45cm}
    \subfigure[$l=0$ On Bottom Face of Tube]{{\includegraphics[scale=0.5]{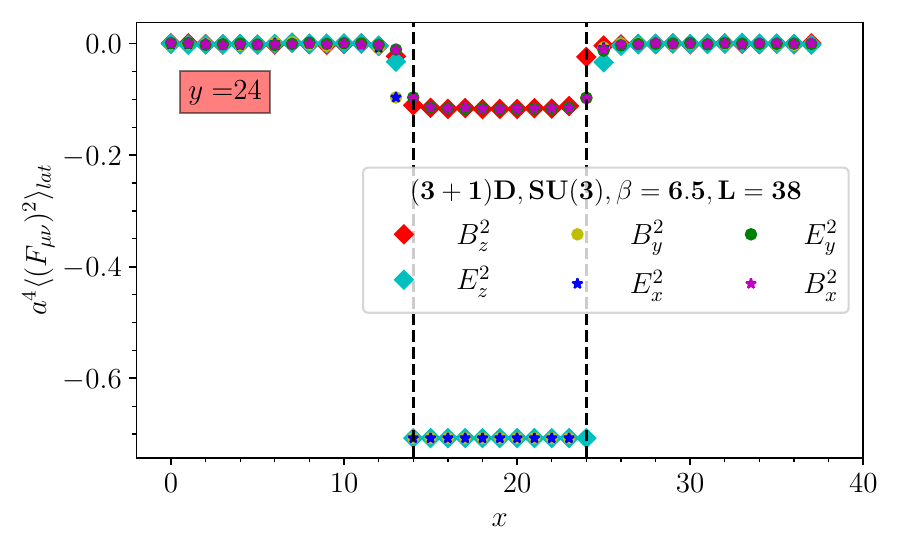} }}%
    \hspace{-0.45cm}
    \subfigure[$l=1$ Below Bottom Face of Tube]{{\includegraphics[scale=0.5]{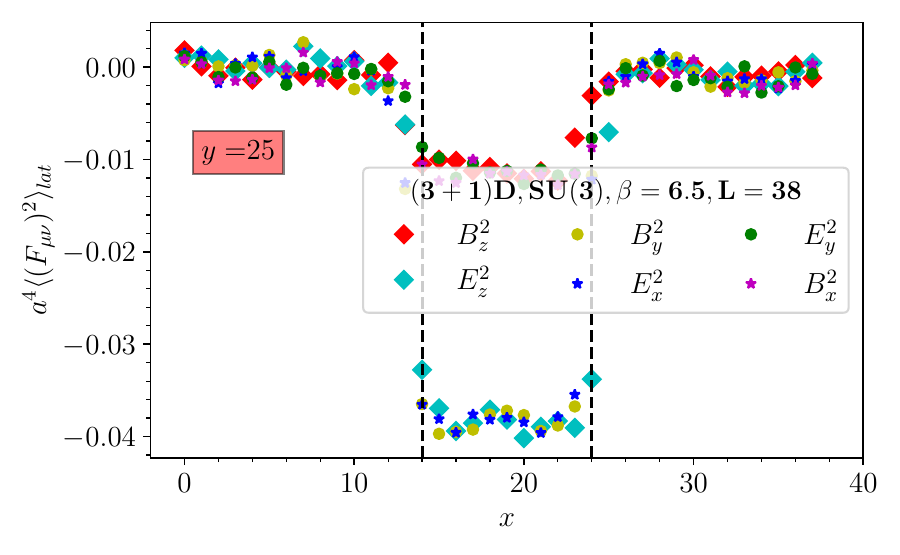} }}%
    \caption{Cross sections of expectation values of the squared field strength tensor components in $(3+1)$D SU(3) for a symmetrical tube with side lengths $R_{lat}=10$ around the bottom face at $y=24$ for all $x$.}%
    \label{fig:fields3D_tube_csin}
\end{figure}

\noindent
We wrap up this section by looking at cross-sectional slices of the 3D field expectation values shown in Fig.\ (\ref{fig:fields3D_tube_all}) along the $\hat{x}$ axis. In Fig.\ (\ref{fig:fields3D_tube_csout}) we take cross-sections as we move from outside to inside the tube, while in Fig.\ (\ref{fig:fields3D_tube_csin}), we move from inside to outside. The qualitative features of these cross-sections are consistent with our observations in the case of parallel plates, and the F3-symmetry is made explicit. The main notable difference is that at a distance $l=1$ away from the face, the degree of suppression of the fields is greater inside than outside... \\

\section{Geometry and Symmetries: Box}
\label{sec:Geometry and Symmetries: Box}

\begin{figure}[!htb]
    \centering
    \subfigure[Symmetric Box]{{\includegraphics[scale=0.48]{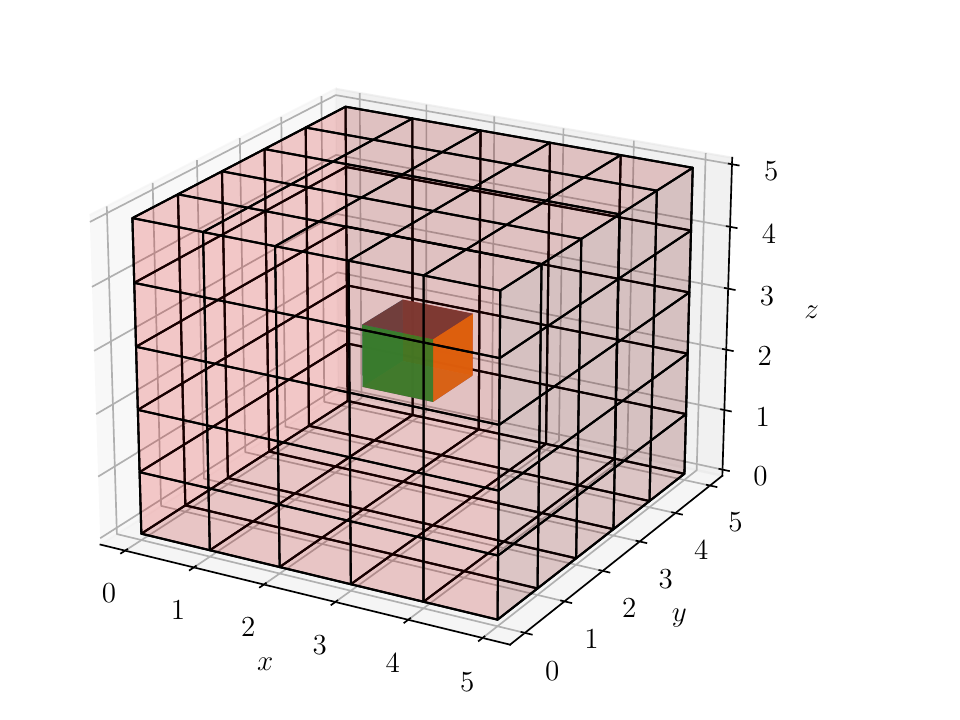} }}%
    \hspace{-1cm}
    \subfigure[Asymmetric Box]{{\includegraphics[scale=0.48]{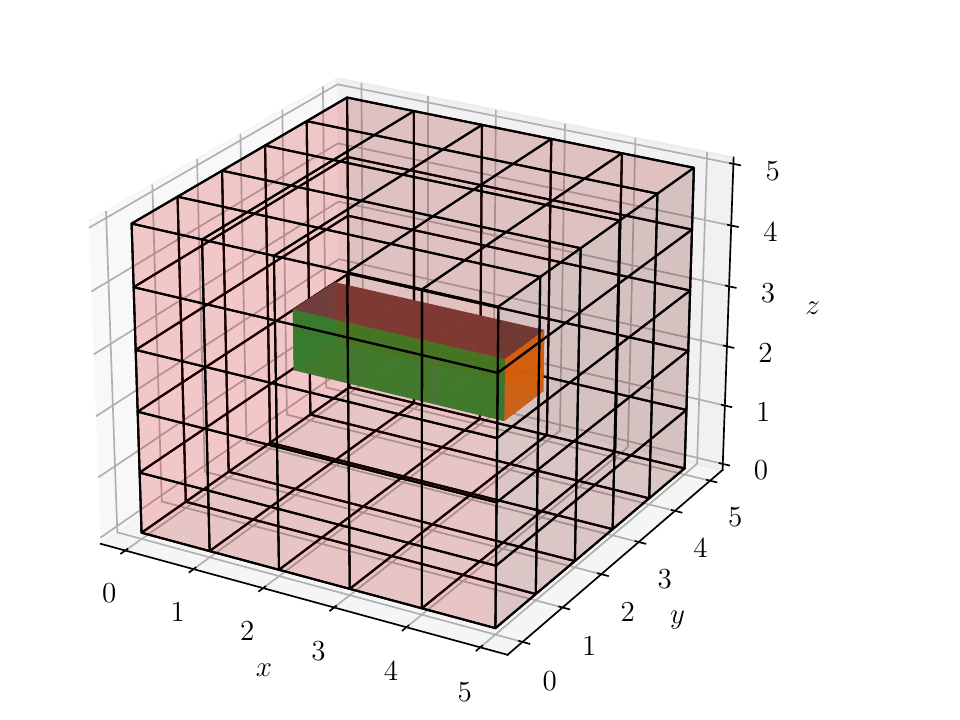} }}%
    \caption{Geometry of chromoelectric boundary conditions on a box in a $(3+1)$D cubic lattice.}%
    \label{fig:geometry_box}
\end{figure}

\noindent
The last geometry that we explore in our Casimir effect studies is a symmetrical and an asymmetrical box shown in Fig.\ (\ref{fig:geometry_box}). The symmetrical box is a cube of side length, $l=R$, while the asymmetric box is a right square prism with two fixed side lengths $R_y=R_z=1$ (square base) and a third side with length, $R_x=R$. In this geometry, all rotational symmetries are now broken because we have fixed the spatial 3D axis, leaving only the temporal direction free. We will only discuss the electromagnetic field configurations for the symmetric box, however, the field components for the asymmetric box contain the same features.\\

\begin{figure}[!htb]
    \centering
    \subfigure[$z=10a$]{{\includegraphics[scale=0.5]{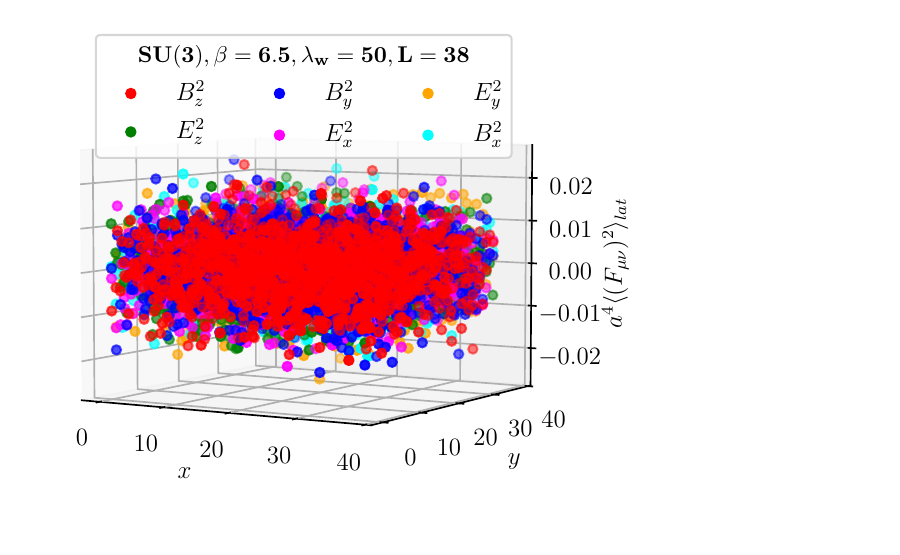} }}%
    \hspace{-2cm}
    \subfigure[$z=13a$]{{\includegraphics[scale=0.5]{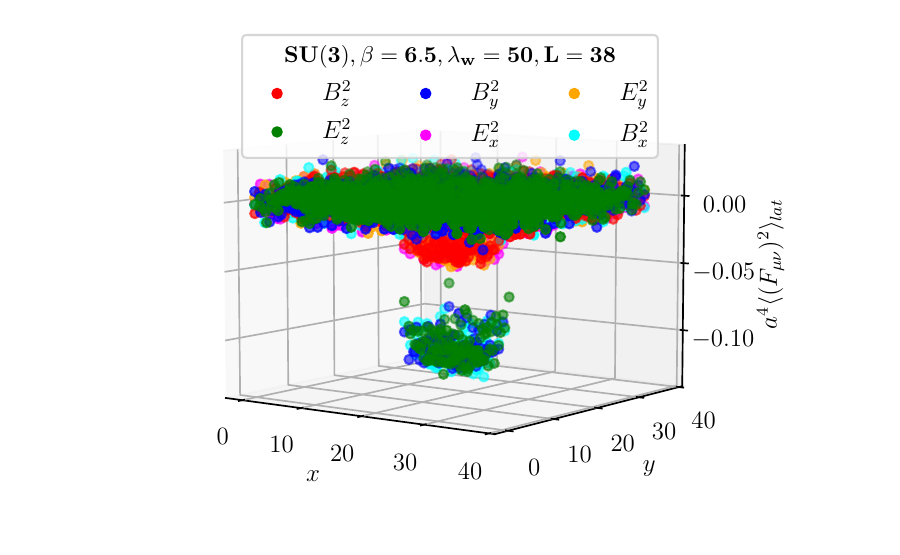} }}%
    \caption{Three dimensional expectation values of the squared field strength tensor components in $(3+1)$D SU(3) outside a symmetrical box with side lengths $R_{lat}=10$ and faces at $x=y=z=14$ and $x=y=z=24$.}%
    \label{fig:fields3D_box_out}%
\end{figure}

\noindent
While it is possible that there exists a nontrivial equivalence of the field expectation values at various localities on the box, we do not find a general analytical expression based on the geometry of the problem. In Fig.\ (\ref{fig:fields3D_box_out}), we show a three dimensional representation of the squared field expectation values in-vacuum\footnote{Far enough outside the box, the surrounding plaquettes do not experience any plaquette contributions from the boundaries of the box.} for a symmetrical box at two cross sections in the $\hat{z}$ direction. On the left frame, a distance $R_{lat}=4$ away from the box, the surrounding gauge fields retain their in-vacuum expectation values and fluctuate around zero in this case because they are vacuum normalised. The various electromagnetic field components are indistinguishable.\\

\noindent
On the right frame of Fig.\ (\ref{fig:fields3D_box_out}), a distance $R_{lat}=1$ away from the lower face of the box, we start seeing a suppression of some of the field components at varying degrees. The intensity of suppression of the individual electromagnetic field components is dependent on the position around the box (i.e., different components are suppressed at varying intensities around the different faces depending on the contributing plaquettes). In this region, the in-vacuum plaquettes are now attached to link variables that are in contact with the boundaries of the box.\\

\begin{figure}[!htb]
    \centering
    \subfigure{{\includegraphics[scale=0.5]{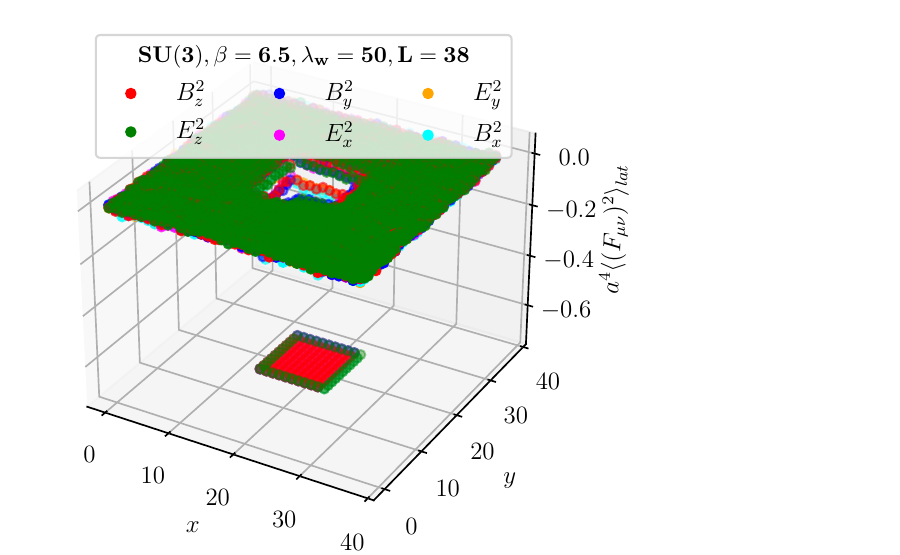} }}%
    \hspace{-2cm}
    \subfigure{{\includegraphics[scale=0.5]{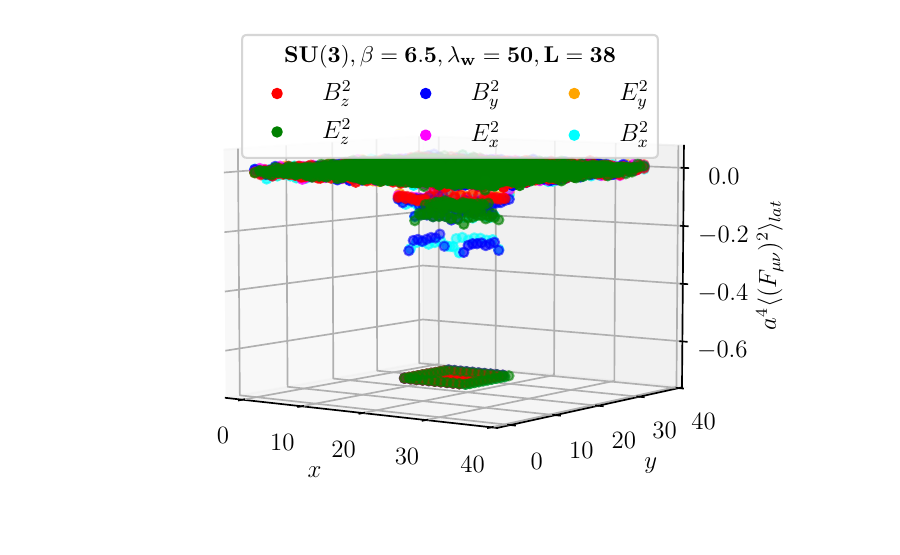} }}%
    \caption{Three dimensional expectation values of the squared field strength tensor components in $(3+1)$D SU(3) on the face (at $z=14$) of a symmetrical box with side lengths $R_{lat}=10$ and faces at $x=y=z=14$ and $x=y=z=24$.}%
    \label{fig:fields3D_box_face}%
\end{figure}

\noindent
In Fig.\ (\ref{fig:fields3D_box_face}), we show a cross section of the electromagnetic field components on the bottom face of the box at $z=14$. As opposed to observing maximally strong field suppression on the walls of the box, the fields are now suppressed equivalently on the whole face. Similarly to the tube geometry, the maximally suppressed field components depend on the face under consideration. In this case, we are looking at a face on the $xy$-plane, therefore the $E_x$, $E_y$ and $B_z$ experience maximal suppression on this face.\\

\begin{figure}[!htb]
    \centering
    \subfigure{{\includegraphics[scale=0.5]{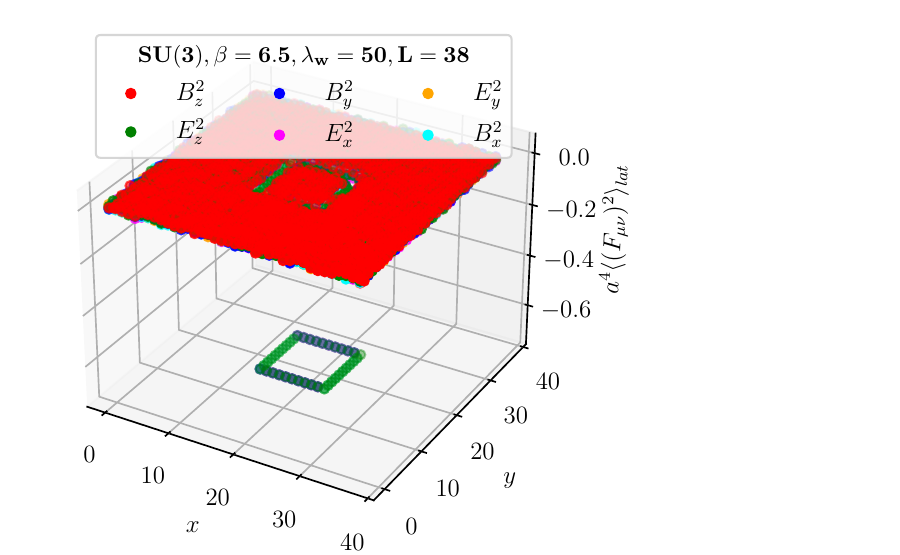} }}%
    \hspace{-2cm}
    \subfigure{{\includegraphics[scale=0.5]{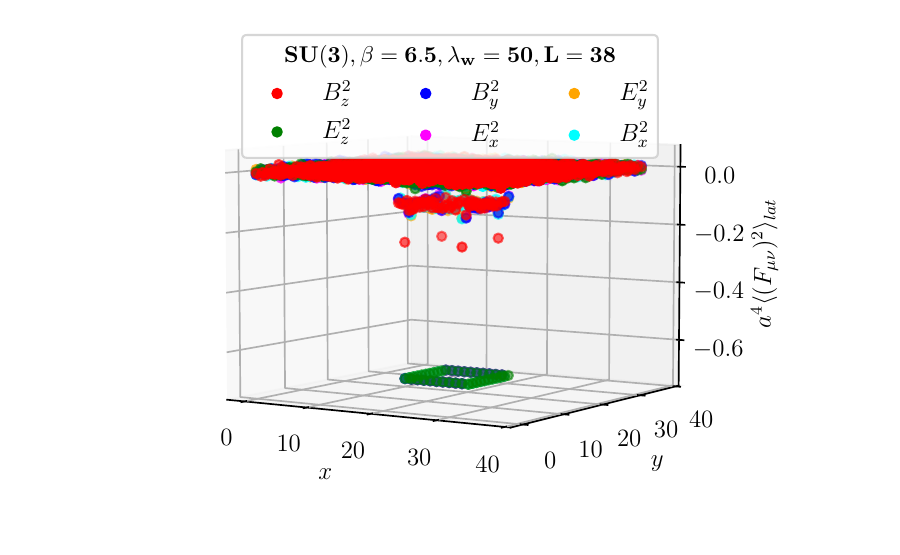} }}%
    \caption{Three dimensional expectation values (at $z=20$) of the squared field strength tensor components in $(3+1)$D SU(3) inside a symmetrical box with side lengths $R_{lat}=10$ and faces at $x=y=z=14$ and $x=y=z=24$.}%
    \label{fig:fields3D_box_in}%
\end{figure}

\begin{figure}[!htb]
    \centering
    \subfigure{{\includegraphics[scale=0.5]{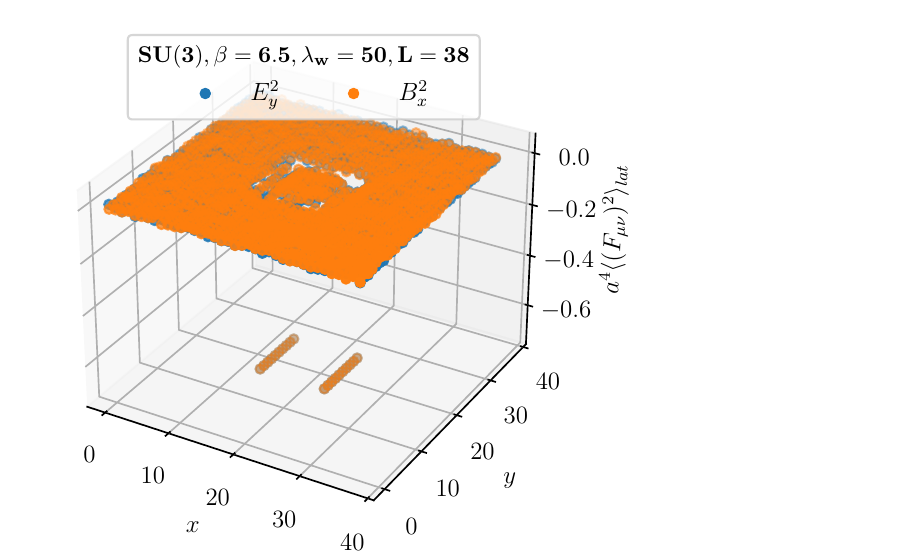} }}%
    \hspace{-2cm}
    \subfigure{{\includegraphics[scale=0.5]{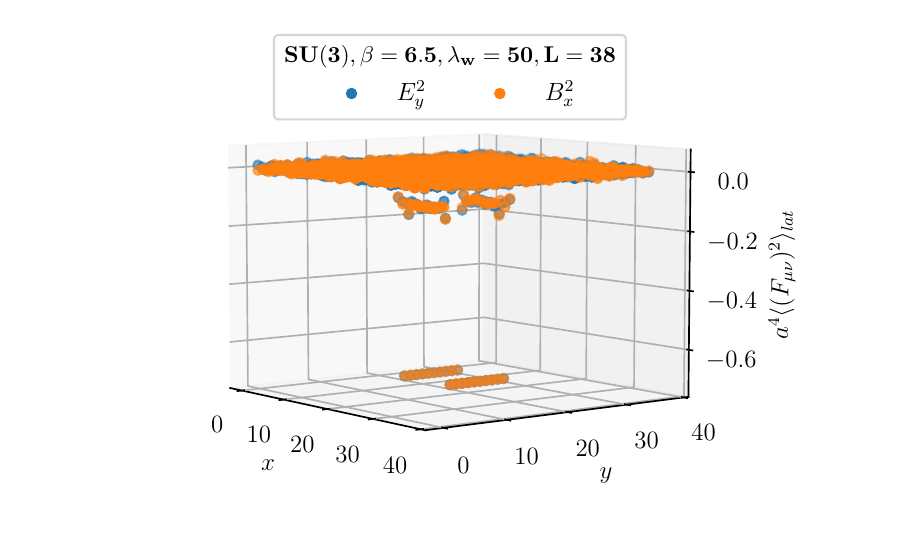} }}%
    \caption{Three dimensional expectation values (at $z=20$) of the $E_y^2$ and $B_x^2$ field components in $(3+1)$D SU(3) inside a symmetrical box with side lengths $R_{lat}=10$ and faces at $x=y=z=14$ and $x=y=z=24$.}%
    \label{fig:fields3D_box_BxEy}%
\end{figure}

\begin{figure}[!htb]
    \centering
    \subfigure{{\includegraphics[scale=0.5]{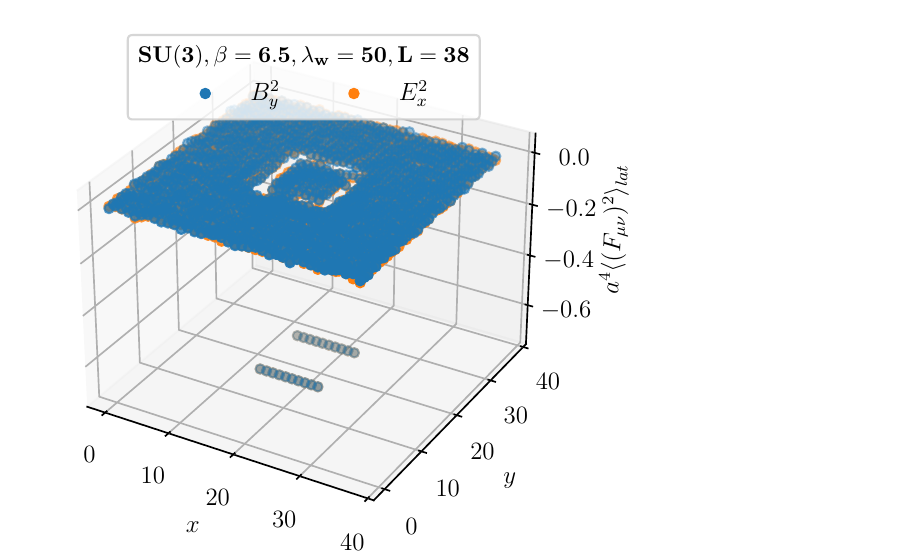} }}%
    \hspace{-2cm}
    \subfigure{{\includegraphics[scale=0.5]{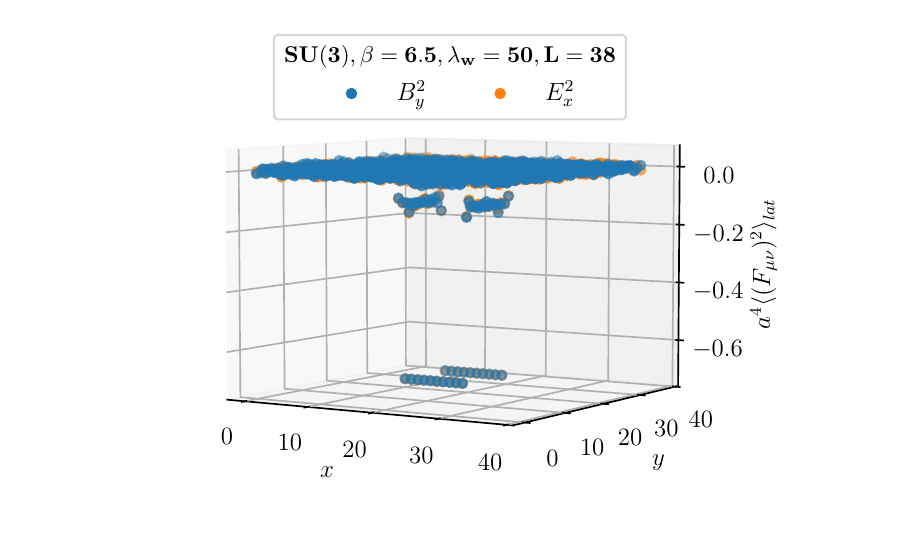} }}%
    \caption{Three dimensional expectation values (at $z=20$) of the $B_y^2$ and $E_x^2$ field components in $(3+1)$D SU(3) inside a symmetrical box with side lengths $R_{lat}=10$ and faces at $x=y=z=14$ and $x=y=z=24$.}%
    \label{fig:fields3D_box_ByEx}%
\end{figure}

\begin{figure}[!htb]
    \centering
    \subfigure{{\includegraphics[scale=0.5]{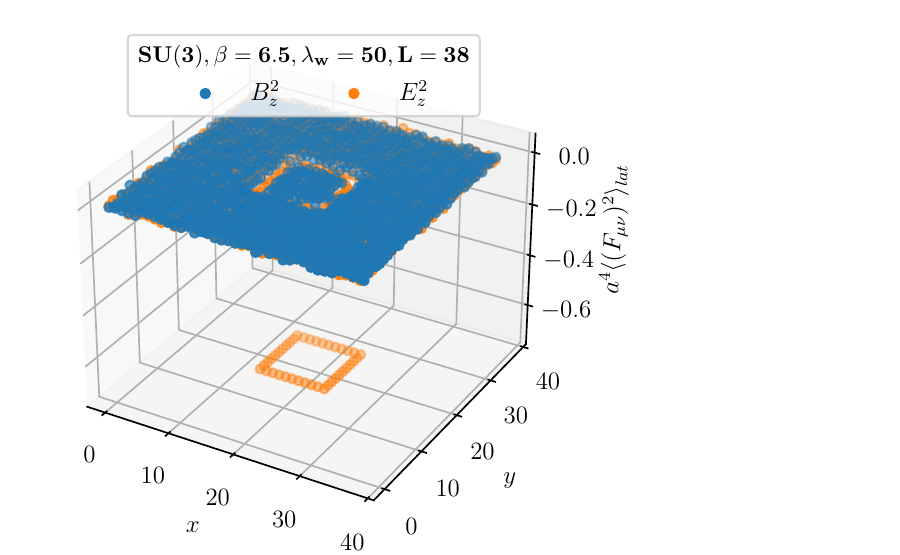} }}%
    \hspace{-2cm}
    \subfigure{{\includegraphics[scale=0.5]{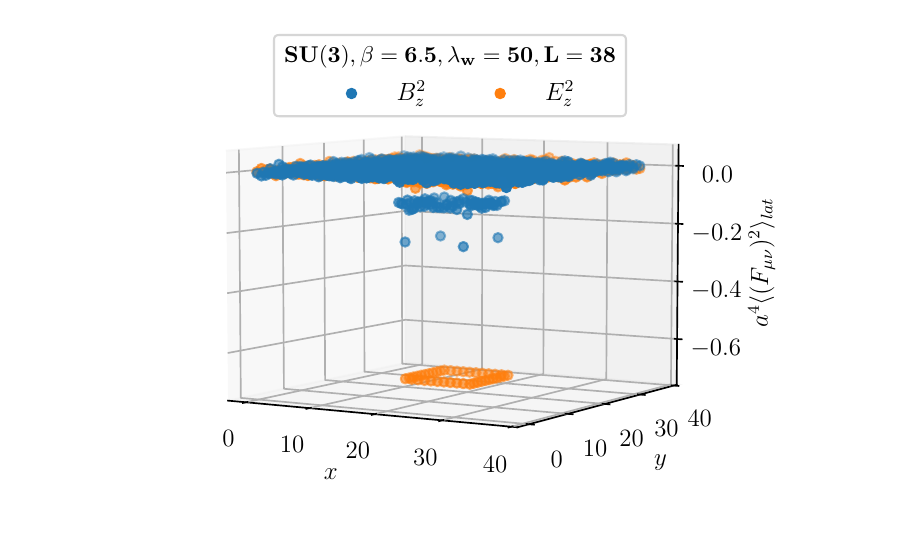} }}%
    \caption{Three dimensional expectation values (at $z=20$) of the $B_z^2$ and $E_z^2$ field components in $(3+1)$D SU(3) inside a symmetrical box with side lengths $R_{lat}=10$ and faces at $x=y=z=14$ and $x=y=z=24$.}%
   \label{fig:fields3D_box_BzEz}%
\end{figure}

\noindent
We then show a cross sectional slice of the field components taken inside the box at $z=14$ in Fig.\ (\ref{fig:fields3D_box_in}). We observe that the field expectation values are reminiscent of the field behaviour in the symmetrical tube geometry\footnote{See Fig.\ (\ref{fig:fields3D_tube_all} - \ref{fig:fields3D_tube_BzEz}) for the electromagnetic fields of a symmetric tube in comparison to Fig.\ (\ref{fig:fields3D_box_in} - \ref{fig:fields3D_box_BzEz}) for the fields on a symmetrical box.}. Such behaviour is expected because in the limit of a large box symmetric box, an observation made on a cross-section inside the box (far enough from the parallel faces) should be indistinguishable from a symmetric tube. At that position, the other two faces are not connected to any nearby plaquettes and their contributions are negligible.\\

\noindent
Similarly, an asymmetric box at large $R$ and far from the faces on the longest side should have fields consistent with the smallest possible symmetric tube (i.e., $R_{lat}=1$). We will discuss the effect of the large $R$ limit for a box on the Casimir pressure of the system in the next chapter. In Fig.\ (\ref{fig:fields3D_box_BxEy} - \ref{fig:fields3D_box_BzEz}) we emphasise the resemblance of the symmetric box and tube at large $R$ by showing that inside the box, the tube's electromagnetic field F3-symmetries in Eq.\ (\ref{eqn:F3}) are recovered.\\

\begin{figure}[!htb]
    \centering
    \subfigure[$l=2$ Outside Front-Bottom Edge of Box]{{\includegraphics[scale=0.5]{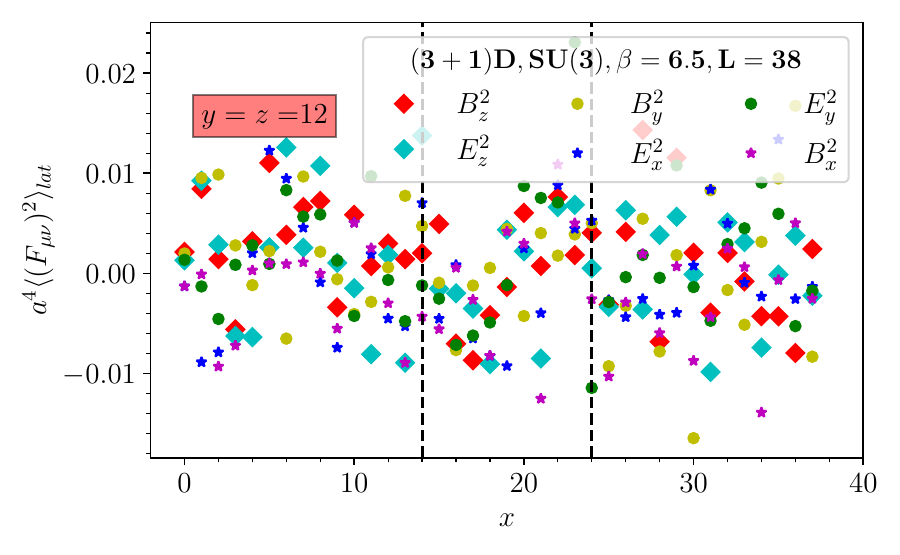} }}%
    \hspace{-0.45cm}
    \subfigure[$l=1$ Outside Front-Bottom Edge of Box]{{\includegraphics[scale=0.5]{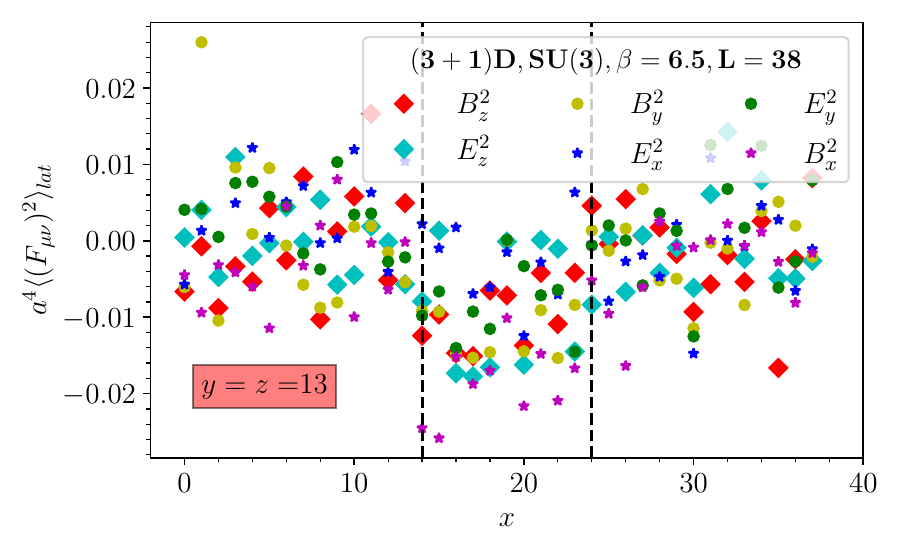} }}%
    \hspace{-0.45cm}
    \subfigure[$l=0$ On Front-Bottom Edge of Box]{{\includegraphics[scale=0.5]{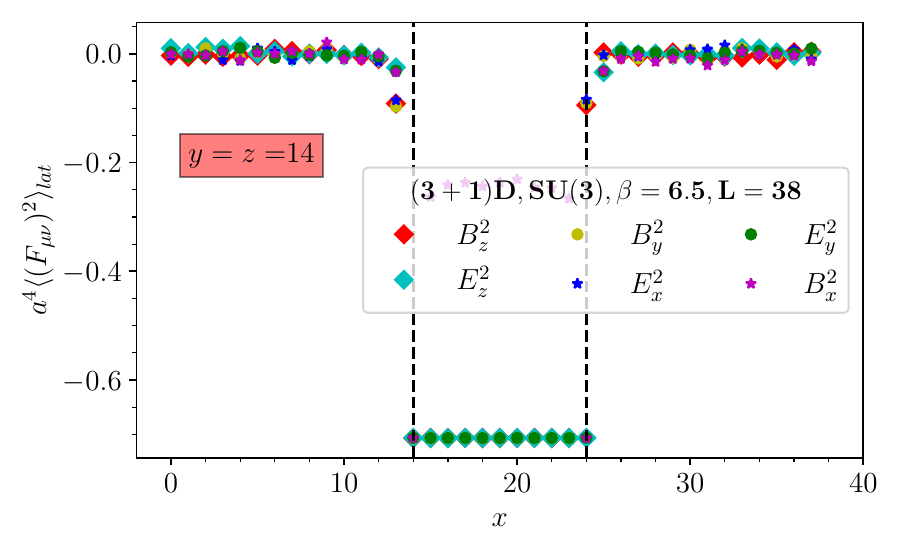} }}%
    \hspace{-0.45cm}
    \subfigure[$l=1$ Inside Front-Bottom Edge of Box]{{\includegraphics[scale=0.5]{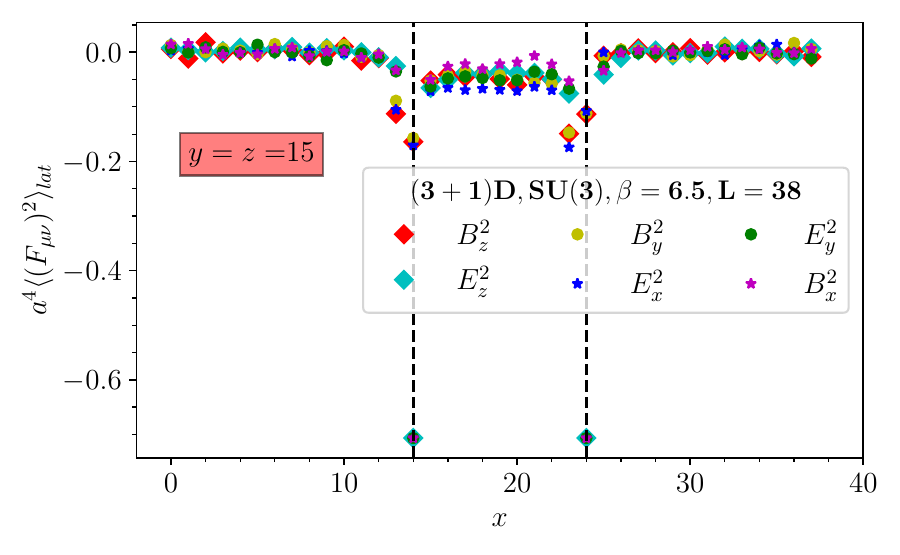} }}%
    \caption{Cross sections of expectation values of the squared field strength tensor components in $(3+1)$D SU(3) for a symmetrical box with side lengths $R_{lat}=10$ around the front and bottom edge at $y=z=14$ for all $x$.}%
    \label{fig:fields3D_box_csout}
\end{figure}

\begin{figure}[!htb]
    \centering
    \subfigure[$l=2$ Inside Back-Top Edge of Box]{{\includegraphics[scale=0.5]{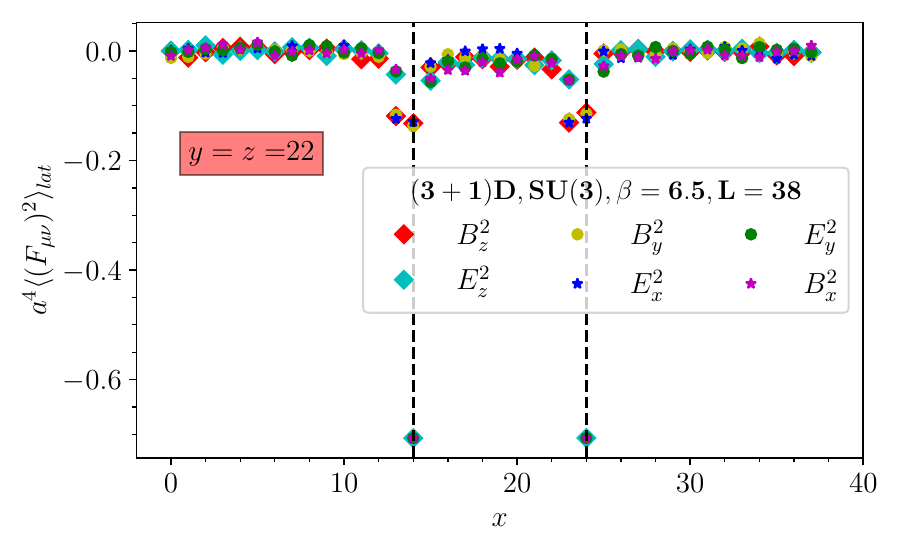} }}%
    \hspace{-0.45cm}
    \subfigure[$l=1$ Inside Back-Top Edge of Box]{{\includegraphics[scale=0.5]{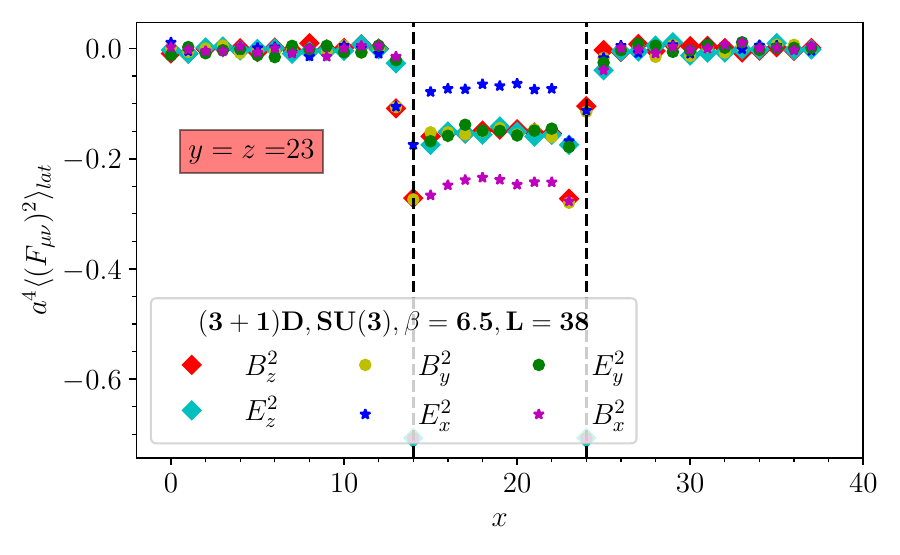} }}%
    \hspace{-0.45cm}
    \subfigure[$l=0$ On Back-Top Edge of Box]{{\includegraphics[scale=0.5]{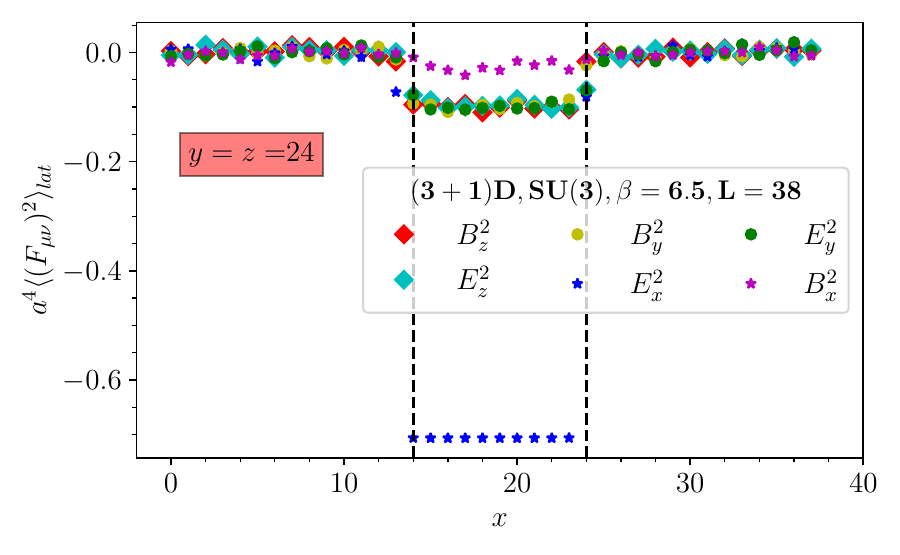} }}%
    \hspace{-0.45cm}
    \subfigure[$l=1$ Outside Back-Top Edge of Box]{{\includegraphics[scale=0.5]{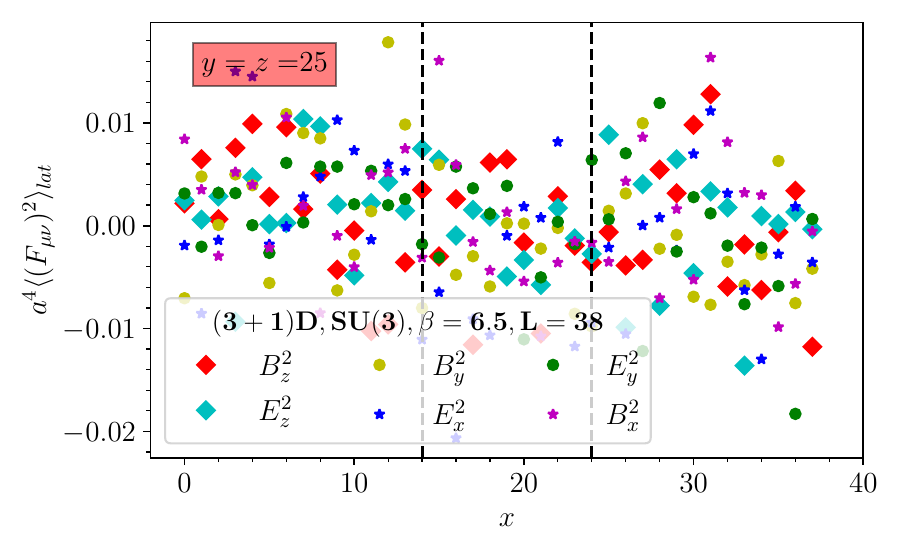} }}%
    \caption{Cross sections of expectation values of the squared field strength tensor components in $(3+1)$D SU(3) for a symmetrical box with side lengths $R_{lat}=10$ around the back and top edge at $y=z=24$ for all $x$.}%
    \label{fig:fields3D_box_csin}
\end{figure}

\noindent
Lastly, we provide the cross-section slices of the electromagnetic field components taken at intersecting points in both the $\hat{y}$ and $\hat{z}$ directions along the $\hat{x}$-axis inside of the symmetric box. Following a similar structure used for the tube, we show the fields as we move from outside to inside the box in Fig.\ (\ref{fig:fields3D_box_csout}), followed by fields moving from inside to outside the box in Fig.\ (\ref{fig:fields3D_box_csin}). The qualitative features are consistent with our expectations from the first two cases, with field fluctuations around zero in-vacuum and varying degrees of suppressed field components on the boundaries. \\

\noindent
We also observe that there are field expectation values that are equivalent \textit{inside the box} as already seen in the three-dimensional representations where, $ \langle B_x^2 \rangle = \langle E_y^2 \rangle$ and $\langle B_y^2 \rangle = \langle E_x^2 \rangle$. This is intuitive because the same field components are equal in the symmetrical tube geometry and there are geometrical similarities at the cross section taken inside the box with no contributions from the perpendicular walls. We do not draw a generalised conclusion on the symmetries of the field components in a box as we have not derived a representative analytical expression. However, in future studies it may be worth investigating whether any other symmetries can be explored in this geometry that directly point out to the observed field behaviour.\\

\chapter{The Casimir Potential}
\label{chapter:The Casimir Potential}

\section{Parallel Wires in (2+1)D}
\label{section:Parallel Wires in (2+1)D}

\noindent
While we aim to reproduce the results of Ref.\ \cite{Chernodub:2018pmt} as a starting point, there is a strong motivation for studying (2+1)D SU($N_c$) gauge theories. Primarily because they have similar characteristics to their (3+1)D counterparts but are also technically easier to explore, and studying one teaches us something about the other. Some similarities include the ultraviolet freedom at short distances, consequently, both theories are strongly coupled at large separation distances and need to be explored non-perturbatively in this regime. The three-dimensional theory is only asymptotically free, whereas in (2+1)D, interactions decay more rapidly with decreasing separation distance.\\

\noindent
This difference in the rate at which the interactions vanish with decreasing distance is attributed to the distinct scaling properties of their gauge couplings. In the three-dimensional theory, the coupling is dimensionless and the interaction strength diminishes logarithmically, $g^2(l)\sim 1/\ln(l\Lambda)$ with decreasing distance (increasing energy), where $l$ is the length scale. Meanwhile in the two-dimensional theory, the coupling constant, $g^2$ has dimensions of mass and the effective dimensionless expansion parameter diminishes more rapidly with as $l\to 0$ \cite{Teper:1998te}. \\ 

\noindent
In addition to having similar coupling behaviour, both theories have propagating degrees of freedom and a similar mass spectra \cite{Teper:1998te}. The Coulomb potential in (2+1)D has a weakly confining logarithmic form, $V(r) \sim g^2\ln(r)$ obtained from Gauss' law analysis. However, at large separation distances, it has been shown in lattice calculations that the large separation confinement behaviour following a nonperturbative linear potential i.e., $V(r)\sim \sigma r$ in (3+1)D gauge theories is also observed in their two-dimensional counterparts \cite{Bagan:2000nc, Gaete:2014esa}. Lastly, the (2+1)D gauge theory has direct implications to the high temperature limit of the three-dimensional theory \cite{Gross:1980br}. An example of this is that the Casimir mass found in the former represents the high temperature magnetic screening mass in the latter \cite{Karabali:2018ael}. \\

\noindent
In the preceding chapter, we provided an expression for the (2+1)D theory energy density in Euclidean space, see Eq.\ (\ref{eqn:2deuclidean_energy_density}). Taking the expectation values of the field components and applying the rotational symmetry condition in Eq.\ (\ref{eqn:fields_F2D}), the energy density is given by
\begin{eqnarray}
    \varepsilon(x)_{\text{wires}} &=& \frac{1}{2} \left[ \langle B_z^2 \rangle - \langle E_x^2 \rangle - \langle E_y^2 \rangle \right]\\
     &=& -\frac{1}{2}\langle E_y^2 \rangle,
\end{eqnarray}
and is a function of $x$, the axis orthogonal to the wires in which the Casimir force acts. On the lattice, this analytical form is evaluated by computing the expectation value of the action contribution from the relevant plaquettes, the $yt$-plaquettes which sit on the worldsurface of the wires in this case,
\begin{eqnarray}
     \varepsilon(x)_{\text{wires}}^{\text{lat}} &=& -\langle S_{P_{yt}} \rangle = \frac{1}{N_{\tau}} \sum\limits_{N_{y}} \sum\limits_{N_{\tau}} S_{P_{yt}}.
\end{eqnarray}
This is similar to how we express the field strength tensor components along the $x$-axis in chapter (\ref{section:Geometry and Symmetries: Wires in $(2+1)$D}). We work on an isotropic lattice and perform an ensemble average over $N_{\tau}$ to get the total energy of the system without introducing any additional length scale. \\

\noindent
Note that the energy density, expressed by the action average, $\langle S_{P_{yt}} \rangle$ does not directly depend on the individual plaquettes, $P_{yt}$, due to the summation, but we use this notation\footnote{This notation is proposed in the work of Chernodub et al.\ in Ref.\ \cite{Chernodub:2018pmt}.} to make explicit that only contributions of selected plaquettes with suitable orientations are used to compute the energy of the system. The corresponding Casimir potential is given by the normalised integral of the energy density,
\begin{eqnarray}
    V^{\text{lat}}_{\text{Cas}}(R) &=& \left[ \int dx\, \varepsilon(x)_{\text{wires}} \right]_{R-R_0} = \left[ \sum\limits_{N_{x}} \varepsilon(x)_{\text{wires}}^{\text{lat}} \right]_{R-R_0}\\
    &=&  \sum\limits_{N_{x}} \left[ \langle S_{P_{yt}} \rangle_R - \langle S_{P_{yt}} \rangle_{R_0} \right] = -\langle \langle S_{P_{yt}} \rangle \rangle,
    \label{eqn:2d_casimir_lattice}
\end{eqnarray}
and depends on the separation distance, $R$, between the wires, where $R_0$ is a renormalisation condition to be discussed shortly. On the lattice, the Casimir potential is equivalent to the normalised expectation value of the total energy of the system,
\begin{eqnarray}
    \langle \langle S_{P_{yt}} \rangle \rangle &=& \left[ \frac{1}{N_{\tau}} \sum\limits_{N_{x}} \sum\limits_{N_{y}} \sum\limits_{N_{\tau}}  S_{P_{yt}} \right]_{R-R_0}.
    \label{eqn:2d_action_lattice}
\end{eqnarray}
We highlight this definition for reasons that will become clear in the subsequent sections. The Casimir potential can be divided by the length of the wires, $N_y$, in order to obtain the Casimir energy per unit length of the wires.\\

\noindent
The normalisation is enforced by subtracting the lattice action expectation value with the parallel wires placed at separation distance, $R_0\to \infty$ apart, which on our finite periodic lattice we choose to be at half the lattice spatial extent, $R_0 = L/2$. Essentially, we compute the total energy of the system with wires placed a finite distance $R$ apart, and from this total energy, we subtract the total energy of the system with the wires placed far apart. At large enough distances between the wires, the Casimir energy will diminish leaving only energy contributions that need to be subtracted. We express the Casimir potential as,
\begin{eqnarray}
    V^{\text{lat}}_{\text{Cas}}(R) &=& E^{\text{Tot}}_R - E^{\text{Tot}}_{R_0}\\
    &=& [E_{\text{Cas}}(R) + E_{\text{Wires}} + E_{\text{Vac}}]_R - [E_{\text{Wires}} + E_{\text{Vac}}]_{R_0},
    \label{eqn:wires_energy_normalisation}
\end{eqnarray}
where the zero-point energy is independent of $R$, but so is the energy contribution from the wires. We will return to the discussion of the energy contributions from imposing boundary conditions in the subsequent sections where it contributes a non-constant value.\\

\noindent
It is clear from Eq.\ (\ref{eqn:wires_energy_normalisation}) that the normalisation condition takes care of two energy contributions to the system, the first is the cancellation of the vacuum ultraviolet divergences such that the energy density describes a local finite quantity. The second is the energy contribution from imposing the chromoelectric boundary condition (geometry of two parallel static wires) on the QCD vacuum, i.e., the act of creating the wires requires energy. Note that if one normalises by performing the vacuum subtraction (in the absence of the wires), the boundary energy contribution remains and the Casimir energy is not isolated.\\

\noindent
The physical Casimir potential is obtained by converting the lattice expression in Eq.\ (\ref{eqn:2d_casimir_lattice}) into physical units. The physical separation distance between the wires is given by $R_{\text{phys}} = aR_{\text{lat}}$, where $a$ is the physical lattice spacing. The corresponding scaling of the Casimir energy density with the physical lattice spacing is
\begin{eqnarray}
    V^{\text{phys}}_{\text{Cas}}(R_{\text{phys}}) &=& \frac{1}{a^2} V^{\text{lat}}_{\text{Cas}}(R_{\text{phys}}/a),
\end{eqnarray}
providing the correct dimensionality for the energy per unit length, $[V^{\text{phys}}_{\text{Cas}}(R_{\text{phys}})] = [1/\text{length}^2]$. A scale-improved expression for the Casimir energy density designed to provide corrections to finite size effects of the inverse coupling on the non-abelian Casimir potential in the continuum limit is proposed in Ref.\ \cite{Chernodub:2018pmt},
\begin{eqnarray}
    V_{\text{Cas}}(R_{\text{phys}})/\sigma &=& -\frac{1}{a^2\sigma } \left( \frac{\beta_I}{\beta} \right)^4 \langle \langle S_{P_{yt}} \rangle \rangle,
    \label{eqn:Vcas2D_formula_scaled}
\end{eqnarray}
where $\sigma$ is the zero-temperature confining Yang-Mills string tension and $a\sqrt{\sigma}$ is dimensionless, hence we express the Casimir potential as a dimensionless quantity. The mean-field-tadpole improved inverse coupling, $\beta_I$, is introduced to reduce finite-size corrections. We refer the reader to chapter (\ref{section:string_tension}) for a discussion on the introduction of physical units to lattice measurements using the string tension to set the physical scale, as well as the improved inverse coupling, $\beta_I$.\\

\begin{figure}[!htb]
\begin{center}
\includegraphics[scale=.7]{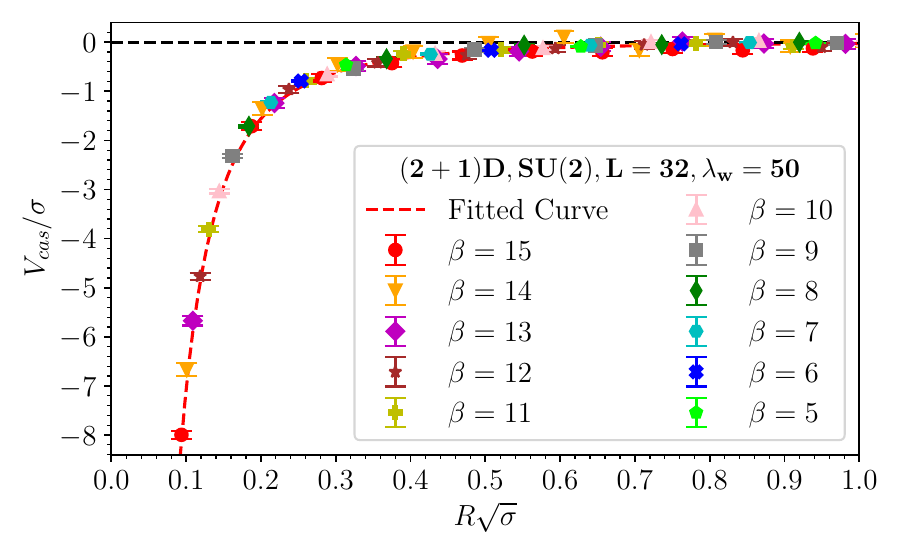}
\caption{The Casimir potential between two parallel wires separated by a physical distance $R\sqrt{\sigma}$ in $(2+1)$D SU(2) for different physical lattice volumes in units of the string tension.}
\label{fig:casimir_potential_2dsu2}
\end{center}
\end{figure}

\noindent
In lattice perturbation theory, the plaquette expectation value acquires additive and multiplicative radiative corrections from higher order contributions of quantum fluctuations of the gauge fields. Additive corrections stem from the ultraviolet divergent vacuum contributions to the energy and are removed through the subtraction scheme in Eq.\ (\ref{eqn:wires_energy_normalisation}). Multiplicative corrections arise from renormalisation effects, which in this case is the scaling of the plaquette expectation value to obtain the physically relevant field strength tensor components, $\beta^4 \langle \text{Tr} U_{\mu\nu} \rangle \sim a^{-4} \langle \text{Tr} U_{\mu\nu} \rangle \sim \langle F^2_{\mu\nu} \rangle_{\text{phys}} $. The factor containing the ratio of the inverse coupling in Eq.\ (\ref{eqn:Vcas2D_formula_scaled}) improves the finite-size scaling of the Casimir potential.  \\

\noindent
Based on Eq.\ (\ref{eqn:Vcas2D_formula_scaled}), we have redone the simulations of Ref.\ \cite{Chernodub:2018pmt}. The resulting dimensionless Casimir potential per unit length between two chromoelectric wires separated by a physical distance $R\sqrt{\sigma}$ in (2+1)D SU(2) non-abelian gauge theory is shown in Fig.\ (\ref{fig:casimir_potential_2dsu2}) expressed in units of the string tension. This result is consistent with the findings of Ref.\ \cite{Chernodub:2018pmt}, where the inverse coupling scaling factor $\lambda_w =50$ tested by Chernodub et al.\ in the in (2+1)D gauge theory has been used in this work.\\ 

\noindent
The different inverse couplings, $\beta$, used correspond to different physical lattice spacings, $a$, and consequently different physical lattice volumes, since the number of grid points on the isotropic lattice, $N=32^3$, is kept constant. The range of inverse couplings, $\beta = 5-15$, corresponds to a range of physical lattice sizes, $L_{\text{phys}} \sim 1.4 - 4.5$ fm, where we have used the fundamental string tension value, $\sqrt{\sigma}=0.485$ GeV, from Ref.\ \cite{Athenodorou:2020ani}. If the physical lattice size is taken to be too small, i.e.\ $L_{\text{phys}}<1$ fm, then finite lattice volume effects can come into play and need to be accounted for.\\

\begin{figure}[!htb]
\begin{center}
\includegraphics[scale=.7]{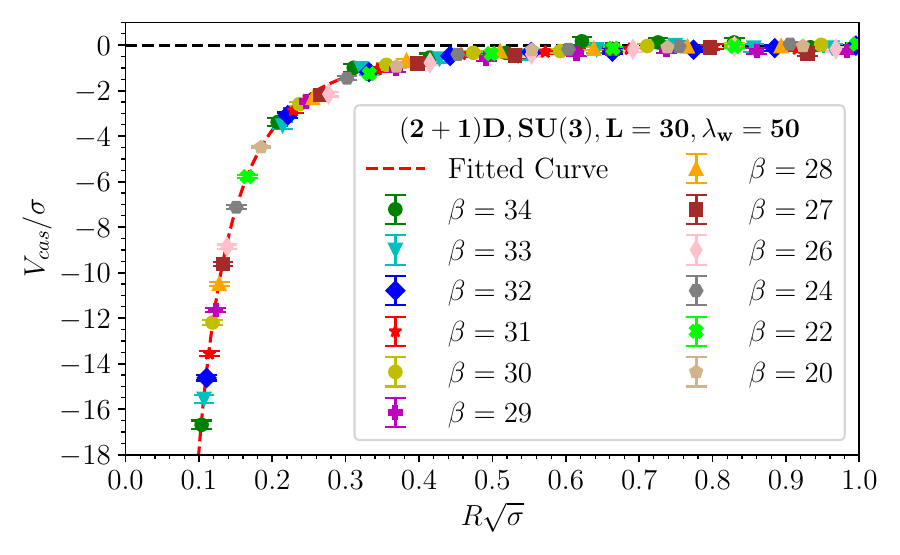}
\caption{The Casimir potential between two parallel wires separated by a physical distance $R\sqrt{\sigma}$ in $(2+1)$D SU(3) for different physical lattice volumes in units of the string tension.}
\label{fig:casimir_potential_2dsu3}
\end{center}
\end{figure}

\noindent
The fitted curve is given in Eq.\ (\ref{eqn:vcas_fit}), $ V_{\text{Cas}}^{\text{fit}} (R) \sim e^{-M_{\text{Cas}}R}/R^{(\nu+2)}\sigma^{(\nu+1)}$ and the relevant parameters are discussed in more detail in the introduction and references therein. The functional form of the fit is an extension of the non-compact abelian gauge theory Casimir potential result in two spatial dimensions. The fit parameters, $(\nu, M_{\text{Cas}})$ have a physical interpretation, where $\nu$ represents the anomalous dimension of the potential at short separation distances. In general, the anomalous dimension indicates how much the scaling dimension deviates from its free-field value in the interacting theory.\\

\noindent
On the other hand, $M_{\text{Cas}}$ represents the effective screening of the potential at large separation distances. The fitted curve confirms that the measured Casimir potential has a good physical scaling as the potential measured at different coupling constants, hence different lattice spacings and physical volumes (for a fixed number of lattice points, $N=32^3$), all follow a single curve. An alternative fitting function for the (2+1)D SU(2) Casimir potential is proposed in Ref.\ \cite{Karabali:2018ael} based on a derivation of the non-abelian theory's Casimir energy by reducing it to that of a massive scalar field and is shown to be quantitatively consistent with Eq.\ (\ref{eqn:vcas_fit}). In the aforementioned paper, the recommended fitting form for the Casimir energy has a mass dependence of $V_{\text{Cas}} \sim e^{-2M_{\text{Cas}}R}$. \\

\noindent
We extend these (2+1)D results to the gauge group SU(3) to test for the effect of increased number of degrees of freedom on the non-abelian gauge theory Casimir potential. The resulting potential is shown in Fig.\ (\ref{fig:casimir_potential_2dsu3}) with the corresponding fit for inverse couplings $\beta = 20-34$ corresponding to physical lattice sizes $L_{\text{phys}} \sim 1.3 - 2.2$ fm. The results that we have provided for the (2+1)D potential are at varying physical lattice volumes for the number of colours under consideration. As we have already shown in the SU(2) case, the potential at varying lattice spacing and slightly different physical volumes follow a single curve. The same is true in SU(3) and a direct comparison is not unfounded.\\ 

\begin{figure}[!htb]
\begin{center}
\includegraphics[scale=.7]{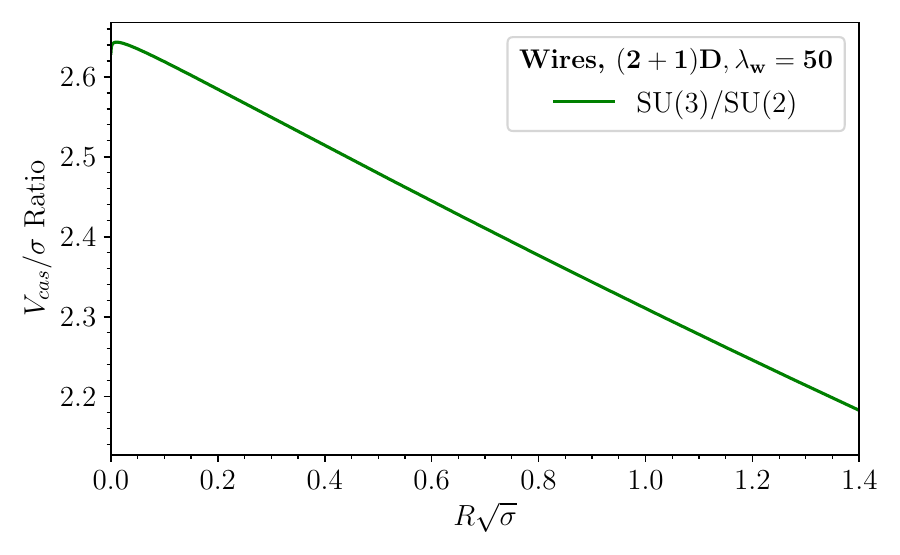}
\caption{Ratio of the Casimir potential between two parallel wires separated by a physical distance $R\sqrt{\sigma}$ in $(2+1)$D with different degrees of freedom.}
\label{fig:casimir_potential_2DRatio}
\end{center}
\end{figure}

\noindent
At short separation distances, qualitatively we observe that the force of attraction experienced by the wires in SU(3) is stronger than that in SU(2). We provide a quantitative comparison which is not continuum extrapolated in Fig.\ (\ref{fig:casimir_potential_2DRatio}) by employing the fitting functions to compute the ratio of the potentials as a function of the physical separation distance. This allows us to make a quantitative statement that increasing the degrees of freedom from $N_c:2\to3$ increases the Casimir force by a factor $\gtrsim 2$, and decreases linearly with increasing separation distance. In the limit, $R\sqrt{\sigma} \to \infty$, this ratio should approach unity as the Casimir effect vanishes. In the weak coupling regime of gauge theories, the static potential scales with the Casimir operator \cite{Greensite:2003bk}, 
\begin{equation}
    C_F = \frac{N_c^2-1}{2N_c},
\end{equation}
in the fundamental representation and the SU(2) theory differs from SU(3) by approximately this factor. The Casimir potential dependence lies between the Casimir operator ratio, $C_F^{SU(3)}/C_F^{SU(2)}=16/9$ and the ratio of the number of colours $N_3/N_2=8/3$.\\

\noindent

\noindent
On the basis of increasing the number of degrees of freedom and resulting increased number of modes, the increased Casimir pressure is somewhat intuitive. In comparison to (2+1)D lattice results of the thermodynamic pressure in pure gauge theory \cite{Caselle:2011mn}, the Casimir pressure scales differently with increased degrees of freedom in the deconfined phase. In the deconfined phase (equivalent to the region between the wires), the thermodynamic pressure lies on the same curve for the gauge groups $N_c=2-6$ \cite{Caselle:2011mn}. In the confined phase where the thermodynamic properties are described by a non-interacting glueball gas, the pressure increases when $N_c:2\to3$ \cite{Caselle:2011fy}. This increase in thermodynamic pressure behaves similarly to the zero-temperature Casimir pressure which increases with the number of colours.\\

\noindent
In terms of the fit parameters, we only provide these for $\lambda_w=50$, as opposed to the optimal perfect conductor condition obtained in the limit $\lambda_w \to \infty$. The result at $\lambda_w=50$ is shown to be sufficiently close to the perfect conductor result \cite{Chernodub:2018pmt}. In SU(2), we find
\begin{equation}
    \nu_{\lambda_w=50} = 0.022(5), \quad M_{\text{Cas}}^{\lambda_w=50} = 1.38(2)\sqrt{\sigma},
\end{equation}
where the value of $\nu$ is reasonably slightly lower than the value of $\nu_{\infty} = 0.05$ and $M_{\text{Cas}}$ is consistent with $M_{\text{Cas}}^{\infty} = 1.38(3)\sqrt{\sigma}$, both obtained through an exponential fitting procedure in Ref.\ \cite{Chernodub:2018pmt}. The Casimir mass is much lower than the lightest glueball mass, $M_{0^{++}} = 4.718(43)\sqrt{\sigma}$ in SU(2) pure gauge theory, motivating once again for the presence of a Casimir-induced deconfined phase in the region between the wires. In SU(3), we find
\begin{equation}
    \nu_{\lambda_w=50} = 0.020(3), \quad M_{\text{Cas}}^{\lambda_w=50} = 1.51(9)\sqrt{\sigma}.
\end{equation}
While the anomalous scaling dimension of the potential at short distances shows no explicit dependence (within errors) on the number of degrees of freedom, the Casimir mass increases as $N_c:2\to3$. However, the Casimir mass is still lower than the mass of the lightest glueball in SU(3), $M_{0^{++}} = 4.329(41)\sqrt{\sigma}$ \cite{Teper:1998te}, which decreases when $N_c:2\to3$. Glueballs are the relevant degrees of freedom in pure gauge theory, but the region between the wires is an induced deconfined phase with interactions governed by particles with a lower mass than the lightest ground-state glueball. Therefore it is plausible that the Casimir mass increases when $N_c:2\to3$ while the glueball mass decreases.\\

\begin{figure}[!htb]
    \centering
    \subfigure[SU(2) with Exponential Fit]{{\includegraphics[scale=0.5]{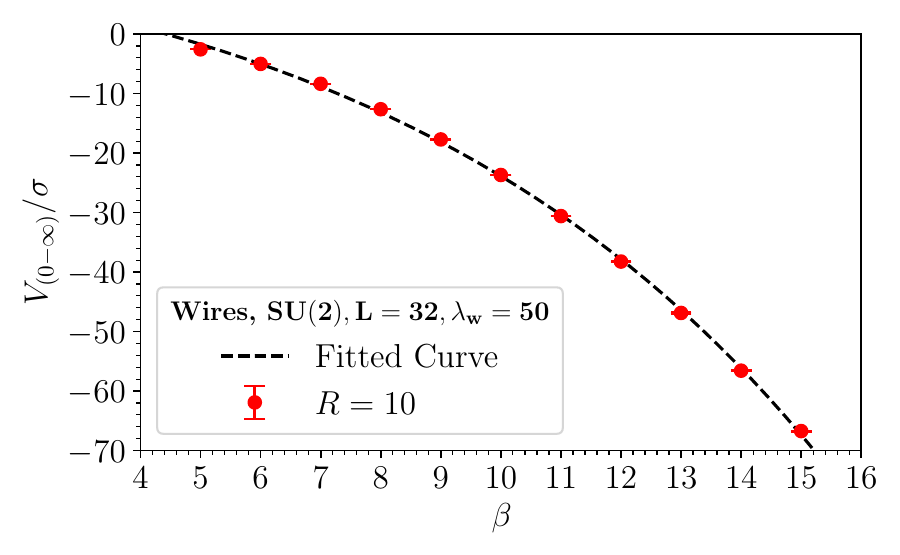} }}%
    \hspace{-0.45cm}
    \subfigure[SU(3) with Linear Fit]{{\includegraphics[scale=0.5]{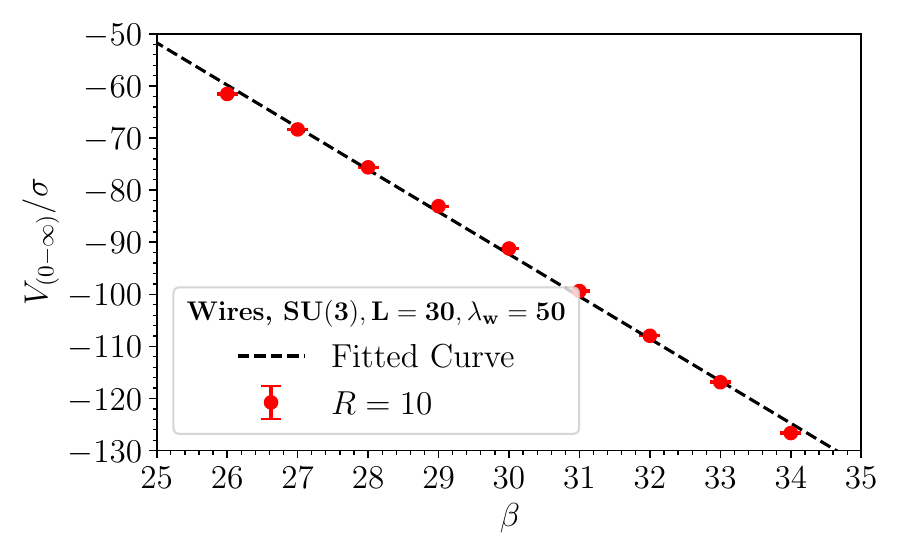} }}%
    \caption{Energy contribution from the chromoelectric boundaries (introducing the wires in the vacuum) on the measured Casimir effect in $(2+1)$D SU(N) as a function of the inverse coupling, $\beta$ at physical separation distance $R=10\sqrt{\sigma}$.}%
    \label{fig:Vcas_wires_Beta}
\end{figure}

\noindent
Lastly, in Fig.\ (\ref{fig:Vcas_wires_Beta}), we show the additional `plaquette potential' that one measures at various inverse couplings provided the energy contribution of imposing the boundary conditions on the lattice is not subtracted. The wire energy contribution is computed according to Eq.\ (\ref{eqn:wires_energy_normalisation}) at physical separation distance $R=10\sqrt{\sigma}$, where we expect an insignificant Casimir energy contribution due to the effective screening of the potential at large separation distances. As mentioned earlier, this energy contribution is independent of the separation distance. \\

\noindent
The two important features in this plot are: the energy contribution of the wires grows with coupling, and that it is also dependent on the number of colours based on the observed functional form. While we do not attempt to provide intricate fitting functions for this energy contribution as it is not important in this geometry, we show that an exponential fit is more appropriate in SU(2) and a linear fit describes the SU(3) data better. In the limit, $\beta \to 0$ corresponding to very large lattice spacings, these curves should approach zero. Understanding how to isolate the boundary energy contribution becomes imperative for the geometries of a tube and box where this contribution depends on the distance between boundaries.\\

\section{Parallel Plates}
\label{section:Parallel Plates}

\noindent
We begin our exploration of the (3+1)D non-abelian gauge theory by considering the geometry of parallel conducting plates separated by a lattice distance $R$. This geometrical set-up is shown in Fig.\ (\ref{fig:geometry_plates}), and the resulting electromagnetic field configurations are discussed in chapter (\ref{sec:Geometry and Symmetries: Plates}). Starting off with the full expression for the energy density in (3+1)D, and employing the rotational symmetries discussed in the preceding chapter, we confirmed that the energy density between the plates takes three equivalent possible forms,
\begin{eqnarray}
     \varepsilon(x)_{\text{plates}} &=& \frac{1}{2}\left[ \langle B_x^2 \rangle + \langle B_y^2 \rangle + \langle B_z^2 \rangle - \langle E_x^2 \rangle - \langle E_y^2 \rangle - \langle E_z^2 \rangle \right] \label{eqn:plates_Edensity_F0}\\
     &=& \frac{1}{2}\left[ \langle B_y^2 \rangle - \langle E_y^2 \rangle \right] \label{eqn:plates_Edensity_F2} \\
     &=& \frac{1}{2}\left[ \langle B_z^2 \rangle - \langle E_z^2 \rangle \right] \label{eqn:plates_Edensity_F3},
\end{eqnarray}
where we have used the F$2$ symmetry to obtain Eq.\ (\ref{eqn:plates_Edensity_F2}) and F$3$ symmetry to obtain Eq.\ (\ref{eqn:plates_Edensity_F3}).\\

\begin{figure}[!htb]
\begin{center}
\includegraphics[scale=.7]{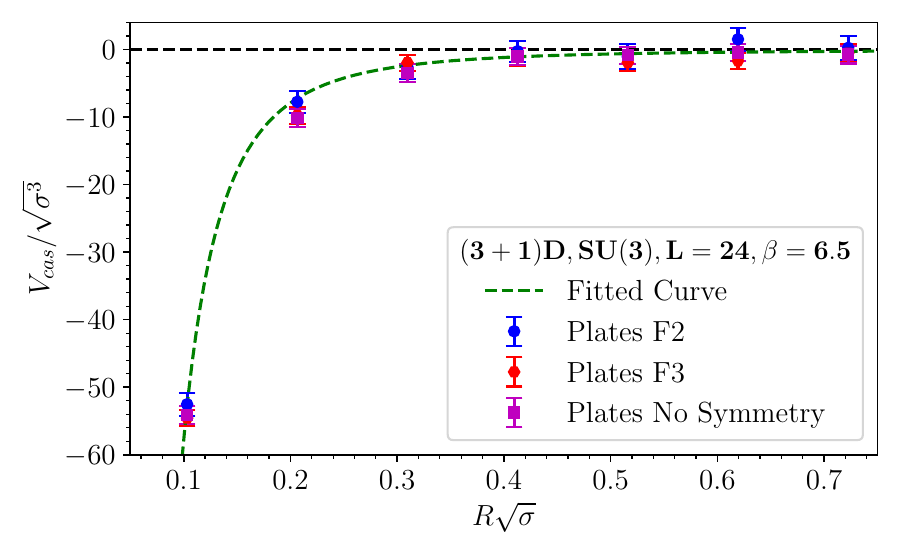}
\caption{The Casimir potential between two parallel plates separated by a distance $R\sqrt{\sigma}$ in $(3+1)$D SU(3) using different symmetry relations.}
\label{fig:symmetry_plates_3dsu3}
\end{center}
\end{figure}

\noindent
The lattice expression assumes a similar form to the (2+1)D case,
\begin{eqnarray}
     \varepsilon(x)_{\text{plates}}^{\text{lat}} &=& \langle S_{P_{ij}} \rangle = \frac{1}{N_{\tau}} \sum\limits_{N_{y}} \sum\limits_{N_{z}} \sum\limits_{N_{\tau}} S_{P_{ij}},
     \label{eqn:Edensity_plates_lat}
\end{eqnarray}
where $P_{ij}$ represents the plaquettes which lie on the worldvolume of the plates. Due to the derived symmetry relations which allow for the simplification of the calculation, we can choose the plaquettes $P_{ij}$ contributing to the total energy density of the system based on the chosen expression in Eq.\ (\ref{eqn:plates_Edensity_F0} - \ref{eqn:plates_Edensity_F3}). As an example, if the full (3+1)D form in Eq.\ (\ref{eqn:plates_Edensity_F0}) is used, then all spatial and temporal plaquettes contribute to the Casimir energy of the plates. However, if one exploits the symmetries of the geometry and employs the form in e.g., Eq.\ (\ref{eqn:plates_Edensity_F2}), then only the plaquettes $P_{xz}$ and $P_{yt}$ need to be considered. \\

\noindent
As discussed in the two-dimensional case, the energy density has no direct dependence on the individual plaquette components $P_{ij}$ contributing to the total energy density due to the summation over the action. We represent the energy density $\langle S_{P_{ij}} \rangle$ with the indices to signal that only contributions from selected plaquettes are considered and for consistency with the notation in the available literature. In similar fashion, we find the three-dimensional normalised Casimir energy by integrating over the normal direction and applying the normalisation condition to remove the energy contribution from the plates,
\begin{eqnarray}
    V^{\text{lat}}_{\text{Cas}}(R) &=& \left[ \int dx\, \varepsilon(x)_{\text{plates}} \right]_{R-R_0} = \left[ \sum\limits_{N_{x}} \varepsilon(x)_{\text{plates}}^{\text{lat}} \right]_{R-R_0}\\
    &=&  \sum\limits_{N_{x}} \left[ \langle S_{P_{ij}} \rangle_R - \langle S_{P_{ij}} \rangle_{R_0} \right] = \langle \langle S_{P_{ij}} \rangle \rangle,
    \label{eqn:3d_casimir_lattice}
\end{eqnarray}
given in lattice units. The lattice expression for the total dimensionless energy of the system,
\begin{eqnarray}
    \langle \langle S_{P_{ij}} \rangle \rangle &=& \left[ \frac{1}{N_{\tau}} \sum\limits_{N_{x}} \sum\limits_{N_{y}} \sum\limits_{N_{z}} \sum\limits_{N_{\tau}}  S_{P_{ij}} \right]_{R-R_0},
    \label{eqn:3d_action_lattice}
\end{eqnarray}
can be compared to the energy density, $T^{00}$, and the total electromagnetic energy, $E(t)$, contained in a spatial volume in classical electrodynamics, 
\begin{eqnarray}
     \langle \langle S_{P_{ij}} \rangle \rangle &=& \frac{1}{N_{\tau}} \sum\limits_{N_{\tau}} \sum\limits_{x,y,z} a^4 T^{00}\\
     &=& \frac{1}{N_{\tau}} \sum\limits_{N_{\tau}} a  \int d^3x\, T^{00}(\tau,x)\\
     &=& \frac{1}{N_{\tau}} \sum\limits_{N_{\tau}} aE(\tau) = a\Bar{E}
    \label{eqn:3d_classical_energy}
\end{eqnarray}
where the prefactor $a^4$ follows directly from our definition of the field strength tensor according to plaquettes in Eq.\ (\ref{eqn:field_components}) and the sum over $N_{\tau}$ is just an ensemble average on the lattice.\\

\noindent
Due to periodic boundary conditions, the plates extend infinitely in the $\hat{y}$ and $\hat{z}$ directions. Since we consider the energy associated with a single lattice volume, we also take as plate area $A=N_z\times N_y$, which is used in Eq.\ (\ref{eqn:Edensity_plates_lat}) to express the energy density per unit area of the plates. The physical potential scales as $V^{\text{phys}}_{\text{Cas}} =  V^{\text{lat}}_{\text{Cas}}/a$, giving a well-defined expression for the total energy,
\begin{eqnarray}
    V_{\text{Cas}}(R_{\text{phys}})/\sigma &=& \frac{1}{a\sqrt{\sigma}} \left( \frac{\beta_I}{\beta} \right)^4 \langle \langle S_{P_{ij}} \rangle \rangle,
    \label{eqn:Vcas3D_total_scaled}
\end{eqnarray}
and the Casimir energy per unit area of the plates scales as $V^{\text{phys}}_{\text{Cas}} =  V^{\text{lat}}_{\text{Cas}}/a^3$, and the resulting potential per unit area is
\begin{eqnarray}
    V_{\text{Cas}}(R_{\text{phys}})/\sqrt{\sigma^3} &=& \frac{1}{a^3 \sqrt{\sigma^3}} \left( \frac{\beta_I}{\beta} \right)^4 \langle \langle S_{P_{ij}} \rangle \rangle,
    \label{eqn:Vcas3D_formula_scaled}
\end{eqnarray}
where we have used the string tension to express the potential in dimensionless units and applied the same scaling with the tadpole improved coupling used in the two-dimensional case.\\

\noindent
In Fig.\ (\ref{fig:symmetry_plates_3dsu3}), we show the Casimir potential obtained from exploring the various symmetry relations in Eq.\ (\ref{eqn:plates_Edensity_F0} - \ref{eqn:plates_Edensity_F3}) in a physical lattice size, $L_\text{phys} \sim 1$ fm. The resulting potential from our numerical simulations on the lattice shows a good quantitative consistency with our analytical results of the symmetry relations. This consistency in the symmetry relations is also discussed for the individual field-strength tensor components in chapter (\ref{fig:symmetry_plates_3dsu3}).\\

\noindent
The fitting function that we have employed for the potential of the plates has a similar functional form and same physical interpretation as the one used for the wires in (2+1)D given in Eq.\ (\ref{eqn:vcas_fit}). It is given by
\begin{equation}
    V_{\text{Cas}}^{\text{fit}} (R) = -(N_c^2-1)\frac{\pi^2}{1440} \frac{1}{R^{(\nu+3)}\sigma^{(\nu+3)/2}} e^{-M_{\text{Cas}}R},
    \label{eqn:vcas_fit_plates}
\end{equation}
where the power of $\sigma$ is chosen to give a dimensionless potential, i.e., $V_{\text{Cas}}/\sqrt{\sigma^3}$. The coefficients are obtained from the expected tree level behaviour of the potential for a non-interacting theory, $M_{\text{Cas}} = \nu =0$ \cite{Ambjorn:1981xw}, 
\begin{equation}
    V_{\text{Cas}}^{\text{tree}} (R) = -(N_c^2-1)\frac{\pi^2}{1440} \frac{1}{R^{3}},
    \label{eqn:vcas_tree_plates}
\end{equation}
describing $(N_c^2-1)$ additive contributions from the non-interacting gluon fields.\\

\begin{figure}[!htb]
    \centering
    \subfigure[Vacuum Subtracted Total Energy]{{\includegraphics[scale=0.5]{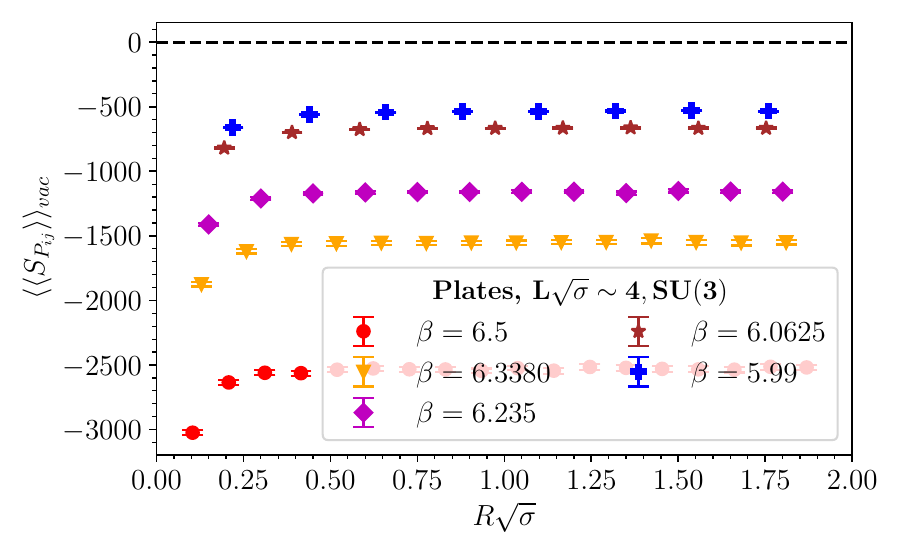} \label{fig:plates_Svac} }}%
    \hspace{-0.45cm}
    \subfigure[$R_{\infty}$ Subtracted Total Energy]{{\includegraphics[scale=0.5]{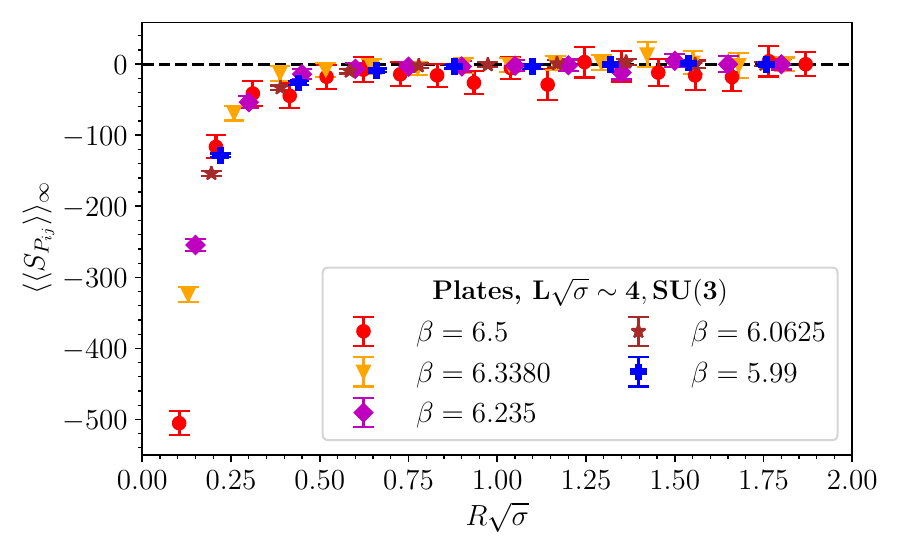} \label{fig:plates_Sinf} }}%
    \caption{Total energy of the system in a parallel plates configuration in $(3+1)$D SU(3) for different couplings,(left) is the vacuum normalisation and (right) is the $R_{\infty}$ normalisation.}%
\end{figure}

\noindent
In the previous section, we mentioned that the energy required to create the wires is independent of the separation distance between the wires. We now return to expand on this discussion using a more easily visualisable example of parallel plates on the three-dimensional axis. Given that both plates extend infinitely in two spatial directions, the number of plaquettes forming each plate is the same within our lattice volume. We refer the reader to chapter (\ref{sec:Geometry and Symmetries: Plates}) where we discuss the field components associated with each plaquette for a visualisation of the differences in the magnitude between the energy contributions at and away from the boundaries.\\

\begin{figure}[!htb]
\begin{center}
\includegraphics[scale=.7]{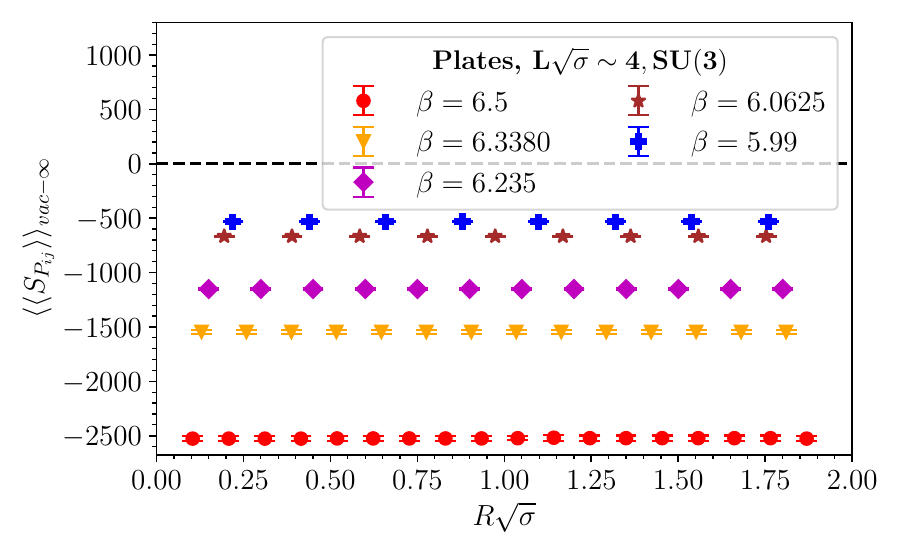}
\caption{Energy contribution from two parallel plates forming the chromoelectric boundaries in $(3+1)$D SU(3) is a constant at each inverse couplings.}
\label{fig:plates_Svac_Sinf}
\end{center}
\end{figure}

\noindent
The energy contribution from the plates can be calculated directly by placing the two plates at a large separation distance where the Casimir effect is insignificant, computing the total energy of the system and subtracting the vacuum contribution. However, we would like to emphasize that the energy contribution from the two plates is independent of $R$ since the area of the plates is the same. Hence we isolate the energy contribution from the plates by comparing the total energy of the system normalised using two different schemes.\\

\noindent
The first one is the \textit{vacuum normalisation} shown in Fig.\ (\ref{fig:plates_Svac}), which corresponds to the total energy of the system less the vacuum energy contribution,
\begin{equation}
    E_{\text{Tot}}^{\text{Vac}} = E_{\text{Cas}} + E_{\text{Plates}},
\end{equation}
where $E_{\text{Vac}}$ is cancelled by the normalisation. The second one is the $R_{\infty}$ \textit{normalisation} corresponding to the subtraction of the total energy of a system consisting of two plates placed far apart ($R_{\infty}=L/2$ on the lattice), where the Casimir energy is negligible. We show the resulting energy in Fig.\ (\ref{fig:plates_Sinf}), with total energy
\begin{equation}
    E_{\text{Tot}}^{\infty} = E_{\text{Cas}},
\end{equation}
because $E_{\text{Plates}}$ and $E_{\text{Vac}}$ are cancelled by the normalisation, and only the Casimir energy remains. The energy difference, $E_{\text{Tot}}^{\text{Vac}} - E_{\text{Tot}}^{\infty} = E_{\text{Plates}}$ is the energy cost from putting in the boundaries, and is dependent on the inverse coupling, but independent of the separation distance as illustrated in Fig.\ (\ref{fig:plates_Svac_Sinf}). Note that this is the total energy described by Eq.\ (\ref{eqn:3d_action_lattice}), and not the energy per unit area of the plates. Most importantly, note that this energy is independent of $R$ because the area of the plates remains fixed.\\ 

\begin{figure}[!htb]
\begin{center}
\includegraphics[scale=.7]{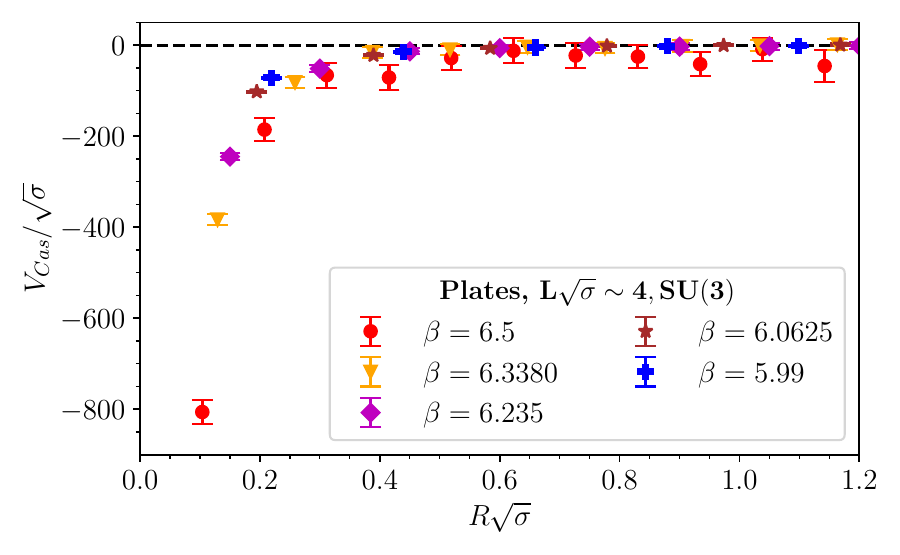}
\caption{The total Casimir potential between two parallel plates separated by a distance $R\sqrt{\sigma}$ in $(3+1)$D SU(3).}
\label{fig:casimir_total_plates_3dsu3}
\end{center}
\end{figure}

\noindent
In Fig.\ (\ref{fig:plates_Sinf}), we observe that the total Casimir energy lies on a single curve for various inverse couplings similarly to the geometry of the wires. The normalisation condition at $R_{\infty}$ ensures that all other energy contributions to the system cancel exactly, leaving only the Casimir energy term. The exponential decay of the potential with separation distance resembles abelian theory results for this geometry. Meanwhile, the total energy in Fig.\ (\ref{fig:plates_Svac}) lies on different curves at different couplings and does not vanish at large separation distances due to the overall constant energy from the boundaries at each coupling. These two scenarios highlight the importance of subtracting correctly the energy contribution from the boundaries.  \\

\begin{figure}[!htb]
\begin{center}
\includegraphics[scale=.7]{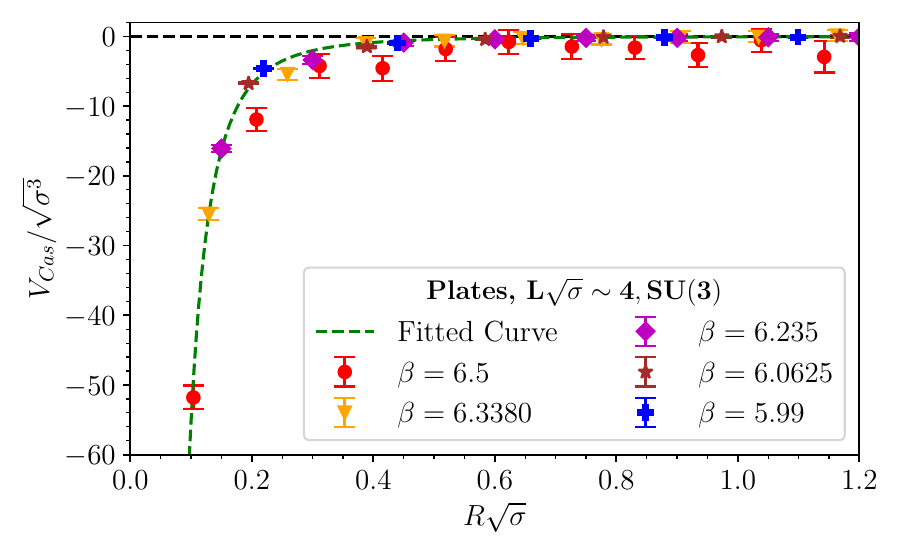}
\caption{The Casimir potential between two parallel plates separated by a distance $R\sqrt{\sigma}$ in $(3+1)$D SU(3) per unit area of the plates.}
\label{fig:casimir_plates_3dsu3}
\end{center}
\end{figure}

\noindent
The dimensionless physical total Casimir energy, $V_{\text{Cas}}/\sqrt{\sigma}$ corresponding to the $R_{\infty}$ normalisation is obtained from applying Eq.\ (\ref{eqn:Vcas3D_total_scaled}). We show this potential in Fig.\ (\ref{fig:casimir_total_plates_3dsu3}) using units of the string tension, while the Casimir potential per unit area of the plates is given in Fig.\ (\ref{fig:casimir_plates_3dsu3}) in SU(3). Similarly to conducting wires in (2+1)D, the potential is negative. Its functional form indicates that the Casimir force, described by the slope of the potential, experienced by the plates in the non-abelian gauge theory is \textit{attractive}. In addition, there is a Casimir \textit{boundary-induced deconfinement} in the region between the plates and we will expand on this discussion in the following chapter when we look at the temperature dependence of the Casimir potential. We also refer the reader to Ref.\ \cite{Chernodub:2023dok} for a recent study of the parallel plate Casimir effect in SU(3) where the boundary induced deconfinement region is discussed in more detail.\\ 

\begin{figure}[!htb]
\begin{center}
\includegraphics[scale=.7]{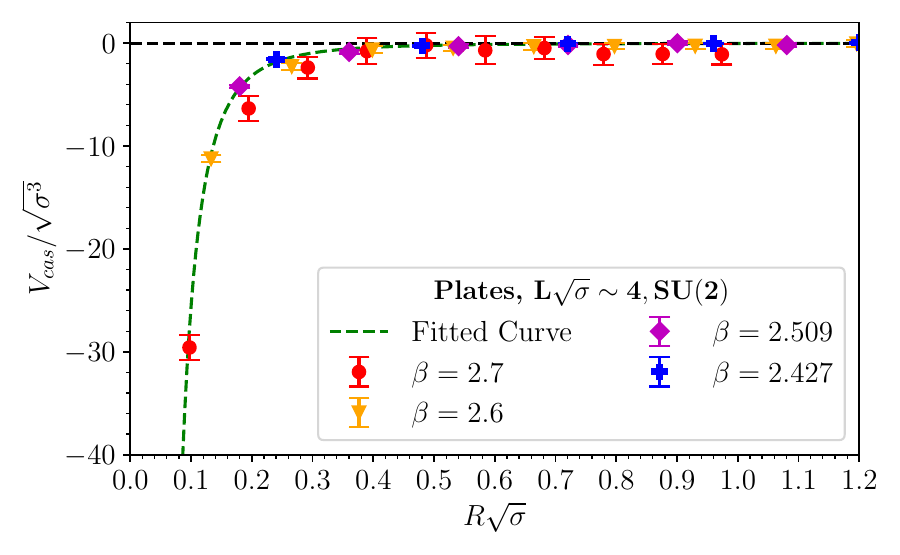}
\caption{The Casimir potential between two parallel plates separated by a distance $R\sqrt{\sigma}$ in $(3+1)$D SU(2) per unit area of the plates.}
\label{fig:casimir_plates_3dsu2}
\end{center}
\end{figure}

\begin{figure}[!htb]
\begin{center}
\includegraphics[scale=.7]{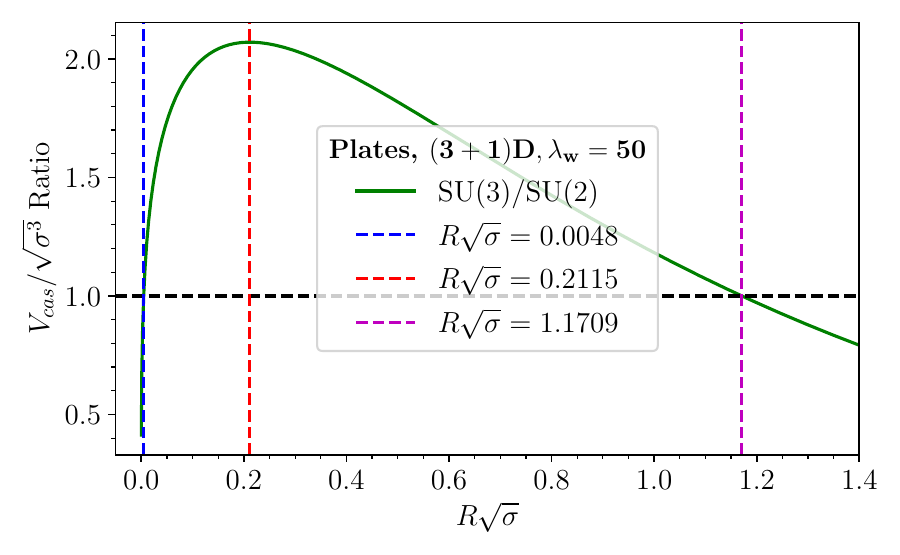}
\caption{Ratio of the Casimir potential between two parallel plates separated by a physical distance $R\sqrt{\sigma}$ in $(3+1)$D with different degrees of freedom.}
\label{fig:casimir_potential_3DRatio}
\end{center}
\end{figure}

\noindent
We also present results for the Casimir potential per unit area of the plates in SU(2) to test for the effect of reducing the number of degrees of freedom on the Casimir effect of a parallel plate configuration in non-abelian gauge theory. The resulting potential is illustrated in Fig.\ (\ref{fig:casimir_plates_3dsu2}) for the same physical lattice volume, $L\sqrt{\sigma} \sim 4$ as our SU(3) studies. This enables us to make a quantitative comparison of these cases as shown by the ratio in Fig.\ (\ref{fig:casimir_potential_3DRatio}).\\

\noindent
Unlike the geometry of parallel wires where the Casimir potential was consistently higher in SU(3) for the separation distances considered, in the (3+1)D theory for parallel plates, we observe a non-linear scaling of the potential when $N_c:2\to3$. At intermediate separation distances, the Casimir pressure experienced by the plates is stronger in SU(3), peaking at approximately twice as strong at $R\sqrt{\sigma}=0.2$. Then when the plates are very close together and at large separation distances where we expect a negligible Casimir force, the pressure is stronger in SU(2). While we do not provide a physical interpretation at this point, we state these results as they point out a need for further investigation of the Casimir effect and relevant fitting functional form between the two theories.\\

\noindent
The dominant higher Casimir pressure in SU(3) is contrary to lattice calculations of thermodynamic quantities in pure gauge theory which suggest that the pressure of the system decreases when $N_c:2\to3$ in the deconfined phase \cite{Giudice:2017dor} (which is the expected behaviour for the region between the plates). Of course, the thermodynamic pressure is different from the Casimir pressure, but also, the SU(2) and SU(3) theories are both quantitatively and qualitatively different. In particular, the deconfinement phase transition is second order in the former and a weak first order transition in the latter. Given the Casimir induced deconfinement transition in the region between the plates, it may be worth investigating the dynamics of this transition between the two theories.\\

\noindent
We conclude this section by looking at the fit-parameters obtained for the anomalous scaling dimension, $\nu$ and the Casimir mass, $M_{\text{Cas}}$ in Eq.\ (\ref{eqn:vcas_fit_plates}). These fits are performed at a finite boundary coupling, $\lambda_w=50$ as opposed to the perfect conductor coupling where $\lambda_w \to \infty$. In the (2+1)D non-abelian theory, the dependence of the fit parameters on the coupling, $\lambda_w$ is provided in Ref.\ \cite{Chernodub:2018pmt}, which shows that as the coupling increases, $\nu$ grows logarithmically then plateaus. On the other hand, $M_{\text{Cas}}$ is an exponential decay such that in the weak coupling boundary limit where $\lambda_w\to1$, the Casimir mass should approach $M_{0^{++}}$.\\ 

\noindent
In the (3+1)D theory, we expect the same dependence of the fit parameters with increasing coupling, $\lambda_w$ to hold. Then our present fit parameters at $\lambda_w=50$ could possibly slightly underestimate the anomalous scaling dimension as seen in the two-dimensional case, while overestimating the Casimir mass. The following fit parameters for the anomalous dimension are obtained,
\begin{equation}
    \nu_{\lambda_w=50} = 
\begin{dcases}
    0.09(4), & \text{SU(2)}\\[5pt]
    0.002(15), & \text{SU(3)},
\end{dcases}
\label{eqn:lambda_fit_3D}
\end{equation}
with corresponding Casimir masses,
\begin{equation}
    M_{\text{Cas}}^{\lambda_w=50} = 
\begin{dcases}
    0.38(12)\sqrt{\sigma}, & \text{SU(2)}\\[5pt]
    0.06(6)\sqrt{\sigma}, & \text{SU(3)}.
\end{dcases}
\label{eqn:MCas_fit_3D}
\end{equation}

\noindent
The anomalous scaling dimension is non-zero in both SU(2) and SU(3), indicating that the gluon self-interactions lead to an anomalous scaling dimension of the Casimir energy at short separation distances between the plates. In addition, similarly to the observations in the two-dimensional theory, the Casimir masses extracted from the fits are much smaller than the ground state lightest glueball masses, $M_{0^{++}} = 3.781(23)\sqrt{\sigma}$ in SU(2) and $M_{0^{++}} = 3.405(21)\sqrt{\sigma}$ in SU(3) \cite{Athenodorou:2021qvs}. We note that the Casimir mass decreases as we move from the $(2\to3)$D theory, also consistent with the behaviour of glueball masses in the two theories.\\ 

\noindent
The SU(3) Casimir potential that we obtain has similar qualitative features to the potential obtained in Ref.\ \cite{Chernodub:2023dok}, however the rate of decay of the potential varies particularly at short separation distances where our measured potential decays more rapidly. In addition, the Casimir mass obtained in Ref.\ \cite{Chernodub:2023dok}, $M_{\text{Cas}} = 1.0(1)\sqrt{\sigma}$ differs substantially from the mass obtained in our work. We highlight this difference in the measured potential and Casimir mass as an aspect that needs to be investigated further in future Casimir effect studies of the parallel plate geometry.\\

\noindent
Given that the Casimir masses measured in the three-dimensional theory, $M_{\text{Cas}} << M_{0^{++}}$, it is worth investigating the exact form of the mass-dependent exponential term in the Casimir potential. In the formulation of the Casimir energy of a scalar field in Ref.\ \cite{Karabali:2018ael}, the exponential term varies as $V_{\text{Cas}} \sim e^{-2M_{\text{Cas}}R}$. If one instead applies this dependence in Eq.\ (\ref{eqn:vcas_fit_plates}), then the resulting Casimir masses would be half the values in Eq.\ (\ref{eqn:MCas_fit_3D}). It is rather interesting that the boundary-induced phase transition results in gluonic excitations with such low masses. Perhaps a lattice approach employing different boundary conditions could improve this current picture of the relevant masses of the exchange particles in Casimir interactions in order to better understand the effect of chromoelectric boundaries on the Casimir effect in non-abelian gauge theories.\\

\section{Symmetrical and Asymmetrical Tube}
\label{section:Symmetric and Asymmetric Tube}

\noindent
The next geometrical set-up that we look at in the (3+1)D non-abelian theory is the hollow tube shown in Fig.\ (\ref{fig:geometry_tube}). The tube has finite extents, $R$ in the $\hat{x}$ and $\hat{y}$-axis, and extends infinitely in the $\hat{z}$ direction, i.e., has a lattice length, $L$ in one direction. We explore two separate cases for this geometry; a symmetrical tube with physical side lengths $R_x=R_y=R\sqrt{\sigma}$ and an asymmetrical tube with $R_x=R\sqrt{\sigma}$ and $R_y=1\sqrt{\sigma}$, both with $R_z=L\sqrt{\sigma}$. Geometrically, the symmetrical tube is a square prism, while the asymmetrical tube is a rectangular prism for $R_{\text{lat}}>1$.\\

\noindent
In the preceding chapter, we looked at the electromagnetic field-strength tensor components and the resulting numerical field configurations for these two geometries. We also showed that due to the symmetry relations, the full expression for the energy density of the system in Euclidean space given by Eq.\ (\ref{eqn:euclidean_energy_density}), reduces to only the $\hat{z}$ components and the energy density in the tube is described by,
\begin{eqnarray}
    \varepsilon_{\text{tube}}(x,y) &=& \frac{1}{2}\left[ \langle B_x^2 \rangle + \langle B_y^2 \rangle + \langle B_z^2 \rangle - \langle E_x^2 \rangle - \langle E_y^2 \rangle - \langle E_z^2 \rangle \right]\\
     &=& \frac{1}{2}\left[ \langle B_z^2 \rangle - \langle E_z^2 \rangle \right]  \label{eqn:tube_Edensity},
\end{eqnarray}
which is now a function of both $x$ and $y$ because the gluon fields are now enclosed in a hollow cavity and the Casimir force acts in two directions.\\

\noindent
On the lattice, the energy density is given by a sum over the plaquettes on the worldvolume of the tube,
\begin{eqnarray}
     \varepsilon(x,y)_{\text{tube}}^{\text{lat}} &=& \langle S_{P_{ij}} \rangle = \frac{1}{N_{\tau}} \sum\limits_{N_{z}} \sum\limits_{N_{\tau}} S_{P_{ij}},
\end{eqnarray}
and these plaquettes contributing to the action are reduced to $P_{xy}$ and $P_{zt}$ according to Eq.\ (\ref{eqn:tube_Edensity}). The expression for the Casimir potential of the tube follows,
\begin{eqnarray}
    V^{\text{lat}}_{\text{Cas}}(R) &=& \left[ \int_{d\mathcal{A}}\, \varepsilon(x,y)_{\text{tube}} \right]_{R-R_0} = \left[ \sum\limits_{N_{x}, N_{y}} \varepsilon(x,y)_{\text{tube}}^{\text{lat}} \right]_{R-R_0}\\
    &=&  \sum\limits_{N_{x}, N_{y}} \left[ \langle S_{P_{ij}} \rangle_R - \langle S_{P_{ij}} \rangle_{R_0} \right] = \langle \langle S_{P_{ij}} \rangle \rangle,
    \label{eqn:tube_casimir_lattice}
\end{eqnarray}
and the total lattice dimensionless energy in the system follows Eq.\ (\ref{eqn:3d_action_lattice}).\\

\begin{figure}[!htb]
    \centering
    \subfigure[Vacuum Subtracted Energy]{{\includegraphics[scale=0.5]{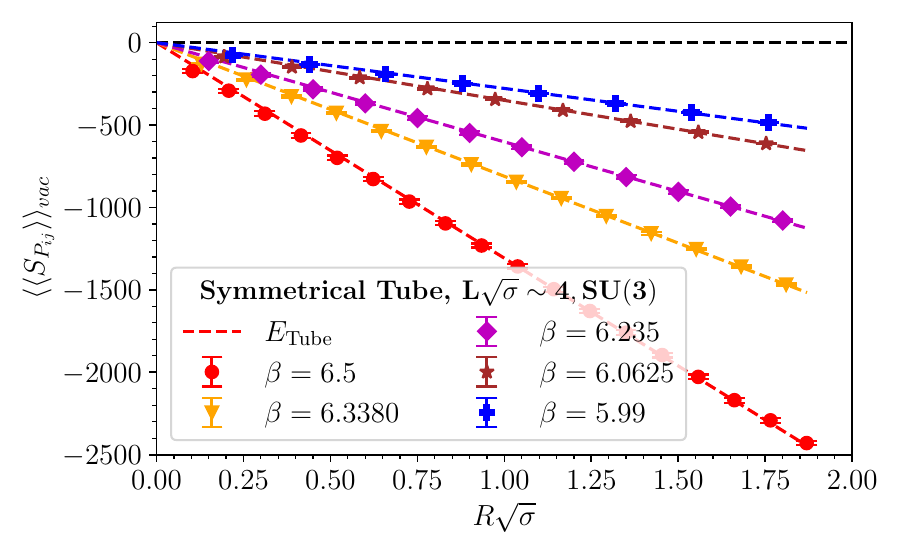} \label{fig:tube_Svac} }}%
    \hspace{-0.45cm}
    \subfigure[$R_{\infty}$ Subtracted Energy]{{\includegraphics[scale=0.5]{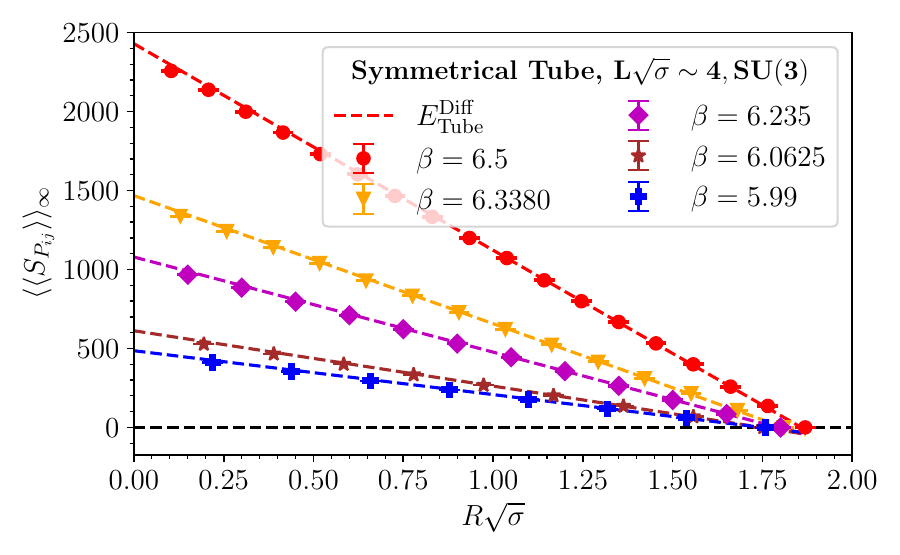} \label{fig:tube_Sinf} }}%
    \caption{Total energy of the system in a symmetrical tube in $(3+1)$D SU(3) for different couplings and different normalisation schemes in lattice units.}%
    \label{fig:tube_total_S}
\end{figure}

\noindent
Now, the next question is, `How can we subtract the energy contribution from the boundaries at $R_0$?' In the previous case of parallel plates, the normalisation condition was well-posed since the area of the plates did not change with $R$. We also showed that the energy contribution from the plates was independent of the separation distance, thus just a constant. However, the energy contribution from the tube is now dependent on the separation distance, $R$ because as $R$ increases, the size of the four rectangular plates forming the tube also increases and their energy contribution changes. We devise two approaches in which this energy can be accounted for, and we discuss both of them for the symmetrical tube and apply the most optimal one for the asymmetrical tube.\\

\subsection{Symmetrical Tube}

\noindent
In order to capture the tube's energy dependence on the separation distance, $R$ and consequently the size of the plates forming the tube, we plot the total energy of a symmetrical tube in Fig.\ (\ref{fig:tube_total_S}) using different normalisation schemes. Let's start by looking at the \textit{first method} shown in Fig.\ (\ref{fig:tube_Svac}) where we plot the \textit{vacuum subtracted} total energy of the tube, $E_{\text{Tot}}^{\text{Vac}}$. It is clear from observation that the total energy,
\begin{equation}
  E_{\text{Tot}}^{\text{Vac}} = E_{\text{Cas}} + E_{\text{Tube}}, 
  \label{eqn:tube_vac_tot}
\end{equation} 
varies linearly with separation distance. However, we also know from extrapolating from findings of the plates, that at large separation distances, the Casimir energy contribution should be negligible.\\

\noindent
We therefore conclude that the energy contribution from creating the tube, $E_{\text{Tube}}$ increases linearly with increasing plate size. Consequently, we apply a linear fit to the total energy for $R\sqrt{\sigma} \gtrsim 1$ at different couplings,
\begin{equation}
    E_{\text{Tube}} = mR\sqrt{\sigma} + c,
    \label{eqn:linear_fit}
\end{equation}
with fit parameters, $m$ capturing the plate size dependence and $c$ providing the overall scaling. In the absence of the tube, i.e., at $R\sqrt{\sigma} =0$, the energy contribution from the tube vanishes. This allows us to set the constant term, $c=0$.\\

\begin{figure}[!htb]
    \centering
    \subfigure[Vacuum Subtracted Potential]{{\includegraphics[scale=0.5]{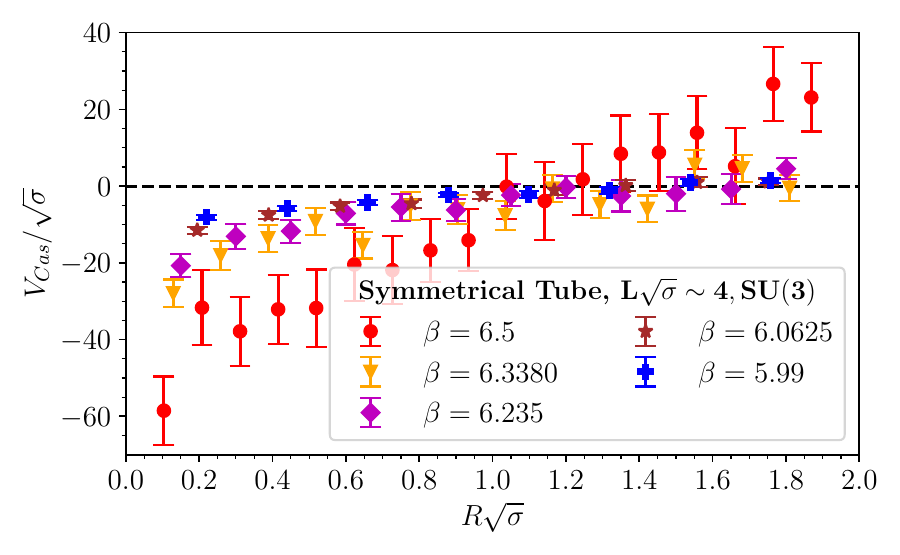} \label{fig:tube_ETotVac} }}%
    \hspace{-0.45cm}
    \subfigure[$R_{\infty}$ Subtracted Potential]{{\includegraphics[scale=0.5]{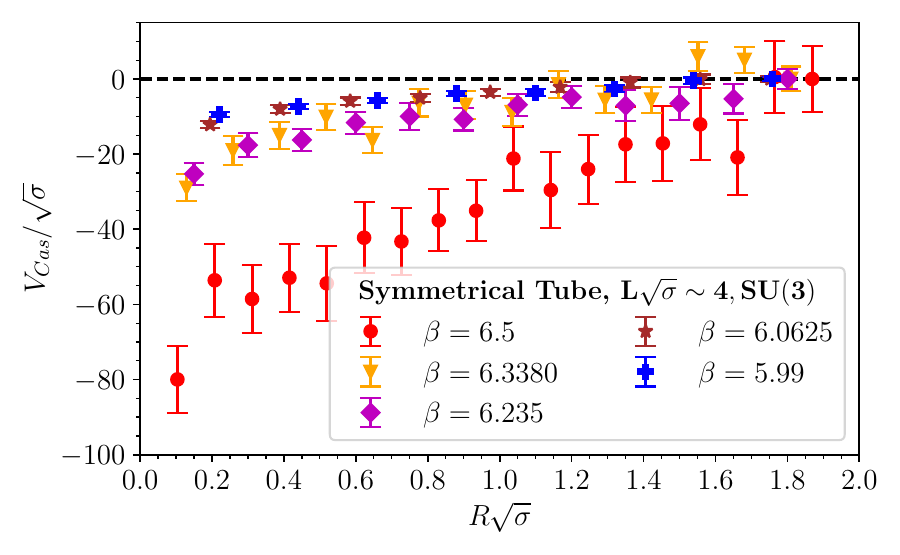} \label{fig:tube_ETotInf} }}%
    \caption{The total Casimir potential for a symmetrical tube with side lengths $R\sqrt{\sigma}$ in $(3+1)$D SU(3).}%
    \label{fig:VcasTubeTotal}
\end{figure}

\noindent
We find, from performing the fitting procedure that, $E_{\text{Tot}}^{\text{Vac}} > E_{\text{Tube}}$, where $E_{\text{Tot}}^{\text{Vac}}$ is represented by coloured points at different inverse couplings. The difference between the two curves captures the Casimir energy, $E_{\text{Cas}}$. To summarise all the steps of this method, the Casimir energy is isolated by:
\begin{enumerate}
    \item Compute the total energy of the system and apply the vacuum subtraction to remove UV divergences, remaining with the total energy in Eq.\ (\ref{eqn:tube_vac_tot}).
    \item Perform a linear fitting procedure on the renormalised total energy at large separation distances to capture the energy contributions from the boundaries (tube). Take the constant term of the fit, $c=0$ because the energy from creating the boundaries vanishes in the absence of boundaries.
    \item Use the fit parameters to perform a second subtraction on the total energy in Eq.\ (\ref{eqn:tube_vac_tot}) in order to remove the energy associated with creating the tube. This isolates the Casimir energy.
\end{enumerate}

\begin{figure}[!htb]
    \centering
    \subfigure[Vacuum Subtracted Potential/Area]{{\includegraphics[scale=0.5]{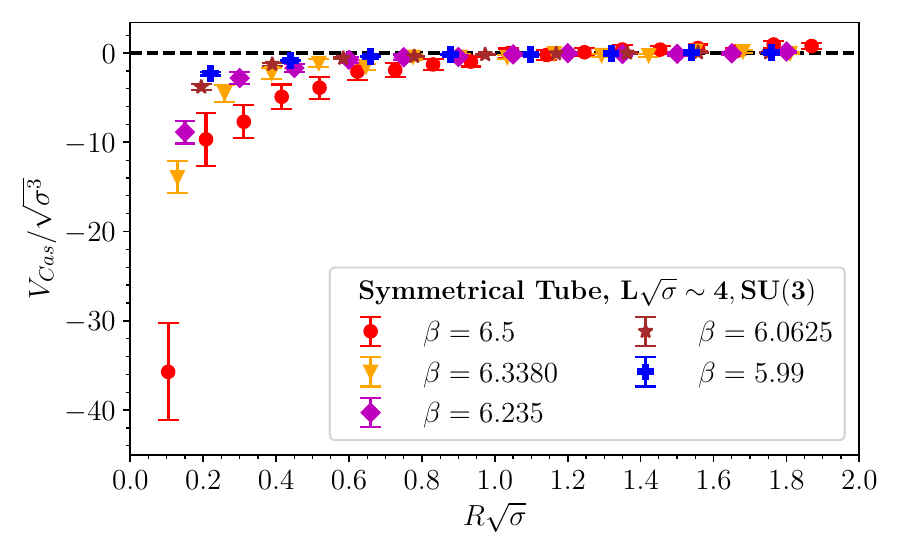} \label{fig:tube_EVac} }}%
    \hspace{-0.45cm}
    \subfigure[$R_{\infty}$ Subtracted Potential/Area]{{\includegraphics[scale=0.5]{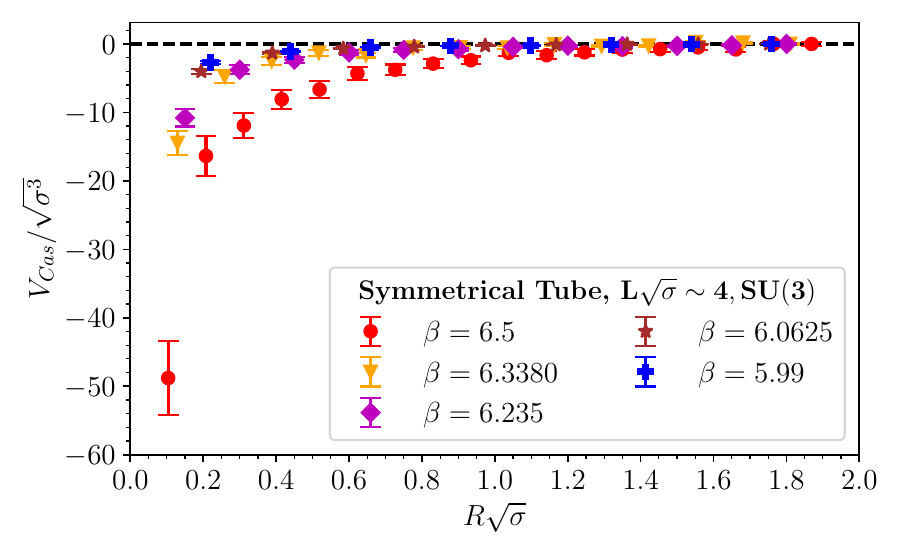} \label{fig:tube_EInf} }}%
    \caption{The Casimir potential for a symmetrical tube with side lengths $R\sqrt{\sigma}$ in $(3+1)$D SU(3) per unit surface area of the tube.}%
    \label{fig:VcasTube}
\end{figure}

\noindent
The resulting physical total Casimir potential for the symmetrical tube as defined in Eq.\ (\ref{eqn:Vcas3D_total_scaled}) is shown in Fig.\ (\ref{fig:tube_ETotVac}). We conclude our discussions on this method by showing the Casimir energy per unit surface area of the symmetrical tube in Fig.\ (\ref{fig:tube_EVac}) using the definition in Eq.\ (\ref{eqn:Vcas3D_formula_scaled}). The corresponding lattice total surface area of the symmetrical tube is, $A_{\text{lat}}=4RL$, where $R$ is the tube side-length and separation distance. Using the total surface area enables us to make a direct comparison of the Casimir potential with other geometries given that asymmetrical geometries consist of faces with different areas. Computing the potential per unit surface area introduces an additional dependence on the side-length of the tube, $R\sqrt{\sigma}$ which scales the diminishing potential with separation distance non-linearly.\\

\noindent
The \textit{second method} of quantifying the energy contributions from the tube's boundaries and ultimately isolating the Casimir energy follows a similar technique as the first. Instead of performing a vacuum normalisation, we use the $R_{\infty}$ \textit{normalisation} where we subtract from the total energy, the energy contribution with a tube of side-length, $R=R_{\infty}=L\sqrt{\sigma}/2$. This is the same normalisation employed in the parallel wires and plates geometries, except, now with a slight twist.\\

\noindent
The primary issue here is that the energy contribution from creating a larger tube of side-length, $R=R_{\infty}=L\sqrt{\sigma}/2$ is greater than the energy required to create a smaller tube with $R<L\sqrt{\sigma}/2$. Hence by subtracting the energy contribution at $R_{\infty}$, we eliminate the vacuum energy but remain with an energy difference between creating a small tube and a large tube according to,
\begin{eqnarray}
    E_{\text{Tot}}^{\infty} &=& E_{\text{Cas}} + E_{\text{Tube}}^{\text{Diff}} , \quad E_{\text{Tube}}^{\text{Diff}} = E_{\text{Tube}}^{\text{R}} -E_{\text{Tube}}^{\text{R}_{\infty}}.
    \label{eqn:tube_Etot_infty}
\end{eqnarray}
Fortunately, the resulting energy difference is linear and shown in Fig.\ (\ref{fig:tube_Sinf}), and collapses to zero at $R=L\sqrt{\sigma}/2$ because the Casimir energy is negligible and the `inner tube' of side-length, $R$ has equal dimensions as the $R_{\infty}$ tube. Accordingly, their energy difference vanishes. Similarly, we employ a linear fit for $R\sqrt{\sigma} \gtrsim 1$, where the Casimir energy contributions are negligible and $E_{\text{Tot}}^{\infty} = E_{\text{Tube}}^{\text{Diff}}$.\\

\begin{figure}[!htb]
\begin{center}
\includegraphics[scale=.7]{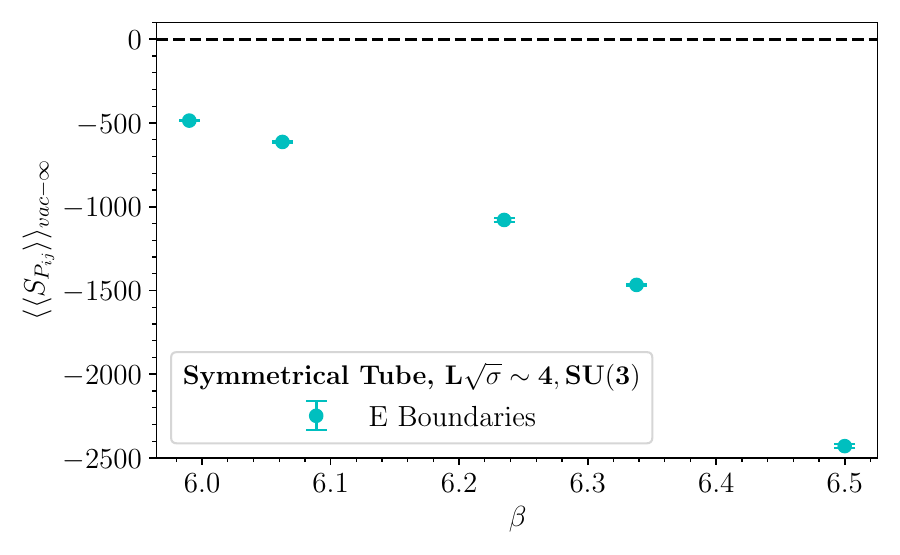}
\caption{The total energy required to create a symmetrical tube with side lengths $L\sqrt{\sigma}/2$ in $(3+1)$D SU(3) at varying lattice couplings.}
\label{fig:tube_Svac_Sinf}
\end{center}
\end{figure}

\noindent
In order to obtain the constant term of the linear fit according to Eq.\ (\ref{eqn:linear_fit}), we note that in the absence of the small tube, the energy difference,
\begin{eqnarray}
    \left. E_{\text{Tube}}^{\text{Diff}} \right|_{\text{R}=0} = \cancelto{0}{E_{\text{Tube}}^{\text{R}=0}} -E_{\text{Tube}}^{\text{R}_{\infty}} = -E_{\text{Tube}}^{\text{R}_{\infty}} = c
\end{eqnarray}
reduces to the energy of creating the larger tube. This calculation is equivalent to subtracting the vacuum energy from the total energy of the system with a tube at $R_{\infty}$ or computing the energy difference as shown in Fig.\ (\ref{fig:tube_Svac_Sinf}). The gradient, $m$ is obtained from the fitting procedure. We summarise these normalisation steps as follows:
\begin{enumerate}
    \item Compute the total energy of the system and apply the energy subtraction using a large tube of side-length, $R=R_{\infty}=L\sqrt{\sigma}/2$. This gives you the remaining total energy defined in Eq.\ (\ref{eqn:tube_Etot_infty}).
    \item Find the total energy required to create the large tube of side-length, $R=R_{\infty}=L\sqrt{\sigma}/2$ either by calculating it directly or easily isolating it according to Fig.\ (\ref{fig:tube_Svac_Sinf}). Use this energy to set the constant term of the fit at $R=0$.
    \item Perform a linear fitting procedure on the total energy at large separation distances to find the energy difference between creating the larger tube and the smaller tubes at varying separation distances. 
    \item Since the energy required to create the `inner tube' is less than the energy required to create the `outer tube', use the resulting fit parameters to add the energy difference back to the total energy in Eq.\ (\ref{eqn:tube_Etot_infty}). This isolates the Casimir energy.
\end{enumerate}

\begin{figure}[!htb]
\begin{center}
\includegraphics[scale=.7]{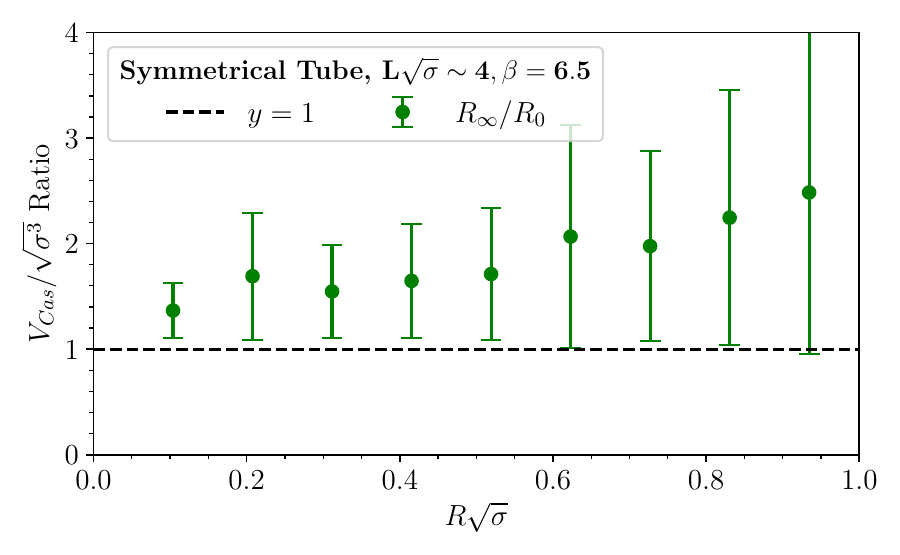}
\caption{The Ratio of the Casimir potential per unit surface area for a symmetrical tube with side lengths $R\sqrt{\sigma}$ in $(3+1)$D SU(3), comparing the two normalisation methods.}%
\label{fig:VcasTube_Fit_Ratio}
\end{center}
\end{figure}

\noindent
We highlight the following important features of the symmetrical tube geometry based on our numerical results:
\begin{itemize}
    \item The Casimir potential is \textit{negative}, which implies that the resulting Casimir force is \textit{attractive}. This result is inconsistent with the Casimir effect of a weakly coupled, massless non-interacting scalar field computed in Ref.\ \cite{Mogliacci:2018oea} using Dirichlet boundary conditions. We highlight that in the lattice formulation, if one does not correctly account for the energy contributions from the boundaries in the symmetrical tube geometry, then one would incorrectly find a \textit{repulsive} Casimir potential as can be seen directly from the slope of the potential Fig.\ (\ref{fig:tube_total_S}).
    \item The functional form of the total Casimir potential (not per unit surface area) shown in Fig.\ (\ref{fig:VcasTubeTotal}) is now linear as compared to a logarithmic form in the case of parallel plates. Hence the Casimir force experienced by the surfaces of the tube diminishes slower with increasing separation distance.
    \item The potential of the tube at varying inverse coupling forms a smooth curve, but it does not collapse to a single curve as was the case for parallel wires and plates. There are no additional energy contributions to the system that could introduce a different scaling, and we note this as a subject for further investigations. 
\end{itemize}

\noindent
In summary, we have proposed two methods in which the energy contributions from creating the tube's boundaries can be accounted for such that the Casimir energy can be isolated in these first principle lattice simulations. We have shown all the relevant quantities for the symmetrical tube side-by-side in Fig.\ (\ref{fig:tube_total_S} - \ref{fig:VcasTube}) to provide a qualitative comparison of the two methods as we expect them to give the same result (within uncertainties) for the Casimir energy. However, there is a clear qualitative difference between the measured Casimir energies, with the $R_{\infty}$ subtraction producing a slightly higher potential.\\

\begin{figure}[!htb]
    \centering
    \subfigure[Vacuum Subtracted Energy]{{\includegraphics[scale=0.5]{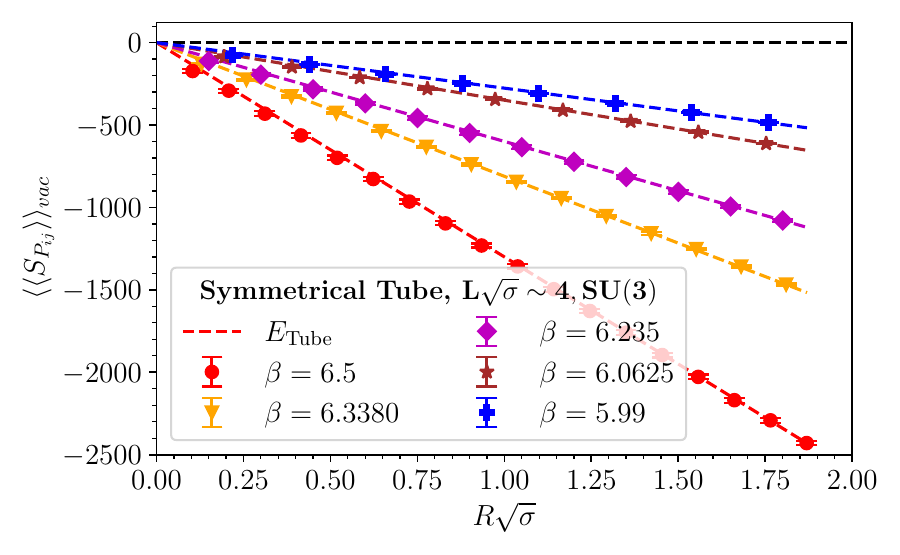} \label{fig:tube_Svac_Slope} }}%
    \hspace{-0.45cm}
    \subfigure[$R_{\infty}$ Subtracted Energy]{{\includegraphics[scale=0.5]{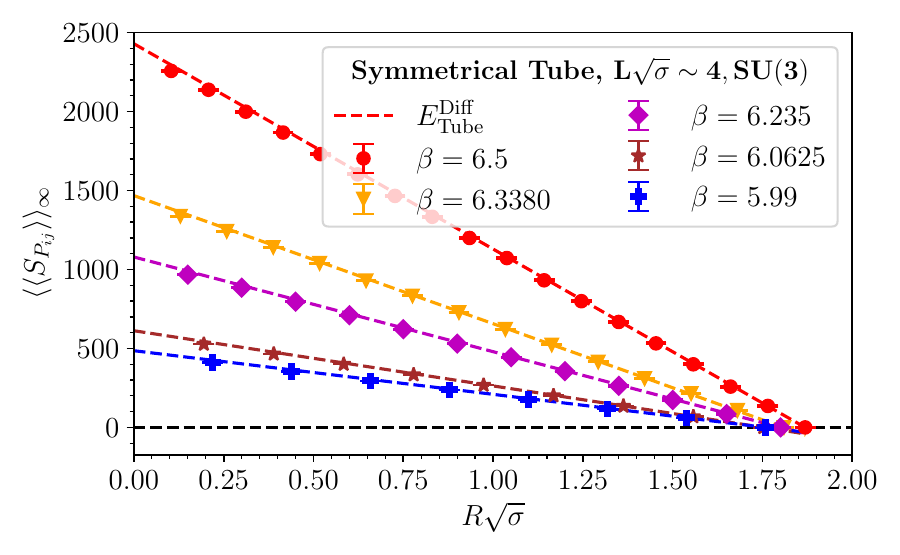} \label{fig:tube_Sinf_Slope} }}%
    \caption{Total energy of the system in a symmetrical tube in $(3+1)$D SU(3) for different couplings and different normalisation schemes in lattice units.}%
    \label{fig:tube_total_S_Slope}
\end{figure}

\noindent
We provide a quantitative comparison of the Casimir potential per surface area for $\beta=6.5$ in Fig.\ (\ref{fig:VcasTube_Fit_Ratio}), where $R_{0}$ and $R_{\infty}$ correspond to the first and second method, respectively. Provided that we have accounted for all the energy contributions to the system, and there are no additional physical parameters to account for this difference in the measured potential, we attribute such differences to the subtleties of the fitting functions. There is a possibility that we are significantly underestimating our errors as we do not include the systematic errors associated with the fit parameters. We leave such an investigation of the errors for possible future explorations and instead, we provide an alternative angle to tackle this problem.\\

\noindent
Let's start by revisiting our second approach shown on the right frames of Fig.\ (\ref{fig:tube_total_S} - \ref{fig:VcasTube}), where we subtracted the energy contribution of the expanding tube's boundaries by comparing it to a larger outer tube with side length, $R=R_{\infty}=L\sqrt{\sigma}/2$. In principle, the fitting function of the energy difference should obey an additional condition that, $\left. E_{\text{diff}} \right|_{\text{R}=L\sqrt{\sigma}/2}=0$ where the size of the `inner tube' is equal to the size of the `outer tube' and the energy difference vanishes. Note that there's an additional implication to this statement, in that, we have two fitting conditions that do not only force the fitted curve to pass through two points, but essentially fix the two endpoints of the fit.\\ 

\noindent
However, the functional form of our fit is linear and knowing the endpoints means we do not need to perform a fit in the first place. We could simply use the two conditions that fix the endpoints of the fitting function and draw a straight line to join them. That straight line describes the energy difference between the `outer tube' at $R=L\sqrt{\sigma}/2$ and the inner tubes at $R<L\sqrt{\sigma}/2$, and all we need to do is to calculate its gradient. The gradient enables us to quantify the energy difference with varying tube size and adding this energy into the system gives us the Casimir energy. We will refer to this approach as the \textit{gradient method} for the remainder of this text, and we apply it to the two methods we have proposed in order to eliminate the fit.\\

\begin{figure}[!htb]
    \centering
    \subfigure[Vacuum Subtracted Potential]{{\includegraphics[scale=0.5]{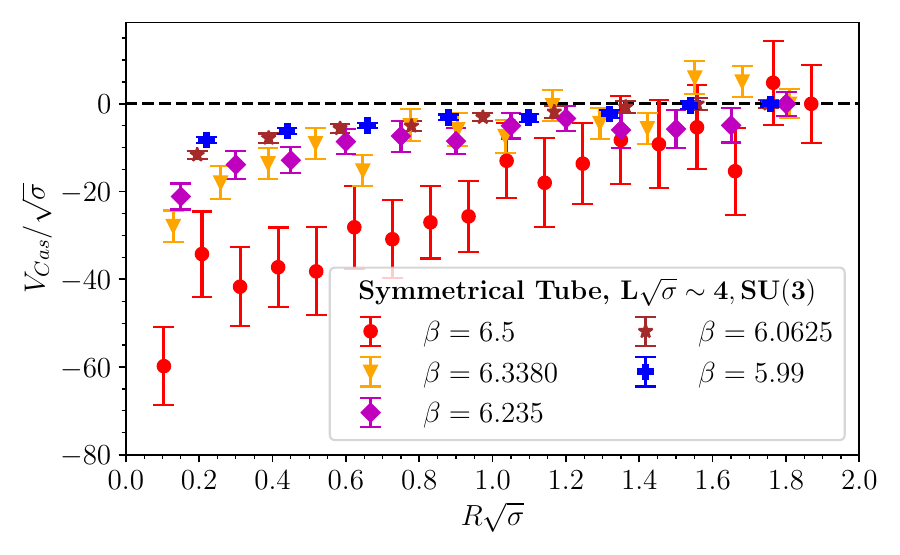} \label{fig:tube_ETotVac_Slope} }}%
    \hspace{-0.45cm}
    \subfigure[$R_{\infty}$ Subtracted Potential]{{\includegraphics[scale=0.5]{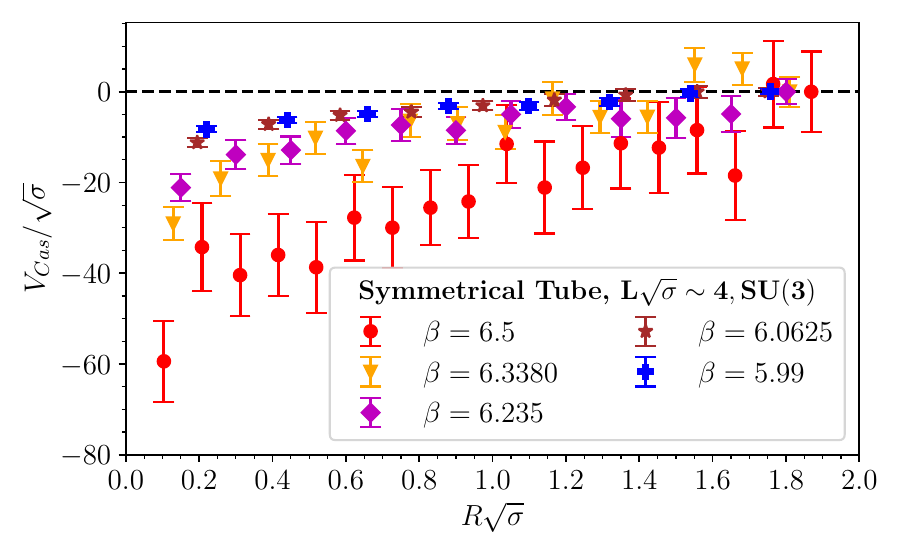} \label{fig:tube_ETotInf_Slope} }}%
    \caption{The total Casimir potential for a symmetrical tube with side lengths $R\sqrt{\sigma}$ in $(3+1)$D SU(3).}%
    \label{fig:VcasTubeTotal_Slope}
\end{figure}

\noindent
Motivated by this trick and taking a closer look at the first approach where we quantified the energy contribution from creating the tube by comparing the energy of the system to the vacuum, one notices that a similar constraint can be imposed. At the endpoint at $R\sqrt{\sigma}=0$, we have already fixed the energy of the system to zero and this fixed the constant term of the fit. Recall that we also computed the energy of the system at $R=L\sqrt{\sigma}/2$ in Fig.\ (\ref{fig:tube_Svac_Sinf}), where the Casimir effect should be negligible.\\

\noindent
We use this result of the energy for the largest possible tube to fix the energy at the second endpoint, and consequently calculate the gradient of the resulting straight line which describes the energy from creating the tube. Subtracting this energy from the system gives us the Casimir potential. The resulting fit parameters for the energy of the walls of a symmetrical tube shown in Fig.\ (\ref{fig:tube_total_S_Slope}) are provided in Table (\ref{tab:tube_fit_params}) at various gauge coupling constants, where $\langle \langle S_{P_{ij}} \rangle \rangle_{vac}$ has the same gradient with $c=0$.\\

\begin{table}[!htb]
    \centering
        {\rowcolors{2}{green!80!yellow!50}{green!70!yellow!40}
        \begin{tabular}{ |P{2.5cm}|P{1.8cm}|P{1.8cm}|P{1.8cm}|  }
        \hline
        \hline
        Energy & $\mathbf{\beta}$ & $m$ & $c$\\
        \hline
         & 6.5 & -1299.96 & 2429.55  \\
         & 6.3380 & -811.02 & 1467.88\\
        $\langle \langle S_{P_{ij}} \rangle \rangle_{\infty}$ & 6.235 & -599.38 & 1079.10 \\
         & 6.0625 & -349.36 & 612.25 \\
         & 5.99 & -276.24 & 485.79\\
        \hline
        \end{tabular}}
    \caption{Linear fit parameters for the energy contributions from the boundaries of a symmetrical tube in SU(3) for lattice size, $L\sqrt{\sigma}\sim 4$.}
    \label{tab:tube_fit_params}
\end{table}

\begin{figure}[!htb]
    \centering
    \subfigure[Vacuum Subtracted Potential/Area]{{\includegraphics[scale=0.5]{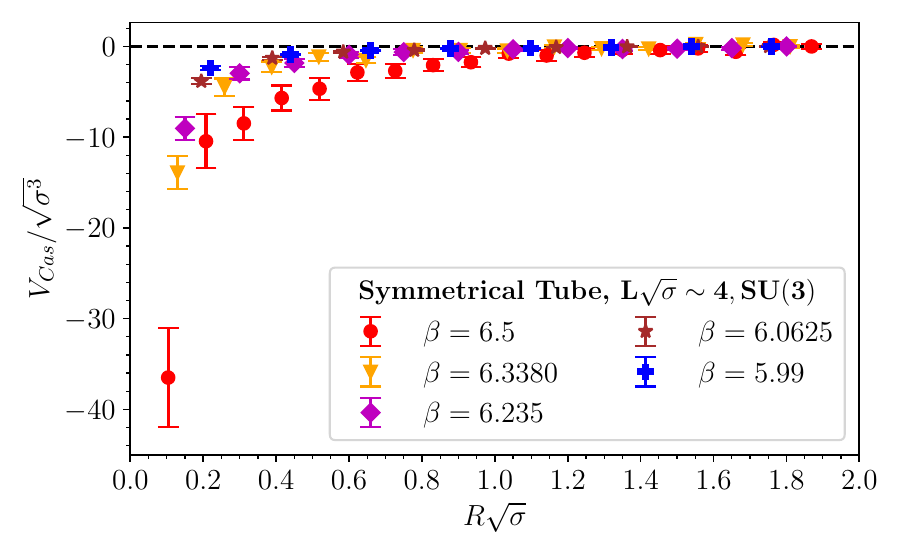} \label{fig:tube_EVac_Slope} }}%
    \hspace{-0.45cm}
    \subfigure[$R_{\infty}$ Subtracted Potential/Area]{{\includegraphics[scale=0.5]{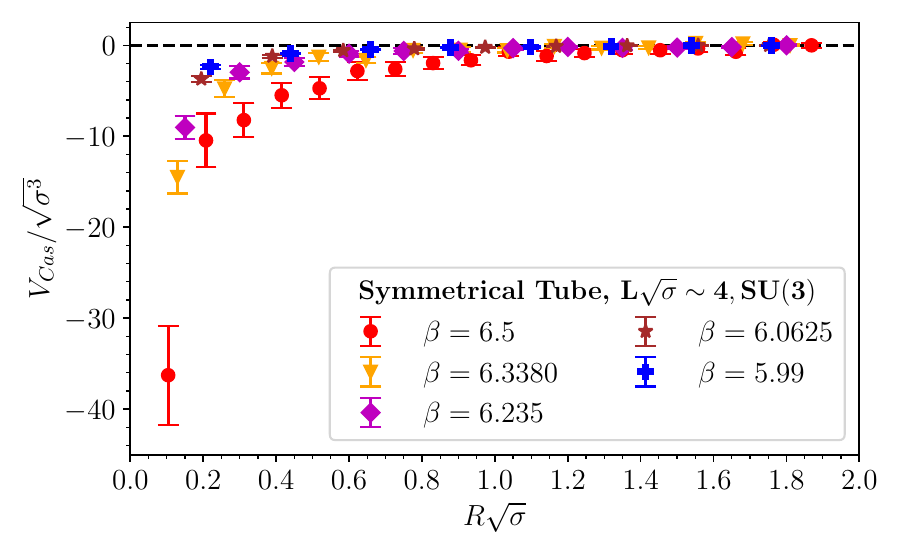} \label{fig:tube_EInf_Slope} }}%
    \caption{The Casimir potential for a symmetrical tube with side lengths $R\sqrt{\sigma}$ in $(3+1)$D SU(3) per unit surface area of the tube.}%
    \label{fig:VcasTube_Slope}
\end{figure}

\begin{figure}[!htb]
\begin{center}
\includegraphics[scale=.7]{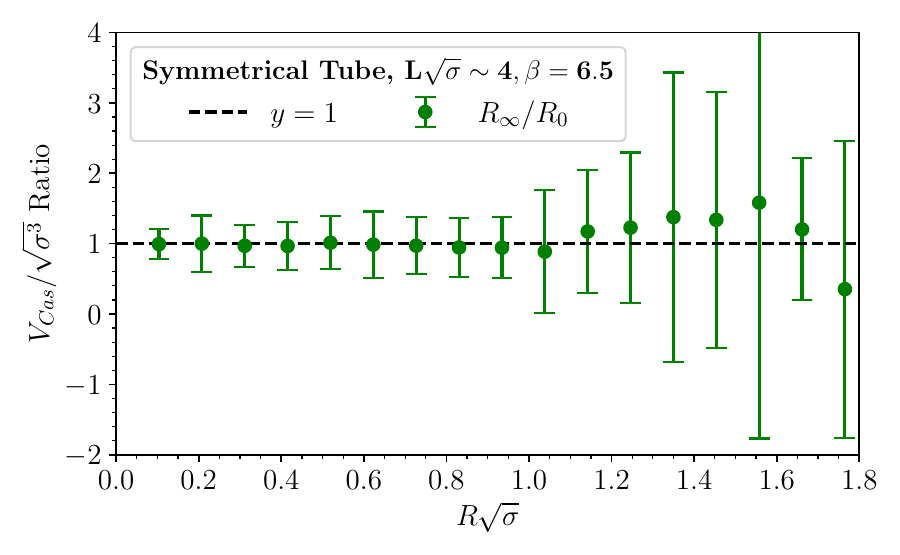}
\caption{The Ratio of the Casimir potential per unit surface area for a symmetrical tube with side lengths $R\sqrt{\sigma}$ in $(3+1)$D SU(3), comparing the two normalisation methods.}%
\label{fig:VcasTube_Fit_Ratio_Slope}
\end{center}
\end{figure}

\noindent
Note that we have essentially reduced the non-trivial computation of the energy contributions from the boundaries of the tube as the tube expands from an intuitive fitting procedure into a simple analytical calculation of the gradient of a straight line. We provide the corresponding results comparing the vacuum subtraction to the $R_{\infty}$ subtraction in Fig.\ (\ref{fig:tube_total_S_Slope} - \ref{fig:VcasTube_Slope}). The measured Casimir potential is also consistent irrespective of the normalisation choice, as opposed to using a fitting function which heavily relies on the goodness of the fit. This indicates that the gradient method captures the energy of creating the walls of the tube in an exact manner. A quantitative comparison of the two methods for the Casimir energy per unit surface area is shown in Fig.\ (\ref{fig:VcasTube_Fit_Ratio_Slope}) for $\beta=6.5$.\\

\subsection{Asymmetrical Tube}

\noindent
The asymmetrical tube is an extension of the symmetrical tube where one side-length is fixed to a constant, $R_y=\sqrt{\sigma}$ while the other side is stretched. Given that the tube also extends infinitely in one direction, $R_z=L\sqrt{\sigma}$ on the lattice, in the limit $R_x \to L\sqrt{\sigma}$ of the expanding side (corresponding to the physical limit, $R_x \to \infty$), the geometric shape of the asymmetric tube resembles that of a parallel plate configuration with $R_x=\sqrt{\sigma}$. Recall that we express our physical units in units of the string tension, so $R_x=\sqrt{\sigma}=R_{phys}=1a$.\\

\begin{figure}[!htb]
\begin{center}
\includegraphics[scale=.7]{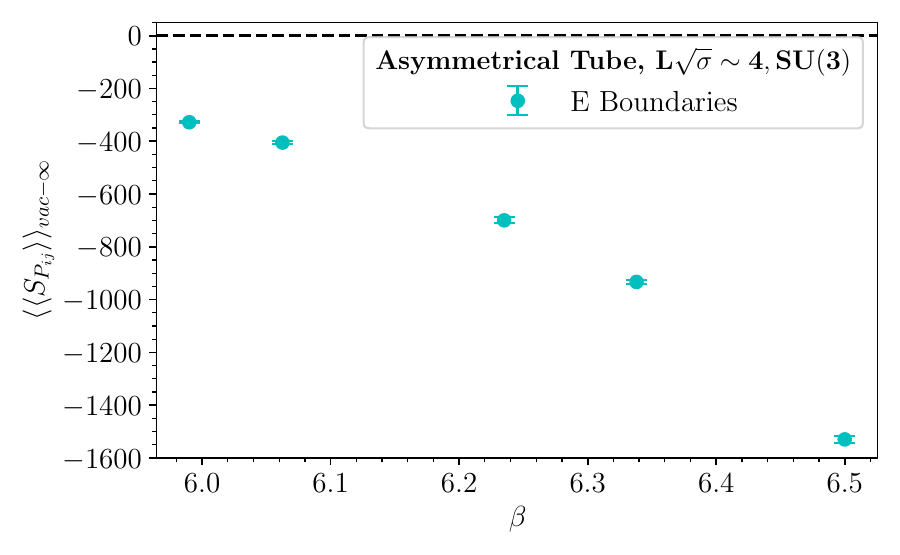}
\caption{The total energy required to create an asymmetrical tube with longest side-length, $L\sqrt{\sigma}/2$ in $(3+1)$D SU(3) at varying lattice couplings.}
\label{fig:recttube_Svac_Sinf}
\end{center}
\end{figure}


\noindent
Studying the asymmetrical tube geometry is more complicated compared to its symmetrical counterpart. This is mostly due to the difficulty in understanding the differences in the contributions from modes that fit into the small fixed side-length of the tube and those that fit into the expanding side. In a formulation of this geometry in the infinite volume limit, one would intuitively expect that in the limit, $R_x \to \infty$ (the one expanding side) then the measured Casimir force should resemble that of the infinitely extending parallel plate geometry with separation distance, $R=\sqrt{\sigma}$. Taking this limit should imply that the number and wavelength of modes that fit into the cavity between the plates or inside the tube should be equivalent.\\

\noindent
In our lattice simulations, we have used periodic boundary conditions, hence the largest asymmetrical tube that we can construct only extends to half the lattice grid-size in one direction before its faces are mirrored. Consequently, we can not geometrically test the equivalence of these two geometries because while the wavelength of modes that fit into the $R_y=\sqrt{\sigma}$ fixed direction in the tube may be the same as those in the direction perpendicular to the plates, the direction parallel to the plates accommodates modes of longer wavelengths than we can fit into the expanding side of the asymmetrical tube. It is unclear at this point how the $R_x \to \infty$ limit Casimir effect can be studied in this geometry in the present formulation employing periodic boundary conditions, however, we do not rule out its possibility.\\  

\begin{figure}[!htb]
    \centering
    \subfigure[Vacuum Subtracted Energy]{{\includegraphics[scale=0.5]{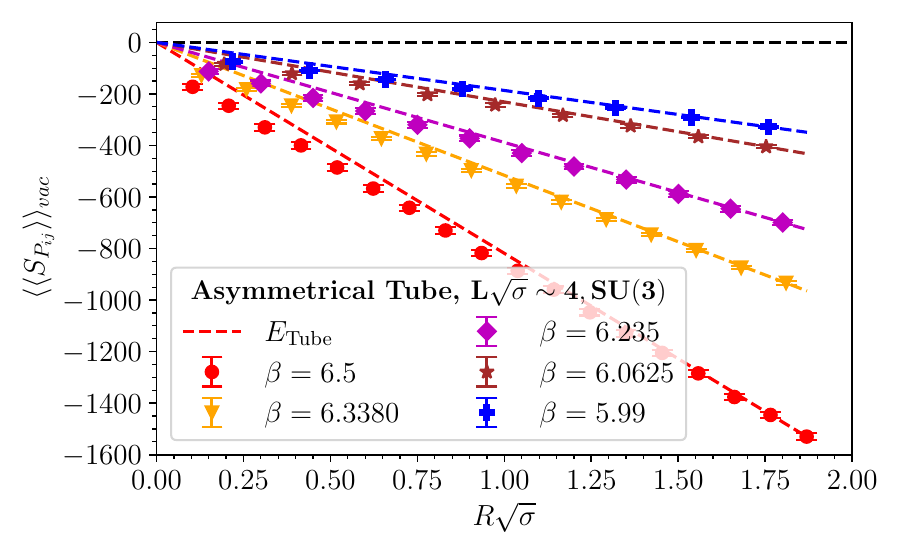} \label{fig:recttube_Svac} }}%
    \hspace{-0.45cm}
    \subfigure[$R_{\infty}$ Subtracted Energy]{{\includegraphics[scale=0.5]{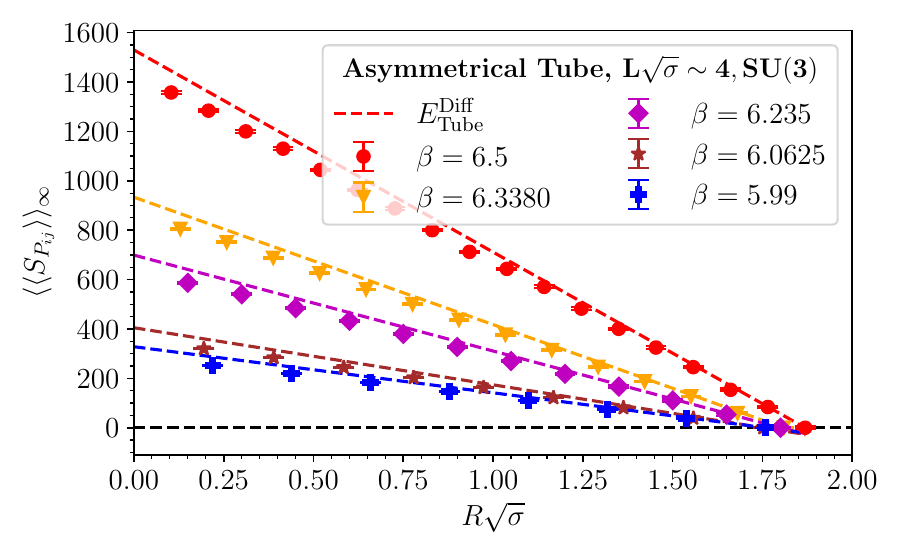} \label{fig:recttube_Sinf} }}%
    \caption{Total energy of the system in an asymmetrical tube in $(3+1)$D SU(3) for different couplings and different normalisation schemes in lattice units.}%
    \label{fig:recttube_total_S}
\end{figure}

\noindent
In the present study, we do not attempt to gain a good understanding of the Casimir effect of the asymmetrical tube in the limit, $R_x\to \infty$ where the modes fitting into the small fixed side-length should continue to contribute to a non-zero Casimir force. This is left for future investigation. Instead, we focus on studying the effect of the asymmetry in the tube configuration at small separation distances where we expect the Casimir effect to be stronger.\\ 

\noindent
We obtain the Casimir energy results for the asymmetrical tube using the same procedures we followed in the symmetrical tube, employing the two methods of accounting for energy contributions from the boundaries, the \textit{vacuum} and $R_{\infty}$ \textit{normalisation}. We only discuss results obtained through the \textit{gradient method} which we have shown to provide optimal results (in terms of consistent results from the two normalisation schemes) compared to fitting functions. The energy required to create the largest possible asymmetrical tube with dimensions, $\sqrt{\sigma}\times L\sqrt{\sigma}/2 \times L\sqrt{\sigma}$ on our finite lattice volume is shown in Fig.\ (\ref{fig:recttube_Svac_Sinf}) and is used to constrain the endpoint with the largest energy in the application of the gradient method.\\

\noindent
This energy is less than the corresponding energy for a symmetrical tube shown in Fig.\ (\ref{fig:tube_Svac_Sinf}) because one side-length is fixed to $\sqrt{\sigma}$. In the case of the parallel plates and symmetrical tube, we have argued that in the limit, $R\to \infty$ or $R\to L\sqrt{\sigma}/2$ on the lattice, the Casimir energy vanishes because of the resulting symmetry on the four-dimensional torus. At half the lattice size, the number of modes inside/outside these geometries should be equal. This is not true for the asymmetrical tube because irrespective of the size of the tube, there is no limit in which the number of modes is equal both inside and outside the tube.\\

\begin{figure}[!htb]
    \centering
    \subfigure[Vacuum Subtracted Potential]{{\includegraphics[scale=0.5]{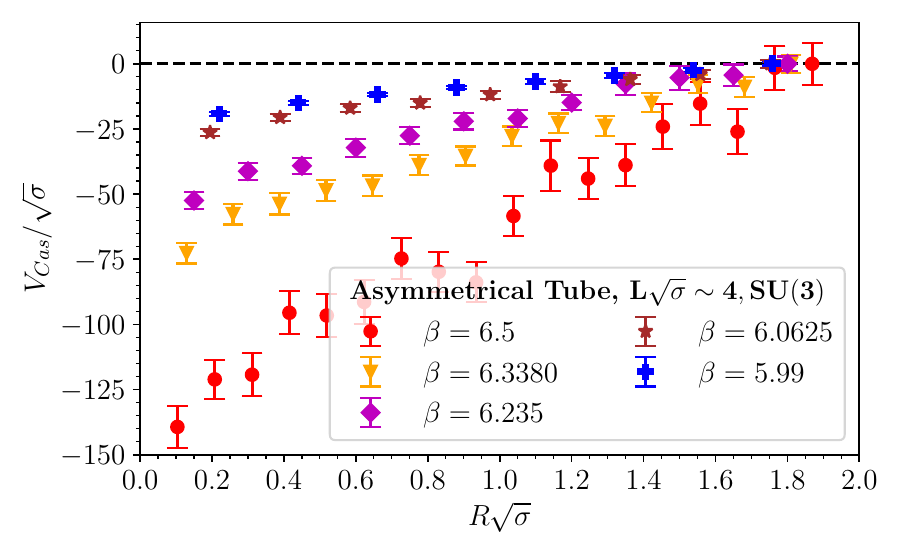} \label{fig:recttube_ETotVac} }}%
    \hspace{-0.45cm}
    \subfigure[$R_{\infty}$ Subtracted Potential]{{\includegraphics[scale=0.5]{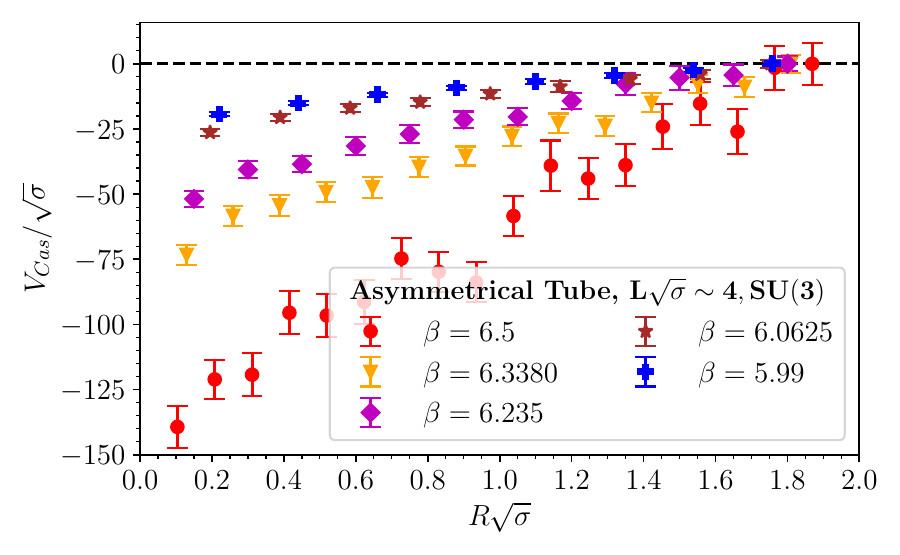} \label{fig:recttube_ETotInf} }}%
    \caption{The total Casimir potential for an asymmetrical tube with side lengths $R\sqrt{\sigma}$ in $(3+1)$D SU(3) at different inverse couplings.}%
    \label{fig:VcasRectTubeTotal}
\end{figure}

\noindent
In principle, because the asymmetrical tube maintains a geometry of two closely spaced parallel plates when one side is stretched, the Casimir energy of the system should not vanish as $R_x\to \infty$, but instead, it should approach the Casimir energy of the parallel plate geometry with separation, $R=\sqrt{\sigma}$. In the vacuum subtraction scheme, we have,
\begin{eqnarray}
     \lim_{R\to\infty} E_{\text{Tot}}^{\text{Vac}} &=& \lim_{R\to L\sqrt{\sigma}/2} \left( E_{\text{Cas}} + E_{\text{Tube}} \right) = E_{\text{Tube}} + corr_1,  
\end{eqnarray}
where the correction term corresponds to the non-vanishing Casimir energy due to the modes perpendicular to the side with fixed-length. Similarly for the $R_x\to \infty$ subtraction,
\begin{eqnarray}
     \lim_{R\to\infty} E_{\text{Tot}}^{\infty} &=& \lim_{R\to L\sqrt{\sigma}/2} \left( E_{\text{Cas}} + E_{\text{Tube}}^{\text{R}} \right) -E_{\text{Tube}}^{\text{R}_{\infty}} - E_{\text{Cas}}^{\text{R}_{\infty}}\\
     &=& E_{\text{Tube}} + corr_1,
\end{eqnarray}
where both the energy of the outer tube and its Casimir energy are just constants. Given that we are not interested in quantifying the correction term, i.e., Casimir energy of the asymmetrical tube at large R, we subtract it with the energy of the tube. One would need to alter this assumption in order to study the $R_x\to \infty$ limit.\\

\noindent
The resulting lattice total energy with the corresponding boundary energy curves for the different normalisation schemes is shown in Fig.\ (\ref{fig:recttube_total_S}) for various gauge couplings. Now, the linear curve not only describes the energy of the boundaries, but also, the Casimir energy at large separation distances given by the correction term. Therefore, by using the linear curve to subtract the energy contribution from the boundaries, we are also subtracting the Casimir energy at large separation distances and consequently enforcing a vanishing Casimir energy as $R\to \infty$. \\

\begin{figure}[!htb]
    \centering
    \subfigure[Vacuum Subtracted Potential/Area]{{\includegraphics[scale=0.5]{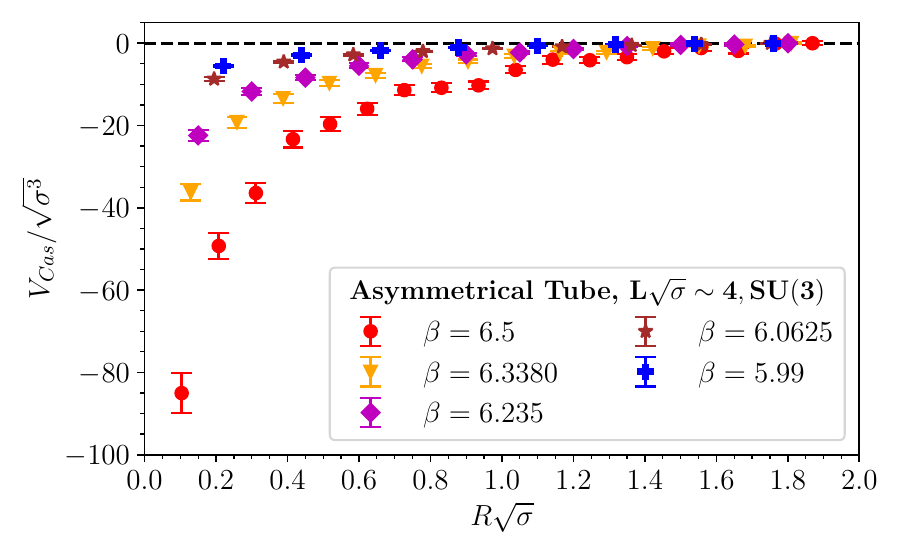} \label{fig:recttube_EVac} }}%
    \hspace{-0.45cm}
    \subfigure[$R_{\infty}$ Subtracted Potential/Area]{{\includegraphics[scale=0.5]{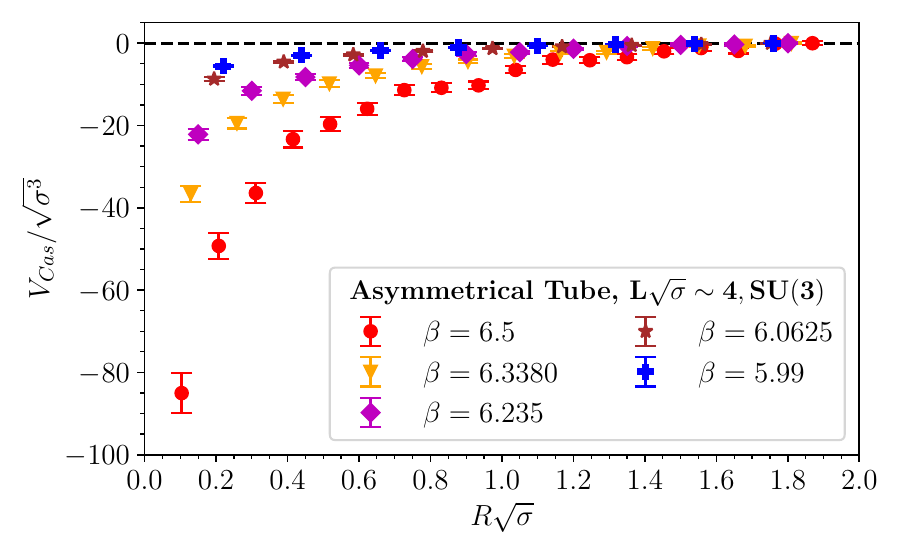} \label{fig:recttube_EInf} }}%
    \caption{The Casimir potential for an asymmetrical tube with side lengths $R\sqrt{\sigma}$ in $(3+1)$D SU(3) per unit surface area of the tube.}%
    \label{fig:VcasRectTube}
\end{figure}

\noindent
As a result of having one fixed side-length in this geometry, the application of the gradient method exactly as we did for the symmetrical tube is inaccurate. We now discuss this and show that an additional correction term is necessary to account for the differences in the symmetrical/asymmetrical geometries. We show the dimensionless physical Casimir energy and the energy per unit surface area in Fig.\ (\ref{fig:VcasRectTubeTotal} - \ref{fig:VcasRectTube}), where the total physical surface area of the asymmetrical tube is, $A=2\sigma(L + R\times L$).\\

\noindent
Qualitatively, it is apparent that the Casimir force on the asymmetrical tube is greater than the corresponding force exerted on the symmetrical tube, particularly in the limit $R\sqrt{\sigma}\to 0$. This is counter-intuitive because in this limit, the Casimir force experienced by the two types of tubes should be approximately equal. This can be easily understood by considering that in the limit $R\sqrt{\sigma}\to 0$, the geometrical shape of the separate tubes should be similar for $R \gtrsim 1$ and \textit{equivalent} for $R=1$. Hence the composition of the number of modes inside/outside the two tubes should also be equivalent.\\

\begin{figure}[!htb]
\begin{center}
\includegraphics[scale=.7]{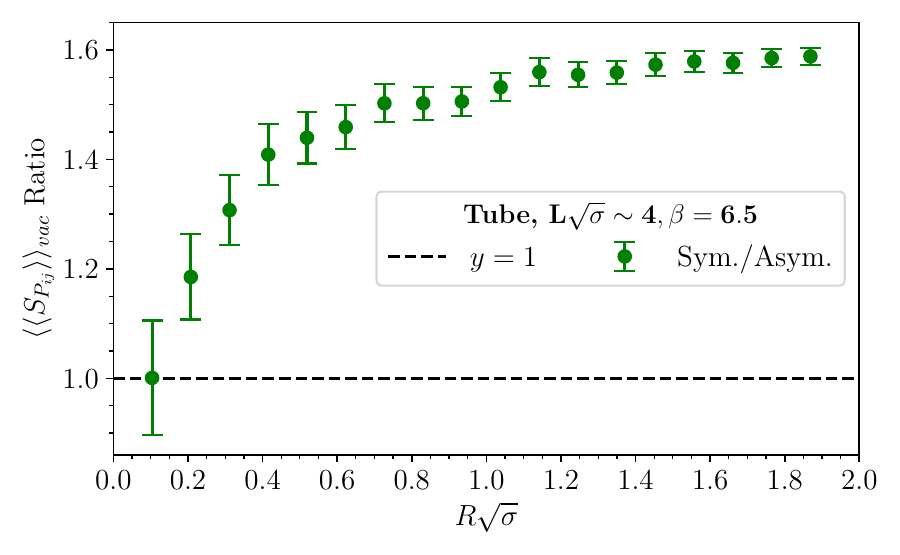}
\caption{Ratio of the vacuum normalised lattice total energy for a symmetrical and asymmetrical tube in $(3+1)$D SU(3).}
\label{fig:tubes_Svac_Ratio}
\end{center}
\end{figure}

\noindent
We emphasize this point by looking at a quantitative comparison of the dimensionless lattice total energy, $E_{\text{Tot}}^{\text{Vac}} = E_{\text{Cas}} + E_{\text{Tube}}$ shown in Fig.\ (\ref{fig:tube_Svac}) for the symmetrical tube and Fig.\ (\ref{fig:recttube_Svac}) for the asymmetrical tube. We show the ratio in Fig.\ (\ref{fig:tubes_Svac_Ratio}) for $\beta=6.5$. As expected, at $R =1$, this ratio is approximately one because the two tubes are geometrically equivalent, hence the energy required to create them and the resulting Casimir energy are equal. Given that this result does not materialise after we subtract the energy of creating the asymmetrical tube, it implies that our subtraction scheme requires modification for the asymmetrical tube. \\

\noindent
This is precisely the caveat we mentioned earlier, which we now discuss. Let's revisit the logic we followed to set the constant of the linear fit to zero at $R=0$ for the vacuum subtraction method. We argued that at $R=0$, we are in-vacuum and there is no tube, hence no boundary contributions to the energy. This argument holds for the symmetrical tube because the linear fit describes the energy contribution from the walls of a symmetrically expanding tube,
\begin{equation}
    E_{\text{Tube}}^{\text{fit}} (R_{x,y}) = E_{x=R} + E_{y=R},
\end{equation}
where both the $\hat{x}$-axis and $\hat{y}$-axis side-lengths change with $R$. In the limit, $R\to0$, both $E_{x=R}\to0$ and $E_{y=R}\to0$, resulting in a vanishing tube energy.\\

\begin{figure}[!htb]
    \centering
    \subfigure[Vacuum Subtracted Energy]{{\includegraphics[scale=0.5]{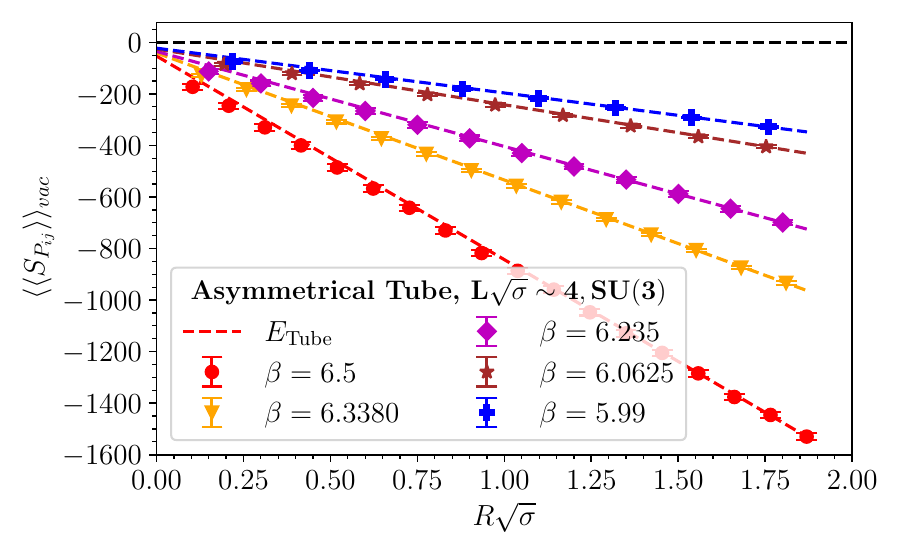} \label{fig:recttube_Svac_norm} }}%
    \hspace{-0.45cm}
    \subfigure[$R_{\infty}$ Subtracted Energy]{{\includegraphics[scale=0.5]{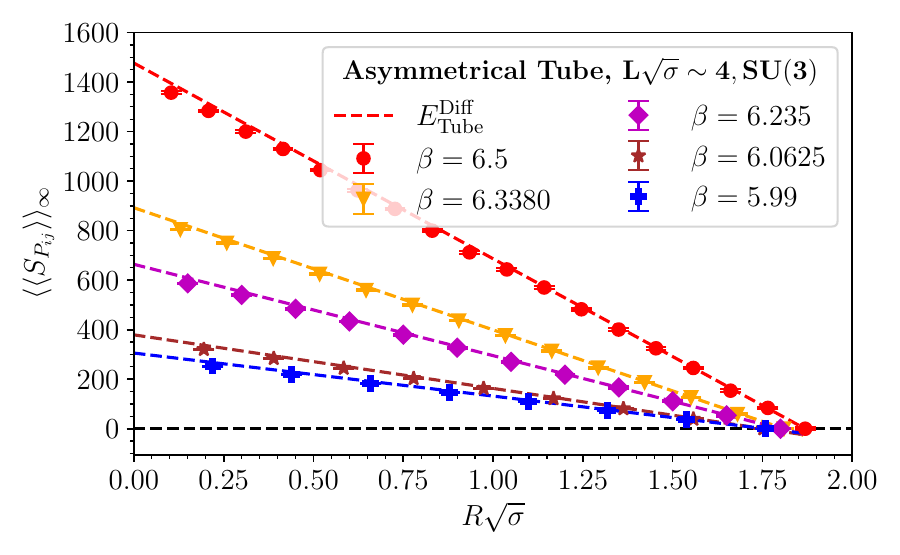} \label{fig:recttube_Sinf_norm} }}%
    \caption{Total energy of the system in an asymmetrical tube in $(3+1)$D SU(3) for different couplings and different normalisation schemes in lattice units.}%
    \label{fig:recttube_total_Snorm}
\end{figure}

\noindent
In the case of the asymmetrical tube, this argument breaks down because the linear fit describes the energy of the walls of the tube with one expanding side with the other side's energy contribution fixed to a constant. Thus we have,
\begin{equation}
    E_{\text{Tube}}^{\text{fit}} (R_{x,y=1}) = E_{x=R} + E_{y=1} = E_{x=R} + c,
\end{equation}
where $c$ is the energy contribution from the two plates with fixed side-lengths, $R_y=1\sqrt{\sigma}$. Therefore, in the limit, $R\to0$,  $E_{x=R}\to0$ but the total energy of the asymmetrical tube does not vanish, $E_{\text{Tube}}\to c$.\\

\begin{figure}[!htb]
    \centering
    \subfigure[Vacuum Subtracted Potential]{{\includegraphics[scale=0.5]{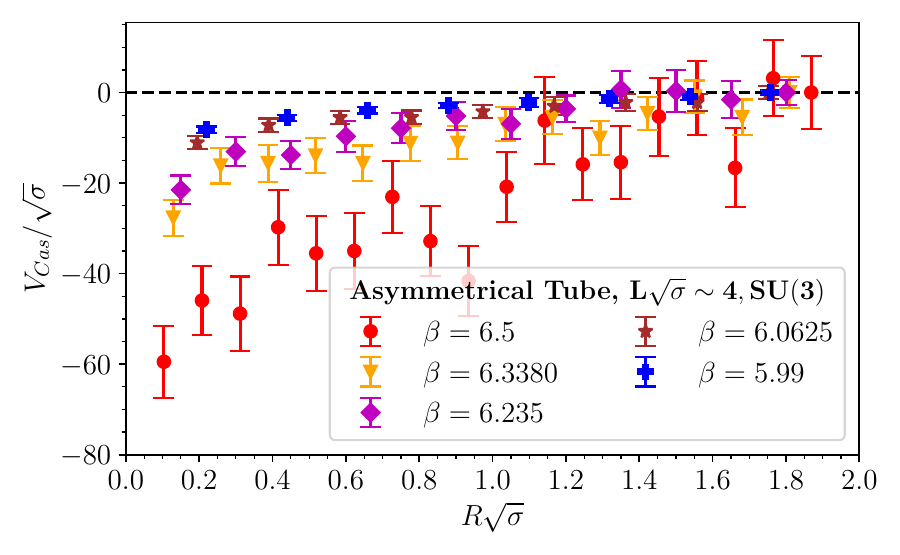} \label{fig:recttube_ETotVac_Norm} }}%
    \hspace{-0.45cm}
    \subfigure[$R_{\infty}$ Subtracted Potential]{{\includegraphics[scale=0.5]{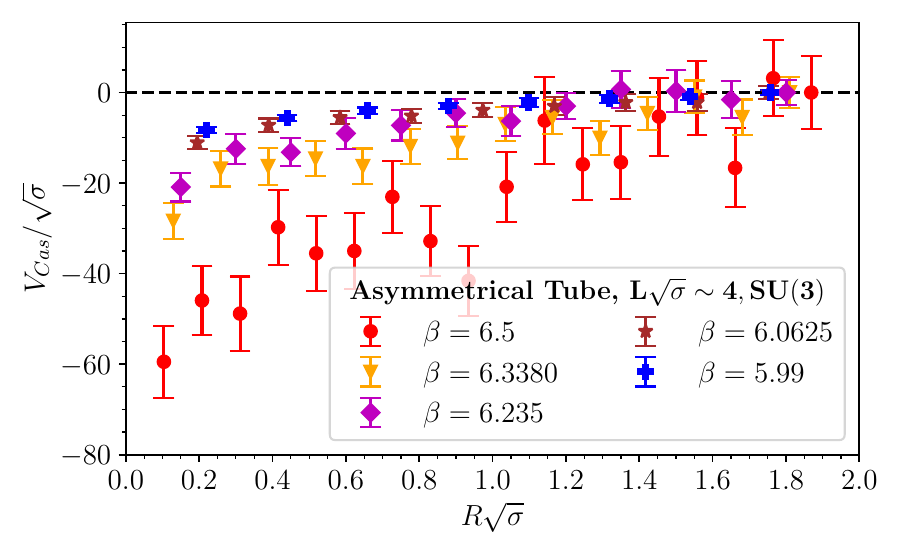} \label{fig:recttube_ETotInf_Norm} }}%
    \caption{The total Casimir potential for an asymmetrical tube with side lengths $R\sqrt{\sigma}$ in $(3+1)$D SU(3) at different inverse couplings.}%
    \label{fig:VcasRectTubeTotal_Norm}
\end{figure}

\noindent
Now, we need to find the constant, $c$ and linear fitting becomes an option, 
but we take a different route. We know that at $R=1$, the energy of creating a symmetrical tube is equal to that of creating an asymmetrical tube. We have already obtained the linear fit parameters describing the energy of the symmetrical tube. Using these fit parameters, we find the energy of the asymmetrical tube at $R=1$,
\begin{eqnarray}
    E_{\text{Tube}}^{\text{fit}} (R_{1,1}) &=& E_{x=1} + E_{y=1} = 2E_{y=1},
\end{eqnarray}
and we use this energy to fix the second endpoint at $R=1$ then apply the gradient method to obtain the slope, $m$ between this point and the other endpoint at $R=L\sqrt{\sigma}/2$. \\

\noindent
Using this gradient to extrapolate to $R=0$, we obtain the constant term,
\begin{eqnarray}
    c &=& \frac{1}{2}E_{\text{Tube}}^{\text{fit}} (R_{1,1}) + corr_2.
\end{eqnarray}
While one may expect the constant term to be half the total energy of the tube at $R=1$ because two parallel plates of side-length, $R_y=1$ remain when $R_x\to 0$, this is not the case and one needs to add a correction term. This correction term can be understood as follows; the basis of our argument is that the Casimir energy is given by the energy difference between the total energy of the system (coloured points) and the linear fitted curve describing the energy of the tube.\\

\noindent
In the case of a symmetric tube, at $R=0$, the Casimir energy should vanish because there is no tube and the energy required to create the tube correspondingly vanishes. However, in the case of the asymmetric tube, at $R=0$, we are left with two parallel plates with side-length, $R_y=1$, but our set-up still requires the Casimir energy to vanish. Hence the correction term ensures that we do not violate energy conservation. In principle, this correction term should be equivalent to the Casimir energy between two parallel plates of side-length and separation distance, $R=1$.\\ 

\begin{table}[!htb]
    \centering
        {\rowcolors{2}{green!80!yellow!50}{green!70!yellow!40}
        \begin{tabular}{ |P{2.5cm}|P{1.8cm}|P{1.8cm}|P{1.8cm}|P{1.8cm}|  }
        \hline
        \hline
        Energy & $\mathbf{\beta}$ & $m$ & $c$ & $corr_2$\\
        \hline
         & 6.5 & -789.62 & -53.05 & -14.47  \\
         & 6.3380 & -493.19 & -41.02 & -11.37\\
        $\langle \langle S_{P_{ij}} \rangle \rangle_{vac}$ & 6.235 & -372.60 & -34.54 & -10.68\\
         & 6.0625 & -216.33 & -25.90 & -8.11 \\
         & 5.99 & -173.79 & -22.52 & -7.84\\
         \hline
          & 6.5 & -789.62 & 1475.75 & -14.47  \\
         & 6.3380 & -493.19 & 892.63 & -11.37\\
        $\langle \langle S_{P_{ij}} \rangle \rangle_{\infty}$ & 6.235 & -372.60 & 670.81 & -10.68 \\
         & 6.0625 & -216.33 & 379.12 & -8.11 \\
         & 5.99 & -173.79 & 305.62 & -7.84\\
        \hline
        \end{tabular}}
    \caption{Linear fit parameters for the energy contributions from the boundaries of an asymmetrical tube in SU(3) for lattice size, $L\sqrt{\sigma}\sim 4$.}
    \label{tab:recttube_fit_params}
\end{table}

\noindent
At the time of this thesis, we have not yet performed a numerical calculation to find the Casimir energy between finitely-sized plates on the lattice, and we leave the confirmation of the correction term for future work. This approach is applied similarly for the $R_{\infty}$ normalisation scheme. The resulting energy contributions from the walls of the asymmetrical tube are shown in Fig.\ (\ref{fig:recttube_total_Snorm}) as linear fits for the two normalisations used and the corresponding fit parameters are given in Table (\ref{tab:recttube_fit_params}).\\

\begin{figure}[!htb]
    \centering
    \subfigure[Vacuum Subtracted Potential/Area]{{\includegraphics[scale=0.5]{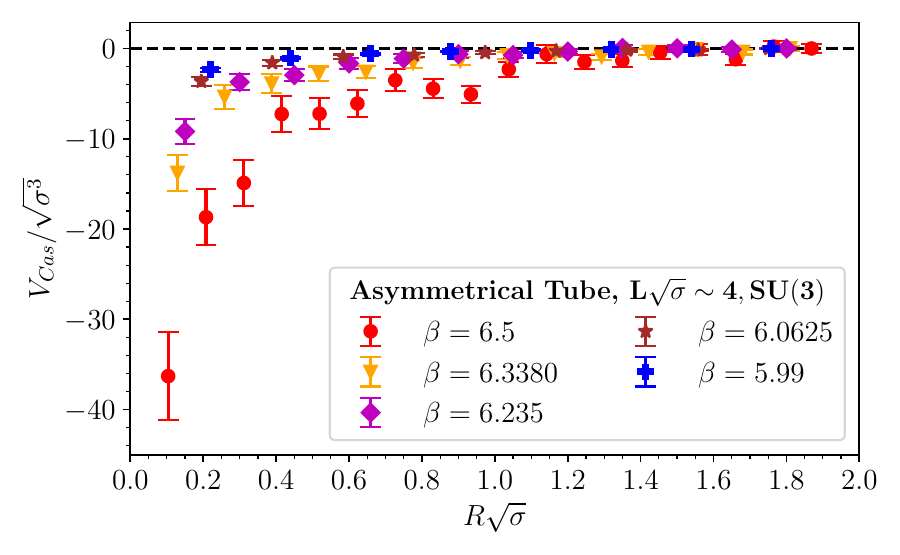} \label{fig:recttube_EVac_Norm} }}%
    \hspace{-0.45cm}
    \subfigure[$R_{\infty}$ Subtracted Potential/Area]{{\includegraphics[scale=0.5]{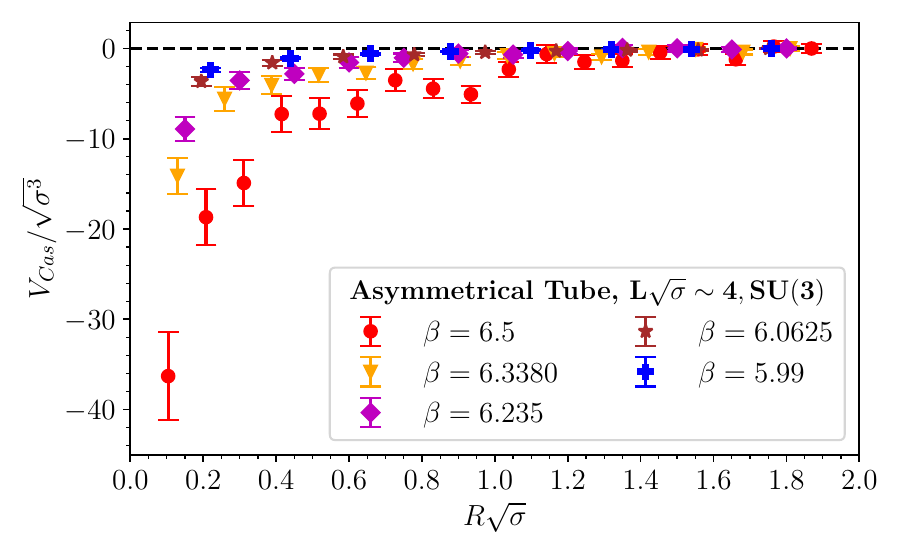} \label{fig:recttube_EInf_Norm} }}%
    \caption{The Casimir potential for an asymmetrical tube with side lengths $R\sqrt{\sigma}$ in SU(3) per unit surface area of the tube.}%
    \label{fig:VcasRectTube_Norm}
\end{figure}

\noindent
The total Casimir energy and the energy per unit surface area of the asymmetrical tube are shown in Fig.\ (\ref{fig:VcasRectTubeTotal_Norm} - \ref{fig:VcasRectTube_Norm}) using the two normalisation schemes. A qualitative comparison of this potential to its symmetrical tube counterpart shows that the magnitudes are now consistent, particularly in the limit $R\to 1$ where the two tubes are geometrically identical. \\

\noindent
Earlier, we provided reasons why one would expect the asymmetrical tube to exhibit similar properties as the parallel plate configuration with separation distance, $R=1$ in the limit, $R\to \infty$. In this limit, the number of modes in the interior cavity of the asymmetrical tube should be the same as those in the cavity between the plates. Intuitively, one would expect that the Casimir effect for these two geometries should have similar characteristics in this limit. The total Casimir energy for the parallel plate configuration is shown in Fig.\ (\ref{fig:casimir_total_plates_3dsu3}), and at $R=1$, the potential is strongly attractive.\\ 

\begin{figure}[!htb]
    \centering
    \subfigure[Vacuum Subtracted Potential]{{\includegraphics[scale=0.5]{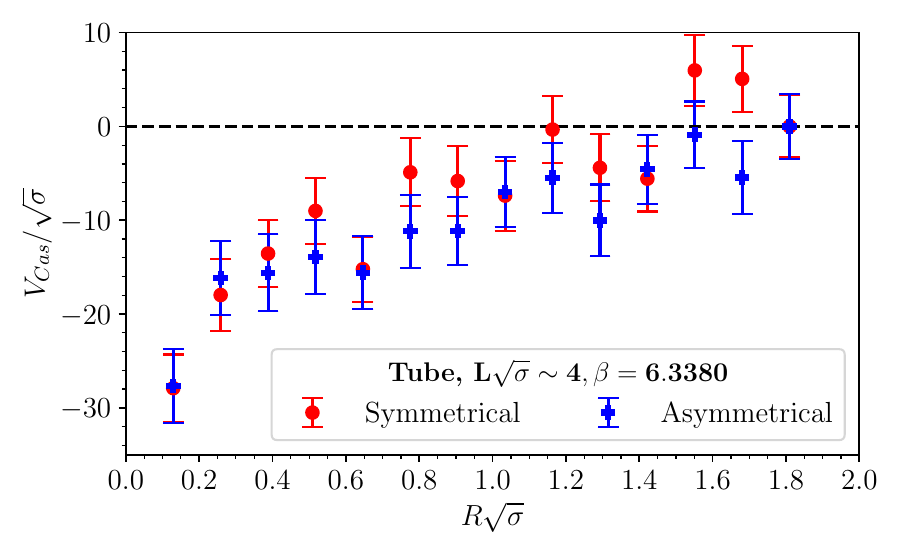} \label{fig:VcasTotalTubes_Compare} }}%
    \hspace{-0.45cm}
    \subfigure[Symmetrical/Asymmetrical Tube Ratio]{{\includegraphics[scale=0.5]{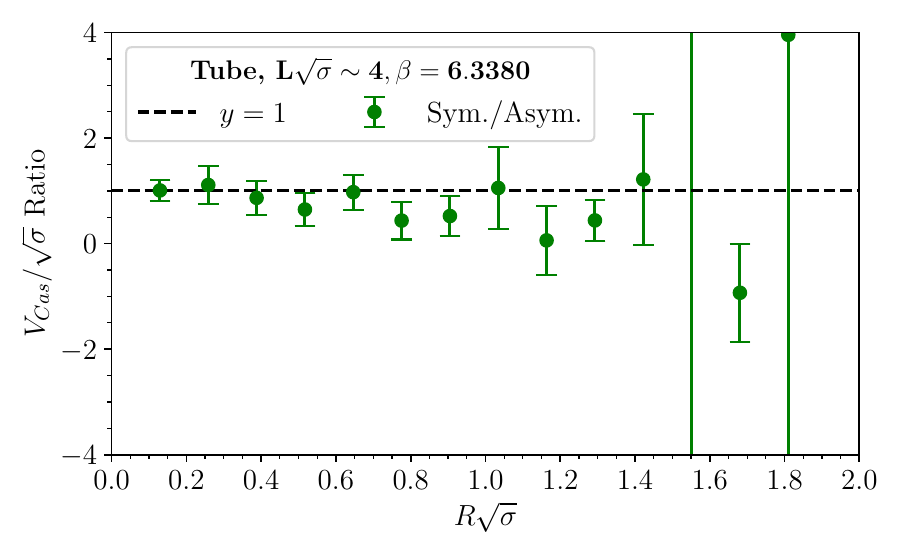} \label{fig:VcasTotalTubes_Ratio} }}%
    \caption{A comparison of the total Casimir potential for a symmetrical and an asymmetrical tube in $(3+1)$D SU(3).}%
    \label{fig:VcasTotalTubes}
\end{figure}

\noindent
On the other hand, the total Casimir energy for the asymmetrical tube is shown in Fig.\ (\ref{fig:VcasRectTubeTotal_Norm}) and is weakly attractive in the limit $R\to\infty$. While we work on a lattice size, $L\sqrt{\sigma}\sim 4 \sim 1.6$ fm where the finite volume nature of the lattice may not allow us to make precise determinations at $R\to\infty$, the primary reason for not seeing any qualitative similarity to the parallel plate geometry is the artificially imposed vanishing energy at $R\to\infty$ which we use to study the effect of the geometrical asymmetry on the Casimir effect. We provide a quantitative comparison of the Casimir effect in the symmetrical and asymmetrical tube in Fig.\ (\ref{fig:VcasTotalTubes}) for the gauge coupling, $\beta=6.3380$. We plot their ratio to draw some insights.\\

\noindent


\noindent
At $R\gtrsim1$, the ratio is approximately one because the two tubes are identical. As the separation distance increases and the tubes assume different geometrical forms, the attractive Casimir force of the asymmetrical tube becomes slightly larger than that of the symmetrical tube. This is consistent with our expectations based on the number of modes inside the two geometries. In the geometry of the tube, there are more modes inside the symmetrical tube than there are in the asymmetrical tube for $R>1$. Hence, the magnitude of the attractive Casimir force should be stronger for the asymmetrical tube.\\

\section{Symmetric and Asymmetric Box}
\label{section:Symmetric and Asymmetric Box}

\noindent
The last geometry that we discuss for the non-abelian gauge theory in SU(3) is a hollow box shown in Fig.\ (\ref{fig:geometry_box}). The box has finite extents on the three-dimensional spatial axis. We consider a symmetrical box with side-lengths $R_x=R_y=R_z=R\sqrt{\sigma}$ and an asymmetrical box with two equal side-lengths, $R_y=R_z=1\sqrt{\sigma}$ and one expanding side with length $R_x=R\sqrt{\sigma}$. We refer the reader to chapter (\ref{sec:Geometry and Symmetries: Box}) for a discussion on the electromagnetic field-strength tensor components and the numerical field configurations results for the symmetrical and asymmetrical box. \\

\noindent
The geometry of a box requires the fixing of all spatial coordinates, leaving only the Euclidean time axis free. Consequently, all rotational symmetries are broken in this geometry because rotations in four-dimensional space require a minimum of two free axes. Therefore, as opposed to all previous geometries that we have discussed, where we explored various symmetry relations to reduce the full expression of the energy density of the system into simpler expressions, such an approach is not possible here. However, in chapter (\ref{sec:Geometry and Symmetries: Box}), we showed that numerical results suggest that some field components are locally equivalent. Hence it would be valuable in future work to explore whether this results in a general analytical simplification of the energy density expression.\\

\noindent
The energy density of the box reads,
\begin{eqnarray}
    \varepsilon_{\text{box}}(x,y,z) &=& \frac{1}{2}\left[ \langle B_x^2 \rangle + \langle B_y^2 \rangle + \langle B_z^2 \rangle - \langle E_x^2 \rangle - \langle E_y^2 \rangle - \langle E_z^2 \rangle \right]
    \label{eqn:box_Edensity}
\end{eqnarray}
and is dependent on $x$, $y$ and $z$ because the gluon fields are enclosed in a finite volume (box) and the Casimir force acts on the walls of the box in all three spatial directions. The lattice expression of the energy density is a sum over the plaquettes on the worldvolume of the box,
\begin{eqnarray}
     \varepsilon(x,y,z)_{\text{box}}^{\text{lat}} &=& \langle S_{P_{ij}} \rangle = \frac{1}{N_{\tau}}\sum\limits_{N_{\tau}} S_{P_{ij}},
\end{eqnarray}
which comprise of all the plaquettes in the spatial and temporal directions according to Eq.\ (\ref{eqn:box_Edensity}). The resulting expression for the Casimir potential of the box is,
\begin{eqnarray}
    V^{\text{lat}}_{\text{Cas}}(R) &=& \left[ \int_{d\mathcal{V}}\, \varepsilon(x,y,z)_{\text{box}} \right]_{R-R_0} = \left[ \sum\limits_{N_{x}, N_{y} , N_{z}} \varepsilon(x,y,z)_{\text{box}}^{\text{lat}} \right]_{R-R_0}\\
    &=&  \sum\limits_{N_{x}, N_{y}, N_{z}} \left[ \langle S_{P_{ij}} \rangle_R - \langle S_{P_{ij}} \rangle_{R_0} \right] = \langle \langle S_{P_{ij}} \rangle \rangle,
    \label{eqn:box_casimir_lattice}
\end{eqnarray}
and the expressions for the physical energies follow similarly to other cases already discussed in four-dimensional space.\\

\begin{figure}[!htb]
    \centering
    \subfigure[Vacuum Subtracted Energy]{{\includegraphics[scale=0.5]{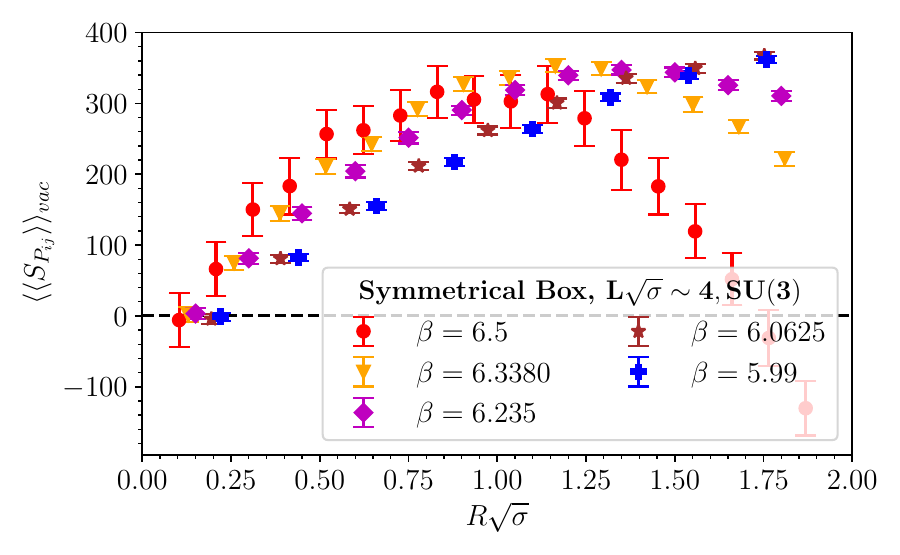} \label{fig:box_Svac} }}%
    \hspace{-0.45cm}
    \subfigure[$R_{\infty}$ Subtracted Energy]{{\includegraphics[scale=0.5]{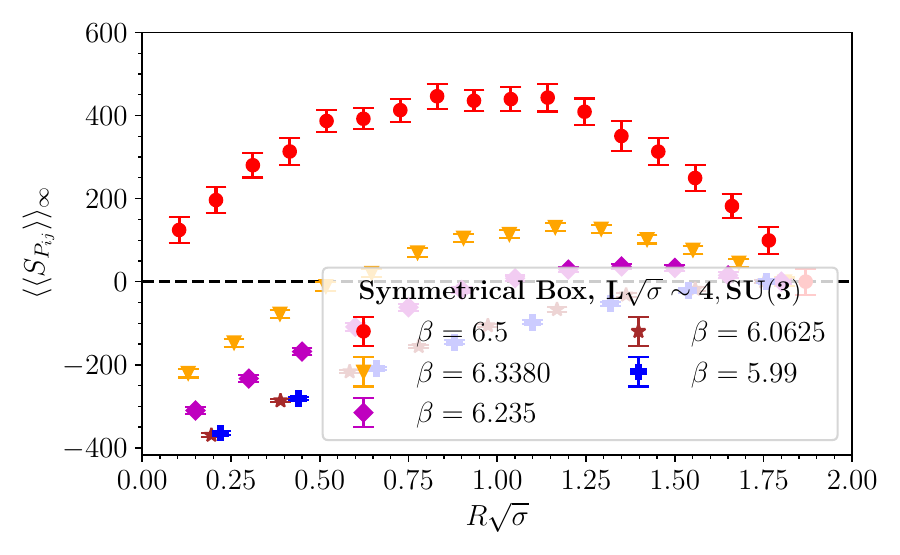} \label{fig:box_Sinf} }}%
    \caption{Total energy of the system in a symmetrical box in $(3+1)$D SU(3) for different couplings and different normalisation schemes in lattice units.}%
    \label{fig:box_total_S}
\end{figure}

\noindent
In the previous section, we went through the intricacies of accounting for the energy contributions from the boundaries of the tube. Such difficulties were overcome by exploiting the linear dependence with respect to tube size of the energy of the walls of the tube. We encounter the same problem in the case of the box geometries, and once again, have to redefine the normalisation condition at $R_0$. We now discuss these boundary energy contributions separately for the symmetrical and asymmetrical box.\\

\subsection{Symmetrical Box}

\noindent
We start by approaching this boundary problem for the symmetrical box in similar fashion as the symmetrical tube. We compute the lattice total energy of the system using Eq.\ (\ref{eqn:box_casimir_lattice}) and compare the result obtained using the vacuum subtraction and it's $R_{\infty}$ counterpart where we subtract the energy of a larger, symmetrical outer box of side-length, $L\sqrt{\sigma}/2$. These results are shown in Fig.\ (\ref{fig:box_total_S}), and we observe that the energy dependence does not vary linearly with increasing box size as was the case for the tubes, allowing for the simplification of the calculation.\\

\noindent
The main concern about the total energy in the symmetric box is that it is non-monotonous. We have already seen in previous sections that the energy contribution from the boundaries increases with the size of the boundary, i.e., the larger the box, the larger the energy contribution from the introduction of its walls should be. Hence this is a puzzling result because intuitively, the Casimir effect should vanish with increasing separation distance or box-size.\\

\noindent
Nothing suggests that any other effects should materialise when the box is increased to a certain size. The same argument is true for thermal fluctuations, which should not fit inside the box. Therefore, we maintain that the observed effect of the total energy in the symmetric box is purely a consequence of the box's boundaries which include non-trivial contributions from the edges and corners. This argument is motivated by the $R_{\infty}$ subtracted energy in Fig.\ (\ref{fig:box_Sinf}), which vanishes at $R=L\sqrt{\sigma}/2$.\\

\noindent
In the case of the symmetrical tube we observed that the functional form of the system's energy was linear for $R\to \infty$ for both normalisation schemes. However, in the case of the symmetrical box, the dependence of the total energy of the system on the separation distance (and consequently the expanding volume of the box) is more intricate. Perhaps a parabolic analytic form would work based on the observation of the energy at coupling, $\beta=6.5$, but we would need to increase our physical lattice volume to validate whether this can be generalised to other couplings. \\

\noindent
Despite the difficulty in assuming a simple analytical form for the system's energy, there some common and consistent features. In the vacuum normalisation, in the limit, $R\sqrt{\sigma} \to 0$, the total energy of the system,
\begin{equation}
    E_{\text{Tot}}^{\text{Vac}} = E_{\text{Cas}} + E_{\text{Box}},
\end{equation}
approaches zero because in this limit, the energy contribution from the walls of the box vanishes and so does the Casimir energy since it has to vanish in the absence of the box. On the other hand, in the $R_{\infty}$ subtraction scheme, in the limit $R\sqrt{\sigma} \to L\sqrt{\sigma}/2$, the total energy of the system given in Eq.\ (\ref{eqn:tube_Etot_infty}) with `Tube' replaced by `Box', vanishes. This is consistent with our expectations because in this limit, the Casimir energy should be negligible and if the `inner box' is the same size as the `outer box', then their energy contributions cancel. \\

\noindent
We propose an alternative normalisation scheme for the symmetrical box. Each box is formed by a combination of six sides/faces, therefore, we start by finding the energy contribution of two finitely extending faces of the same size placed a distance, $R= L\sqrt{\sigma}/2$ apart. As discussed in previous sections, the energy contribution of the faces is independent of their separation distance. We merely place these faces far apart to ensure that there are no Casimir energy contributions to the total energy of the system.\\

\begin{figure}[!htb]
\begin{center}
\includegraphics[scale=.7]{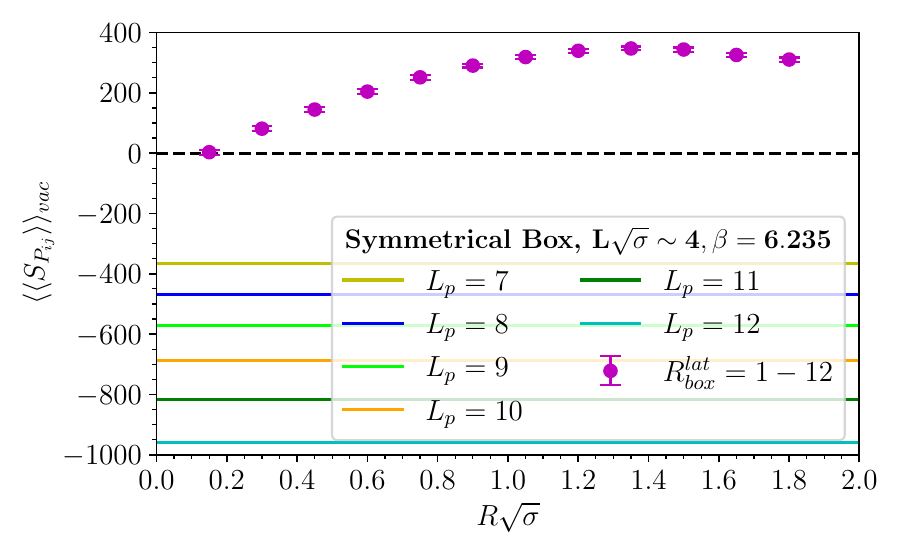}
\caption{Vacuum normalised energy of six faces compared to the vacuum normalised energy of a box with side-lengths equal to the size of the faces.}
\label{fig:Svac_box_plates}
\end{center}
\end{figure}

\noindent
The box system has total energy,
\begin{eqnarray}
  E_{\text{Box}}^{\text{Tot}}  &=& E_{\text{Cas}}^{\text{Box}} + E_{\text{Box}} + E_{\text{Vac}},
\end{eqnarray}
whereas the two finite parallel faces system has total energy,
\begin{eqnarray}
  E_{\text{Faces}}^{\text{Tot}}  &=& \cancelto{0}{E_{\text{Cas}}^{\text{Faces}}} + E_{\text{Faces}} + E_{\text{Vac}}.
\end{eqnarray}
We numerically compute the energy contribution from two finite faces placed far apart. In the end we are interested in the six-faces configuration in order to compare the total energy of this configuration to the energy of the box system (whose geometry is formed by a total of six faces). Thus, we multiply $E_{\text{Faces}}^{\text{Tot}}$ by three to obtain,
\begin{eqnarray}
  E_{\text{Faces}}^{\text{Tot}_2}  &=& 3E_{\text{Faces}}^{\text{Tot}} = 3(E_{\text{Faces}} + E_{\text{Vac}}) = \Tilde{E}_{\text{Faces}} + 3E_{\text{Vac}}.
\end{eqnarray}
Performing a vacuum subtraction on the energies of the two systems we obtain,
\begin{eqnarray}
  \Tilde{E}_{\text{Box}}^{\text{Tot}}  &=& E_{\text{Box}}^{\text{Tot}} - E_{\text{Vac}} = E_{\text{Cas}}^{\text{Box}} + E_{\text{Box}} \label{eqn:box_energy_vacnorm}\\
  \Tilde{E}_{\text{Faces}}^{\text{Tot}}  &=& E_{\text{Faces}}^{\text{Tot}_2} - 3E_{\text{Vac}} = \Tilde{E}_{\text{Faces}},
\end{eqnarray}
and we show the vacuum normalised energy for the box in Fig.\ (\ref{fig:Svac_box_plates}) as coloured points, along with the corresponding energy of the six faces forming a box as coloured lines, where $R=L_p$ is the lattice side-length of each face. It is reasonable that the energy contributions from the faces increases with the size of the faces.\\


\noindent
Note that the energy contribution from six faces is not equal to the energy contribution of the box (composed of six faces). Hence, this is not a direct comparison of the energy contributions in the two systems because the geometrical set-up of finite parallel faces is different from that of a single box with the same side-lengths. The differences in energies can be attributed to the presence of corners in the box (see the field contributions from the corners in the preceding chapter) and varying contributions from the edges in the two geometries.\\

\begin{figure}[!htb]
\begin{center}
\includegraphics[scale=.7]{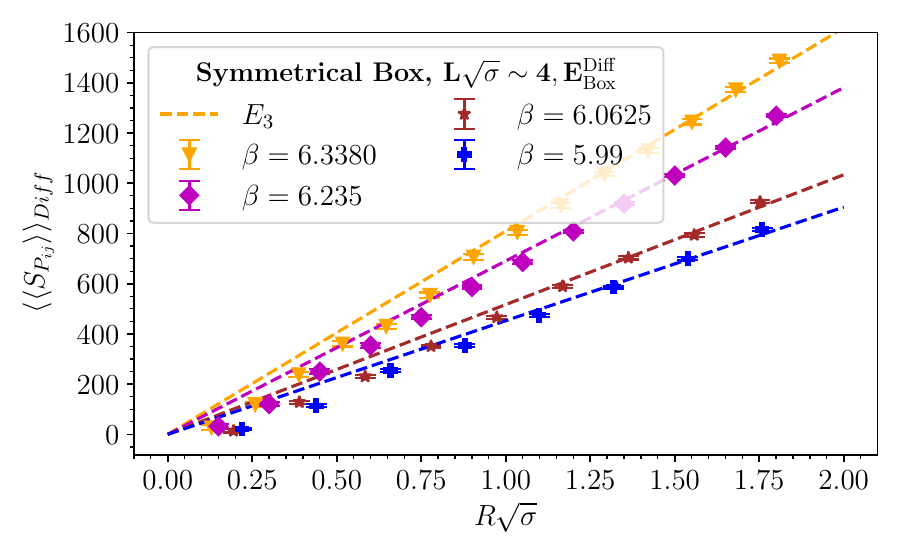}
\caption{Energy difference between the creation of six faces and a box with side-lengths equal to the size of the faces.}
\label{fig:Sdiff_box_plates}
\end{center}
\end{figure}

\noindent
This energy difference between the two systems is given by 
\begin{equation}
    E_{\text{Box}}^{\text{Diff}}(R) = \Tilde{E}_{\text{Box}}^{\text{Tot}} - \Tilde{E}_{\text{Faces}}^{\text{Tot}} = E_{\text{Cas}}^{\text{Box}} +  E_{\text{Box}} - \Tilde{E}_{\text{Faces}},
\end{equation}
and shown in Fig.\ (\ref{fig:Sdiff_box_plates}) for different gauge couplings. The Casimir energy contribution for the geometry with two finite faces is negligible irrespective of the size of the faces because the faces are placed far apart at a distance, $R= L\sqrt{\sigma}/2$. At large separation distances (large enough box-size), the Casimir energy for the box becomes negligible, leaving only the energy difference between the boundaries of the two geometries. Therefore, for $R\gtrsim 1$, this energy difference should describe,
\begin{equation}
    E_3 = \left . E_{\text{Box}}^{\text{Diff}}\right|_{R\gtrsim 1} = E_{\text{Box}} - \Tilde{E}_{\text{Faces}}.
\end{equation}

\noindent
We find that this energy difference varies linearly with the size of the box and faces. Therefore, instead of finding intricate functions to extract the energy of the box's boundaries from the observed behaviour in Fig.\ (\ref{fig:box_total_S}), we have reduced the problem to finding the fitting parameters of a linear function. We find the constant of the fit by fixing the energy difference at the $R=0$ endpoint to zero since the boundary contributions from both geometries vanish. The resulting fit parameters are provided in Table (\ref{tab:Sdiff_fit_params}) at different couplings.\\

\begin{table}[!htb]
    \centering
        {\rowcolors{2}{green!80!yellow!50}{green!70!yellow!40}
        \begin{tabular}{ |P{2.5cm}|P{1.8cm}|P{1.8cm}|P{1.8cm}|  }
        \hline
        \hline
        Energy & $\mathbf{\beta}$ & $m$ & $c$\\
        \hline
         & 6.3380 & 809.36 & 0\\
        $\langle \langle S_{P_{ij}} \rangle \rangle_{Diff}$ & 6.235 & 690.41 & 0 \\
         & 6.0625 & 516.86 & 0 \\
         & 5.99 & 452.55 & 0\\
        \hline
        \end{tabular}}
    \caption{Linear fit parameters for the energy difference between the boundaries of six faces and those of a symmetrical box with equal side-lengths as the faces.}
    \label{tab:Sdiff_fit_params}
\end{table}

\begin{figure}[!htb]
\begin{center}
\includegraphics[scale=.7]{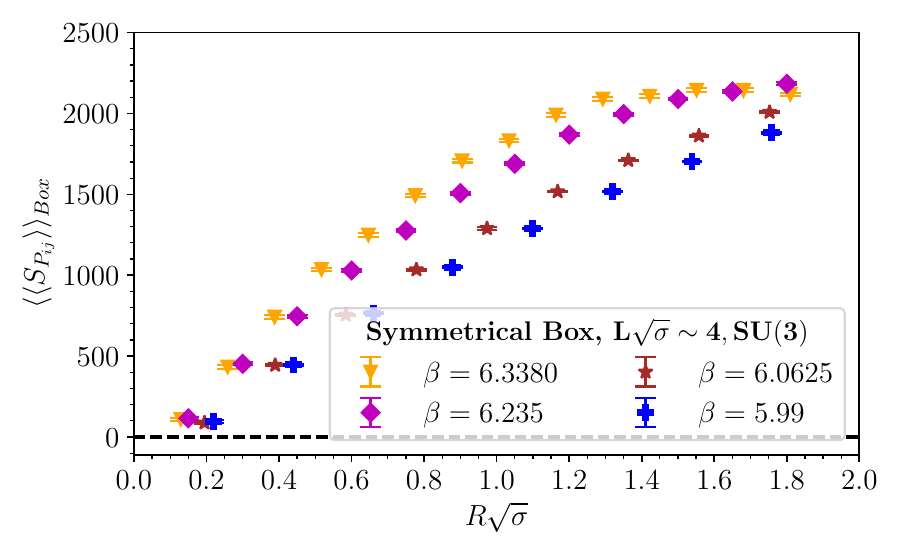}
\caption{The total energy contribution from the boundaries of a symmetrical box of different sizes.}
\label{fig:Sbox_creation}
\end{center}
\end{figure}

\noindent
Once we have obtained this energy difference, we can normalise the energy of a symmetrical box using the energy from the boundaries of a finite parallel faces configuration, accounting for six faces that form a box. We then use the linear fit to remove the expected energy difference between the boundaries of a box and the parallel faces set-up,
\begin{eqnarray}
  E_{\text{Cas}} &=& \Tilde{E}_1 - \Tilde{E}_2 - E_3.
  \label{eqn:box_cas_energy}
\end{eqnarray}
This approach allows us to isolate the Casimir energy of the symmetrical box, as well as the approximate energy from the walls of the box,
\begin{eqnarray}
  E_{\text{Box}} &=& \Tilde{E}_2 + E_3,
  \label{eqn:box_walls_energy}
\end{eqnarray}
where $E_1$ is the total energy of the system with a box present, $E_{\text{Faces}}^{\text{Tot}}$ is the total energy of the system with a two parallel faces placed far apart (times three) and $E_3$ is the energy difference of the respective boundaries.\\

\noindent
Most importantly, we have not only isolated the Casimir energy for the box, but also, the energy contributions from the walls of the box given by Eq.\ (\ref{eqn:box_walls_energy}). Despite not having an analytical formula to describe the energy from creating the box, we provide an approach to numerically extract it from lattice simulations data and we show this energy in Fig.\ (\ref{fig:Sbox_creation}). While our result acquires a dependence on the goodness of the fit, it still provides a good estimate for the Casimir effect in this geometry. \\

\begin{figure}[!htb]
    \centering
    \subfigure[Total Casimir Energy]{{\includegraphics[scale=0.5]{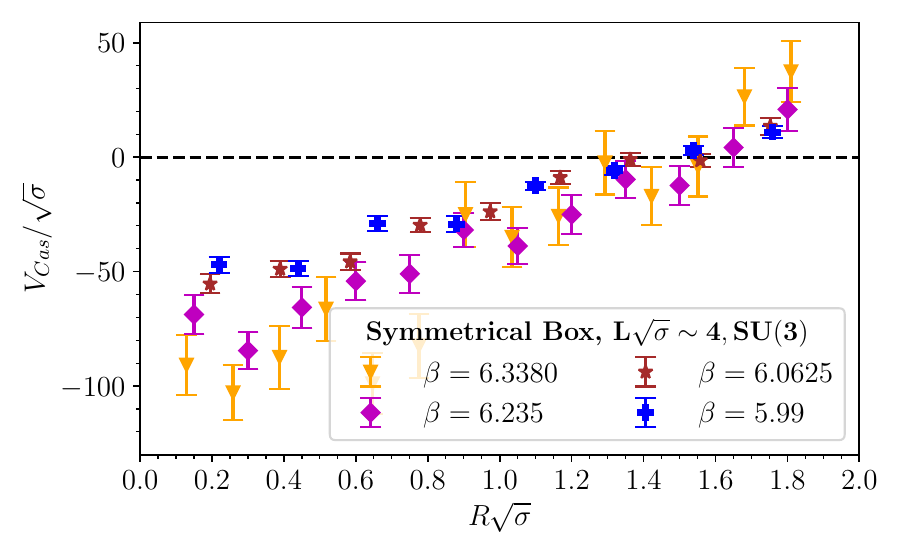} \label{fig:VcasTotal_box} }}%
    \hspace{-0.45cm}
    \subfigure[Casimir Energy/Surface Area]{{\includegraphics[scale=0.5]{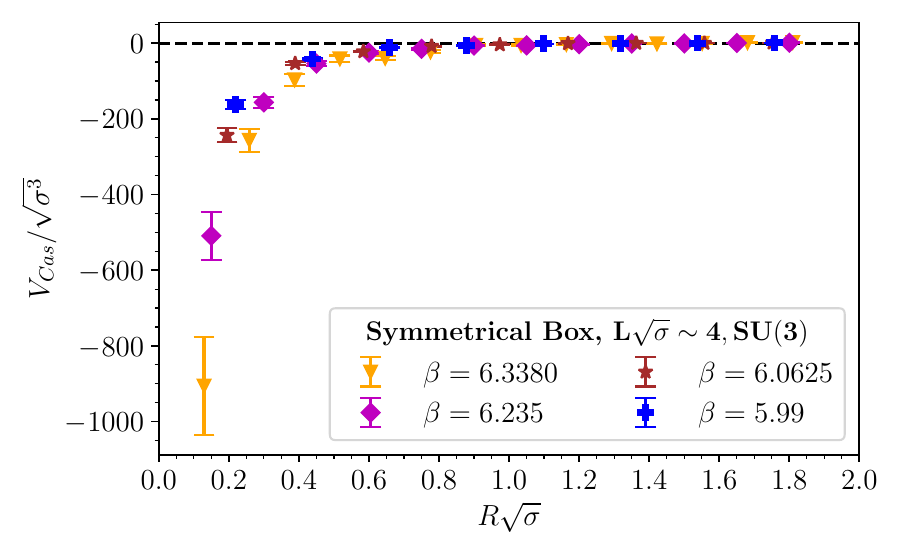} \label{fig:Vcas_box_area} }}%
    \caption{The Casimir potential for a symmetrical box with side lengths $R\sqrt{\sigma}$ in SU(3). (left) raw, (right) normalised per surface area.}%
    \label{fig:Vcas_box}
\end{figure}

\noindent
The resulting total Casimir potential and the potential per unit surface area of a symmetrical box is shown in Fig.\ (\ref{fig:Vcas_box}), where the area is given by $A=6R^2$. One key feature of this potential is that the Casimir force experienced by the symmetrical box is \textit{attractive}. This result is consistent with the Casimir effect of a weakly coupled, massless non-interacting scalar field computed in Ref.\ \cite{Mogliacci:2018oea}. As shown in Fig.\ (\ref{fig:box_Sinf}), incorrectly accounting for the boundary energy contributions for the box would result in a repulsive Casimir effect for some couplings and an attractive potential for others, with a sign flip in-between.\\ 

\subsection{Asymmetrical Box}

\noindent
We conclude our studies of the Casimir effect in non-abelian gauge theory by considering the geometry of an asymmetrical box in SU(3). In this geometry, two side-lengths are kept fixed while the remaining side with length, $R_x=R\sqrt{\sigma}$ expands. In the limit, $R_x\to \infty$, the geometry of an asymmetrical box resembles that of a symmetric tube with sides, $R_x=R_y=\sqrt{\sigma}$. In the infinite lattice volume limit, we expect that the Casimir effect of the large asymmetric box to to be equivalent to that of the symmetric tube. This expectation is also supported by numerical evidence of the field components inside the asymmetric box, which resemble those of a symmetric tube. Refer to the discussion in chapter (\ref{sec:Geometry and Symmetries: Box}).\\

\begin{figure}[!htb]
\begin{center}
\includegraphics[scale=.7]{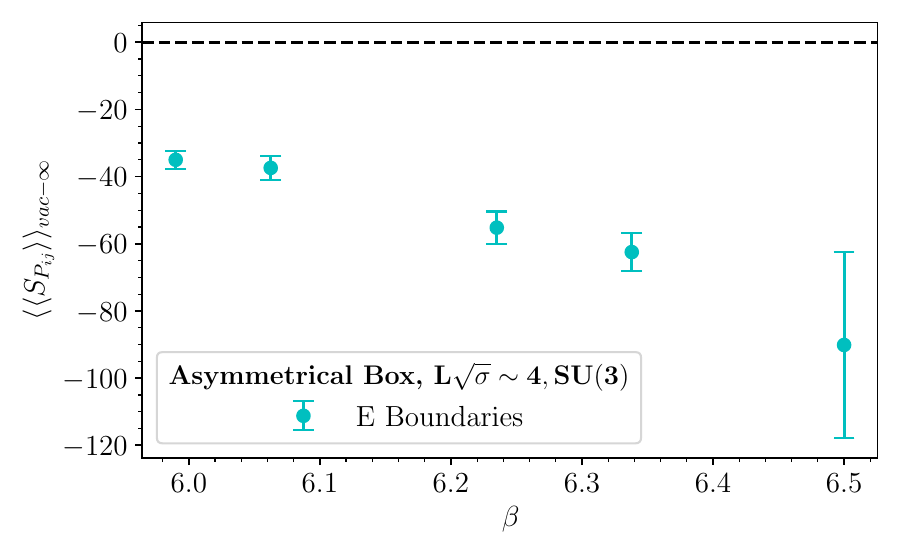}
\caption{The total energy required to create an asymmetrical box with longest side-length, $L\sqrt{\sigma}/2$ in $(3+1)$D SU(3) at varying lattice couplings.}
\label{fig:rectbox_Svac_Sinf}
\end{center}
\end{figure}

\noindent
As was the case for the asymmetrical tube, in this work we do not attempt to investigate the large $R$ limit of this geometry, and we do not provide a comparison of the asymmetrical box to the small symmetrical tube. This is due to our formulation of the asymmetrical box geometry in a Euclidean lattice with periodic boundary conditions. The expanding side, $R_x$ can maximally be extended to half the lattice size, thus allowing modes of different wavelengths to fit inside the asymmetrical box as compared to those that fit inside the infinitely extending length of the small symmetrical tube. \\

\noindent
At $R=L\sqrt{\sigma}/2$, we show the total energy present in the system after the vacuum subtraction in Fig.\ (\ref{fig:rectbox_Svac_Sinf}). Note that in the symmetrical box, this energy would be equivalent to the energy required to create the large box because the Casimir energy is negligible at large separation distances. In the asymmetrical box, this energy is given by
\begin{eqnarray}
     \lim_{R\to\infty} E_{\text{Tot}}^{\text{Vac}} &=& \lim_{R\to L\sqrt{\sigma}/2} \left( E_{\text{Cas}} + E_{\text{Box}} \right) = E_{\text{Box}} + corr_1,  
     \label{eqn:large_rectbox_energy}
\end{eqnarray}
where the correction term quantifies the non-vanishing Casimir energy contribution which should resemble that of the smallest symmetrical tube in the limit, $R\to\infty$. We will also subtract this correction term when we subtract the energy from creating the box.\\

\begin{figure}[!htb]
    \centering
    \subfigure[Vacuum Subtracted Energy]{{\includegraphics[scale=0.5]{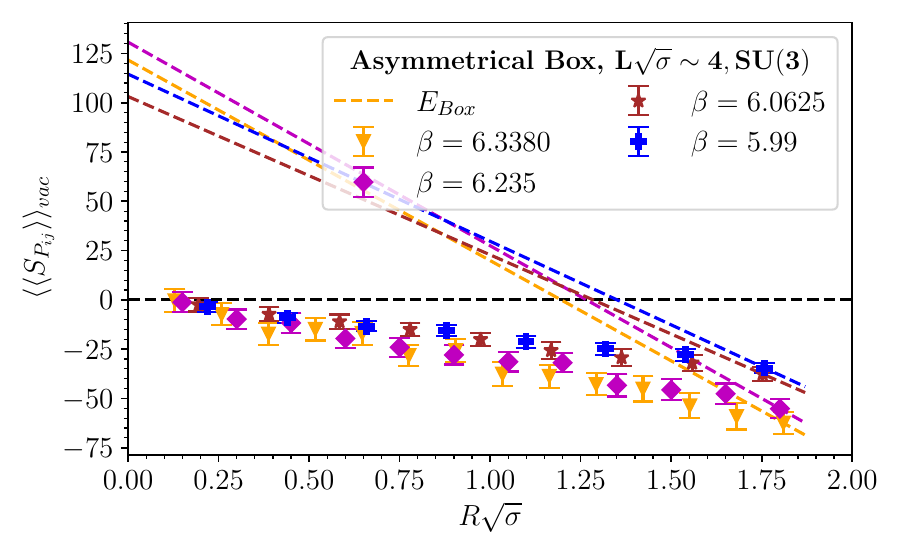} \label{fig:rectbox_Svac} }}%
    \hspace{-0.45cm}
    \subfigure[$R_{\infty}$ Subtracted Energy]{{\includegraphics[scale=0.5]{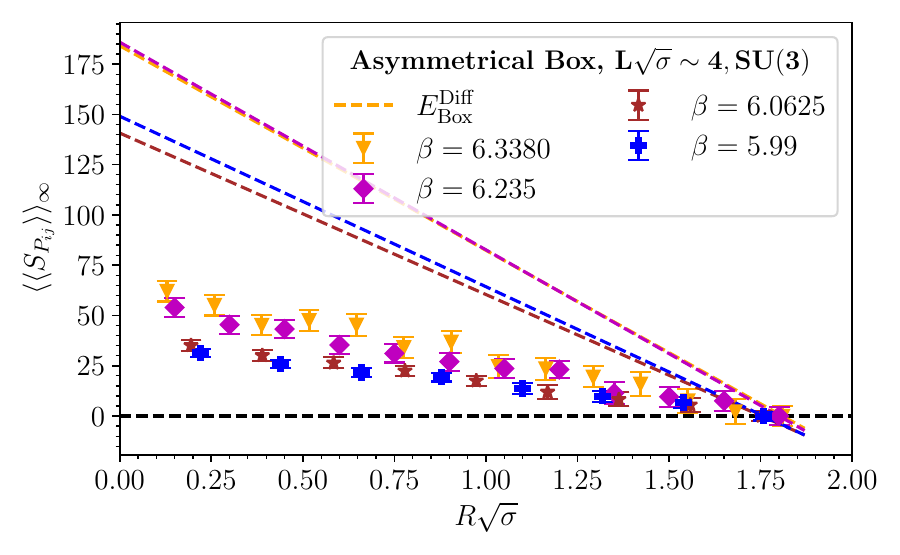} \label{fig:rectbox_Sinf} }}%
    \caption{Total energy of the system in an asymmetrical box in SU(3) for different couplings and different normalisation schemes in lattice units.}%
    \label{fig:rectbox_total_S}
\end{figure}

\noindent
Given that we have already tackled a similar geometry in the asymmetrical tube where we have a fixed side and an expanding side, we will approach this geometry in similar fashion. We start by considering the total energy of the asymmetrical box system defined in Eq.\ (\ref{eqn:box_casimir_lattice}). Similarly to the asymmetrical tube geometry, we consider the two normalisation schemes (vacuum and $R_{\infty}$ normalisation) and show the resulting system's energy in Fig.\ (\ref{fig:rectbox_total_S}).\\

\noindent
We will only focus on the vacuum normalisation shown oin Fig.\ (\ref{fig:rectbox_Svac}) and described by Eq.\ (\ref{eqn:box_energy_vacnorm}), but one can also employ the $R_{\infty}$ subtraction which gives the same result. By qualitatively comparing the vacuum subtracted energy of the asymmetrical box in Fig.\ (\ref{fig:rectbox_Svac}) to that of the symmetrical box in Fig.\ (\ref{fig:box_Svac}), we observe that at $R=1a$ the two energies are consistent because the two boxes are geometrically equivalent.\\

\noindent
The general behaviour of the total energy in the asymmetrical box system is linearly increasing with box size. In comparison to Fig.\ (\ref{fig:recttube_Svac_norm}), the magnitude and linear form of the total energy of the two systems is consistent with their geometrical similarities. In the case of the asymmetrical box, as $R$ increases four out of the six faces experience a change in size/area, but one out of three sides expand. Drawing insights from the asymmetrical tube, we also conclude that the observed linearly varying energy describes the energy change of the expanding side only. Therefore, for the vacuum normalisation, the energy contribution described by the linear curve does not vanish at $R=0$.\\


\noindent
In order to find the constant of the linear curve (or the energy contribution from the expanding faces only), we can use our knowledge of the energy of the smallest possible box that we can create on the lattice with physical volume, $V=(\sqrt{\sigma})^3$. At $R=1a$, the energy contributions from the boundaries of an asymmetrical box is the same as that of a symmetrical box shown in Fig.\ (\ref{fig:Sbox_creation}). This energy is given by
\begin{eqnarray}
    E_{\text{Box}}^{\text{fit}} (R_{1,1,1}) &=& E_{x=1} + E_{y=1} + E_{z=1} = 3E_{x=1}\\
    E_{x=1} &=& \frac{1}{3}E_{\text{Box}}^{\text{fit}} (R_{1,1,1}),
\end{eqnarray}
where $E_{x=1}$ is the energy contribution from the two parallel faces along the $\hat{x}$-axis. Once $R$ is increased, our linear curve with varying $R$ only describes energy changes due to the one expanding side only. Accordingly, the energy at $R=0$ collapses to the energy of the two remaining fixed side-lengths with a correction term,
\begin{eqnarray}
    c &=& \frac{2}{3}E_{\text{Box}}^{\text{fit}} (R_{1,1,1}) + corr_2,
\end{eqnarray}
because only one side-length (of the possible three) is increasing. \\

\noindent
Similarly to the asymmetrical tube, the correction term ensures that we do not violate energy conservation. In this case, we expect this correction term to describe the Casimir energy of a small symmetrical hollow tube with dimensions, $\sqrt{\sigma} \times \sqrt{\sigma} \times \sqrt{\sigma}$. We obtained the energy of the symmetrical box in the preceding subsection, and we use it to fix the energy of the asymmetrical box at $R=\sqrt{\sigma}$. Lastly, we draw a straight line connecting the energy at $R=\sqrt{\sigma}$ to the energy at $R=L\sqrt{\sigma}/2$ given in Eq.\ (\ref{eqn:large_rectbox_energy}) and extrapolate to $R=0$. The resulting parameters of the linear curve are given in Table (\ref{tab:rectbox_fit_params}) and describe the energy from the walls of the asymmetric box.\\

\begin{table}[!htb]
    \centering
        {\rowcolors{2}{green!80!yellow!50}{green!70!yellow!40}
        \begin{tabular}{ |P{2.5cm}|P{1.8cm}|P{1.8cm}|P{1.8cm}|P{1.8cm}|  }
        \hline
        \hline
        Energy & $\mathbf{\beta}$ & $m$ & $c$ & $corr_2$\\
        \hline
         & 6.3380 & -101.75 & 121.68 & -49.33\\
        $\langle \langle S_{P_{ij}} \rangle \rangle_{vac}$ & 6.235 & -103.27 & 130.72 & -53.90\\
         & 6.0625 & -80.30 & 103.13 & -44.80 \\
         & 5.99 & -84.79 & 114.50 & -50.59\\
        \hline
        \end{tabular}}
    \caption{Linear curve parameters for the energy contributions from the boundaries of an asymmetrical box in SU(3) for lattice size, $L\sqrt{\sigma}\sim 4$.}
    \label{tab:rectbox_fit_params}
\end{table}

\begin{figure}[!htb]
    \centering
    \subfigure[Total Casimir Energy]{{\includegraphics[scale=0.5]{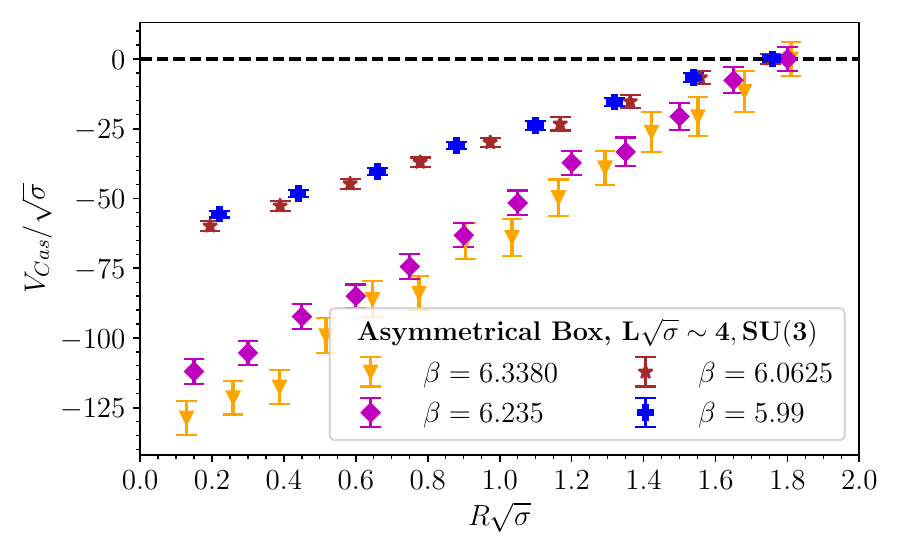} \label{fig:VcasTotal_rectbox} }}%
    \hspace{-0.45cm}
    \subfigure[Casimir Energy/Surface Area]{{\includegraphics[scale=0.5]{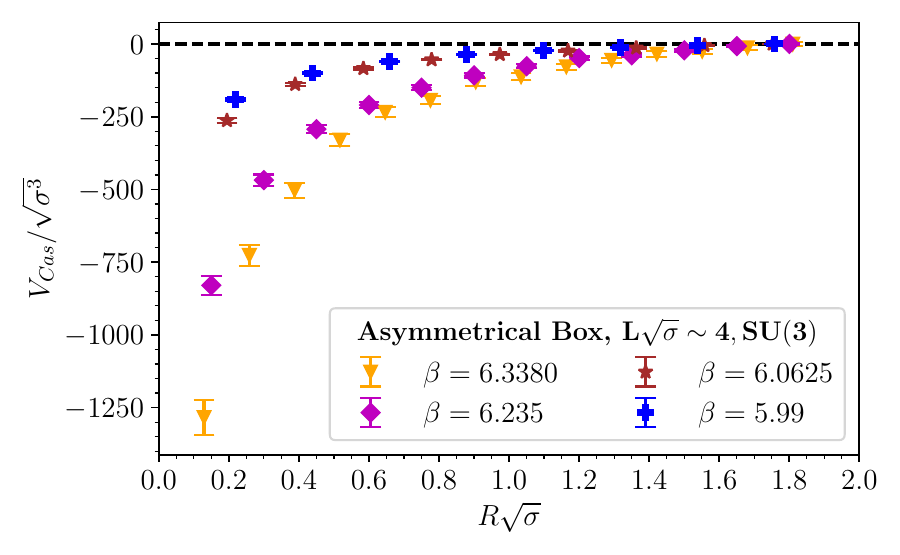} \label{fig:Vcas_rectbox_area} }}%
    \caption{The Casimir potential for an asymmetrical box with side lengths $R\sqrt{\sigma}$ in SU(3).}%
    \label{fig:Vcas_rectbox}
\end{figure}

\noindent
We show the total Casimir potential and the potential per unit surface area of the asymmetrical box in Fig.\ (\ref{fig:Vcas_rectbox}), where the physical surface area is, $A=2\sigma(2R + 1)$. We note that the Casimir potential in the asymmetrical box appears to be overestimated, particularly in the limit, $R\to1$ where the magnitude of the attractive force should coincide with that of a symmetrical box. We attribute this stark difference to boundary energy contributions where our linear fit does not seem to capture correctly the exact functional form of the asymmetrical box's boundaries. \\

\noindent
Despite the overestimation of the energy contributions from the box's boundaries, we can safely conclude that the Casimir force experienced by an asymmetrical box geometry is \textit{attractive}. Therefore, all the geometries that we have explored have resulted in an attractive Casimir potential in non-abelian gauge theory. In the case of a tube, this is contrary to Casimir measurements in a weakly coupled, massless non-interacting scalar field presented in Ref.\ \cite{Mogliacci:2018oea}. The observed attractive potential in our geometries is explained by the mode exclusion in the cavity inside the geometry.\\

\begin{figure}[!htb]
    \centering
    \subfigure[Box vs. Faces Energy]{{\includegraphics[scale=0.5]{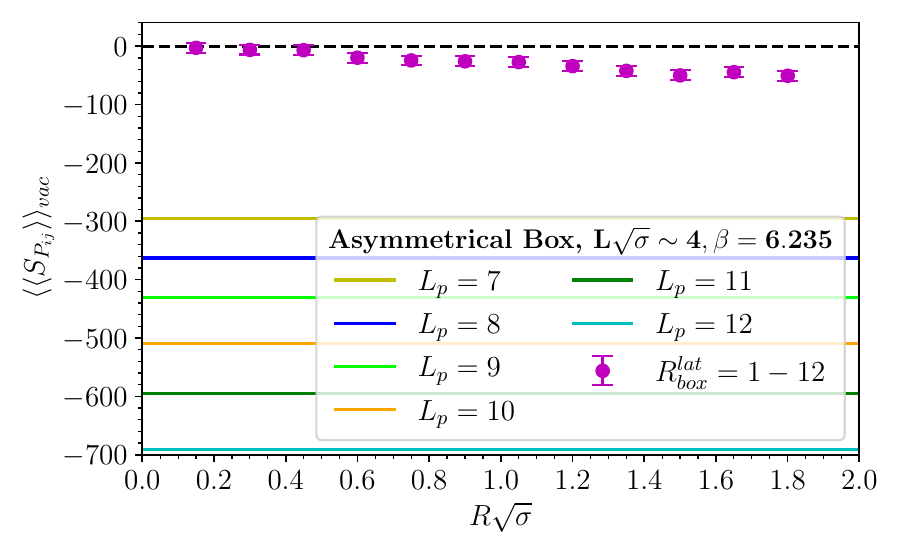} \label{fig:Svac_rectbox_plates} }}%
    \hspace{-0.45cm}
    \subfigure[Box vs. Faces Energy Difference]{{\includegraphics[scale=0.5]{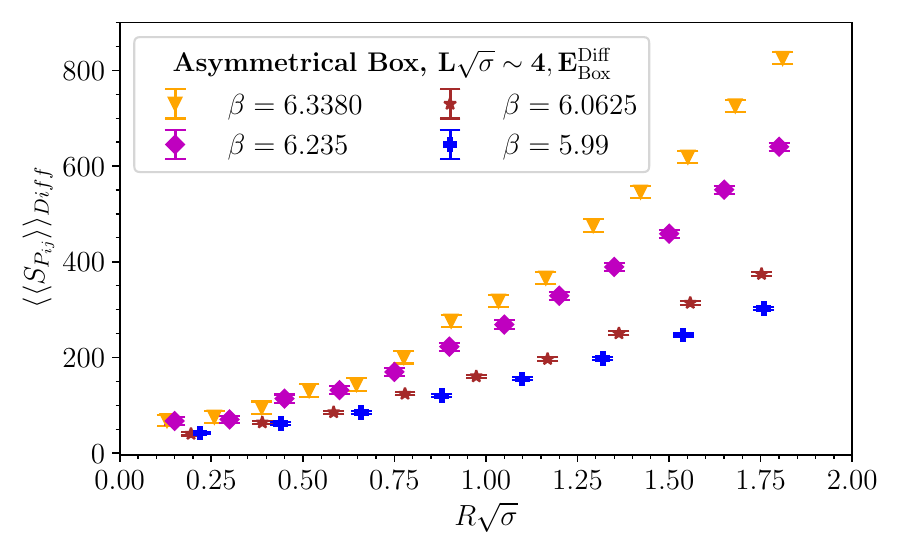} \label{fig:Sdiff_rectbox_plates} }}%
    \caption{(left)Vacuum normalised energy of six faces compared to the vacuum normalised energy of an asymmetrical box.\ (right)Energy difference between the creation of the six faces and an asymmetrical box.}%
    \label{fig:rectbox_plates}
\end{figure}

\noindent
In our discussion of the symmetrical box, we showed that it is possible to account for the energy of creating the walls of the box by considering the energy required to create the six faces that form part of the box. This approach worked because we found that the energy difference between creating a full box and creating six faces varied linearly. In Fig.\ (\ref{fig:Svac_rectbox_plates}), we show the vacuum normalised energy of creating the asymmetric box compared to the vacuum normalised energy of creating six faces that form the box (where two of the faces have a fixed area). Whereas in Fig.\ (\ref{fig:Sdiff_rectbox_plates}) we show the energy difference between creating a box and creating six faces and we observe that this energy difference does not vary linearly for the asymmetric box. However, we highlight this method as a possible approach for correcting the energy contributions from the boundaries of the asymmetrical tube.\\

\noindent
In summary, in this chapter we have presented Casimir potential measurements in (2+1)D and (3+1)D non-abelian gauge theories for gauge groups SU(2) and SU(3) in the lattice formalism for various geometries. We have discussed various methods in which the Casimir energy can be isolated from the energy contributions from creating the boundaries in lattice numerical results of varying geometries. We have also shown that incorrectly accounting for these chromoelectric boundary energy contributions could result in misinterpretation of the measured Casimir potential such that it becomes repulsive.\\

\chapter{The Polyakov Loop and Deconfinement}
\label{chapter:The Polyakov Loop and Deconfinement}

\noindent
In chapter (\ref{section:string_tension}), we introduced the string tension, $\sqrt{\sigma}$, which we have (until now) used to set our physical energy scale. In this chapter, we discuss the temperature dependence of the Casimir potential and compare the potential in the parameter regimes associated with the confined and deconfined phase. We have insofar presented our physical results in units of the string tension; in order to provide a well-defined temperature scale, we now express our physical energy scale in MeV units. In order to do so, we need to specify the value of the numerical value of `zero temperature' string tension. \\

\noindent
A popular approach to scale setting in QCD is to find a physical distance $r$ between two static \textit{heavy} quarks at which the force, $F(r)$, between them assumes a certain value. See Eq.\ (\ref{eqn:hq_potential}, \ref{eqn:hq_force}) on how the heavy quark potential and force are defined. Sufficiently heavy quarks are used because in the limit of large (compared to typical QCD scales) heavy quark mass, $q\Bar{q}$ bound states are described by an effective non-relativistic Schrodinger equation \cite{Eichten:1980mw}. Through the comparison of phenomenological models to experimental data of $b\Bar{b}$ and $c\Bar{c}$ spectra, one can define a distance $r_0$, the so called Sommer parameter where \cite{Sommer:1993ce}
\begin{equation}
    r^2F(r)|_{r=r_0} = 1.65.
\end{equation}
While rigorously defined in QCD, it offers an analogous way to define a physical scale in quenched QCD and has the value \cite{Sommer:2014mea},
\begin{eqnarray}
    r_0 = 0.472(5) \text{ fm} = \frac{1}{418(5) \text{ MeV}},
\end{eqnarray}
determined from lattice calculations with $N_f >2$. The string tension is then obtained using the Sommer parameter \cite{Athenodorou:2020ani},
\begin{eqnarray}
    \sqrt{\sigma} =\frac{1.160(6)}{r_0} = 485(6) \text{ MeV}.
    \label{eqn:rootsig}
\end{eqnarray}
On the other hand, phenomenological models give a string tension, $\sqrt{\sigma} = 440$ MeV \cite{Lucini:2013qja}. Once the numerical value of the string tension is known, the physical lattice spacing can be determined and all other quantities in physical units follow.\\

\noindent
The temperature is defined through the physical lattice spacing and the lattice temporal extent,
\begin{equation}
    T = \frac{1}{N_{\tau}a} = \frac{1}{\beta}.
    \label{eqn:temperature}
\end{equation}
This definition is consistent with the definition of the inverse temperature, $\beta = 1/(k_BT)$ in quantum mechanics, with $k_B=1$. On large isotropic lattices (i.e., $N_s = N_{\tau}$), in the Euclidean formulation where one cannot distinguish between the spatial and temporal directions, we assume \textit{zero temperature}. In principle, these are low (but still finite) temperatures because of computational limitations. However, since the theory is confined at these low temperatures, the dynamics of the relevant degrees of freedom should remain unaffected, and this provides a good estimate of the zero temperature theory.\\

\noindent
Using the temperature definition given in Eq.\ (\ref{eqn:temperature}), the zero temperature theory corresponds to the limit $T \to 0$ as $\beta \to \infty$. Thus for a fixed value of $\beta$, which gives us a fixed value of the physical lattice spacing, $a$, fixing $N_{\tau}$ fixes the temperature of the system. Increasing the temperature means decreasing the lattice temporal extent, and lattice measurements have shown that finite volume effects are reduced as the aspect ratio $N_s/N_{\tau}$ increases in both pure gauge and full QCD. One reason for the reduced finite volume effects in pure gauge theory is that a larger $N_s/N_{\tau}$ leads to a larger ratio of the available spatial extent and spatial correlation length of gluon fields, which characterises the typical size of gluonic excitations.\\


\noindent
Yang-Mills theories are confined; only colour neutral excitations propagate at low temperatures, and, as the temperature is increased (or at high densities), the YM theories reach a thermodynamic phase transition where the theory deconfines. The presently accepted values of the critical temperature at which the phase transition occurs in SU($N_c$) pure gauge theory in the absence of boundaries in (3+1)D can be approximated by the empirical formula \cite{Fingberg:1992ju, Boyd:1996bx},
\begin{eqnarray}
    T_c^{SU(2)} &=& 0.69(2)\sigma_{SU(2)}^{1/2} \approx 335 \text{ MeV}\\
    T_c^{SU(3)} &=& 0.629(3)\sigma_{SU(3)}^{1/2} \approx 305 \text{ MeV},
\end{eqnarray}
where we have used the string tension values in Eq.\ (\ref{eqn:rootsig}). These formulae overestimate the critical temperature based on the lattice extracted string tension value in comparison to the phenomenological string tension value which yields, 
\begin{eqnarray}
    T_c^{SU(2)} &\approx& 304 \text{ MeV}\\
    T_c^{SU(3)} &\approx& 277 \text{ MeV},
\end{eqnarray}
consistent with the pure Yang-Mills SU(3) result, $T_c \sim 270$ MeV obtained in various lattice calculations. We use these formulae to show the resulting $T_c$ value for consistency because in our calculation of the temperature, we have used the result of the string tension in Eq.\ (\ref{eqn:rootsig}). In pure gauge theory, the phase transition is a second-order (continuous crossover) for $N_c=2$ and first-order for $N_c\geq3$ \cite{Boyda:2020nfh}.\\

\begin{figure}[!htb]
    \centering
    \subfigure[Casimir Potential/Area]{{\includegraphics[scale=0.5]{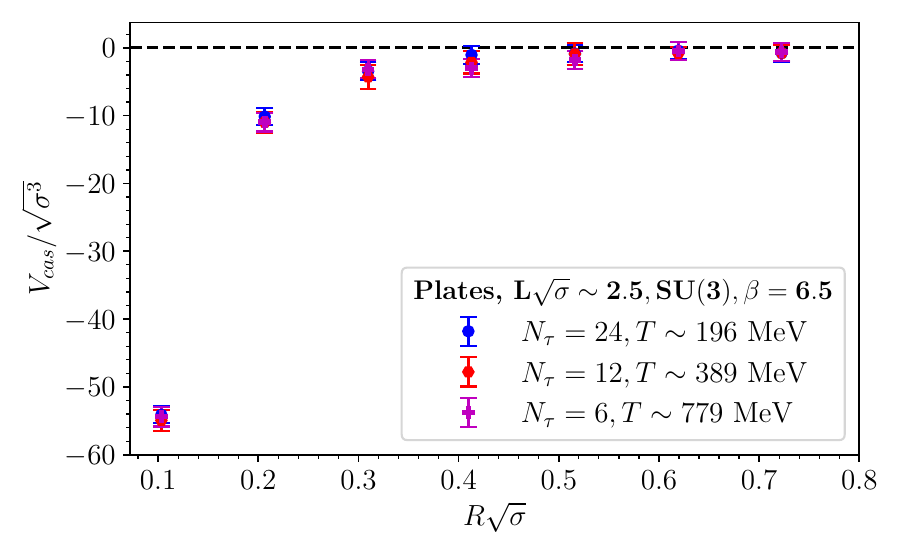} \label{fig:Vcas_temp} }}%
    \hspace{-0.45cm}
    \subfigure[Deconfined/Confined Phase Ratio]{{\includegraphics[scale=0.5]{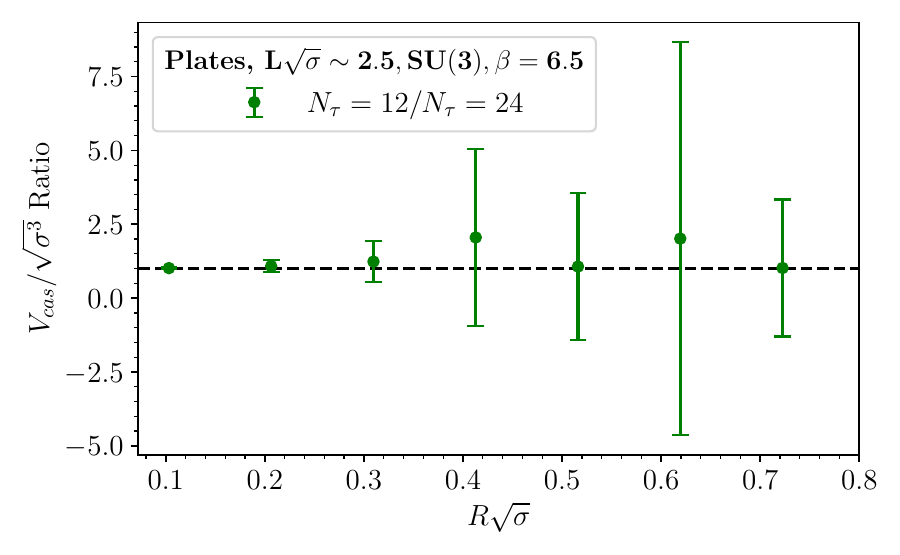} \label{fig:Vcas_temp_ratio} }}%
    \caption{The Casimir potential per unit area between two parallel plates separated by a distance $R\sqrt{\sigma}$ in SU(3) at different temperatures.}%
    \label{fig:Vcas_plates_temp}
\end{figure}

\noindent
It is suggested in Ref.\ \cite{Chernodub:2018pmt} that in the two-dimensional non-abelian gauge theory, where the Casimir effect is studied for parallel wires in SU(2), the region between the wires is a Casimir-induced deconfined phase. As a result, the gluons in this finite region exhibit similar behaviour to thermal glueballs in a heat-bath at finite temperature. This is attributed to the absence of the periodicity of gluonic fields in the presence of Casimir boundaries. The same behaviour is observed for parallel plates in SU(3) \cite{Chernodub:2023dok} where these thermal glueballs are interpreted as colourless states of gluons bound to their negatively coloured images on the chromoelectric boundary. \\

\noindent
This suggestion is also supported by evidence of the measured mass of the relevant degrees of freedom in the Casimir interaction, $M_{\text{Cas}}$ in Eq.\ (\ref{eqn:vcas_fit}). The Casimir mass was shown to assume a lower value than the lightest ground state glueball in SU(2) \cite{Chernodub:2018pmt} and SU(3) \cite{Chernodub:2023dok}, which should be the lowest mass in the pure gluonic system. The reduced glueball mass in the `heat-bath' Casimir cavity is consistent with finite temperature lattice measurements of thermal glueball properties, which show that the mass of the lightest glueball, $M_{0^{++}}$, decreases with temperature, even in the confined phase, with a mass reduction of up to $\sim 20\%$ around $T_c$ in SU(3) \cite{Ishii:2002ww,Meng:2009hq}.\\

\noindent
We have expanded the discussion of the Casimir mass in the non-abelian theory for parallel wires in (2+1)D and plates in (3+1)D for both SU(2) and SU(3). These results are provided in chapter (\ref{section:Parallel Wires in (2+1)D}) and (\ref{section:Parallel Plates}), respectively, where we have also obtained Casimir masses lower than the lightest glueball, $M_{0^{++}}$, in the corresponding theories at temperatures corresponding to the confined phase. While we have studied the Casimir effect for the tube and box geometries, at the present stage, we have not performed Casimir mass calculations to test for any geometry dependence in the three-dimensional theory due to time limitations. \\

\noindent
In Fig.\ (\ref{fig:Vcas_plates_temp}), we show the temperature dependence of the measured Casimir effect of the parallel plate geometry as we move from the number of temporal lattice grid points $N_{\tau}$ associated with the confined to deconfined phase. Because the region between the plates is already a boundary induced deconfined phase where the relevant degrees of freedom have different properties compared to those measured for pure gluodynamics, decreasing $N_{\tau}$ (increasing the temperature) from the confined to deconfined phase does not alter the Casimir effect. The non-abelian Casimir effect is a deconfined phase phenomenon due to the boundaries and is insensitive to temperature changes across $T_c$. However, it may be valuable in future studies to explore the limit $T\to 0$, which is resource intensive on the lattice.\\

\noindent
In order to better understand gluodynamics along the temperature axis, we look at the finite temperature deconfinement order parameter described by the Polyakov loop (or \textit{thermal Wilson line}) \cite{Polyakov:1976fu}, 
\begin{equation}
    L_{\bm x} = \frac{1}{N_c} \text{Tr} \prod_{\tau=0}^{N_{\tau -1}} U_{\mu=4}(\bm x,\tau),
\end{equation}
which is a straight-line product of $N_{\tau}$ link variables in the Euclidean time direction, all situated at a single spatial point $\bm x \coloneqq (x_1, x_2, x_3)$. The Polyakov loop is a trace over a closed loop (through periodic boundary conditions) and is therefore gauge invariant. In lattice terminology, the term under the trace is also referred to as the \textit{temporal transporter} and appears in the definition of the Wilson loop.\\

\begin{figure}[!htb]
\begin{center}
\includegraphics[scale=.7]{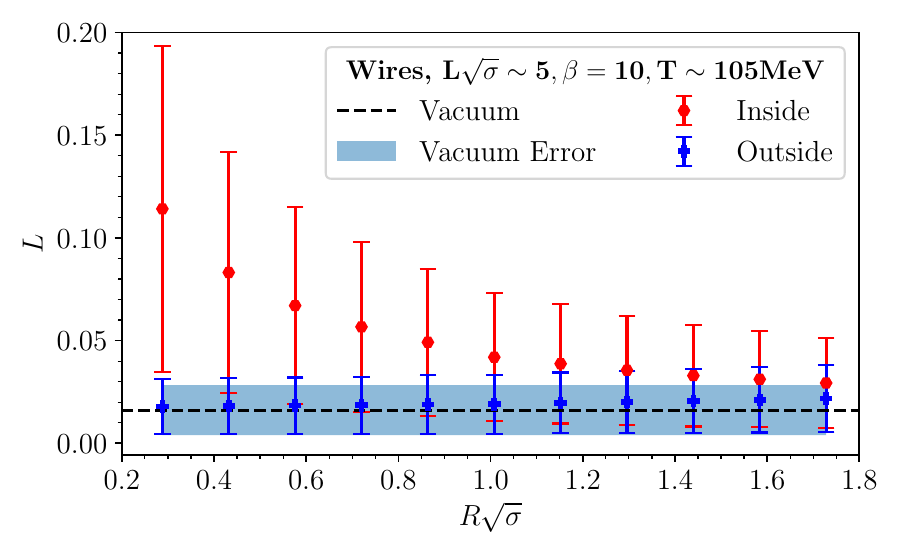}
\caption{The Polyakov loop expectation value in the region between and outside the parallel wires in SU(2) in the confined phase at $T\sim 195$ MeV.}
\label{fig:poly_wires_T0}
\end{center}
\end{figure}

\noindent
The spatial position of the Polyakov loop is irrelevant due to translational invariance in the absence of the non-trivial boundaries, and the deconfinement order parameter is given by the spatial average
\begin{equation}
    L \equiv \frac{1}{V_r} \left\langle \sum_{\bm x \in V_r} L_{\bm x}  \right\rangle,
    \label{eqn:polyakov_loop_avg}
\end{equation}
where $V_r=N_s^3$ is the physical volume. Given that we impose boundaries, note that we use the subscript $r$ to specify between the regions in our Casimir effect geometries, where one region is inside (e.g., between the plates) and the other is outside. Thus we compute the Polyakov loop separately between these two regions.\\ 

\noindent
The expectation value of a single Polyakov loop $L_{\bm x}$ is interpreted as the probability to observe a single static charge and is approximated by the exponential of the \textit{free energy} \footnote{The energy required to separate a quark-antiquark pair to an infinite distance apart.}, $F_q$ \cite{Gattringer:2010zz},
\begin{equation}
    \frac{1}{V}\left| \left\langle \int d^3\, x L_{\bm x} \right\rangle \right| \sim e^{-F_q/T},
\end{equation}
of a single static charge. That is, the theory is confined in the limit $F_q \to \infty$, implying that isolated quarks cannot exist as free particles in this phase, and deconfined for finite free energy.\\

\noindent
The following conditions tell us about the phases of the theory,
\begin{equation}
    L \Leftrightarrow 
\begin{dcases}
    \text{Confined phase}, & \langle L \rangle = 0\\
    \text{Deconfined phase}, & \langle L \rangle \neq 0.
\end{dcases}
\label{eqn:poly_condition}
\end{equation}
In quenched QCD, performing a center transformation i.e.\ multiplying all temporal links at the same time slice by an element $z$\footnote{$z\in Z_3=\{e^{-i2\pi/3}, 1, e^{i2\pi/3}\}$.} of the center group $Z_3$ of SU(3) leads to the transformation rule $L \to zL$ for the Polyakov loop which leaves the Wilson gauge action invariant. In the confined phase, the expectation value of the Polyakov loop vanishes as the local phase is equally distributed among all possible values and \textit{center symmetry}\footnote{The expectation value of the Polyakov loop may be expressed through elements of the center group as, $\langle L \rangle = \frac{1}{3} \langle L+zL+z^2L \rangle = \frac{1}{3} \left( 1+e^{i2\pi/3}+e^{-i2\pi/3} \right)\langle L \rangle=0$ due to the action symmetry.} is preserved. In the deconfined phase, individual colour charges may be observed, resulting in a non-vanishing Polyakov loop due to the presence of nontrivial thermal fluctuations of the gauge fields where possible phases are not equally populated and center symmetry is spontaneously broken above $T_c$. The thermal fluctuations indicate that the colour charges are free to move over long distances.\\

\noindent
Since we have redone the simulations Ref.\ \cite{Chernodub:2018pmt}, we show in Fig.\ (\ref{fig:poly_wires_T0}), the resulting expectation value of the Polyakov loop for parallel wires in SU(2) for completeness. Our result is consistent with that of Chernodub et al.\ where we have used their preferred definition\footnote{We only use this definition for the case of the parallel wires in Fig.\ (\ref{fig:poly_wires_T0}) in order to make a direct comparison.} of the Polyakov loop which requires taking the magnitude before computing the average in Eq.\ (\ref{eqn:polyakov_loop_avg}). The increasing Polykov loop expectation value with decreasing separation distance indicates deconfinement in the region between the wires barring the large errors.\\

\begin{figure}[!htb]
\begin{center}
\includegraphics[scale=.7]{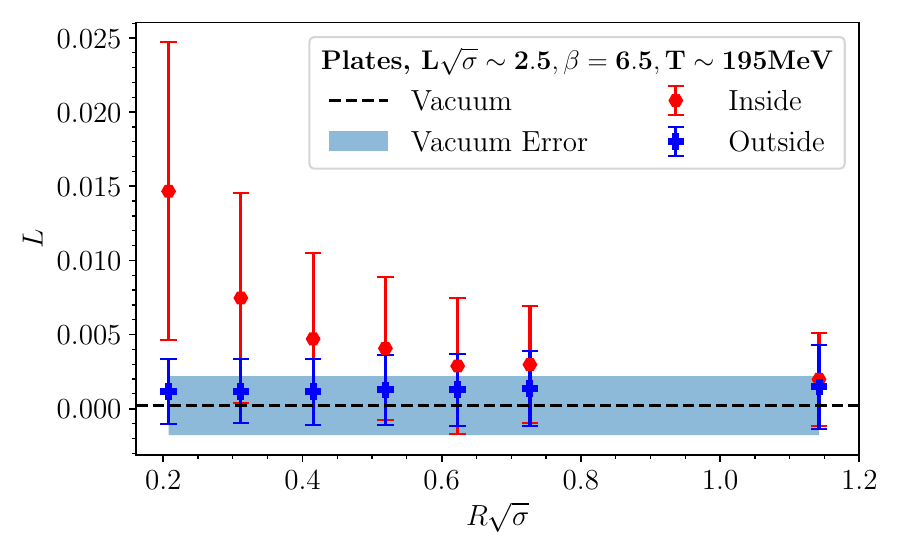}
\caption{The Polyakov loop expectation value in the region between and outside the parallel plates in SU(3) in the confined phase at $T\simeq 195$ MeV.}
\label{fig:poly_confined}
\end{center}
\end{figure}

\noindent
We have shown in Fig.\ (\ref{fig:Vcas_plates_temp}) that increasing the temperature from the confined to deconfined phase has no observable effect on the Casimir effect between parallel plates in SU(3). We now look at the corresponding Polyakov loop expectation values as $N_{\tau}$ is decreased from the confined to deconfined phase. In Fig.\ (\ref{fig:poly_confined}), we show the Polyakov loop expectation value in the confined phase as defined in Eq.\ (\ref{eqn:polyakov_loop_avg}). In the absence of the plates, $L\approx 0$ because the pure glueball system is in the confined phase and center symmetry is preserved.\\

\noindent
In the region between the plates, the Polyakov loop expectation value is non-zero and increases with decreased separation distance between the plates, supporting the idea of a deconfined phase in this region. While there is a clear trend in the Polykov loop expectation value, the error bars are large resulting in data-points for $R>1$ being consistent with zero, which would suggest the system remains confined at these distances. Improvement of the statistics is necessary to draw quantitative conclusions. Meanwhile the Polyakov loop, $L\sim 0$ in the large volume outside the plates where the gluonic system remains confined and increases only slightly with separation distance. As the separation distance between the plates, $R\to L\sqrt{\sigma}/2$ where the boundaries are mirrored, the expectation value of the Polykov loop inside and outside the plates approaches the same value because the two regions become indistinguishable.\\

\begin{figure}[!htb]
    \centering
    \subfigure[$T\simeq 389$ MeV]{{\includegraphics[scale=0.5]{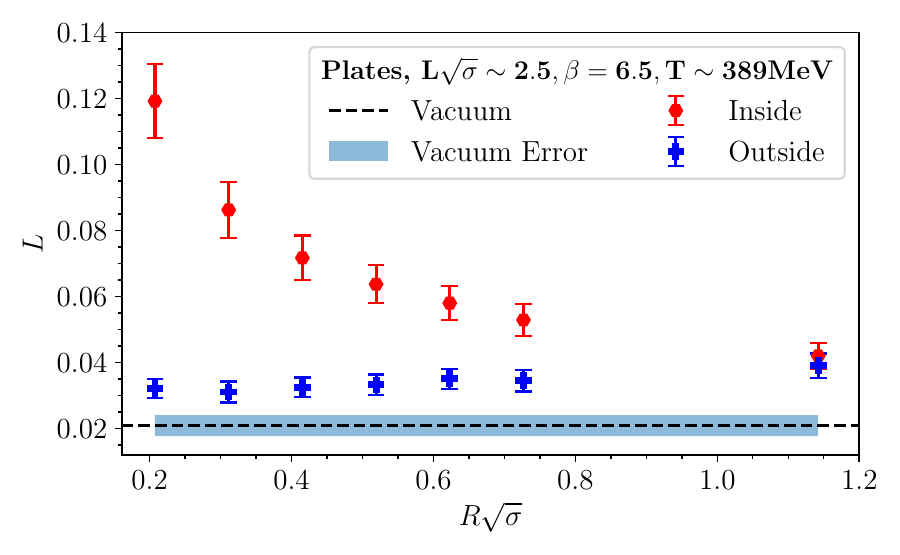} \label{fig:poly_Nt12} }}%
    \hspace{-0.45cm}
    \subfigure[$T\simeq 779$ MeV]{{\includegraphics[scale=0.5]{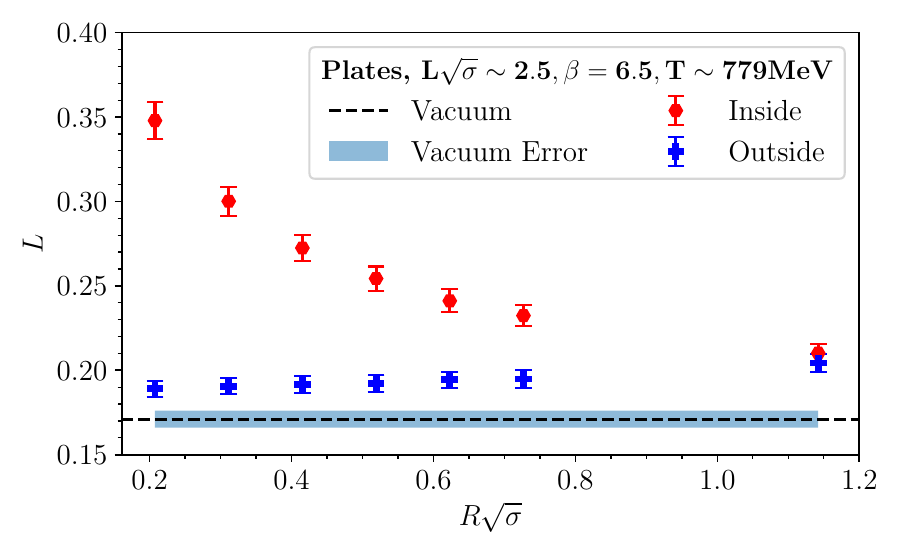} \label{fig:poly_Nt6} }}%
    \caption{The Polyakov loop expectation value in the region between and outside the parallel plates in SU(3) in the deconfined phase.}%
    \label{fig:poly_deconfined}
\end{figure}

\noindent
As the temperature is increased and the system moves into a deconfined phase, the vacuum (in the absence of the plates) Polyakov loop expectation value in the pure gluonic system becomes non-zero, thus signalling deconfinement. While in the deconfined phase, this vacuum expectation value continues to increase with temperature as shown in Fig.\ (\ref{fig:poly_deconfined}), however $L_{vac} < L_{in}$. In the confined phase, we observed that in the region outside the plates, $L\approx 0$ and equal to the vacuum expectation value. This is no longer the case in the deconfined phase, $L_{vac} < L_{out} < L_{in}$ and there is an observable linear increase in $L_{out}$ with separation distance. \\

\noindent
On the other hand, in the region between the plates, the magnitude of the deconfinement order parameter increases with temperature, however we have observed that this does not effect the Casimir effect. Given the observable change in the Polyakov loop with increasing temperature, it is surprising that the Casimir effect remains insensitive to the temperature change. Thermal glueballs studies \cite{Meng:2009hq} show that the glueball mass decreases with increasing temperature above $T_c$, hence the relevant degrees of freedom in the Casimir interaction should change with increased temperature. The temperature independence of the Casimir effect requires further investigation.\\

\begin{figure}[!htb]
    \centering
    \subfigure[Symmetrical Tube]{{\includegraphics[scale=0.5]{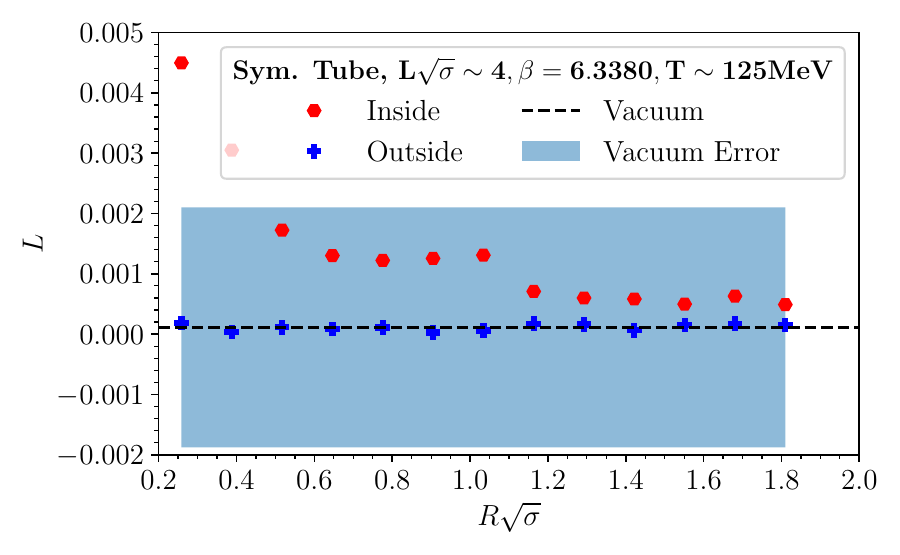} \label{fig:poly_tube} }}%
    \hspace{-0.45cm}
    \subfigure[Asymmetrical Tube]{{\includegraphics[scale=0.5]{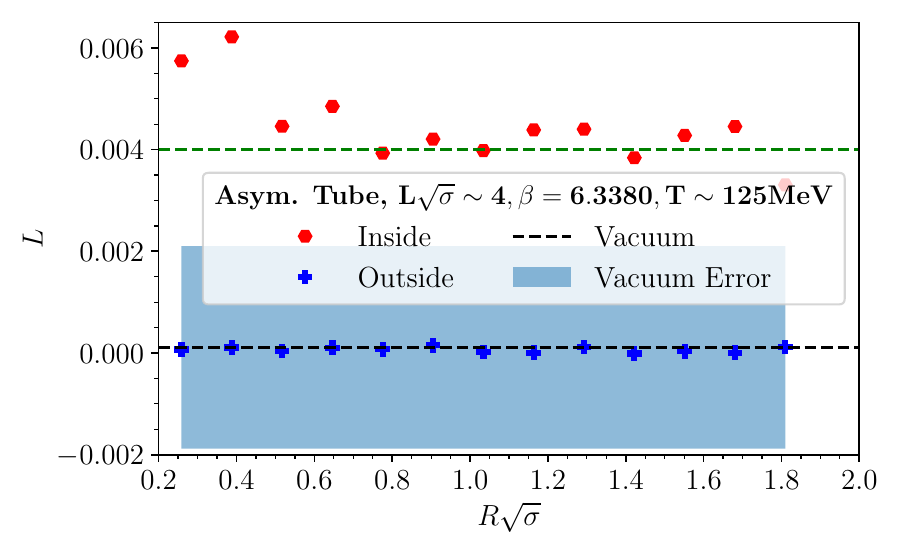} \label{fig:poly_recttube} }}%
    \caption{The Polyakov loop expectation value in the region inside and outside a tube in SU(3) in the confined phase.}%
    \label{fig:poly_tube_confined}
\end{figure}

\noindent
Lastly, we show the expectation value of the Polyakov loop in the confined phase for the symmetrical and asymmetrical tube in Fig.\ (\ref{fig:poly_tube_confined}), omitting the large error-bars for neatness. We emphasize that these results are only preliminary and still require error-reduction. In both geometries, the Polyakov loop fluctuates around the vacuum expectation value outside the tube. Inside the symmetrical tube, the Polyakov loop behaves similarly to the parallel plate geometry and approaches zero with increased tube-size. \\

\begin{figure}[!htb]
\begin{center}
\includegraphics[scale=.7]{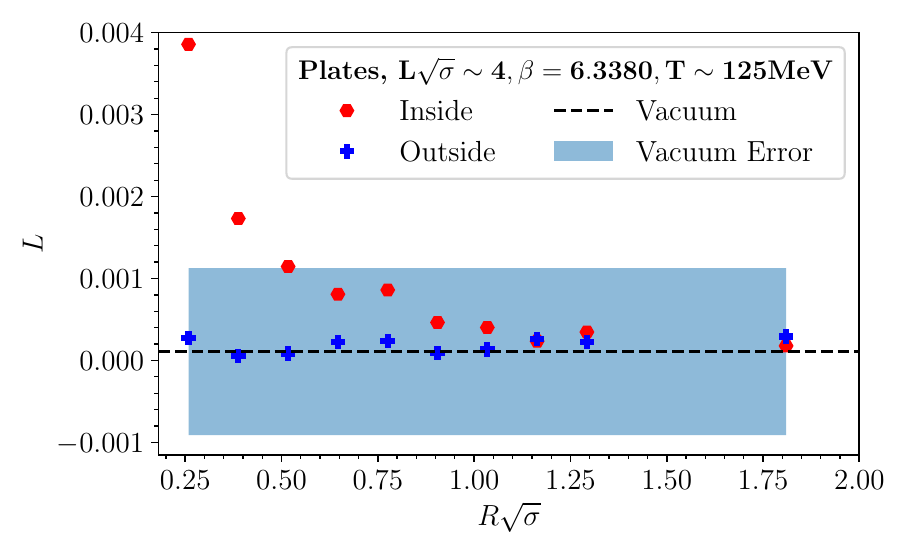}
\caption{The Polyakov loop expectation value in the region between and outside the parallel plates in SU(3) in the confined phase at $T\sim 125$ MeV.}
\label{fig:poly_confined_Nt30}
\end{center}
\end{figure}

\noindent
Inside the asymmetrical tube, the expectation value of the Polyakov loop does not vanish as the tube expands due to the geometrical asymmetry, but instead fluctuates around $L\sim 0.004$. In the previous chapter, we argued that the Casimir effect for the asymmetrical tube should not vanish in the limit, $R\to \infty$, but should approach the parallel plate potential at separation distance, $R=\sqrt{\sigma}$. We make a direct comparison of the Polykov loop of the asymmetrical tube to that of parallel plates at the same temperatures and physical lattice volume in Fig.\ (\ref{fig:poly_confined_Nt30}). We observe that,
\begin{eqnarray}
    L_{\text{Plates}}|_{R=1a}\sim L_{\text{Tube}}|_{R\to L\sqrt{\sigma}/2}. 
\end{eqnarray}
This non-vanishing Polyakov loop for the asymmetrical tube and the similarity to the plates supports our claim that their Casimir effect should be equivalent in the limit $R\to \infty$ for the asymmetrical tube. We have not performed these calculations for the box geometries to good accuracy, however this analysis should similarly help to identify whether there is an equivalence between the Casimir effect of the asymmetrical box in the limit, $R\to \infty$ to the symmetrical tube with sides, $R=\sqrt{\sigma}$.\\

\chapter{Conclusions}
\label{chapter:Conclusions}
Vacuum fluctuations of quantum fields manifest themselves as virtual particles which momentarily appear and disappear in empty-space with lifetimes constrained by the energy-time uncertainty principle. The energy of these fluctuating fields, the zero-point energy in quantum field theory, corresponds to an infinite sum over all possible frequencies and is an ultraviolet divergence. While this infinite energy in quantum field theory is not a physically measurable quantity, we are able to measure the zero-point energy differences in-vacuum by imposing specific boundary conditions on the fields which modify the vacuum properties. These energy differences explain the Casimir effect which we have studied in this thesis in non-abelian gauge theory. \\

\noindent
While the Casimir effect has been successfully studied primarily in the abelian gauge theory using perturbative analytic methods. However, complex geometries require simplifications such as fixed boundary conditions such as Dirichlet, Neumann etc. These simplifications in turn lead to inaccuracies in the measured Casimir effect mostly due to thermal corrections \cite{Mitter:1999hu}. References to comprehensive discussions are provided in the introduction. These difficulties encountered in perturbative techniques provide a strong motivation for the exploration of the Casimir effect using non-perturbative methods such as lattice QED/QCD. We draw inspiration from the work of Chernodub et al.\ \cite{Chernodub:2018pmt} where the Casimir effect in non-abelian gauge theory is studied on the lattice for perfectly conducting static parallel wires in (2+1)D SU(2) at zero temperature using chromoelectric boundaries.\\

\noindent
In order to better understand the implication of these boundary conditions on the gauge fields, we have studied the effect of the chromoelectric boundary conditions on the individual field components in chapter (\ref{chapter:The Casimir Effect: Fields and Symmetries}). We find different degrees of suppression of the individual field components at the boundaries and the surrounding region. Analytically applying rotational matrices on the field-strength tensor, we find that there exists an equivalence in some of the electric and magnetic field components based on the geometrical set-up due to the rotational symmetries. Using these rotational symmetries, we numerically show the equivalent field expectation values for the configurations of the parallel wires/plates and the tube. This, in turn allows us to simplify the expressions of the energy density in these geometries due to cancellations in the equivalent terms. Meanwhile, there are no rotational symmetries to explore in the box geometry.\\

\noindent
The main results of this thesis are the Casimir potentials in non-abelian gauge theories for different geometries. We start by reproducing literature results for parallel wires in (2+1)D SU(2) shown in Fig.\ (\ref{fig:casimir_potential_2dsu2}) and extending these studies to SU(3) in Fig.\ (\ref{fig:casimir_potential_2dsu3}). We show that increasing the number of degrees of freedom from $N_c:2\to3$ results in an increased Casimir potential by a factor $\gtrsim 2$ which decreases linearly with separation distance. We attribute this increase in pressure to the increase in the number of modes in the region outside the wires, which is greater than the increase in the number of modes in the region inside the wires. Using an empirically chosen fitting function for the potential, we show that the anomalous scaling dimension $\nu \approx 0.02$ and is independent of the number of colours. On the other hand, the Casimir mass, $M_{\text{Cas}} = 1.38(2)\sqrt{\sigma}$ in SU(2), consistent with the results of Ref.\ \cite{Chernodub:2018pmt}; and increases with increased degrees of freedom to $M_{\text{Cas}} = 1.51(9)\sqrt{\sigma}$ in SU(3).\\

\noindent
We then provide new results in (3+1)D for the geometries of a parallel plate configuration in SU(2) shown in Fig.\ (\ref{fig:casimir_plates_3dsu2}) and SU(3) in Fig.\ (\ref{fig:casimir_plates_3dsu3}). In the case of parallel plates, increasing the number of degrees of freedom from $N_c:2\to3$ generally results in an increased Casimir potential for intermediate separation distances. However, we observe that at very small and large separation distances, the potential is greater in SU(2). At this point, we have no physical interpretation of this phenomena, and the observed behaviour could instead point out the need to better understand the empirical Casimir fitting functions. \\

\noindent
In the parallel plate geometry, we find an anomalous scaling dimension, $\nu \approx 0.09$ in SU(2) and $\nu \approx 0.002$ in SU(3). Whereas the Casimir mass, $M_{\text{Cas}} = 0.38(12)\sqrt{\sigma}$ in SU(2) $M_{\text{Cas}} = 0.06(6)\sqrt{\sigma}$ in SU(3). The differences in the orders of magnitude from $N_c:2\to3$ is puzzling similarly to the behaviour of the ratio of the potential in Fig.\ (\ref{fig:casimir_potential_3DRatio}) and calls for further investigation. Overall, the measured Casimir mass in both the two and three-dimensional theory is lower than the mass of the lightest ground state glueball, $M_{0^{++}}$ which is the lowest mass in a pure gauge non-abelian system. In the two-dimensional theory, $M_{0^{++}} = 4.718(43)\sqrt{\sigma}$ in SU(2) and $M_{0^{++}} = 4.329(41)\sqrt{\sigma}$ in SU(3). Whereas in the three-dimensional theory, $M_{0^{++}} = 3.781(23)\sqrt{\sigma}$ in SU(2) and $M_{0^{++}} = 3.405(21)\sqrt{\sigma}$ in SU(3). The low Casimir mass indicates that the relevant degrees of freedom in the Casimir interaction are not glueballs, but rather lighter particles exhibiting similar behaviour (such as a decreased mass in the deconfined phase) to thermal glueballs in a heat-bath.\\

\noindent
The interpretation of the low Casimir mass defined in Eq.\ (\ref{eqn:vcas_fit_plates}) as a boundary induced deconfined phase in the region between the boundaries is supported by our studies of the Polyakov loop. However, in Ref.\ \cite{Meng:2009hq}, it is shown that the pole mass of the lowest lying glueball at finite temperature decreases with temperature above $T_c$ when employing a fit with a single hyperbolic cosine (corresponding to a Dirac-delta like spectral function). On the other hand, a Breit-Winger ansatz for the spectral function gives an approximately constant glueball mass (i.e., peak position in the spectral function), with a width that increases with the temperature (consistent with our understanding of thermal spectral functions in other cases such as quarkonia). On the basis of the latter interpretation, where the spectral function width increases at high temperatures and not the glueball mass, it is plausible that something similar is happening in the case of the Casimir mass. It is possible that there exists a spectral decomposition of the Casimir energy, with an appropriately defined
spectral function, and an analogue of thermal broadening which makes the single exponential ansatz inappropriate. We highlight this as a possible avenue for future investigations. \\

\noindent
We extend our Casimir potential studies in SU(3) to the geometry of a symmetrical and asymmetrical tube in chapter (\ref{section:Symmetric and Asymmetric Tube}), followed by the geometry of a symmetrical and asymmetrical box in chapter (\ref{section:Symmetric and Asymmetric Box}). We show that the resulting Casimir potential is attractive in all these geometries. Due to the expanding nature of the tube and box (thus a changing area of the plates) with separation distance as opposed to the parallel plate geometry where the area was fixed, we show that a careful treatment of the energy contribution from the boundaries is necessary, else one could incorrectly measure a repulsive potential for the tubes and asymmetrical box. Not capturing the boundary energy contributions correctly for the box results in a potential that moves from attractive to repulsive depending on the lattice coupling and box-size.\\

\noindent
We have proposed two methods in which this energy of creating the boundaries could be accurately accounted for in the cases of both tube geometries and the asymmetrical box. The first method is the vacuum normalisation where one subtracts the vacuum energy contribution from the total energy of the system, then uses a linear function to describe the remaining energy in the system, $E_{\text{Cas}} + E_{\text{Tube}}$. The energy from creating the boundaries is found by fitting this linear function at large separation distances where the Casimir energy contribution is insignificant. The second method is the $R_{\infty}$ normalisation where we subtract the energy contribution from creating a large `outer box/tube' (similarly to the parallel plate geometry) from the total energy of the system, then use a linear fit to describe the resulting energy difference from creating the `outer tube/box' and the `inner tube/box'. The resulting linear fit parameters are used to renormalise the total energy in order to isolate the Casimir energy. We also show that both methods can be improved by using our knowledge of the system's energy at $R=0$ and $R=L\sqrt{\sigma}/2$.\\

\noindent
The energy contribution from the boundaries of a symmetrical box is nonlinear and unique in that it is non-monotonous. Thus the normalisation schemes that we have used in other geometries do not apply here. Instead, given that the box is composed of six faces, we initially normalise by subtracting the energy of creating the six faces. We place two finite-size faces far apart (where their Casimir energy is negligible) and we find the energy of creating these six faces. We compare the energy difference of creating large boxes with an insignificant Casimir contribution to the energy of creating the six faces. We find that this energy varies linearly and we use a linear fitting form to extract the corresponding parameters which we use to renormalise the total energy of the box and isolate the Casimir energy.\\

\noindent
In the last chapter, we show that the Casimir potential is independent of a temperature increase from the confined to the deconfined phase. We provide results of the expectation value of the Polyakov loop for the wires, plates, symmetrical and asymmetrical tube. These results confirm that at temperatures below $T_c$, the Polyakov loop is zero outside the plates (consistent with a confined system) but it is non-zero in the region between the plates. This observation confirms that the region between each geometry is a boundary-induced deconfined phase, hence the measured Casimir mass of the relevant degrees of freedom is lower than the lightest glueball masses. \\

\noindent
Our study has only been performed at microscopic separation distances between the boundaries, which allows for the interpretation of the matter inside the boundaries as a boundary induced deconfined phase. However, in the case of macroscopic separation distances between the boundaries, the matter in the region between the boundaries has no knowledge of the presence of the boundaries, as is the case for the region outside the boundaries even for microscopic distances considered in this study. Hence at low temperatures, a plausible scenario would be the Polyakov indicating confined matter in both the region between and outside the boundaries, while showing signs of deconfinement close to the boundaries (i.e., a deconfined `boundary layer'). This suggests a possible alternative to the interpretation of the region inside the boundaries and we highlight it as an idea for future investigations. \\

\noindent
Lastly, in our exploration of the Casimir effect in the asymmetrical geometries, we discussed that the Casimir effect for the asymmetrical tube in the limit $R\to\infty$ should resemble that of the parallel plate geometry with $R=\sqrt{\sigma}$. On the other hand, the Casimir effect for the asymmetrical box in the limit $R\to\infty$ should resemble that of the symmetrical tube with $R=\sqrt{\sigma}$. In our studies, we use a lattice volume with periodic boundary conditions and the geometry gets mirrored at $R=L\sqrt{\sigma}/2$, making it difficult to study the $R\to\infty$ limit for these geometries. However, in Fig.\ (\ref{fig:poly_tube_confined} - \ref{fig:poly_confined_Nt30}), we show that the expectation value of the asymmetrical tube Polykov loop at large $R$ fluctuates around $L\approx 0.004$, which is equivalent to the Polykov loop of the parallel plate geometry at $R=\sqrt{\sigma}$. This observation is consistent with our expectation, and we highlight the need to perform a similar study for the asymmetrical box. \\

\noindent
In this thesis, we have studied the Casimir effect in non-abelian gauge theories in SU(2) and SU(3). In (2+1)D, we studied the geometry of parallel conducting wires, whereas in (3+1)D, we studied the geometry of parallel plates as well as a symmetrical and asymmetrical tube and box, respectively.  We performed these calculations on a periodic lattice volume and used chromoelectric boundary conditions on the walls of our geometries. In the geometries of the tube and box where the plate-size expands with increasing separation distance, we discussed techniques that can be used to correctly describe the energy from creating the boundaries. We show that the resulting Casimir potential is attractive for all the geometries considered.\\


\printbibliography


\appendix
\addcontentsline{toc}{chapter}{Appendices}
\chapter{Generators of the Lie Algebra}
\label{appendix:generators}


\section{Pauli Matrices}
The special unitary group SU(N) has $N^2-1$ independent parameters given by the generators of the Lie group. In the fundamental representation of SU(2), the generators $T_k$ are proportional to the Pauli matrices $\sigma_k$,
\begin{equation}
    T_k = \frac{1}{2}\sigma_k, \quad k=1,2,3
\end{equation}

\begin{equation}
\sigma_1 = \begin{pmatrix}
0 & 1 \\
1 & 0
\end{pmatrix}, \quad
\sigma_2 = \begin{pmatrix}
0 & -i \\
i & 0
\end{pmatrix}, \quad
\sigma_3 = \begin{pmatrix}
1 & 0 \\
0 & -1
\end{pmatrix}.
\end{equation}

\section{Gell-Mann Matrices}
In SU(3), the generators $T_k = \frac{1}{2}\lambda_k$, where $k\in[1,8]$ are represented by the Gell-Mann matrices $\lambda_k$,

\begin{equation}
\begin{aligned}
\lambda_1 &= 
\begin{pmatrix}
0 & 1 & 0 \\
1 & 0 & 0 \\
0 & 0 & 0
\end{pmatrix}, \quad
\lambda_2 =
\begin{pmatrix}
0 & -i & 0 \\
i & 0 & 0 \\
0 & 0 & 0
\end{pmatrix}, \quad
\lambda_3 =
\begin{pmatrix}
1 & 0 & 0 \\
0 & -1 & 0 \\
0 & 0 & 0
\end{pmatrix}, \\
\lambda_4 &= 
\begin{pmatrix}
0 & 0 & 1 \\
0 & 0 & 0 \\
1 & 0 & 0
\end{pmatrix}, \quad
\lambda_5 =
\begin{pmatrix}
0 & 0 & -i \\
0 & 0 & 0 \\
i & 0 & 0
\end{pmatrix}, \quad
\lambda_6 =
\begin{pmatrix}
0 & 0 & 0 \\
0 & 0 & 1 \\
0 & 1 & 0
\end{pmatrix}, \\
\lambda_7 &= 
\begin{pmatrix}
0 & 0 & 0 \\
0 & 0 & -i \\
0 & i & 0
\end{pmatrix}, \quad
\lambda_8 = \frac{1}{\sqrt{3}}
\begin{pmatrix}
1 & 0 & 0 \\
0 & 1 & 0 \\
0 & 0 & -2
\end{pmatrix}.
\end{aligned}
\end{equation}

\chapter{Rotation Transformations}
\label{appendix:rotation_matrices}
Rotation matrices describe transformations that preserve the structure of Minkowski spacetime under the Lorentz group. These transformations include spatial rotations, which preserve spatial distances, and Lorentz boosts, which relate the coordinates of observers in relative motion and together, they maintain the spacetime interval invariant. This work is performed on a Euclidean lattice and we omit boosts. 

\begin{multicols}{2}
\begin{equation}
R_{xy} = 
\begin{bmatrix}
\cos \theta & -\sin \theta & 0 & 0\\
\sin \theta & \cos \theta & 0 & 0\\
0 & 0 & 1 & 0\\
0 & 0 & 0 & 1
\end{bmatrix} 
\end{equation}
\break

\begin{equation}
R_{xz} = 
\begin{bmatrix}
\cos \theta & 0 & -\sin \theta & 0\\
0 & 1 & 0 & 0\\
\sin \theta & 0 & \cos \theta & 0\\
0 & 0 & 0 & 1
\end{bmatrix} 
\end{equation}
\end{multicols}

\begin{multicols}{2}
\begin{equation}
R_{xt} = 
\begin{bmatrix}
\cos \theta & 0 & 0 & -\sin \theta\\
0 & 1 & 0 & 0\\
0 & 0 & 1 & 0\\
\sin \theta & 0 & 0 & \cos \theta
\end{bmatrix} 
\end{equation}
\break

\begin{equation}
R_{yz} = 
\begin{bmatrix}
1 & 0 & 0 & 0\\
0 & \cos \theta & -\sin \theta & 0\\
0 & \sin \theta & \cos \theta &  0\\
0 & 0 & 0 & 1
\end{bmatrix} 
\end{equation}
\end{multicols}

\begin{multicols}{2}
\begin{equation}
R_{yt} = 
\begin{bmatrix}
1 & 0 & 0 & 0\\
0 & \cos \theta & 0 & -\sin \theta\\
0 & 0 & 1 & 0\\
0 & \sin \theta & 0 & \cos \theta
\end{bmatrix} 
\end{equation}
\break

\begin{equation}
R_{zt} = 
\begin{bmatrix}
1 & 0 & 0 & 0\\
0 & 1 & 0 & 0\\
0 & 0 & \cos \theta & -\sin \theta\\
0 & 0 & \sin \theta & \cos \theta
\end{bmatrix} 
\end{equation}
\end{multicols}

\end{document}